\documentclass[runningheads,openany]{svmult}

\usepackage{palatino}
\usepackage{euler}
\usepackage{helvetic}
\usepackage{graphicx}  
\usepackage{physprbb}  

\usepackage{fest}  
\usepackage{epsfig}

\usepackage{amssymb}

\makeindex             

\usepackage{amsmath}   
\let\tocauthor\authorrunning

\begin{document}

\frontmatter

\thispagestyle{empty}
\parindent=0pt

{\Large\sc Blejske delavnice iz fizike \hfill Letnik~2, \v{s}t. 2}

\smallskip

{\large\sc Bled Workshops in Physics \hfill Vol.~2, No.~2}

\smallskip

\hrule

\hrule

\hrule

\vspace{0.5mm}

\hrule

\medskip
{\sc ISSN 1580--4992}

\vfill

\bigskip\bigskip
\begin{center}

{\bfseries 
{\Huge  Proceedings to the workshops\\
What comes beyond the Standard model 2000, 2001}

\vspace{5mm}
\centerline{\Large  Volume 1} 
\vspace{5mm}
\centerline{\Huge Festschrift}
\vspace{5mm}
\centerline{\Large dedicated to the 60th birthday of}
\vspace{5mm}
\centerline{\Huge Holger Bech Nielsen} }

\vfill

{\bfseries\large
Edited by

\vspace{5mm}
Norma Manko\v c Bor\v stnik\rlap{$^{1,2}$}

\smallskip

Colin D. Froggatt\rlap{$^{3}$}

\smallskip

Dragan Lukman\rlap{$^2$}

\bigskip

{\em\normalsize $^1$University of Ljubljana, $^2$PINT, $^3$Glasgow University}

\vspace{12pt}

\vspace{3mm}

\vrule height 1pt depth 0pt width 54 mm}

\vspace*{3cm}

{\large {\sc  DMFA -- zalo\v{z}ni\v{s}tvo} \\[6pt]
{\sc Ljubljana, december 2001}}
\end{center}
\newpage

\thispagestyle{empty}
\parindent=0pt

\begin{flushright}
{\parskip 6pt
{\bfseries\large
                  Publication of \textit{Festschrift in Honor of the 60th 
                  Birthday of Holger Bech Nielsen}}

\bigskip

{\bf was sponsored by}

{\parindent8pt
\textit{Ministry of Education, Science and Sport of Slovenia}

\textit{Department of Physics, Faculty of Mathematics and Physics,
University of Ljubljana}

\textit{Primorska Institute of Natural Sciences and Technology, Koper}

\textit{Society of Mathematicians, Physicists and Astronomers
of Slovenia}}}
\bigskip
\medskip
\end{flushright}

\setcounter{tocdepth}{0}
\tableofcontents
\cleardoublepage
\chapter*{Preface}
\addcontentsline{toc}{chapter}{Preface}

One of the most important things in one person's life is to have good
friends, whom one can exchange thoughts and ideas with.  The present
volume is a collection of contributions by friends of Holger Bech
Nielsen for his 60 th birthday. The collection of papers covers a
broad range of physics, which shows that Holger Bech Nielsen has been
working and is working on many topics --- also because different fields
of physics are much more connected as one would believe, at least when
fundamental questions of physics are concerned.  The most fascinating
works in physics bring new understanding of known phenomena, since
they lead to theories beyond the known ones.  Holger Bech Nielsen is
one of (very rare and because of that additionally precious )
scientists who is always willing to play with new thoughts and
progressive ideas, trying to formulate them into formal proofs built
on well defined assumptions and into formulas, pointing out in such a
way that mathematics is a part of Nature.

It is my pleasure and my privilege to have Holger as a real friend in
the last five years, since we have started ---\,together with Colin
Froggatt\,--- the annual workshop at Bled, Slovenia, which is entitled
``What comes Beyond the Standard Model''.  During the workshop the
ideas connected with the open problems of (elementary particle)
physics and cosmology are very vividly and openly discussed.

The first volume of this Proceedings is a Festschrift dedicated to Holger. 
I wish to thank all his friends who are contributing to
this volume and also to those friends, who have sent their
contributions too late to be included.

The second volume is collecting contributions and discussions from the
last two workshops and will appear with a little delay.

The editors would like to thank Yasutaka Takanishi for a lot of work,
which contributed to make the Volume 1 of the Proceedings see the light 
of day. \\[2mm]

Dear Holger: I am honoured  to congratulate you for your 60th jubilee
in the name of all your friends contributing to this volume and also
in the name of the others, wishing you many additional fruitful years,
health, wealth and success.

In the name of your friends:\\[2mm]
\textit{Norma Susana Manko\v c Bor\v stnik}
\qquad\qquad\qquad\qquad\qquad
\textit{Ljubljana, December 2001}

\newpage
\thispagestyle{empty}

{\bf Workshops organized at Bled}

\begin{description}
\item[$\triangleright$]
What Comes beyond the Standard Model (June 29--July 9, 1998)
        \item[$\triangleright$]
Hadrons as Solitons (July 6-17, 1999)
        \item[$\triangleright$]
What Comes beyond the Standard Model (July 22--31, 1999)
        \item[$\triangleright$]
Few-Quark Problems (July 8-15, 2000)
        \item[$\triangleright$]
What Comes beyond the Standard Model (July 17--31, 2000)
        \item[$\triangleright$]
Statistical Mechanics of Complex Systems (August 27--September 2, 2000)
        \item[$\triangleright$]
What Comes beyond the Standard Model (July 17--27, 2001)
        \item[$\triangleright$]
Studies of Elementary Steps of Radical Reactions in Atmospheric Chemistry
(August 25--28, 2001)

\end{description}


\mainmatter
\parindent=20pt
\setcounter{page}{5}
\setcounter{page}{1}
%
\newcommand{\SO}{\mathrm{SO}}
\newcommand{\SU}{\mathrm{SU}}
\newcommand{\Qed}{\rule{2.5mm}{3mm}}
\newcommand{\balpha}{\mbox{\boldmath {$\alpha$}}}
\newcommand{\unit}{\mathrm{U}}

\title*{Unified Internal Space of Spins and Charges}
\author{N. Manko\v c Bor\v stnik\thanks{e-mail:norma.s.mankocs@ijs.si}}
\institute{%
Faculty of Mathematics and Physics, University of
Ljubljana, Jadranska 19, Ljubljana
and Primorska Institute for Natural Sciences and Technology, 
C. Mare\v zganskega upora 2, Koper 6000, Slovenia} 

\authorrunning{N. Manko\v c Bor\v stnik}
\titlerunning{Unified Internal Space of Spins and Charges}
\maketitle

\begin{abstract}
Can the assumption that the spin and all the charges of either fermions or bosons unify, help to 
find the answers to the open questions of the electroweak Standard model? The approach is presented 
in which polynomials in Grassmann space are used to describe all the internal degrees of freedom 
of spinors, scalars and vectors, that is their spins and charges \cite{mankoc1992,mankoc1999}. The same 
can be achieved \cite{mankocandnielsen1999} also by polynomials of differential forms. If the space - 
ordinary and anti-commutative - has 14 dimensions or more, the appropriate spontaneous breaking of 
symmetry leads gravity in $d$ dimensions to manifest in four-dimensional subspace as ordinary 
gravity and all the gauge fields as well as the Yukawa couplings. The approach manifests 
four generations of massless fermions, which are 
left handed $SU(2)$ doublets and right handed $SU(2)$ singlets.
\end{abstract}
 

\section{  Introduction}
 \label{intr}

It can not be any doubt that the internal space of spins and charges plays an important role in our world, 
so important as the ordinary space of coordinates and momenta does. Without the internal space of spins and charges 
no spinors (fermions), no vectors (gauge fields), no tensors (gravity) would exist, scalars (if there are any 
elementary scalars at low energies)  would not interact and accordingly no matter could exist.
We have shown\cite{mankoc1992,mankoc1993,mankoc1994,mankoc1995,%
mankoc1997,borstnikandmankoc1999} how a space of anti-commuting 
coordinates can be used to describe spins and charges of not only fermions but also of bosons,
unifying spins and charges for either fermions or  bosons and that spins in d-di\-men\-sional space manifest
(at low energies) as spins and charges in four-dimensional space-time, while accordingly 
gravity in d-dimen\-sional space - after the appropriate breaking of
symmetry, such as 
\[
\begin{array}{c}
\underbrace{%
\begin{array}{rrcll}
 & & \SO(1,13) &&\\
 & & \downarrow && \\
 & & \SO(1,7) \otimes \SO(6) &&\\
 & \swarrow & &  \searrow &\\
 \SO(1,7) & & & \SU(3)\otimes\unit(1)& \\
 \downarrow\qquad & & & \qquad\downarrow& \\
\SO(1,3)\otimes\SU(2)\otimes\unit(1) & & & \SU(3)\otimes\unit(1)&
\end{array}} \\
\SO(1,3)\otimes\SU(2)\otimes\unit(1)\otimes\SU(3)
\end{array}
\]
- manifests in four-dimensional subspace as the ordinary gravity and all the (known)  gauge fields.

The knowledge of the way how our universe (or universes) ``has (or have) made a choice'' of the signature 
of space-time and of the way of breaking symmetries of (external and internal) space-time from $SO(q,d-q)$ 
down to $SO(1,3) \times SU(3) \times SU(2) \times U(1)$ can help to answer the open questions of the Standard Model,
such as:

i) Why only left handed spinors carry the weak charge while right handed spinors are weak charge-less?

ii) Why besides the Planck scale (at least) the weak scale occurs?

iii) Why there are spin ($SO(1,3)$) and charge ($SU(3), SU(2), U(1)$) internal degrees of freedom?

iv) Why there are besides the gravity (gauge) field also the charge gauge fields?

v) Where do the families come from?

vi) Where the Higgs and the Yukawa couplings come from?

and many others.

If the break of let us say $SO(1,13)$ occurred in the way presented in the above diagram, then
we could say why left handed weak charge doublets and right handed weak charge singlets appear in the same
multiplet ( in this case not only the weak charge scale but also the intermediate scale  before the Planck scale) 
exists, and why there are (four rather than three) families of quarks and leptons, as well as that there is 
gravity in 
$d$-dimensional space, which manifests at low energies as the ordinary gravity and all the (known) gauge 
fields, the Higgs and the Yukawa couplings.

In this article, we briefly present the approach, which unifies spins and 
charg\-es, gravity and gauge fields,
leads to multiplets of left handed weak charged spinors and of right handed weak chargless spinors, to families
of quarks and leptons, pointing out to questions like why and how symmetries break, or
why and how signatures are chosen.

We present a possible Lagrange function for a free particle
and the  quantization of anti-commuting
coordinates  \cite{mankoc1992,mankoc1994,mankoc1995,mankoc1997,mankoc1999}.  
Introducing vielbeins and spin connections,
we demonstrate on the level of a covariant momentum how the spontaneous breaking of symmetry might lead to
the symmetries of the Standard model. This
part of the work has  been started together with 
A. Bor\v stnik \cite{borstnikandmankoc1999} and is continuing with Holger Bech Nielsen.  
We show how the symmetry of the group $SO(1,13)$ breaks to $SO(1,7)$ (leading to multiplets with left handed 
$SU(2)$ doublets and right handed $SU(2)$ singlets) and to $SO(6)$, which then leads to the $SO(1,3) \times SU(2) 
\times U(1) \times SU(3) \times U(1)$. The two $U(1)$ symmetries enable besides the
hyper-charge, needed in the Standard Model, an additional
hyper-charge, which is nonzero for a right handed $SU(2)$
and $SU(3)$ singlet, like it is a right handed neutrino in the
Standard model. 
In the last four years, the author of the paper organized together with Holger Bech Nielsen and Colin Froggatt 
 annual  workshops  at Bled, Slovenia, entitled ``What comes beyond the Standard Model?'', 
 the real workshops in which we discuss all the open questions of the Standard Model, 
 as well as the proposed 
approaches which might lead to physics beyond the Standard Model. 
Very open discussions leaded to many new ideas and suggestions, so that we all profit from these
workshops a lot. Not only one sees connections and correspondences between different approaches, 
but we also open new questions, new problems, trying to find new solutions to the problems.

Thoughts and ideas, presented in this paper, although started and developed by the approach
of unification of spins and charges\cite{mankoc1992}, have been enriched and developed under 
the influence of these discussions. Some of the discussions led to common papers with the main 
opponent Holger Bech Nielsen, some papers are in preparation, with the main opponent and with others.

\section{Dirac equation in ordinary space and in space of anti-commutative coordinates }
\label{dirac}

What we call quantum mechanics in Grassmann space 
is the model for going beyond the Standard Model with extra
dimensions of ordinary and anti-commuting coordinates,
describing spins and charges of either fermions or bosons in a
unique way \cite{mankoc1992,mankoc1993,mankoc1994,mankoc1995,%
mankoc1997,borstnikandmankoc1999,mankocandnielsen1999}.\\
In a $d$-dimensional space-time the internal degrees of freedom
of either spinors or vectors and scalars come from the Grassmann
odd variables 
\begin{equation*}
\theta^a, \qquad a \in \{ 0,1,2,3,5,\cdots ,d \}.  
\end{equation*}
We write wave functions describing either spinors or vectors 
in  the form
\begin{equation}
<\theta^a|\Phi>  = \sum_{i=0,1,..,3,5,..,d} \quad \sum_{\{ a_1<
a_2<...<a_i\}\in \{0,1,..,3,5,..,d\} }
\alpha_{a_1, a_2,...,a_i}\theta^{a_1} 
\theta^{a_2} \cdots \theta^{a_i},
\label{phi}
\end{equation} 
where the coefficients $ \alpha_{a_1, a_2,...,a_i}$ depend on
 commuting coordinates $ x^a, \; a \in \{0,1,2,$ $3,5,..,d \}. $
The wave function space spanned over Grassmannian coordinate
space has the dimension $2^d$. Completely analogously to usual
quantum mechanics we have the operator for the conjugate
variable $\theta^a$ to be 
\begin{equation}
p^{\theta}_a = -i\overrightarrow{\partial}_a.
\label{pt}
\end{equation} 
The right arrow tells, that the derivation has to be performed
from the left hand side. These operators then obey the odd
Heisenberg algebra, which written by means of the generalized
commutators 
\begin{equation}
\{ A, B \}: = AB - ( -1)^{n_{AB}} BA,
\label{gc}
\end{equation}
where 
\begin{equation}
n_{AB}=\left\{ \begin{array}{rl} +1, \quad &\hbox{if A and B have
Grassmann odd character}\\ 0, \quad &\hbox{otherwise,}\end{array} \right.
\nonumber
\end{equation}   
takes the form
\begin{equation}
\{p^{\theta a},p^{\theta b}\} = 0 = \{\theta^a, \theta^b \} , \quad
\{ p^{\theta a}, \theta^b \} = -i \eta^{ab}.
\label{pptt}
\end{equation}
Here $\eta^{ab}$ is the flat metric $\eta = diag\{1,-1,-1,...\}$.
\\
We may  define the operators 
\begin{equation}
\tilde{a}^a := i(p^{\theta a}-i\theta^a), \quad\tilde{
\tilde{a}}{ }^a :=
-(p^{\theta a}+i\theta^a),
\label{at}
\end{equation}
for which we can show that the $\tilde{a}^a$'s among themselves 
fulfill the Clifford 
algebra as do also the $\tilde{\tilde{a}}{ }^a$'s, while they
mutually anti-commute:
\begin{equation}
\{ \tilde{a}^a, \tilde{a}^b\} = 2\eta^{ab} = \{\tilde{\tilde{a}}{
}^a, 
\tilde{\tilde{a}}{ }^b\}, \quad \{\tilde
{a}^a, \tilde{\tilde{a}}{ }^b\} = 0.
\label{ateta}
\end{equation}
We could recognize formally
\begin{equation}
{\mathrm either} \quad \tilde{a}^a p_a|\Phi> = 0, \qquad {\mathrm or} \quad 
\tilde{\tilde{a}}{ }^a p_a|\Phi> = 0 
\label{d}
\end{equation}
as the Dirac-like equation, because of the 
above generalized 
commutation relations.
Applying either the operator $\tilde{a}^a p_a$ on the left-hand
side equation or $\tilde{\tilde{a}}{
}^a p_a$ on the right-hand side equation we get the Klein-Gordon equation
$p^ap_a|\Phi> = 0$, where 
we define  $p_a = i\frac{\partial}{\partial x^a}$.
\\
One can check that none of the two equations
(\ref{d}) have solutions which would transform as spinors with
respect to the  generators of the Lorentz transformations,
when taken in analogy with the generators of the Lorentz
transformations in ordinary space ($L^{ab} = x^a p^b - x^b p^a $)
\begin{equation}
S^{ab}:= \theta^a p^{\theta b} - \theta^b p^{\theta a}.
\label{vecs}
\end{equation}
But we can write these generators as the sum
\begin{equation}
S^{ab} = \tilde{S}^{ab} + \tilde{\tilde{S}}{ }^{ab},
\quad \tilde{S}^{ab} := \frac{i}{4} [\tilde{a}^a, \tilde{a}^b], \quad
\tilde{\tilde{S}}{ }^{ab} := \frac{i}{4}[\tilde{\tilde{a}}{
}^a,\tilde{\tilde{a}}{ }^b],  
\label{vecsp}
\end{equation}
with $[A,B]:=AB-BA$ and recognize that the solutions of the two
equations (\ref{d}) now transform as spinors with respect to
either $ \tilde{S}^{ab}$ or $ \tilde{\tilde{S}}{ }^{ab}.$

One also can easily see that the untilded, the single tilded and
the double tilded  
$S^{ab}$ obey the $d$-dimensional Lorentz generator algebra
$\{ M^{ab}, M^{cd} \} = -i(M^{ad} \eta^{bc} + M^{bc} \eta^{ad} 
- M^{ac} \eta^{bd} - M^{bd} \eta^{ac}),$
 when inserted for
$M^{ab}$.

K\" ahler formulated spinors \cite{kahler1962} in terms of wave
functions which are superpositions of the p-forms in the ($d=4$)-
dimensional space.  A general linear combination of p-forms follows from
Eq.(\ref{phi}) if replacing $\theta^a$ by 
$dx^{a}\wedge $.

We presented in ref. \cite{mankocandnielsen1999} the
parallelism between our approach and the K\" ahler 
approach. We also presented in the same reference the generalization
of the K\" ahler approach, suggested by our approach. In both approaches
two types of operators fulfilling the Clifford algebra (the ones
of our approach are presented in 
Eqs.(\ref{at},\ref{ateta})) as well 
as the two Dirac-like equations (Eq.(\ref{d}) represent our
Dirac-like equation) can be obtained.
Both approaches offer the generators of the Lorentz
transformations, describing not only spinors but also vectors
(Eq.(\ref{vecs})).  In both approaches the $\gamma^a$ matrices,
fulfilling the Clifford algebra and having an Grassmann even
characters (which assures that $\gamma^a$'s transform Grassmann
odd object into Grassmann odd objects and accordingly do not
change the Grassmann character of spinors) can be defined
\begin{equation}
\tilde{\gamma}^a = i \; \tilde{\tilde{a}}{ }^0 \;
\tilde{a}^a, \qquad {\mathrm with} \qquad \{ \tilde{\gamma}^a,
\tilde{\gamma}^b \} = 2 \eta^{ab}. 
\label{gamat}
\end{equation}
The ''naive'' definition of gamma-matrices ($\gamma^a_{naive} :=
\tilde{a}^a$), which 
changes the Grassmann character of spinors, differs from the
Grassmann even definition of gamma-matrices, presented in
Eq.(\ref{gamat}), which keeps 
the Grassmann character of 
spinors (both fulfilling the Clifford algebra), only when
$\gamma^0$-matrix has to 
simulate the parity reflection which is 
$ \vec{\theta}\rightarrow - \vec{\theta}.$
In all physical applications (such as construction of currents)
the two definitions can not be
distinguished among themselves, since $\gamma^a$'s always
appear in pairs. 
We can check that the $\tilde{\gamma}^a$ (Eq.(\ref{gamat})) indeed
perform the operation of the parity reflection. 
\\

\subsection{Scalar product}
 \label{sptk}

In our approach \cite{mankoc1992,mankoc1993,mankoc1994,mankoc1995,%
mankoc1997} the scalar
product between 
the two functions $<\theta^a |\Phi_1>$  and $<\theta^a |\Phi_2>$ is
defined 
\begin{equation}
<\Phi_1|\Phi_2> = \int
d^d\;\theta\;\;(\omega\;<\theta^a|\Phi_1>)<\theta^a |\Phi_2> 
\label{scptk}
\end{equation}
and $\omega $ is a weight function
\begin{equation}
\omega = \prod_{i=0,1,..,d}\;\;(\theta^i +
\overrightarrow{\partial}^i), 
\nonumber
\end{equation}
which operates on only the first function $<\theta^a|\Phi_1>$
and 
\begin{equation}
\int \; d\theta^a = 0, \quad \int \; d^d\theta \theta^0
\theta^1...\theta^d = 1,\quad d^d\theta = \theta^d
...\theta^1 \theta^0.
\nonumber
\end{equation}
According to the above definition and Eq.(\ref{phi}) it follows
\begin{equation}
<\Phi^{(1)}|\Phi^{(2)}> = \sum_{0,d}\;\;
\sum_{\alpha_1<\alpha_2<..<\alpha_d}
\alpha^{(1)*}_{\alpha_1..\alpha_i}\;
\alpha^{(2)}_{\alpha_1..\alpha_i}. 
\label{scpt}
\end{equation}

\subsection{ Four copies of Weyl bi-spinors }\label{fws}

We present in this subsection four copies of two-Weyl spinors.

\begin{center}
\begin{tabular}{||c|c||c||c|c|c|c|c||}
\hline
\hline 
a & i &$<\theta|{}^a \Phi_i>$& $\tilde{S}^{12}$ & $
\tilde{S}^{03}$ & $\tilde{\Gamma}^{(4)}$& family & Grass. cha.\\ 
\hline
1 & 1 &$\frac{1}{2}(\tilde{a}^1 + i \tilde{a}^2)(\tilde{a}^0 +
\tilde{a}^3)$ &$ \frac{1}{2}$ &$ -\frac{i}{2}$&-1&&\\
1 & 2 &$\frac{1}{2}(1 - i\tilde{a}^1 \tilde{a}^2)(1+
\tilde{a}^0 \tilde{a}^3) $ &$ -\frac{1}{2} $&$ \frac{i}{2}$&-1&&\\
&&&&&&I& even\\
2 & 1 &$\frac{1}{2}(\tilde{a}^1 + i \tilde{a}^2)(\tilde{a}^0 -
\tilde{a}^3)$ &$ \frac{1}{2}$ &$ \frac{i}{2}$&1&&\\
2 & 2 &$\frac{1}{2}(1 - i\tilde{a}^1 \tilde{a}^2)(1-
\tilde{a}^0 \tilde{a}^3)$  &$ -\frac{1}{2} $&$ -\frac{i}{2}$&1&&\\
\hline
\hline
3 & 1 &$\frac{1}{2}(\tilde{a}^1 + i \tilde{a}^2)(1 + \tilde{a}^0 
\tilde{a}^3)$ &$ \frac{1}{2}$ &$ \frac{i}{2}$&1&&\\
3 & 2 &$\frac{1}{2}(1 - i\tilde{a}^1 \tilde{a}^2)(\tilde{a}^0 +
 \tilde{a}^3) $ &$ -\frac{1}{2} $&$ -\frac{i}{2}$&1&&\\
&&&&&&II& odd\\
4 & 1 &$\frac{1}{2}(\tilde{a}^1 + i \tilde{a}^2)(1 - \tilde{a}^0 
\tilde{a}^3)$ &$ \frac{1}{2}$ &$ -\frac{i}{2}$&-1&&\\
4 & 2 &$\frac{1}{2}(1 - i\tilde{a}^1 \tilde{a}^2)(\tilde{a}^0 -
\tilde{a}^3)$  &$ -\frac{1}{2} $&$ \frac{i}{2}$&-1&&\\
\hline
\hline
5 & 1 &$\frac{1}{2}(1 +i \tilde{a}^1 \tilde{a}^2)(\tilde{a}^0 +
\tilde{a}^3)$ &$ \frac{1}{2}$ &$ -\frac{i}{2}$&-1&&\\
5 & 2 &$\frac{1}{2}(\tilde{a}^1 -i \tilde{a}^2)(1 +
\tilde{a}^0 \tilde{a}^3) $ &$ -\frac{1}{2} $&$ \frac{i}{2}$&-1&&\\
&&&&&&III&odd\\
6 & 1 &$\frac{1}{2}(1+ i\tilde{a}^1 \tilde{a}^2)(\tilde{a}^0 -
\tilde{a}^3)$ &$ \frac{1}{2}$ &$  \frac{i}{2}$&1&&\\
6 & 2 &$\frac{1}{2}(\tilde{a}^1 -i \tilde{a}^2)(1-
\tilde{a}^0 \tilde{a}^3)$  &$ -\frac{1}{2} $&$ -\frac{i}{2}$&1&&\\
\hline
\hline
7 & 1 &$\frac{1}{2}(1 +i\tilde{a}^1 \tilde{a}^2)(1+ \tilde{a}^0 
\tilde{a}^3)$ &$ \frac{1}{2}$ &$ \frac{i}{2}$&1&&\\
7 & 2 &$\frac{1}{2}(\tilde{a}^1 -i\tilde{a}^2)(\tilde{a}^0 +
 \tilde{a}^3) $ &$ -\frac{1}{2} $&$ -\frac{i}{2}$&1&&\\
&&&&&&IV&even\\
8 & 1 &$\frac{1}{2}(1+i\tilde{a}^1 \tilde{a}^2)(1 - \tilde{a}^0 
\tilde{a}^3)$ &$ \frac{1}{2}$ &$ -\frac{i}{2}$&-1&&\\
8 & 2 &$\frac{1}{2}(\tilde{a}^1 -i \tilde{a}^2)(\tilde{a}^0 -
\tilde{a}^3)$  &$ -\frac{1}{2} $&$ \frac{i}{2}$&-1&&\\
\hline
\hline
\end{tabular}
\end{center}

\noindent
Table I: The polynomials of $\theta^m$, representing the four
times two Weyl spinors, are presented. For each state the
eigenvalues of $\tilde{S}^{12}, \tilde{S}^{03},\;
\tilde{\Gamma}^{(4)}: = i \tilde{a}^0 \tilde{a}^1 \tilde{a}^2 \tilde{a}^3$ are
written. The Roman numerals tell the possible family number.
We use the relation $\tilde{a}^a |0> = \theta^a$.

We present here for $d = 4$ the $2^d $ vectors, which we arrange
 into four copies of 
two Weyl spinors, one left 
( $<\tilde{\Gamma}^{(4)}> = - 1,\;\;\; 
           \Gamma^{(4)} = i \frac{(-2i)^2}{4!} \epsilon_{abcd} {\mathcal S}^{ab} {\mathcal S}^{cd}$
) and one right ( $< \tilde{\Gamma}^{(4)}> = 1$) handed 
in such a way that they are at the same time also the eigenvectors
of the operators $\tilde{S}^{12}$ and the
$\tilde{S}^{03}$ and have  
either an odd or an even Grassmann character. We
have made a choice of $(\tilde{\; })$ operators, putting the
operators of the type $(\tilde{\tilde{\; }}) $ equal to zero. We
present these vectors as 
polynomials of $\theta^m$'s, $m \in (0,1,2,3)$. The
corresponding K\" ahler's p-forms follow if $\theta^a$'s are
replaced by $dx^a \wedge $.  The two Weyl
vectors of one copy of the Weyl bi-spinors are connected by the
$\tilde{\gamma}^m$ 
(Eq.(\ref{gamat})) operators, while the two copies of different
Grassmann character are connected by $\tilde{a}^a$,
respectively. The two copies of an even 
Grassmann character are connected by the ( a kind of a time
reversal operation) $ \theta^0 \; \rightarrow \; - \theta^0 $ (or
equivalently $dx^0 \; \rightarrow \; - dx^0 $), if differential
forms are concerned.

We present in Table I four copies of the Weyl two spinors as
polynomials of $\theta^a$.
 Eigenstates are
orthonormalized according to the scalar product of Eq.(\ref{scpt})\\

Analyzing the irreducible representations of the group $SO(1,3)$
with respect to the generator of the Lorentz transformations of
the vectorial type \cite{mankoc1993,mankoc1994,mankoc1995,%
mankoc1997,borstnikandmankoc1999} (Eqs.(
\ref{vecs})) one finds 
for d = 4 two scalars (a scalar and 
a pseudo scalar), two three vectors (in the complex version of
the $SU(2) \times SU(2) 
$ representation of $SO(1,3)$ denoted by $(1,0) $ and $(0,1)$
representation, respectively, with $<\Gamma^{(4)}> = \pm 1$) and
two four vectors. One can find the polynomial 
representation for this case in ref.\cite{mankoc1993}.

\subsection{Generalization to extra dimensions}
\label{ged}

 It has  been
suggested \cite{mankoc1994} that the Lorentz
transformations in the 
space of $ \theta^a$'s in  ($d-4$)-dimensions manifest
themselves as generators for charges observable  for
the four dimensional particles. Since both, the
extra dimensional spin degrees of freedom and the ordinary spin
degrees of freedom, originate from the $\theta^a$'s (or the
forms) we have a unification of these internal degrees of freedom.
\\
Let us take as an example the model \cite{mankoc1995,borstnikandmankoc1999} which has 
$d=14$ and at first - at the high energy
level - $SO(1,13)$ Lorentz group, but which should be broken 
( in two steps ) to first $SO(1,7)\times SO(6)$ and then to
$SO(1,3)\times SU(3)\times SU(2) $. We shall comment on this
model in section \ref{bltsm}.

\subsection{ Appearance of spinors} 
\label{as}

By   exchanging the Lorentz 
generators ${\mathcal S}^{ab}$ by the $\tilde{S}^{ab}$ say ( or the
$\tilde{\tilde{S}}{ }^{ab}$ if we 
choose them instead), of Eq.(\ref{vecsp}), a  spinor field appears out of models with only scalar,
vector and tensor objects.
One of the two kinds of operators
fulfilling the Clifford 
algebra and anticommuting with the other kind - it has been made a
choice of  
$\tilde{\tilde{a}}{ }^a $ in our approach and similarly one also
can proceed  in the K\" ahler case  - are put
to zero in the operators of the Lorentz transformations; as well as in
all the operators representing physical quantities. The  use
of $ \tilde{\tilde{a}}{ }^0 $ in
the operator $\tilde{\gamma}^0$ (and equivalently also in the
Dirac case) is the exception,  only used to
simulate the Grassmann even  parity operation 
$ \vec{\theta} \to - \vec{\theta} $ (or for p-forms $\vec{dx}^a
\to -\vec{dx}^a $).

\section{Lagrange function for a free massless particles in ordinary
and in Grassmann space and canonical quantization} 
\label{lfmp}

We  present in this section the Lagrange function for a
particle which lives in a d-dimensional ordinary space
of commuting coordinates and in a d-dimensional Grassmann space
of anti-commuting coordinates $ X^a \equiv \{ x^a, \theta^a \} $
and has its geode\-sics parameterized by an ordinary Grassmann even
parameter ($\tau$) and a Grassmann odd parameter($\xi$).  We
derive the Hamilton function and the corresponding Poisson
brackets and perform 
the canonical quantization, which leads to the Dirac equation with
operators presented in section \ref{dirac}.

The coordinates $ X^a = X^a(x^a,\theta^a,\tau,\xi)$ are called the
super-coordinates. 
We define the dynamics of a particle by choosing the 
action (in complete analogy with the usual definition of the scalar
product in ordinary space \cite{mankoc1992,ikemori1987} 
\begin{equation*}
I= \frac{1}{2} \int d \tau d \xi E E^i_A \partial_i X^a E^j_B
\partial_j X^b  \eta_{a b} \eta ^{A B},
\end{equation*}
where $ \partial _i : = ( \partial _ \tau , {\overrightarrow
{\partial}} _\xi ), \tau^i = (\tau, \xi) $, while $ E^i _A
$ determines a metric on a two dimensional super-space $ \tau ^i
$ , $ E = det( E^i _A )$ . We choose $ \eta _{A A} = 0,   
\eta_{1 2} = 1 = \eta_{2 1} $, while $ \eta_{a b} $ is the
Minkowski metric with the diagonal elements $
(1,-1,-1,-1,$ $...,-1) $. The action is invariant under the Lorentz
transformations of super-coordinates: $X'{ }^a = \Lambda^a{ }_b X^b $.
Since a super-matrix $ E^i{ }_A $ transforms as a vector in a
two-dimen\-sional super-space $\tau^i$ under general coordinate
transformations of $\tau^i$, $ E^i{ }_A \; \tau_i $ is invariant
under such transformations and so is $d^2 \; \tau E$. 
The action is locally super-symmetric.
 The inverse matrix $ E^A{ }_i$ 
is defined as follows: $E^i{ }_A E^B{ }_i = \delta^B{ }_A$.
 
Taking into account that either $ x^a $ or $ \theta^a $ depend
on an ordinary time parameter $ \tau $ and that $ \xi^2 = 0 $ ,
the geodesics can be described  as a
polynomial of  $ \xi $ as follows: 
$ X^a = x^a + \varepsilon \xi \theta^a $. We choose $
\varepsilon^2 $ 
to be equal either to $ +i $ or to $ -i $ so that it defines two
possible combinations of super-coordinates. Accordingly we
also choose  the metric  $ E^i { }_A $ : $ E^1{ }_1 = 1, \; E^1{
}_2 = - \varepsilon M,\; E^2{ }_1 = \xi, E^2{ }_2 = N -
\varepsilon \xi M $, with $ N $ and $ M $  Grassmann even and
odd parameters, respectively. We write $ \dot{A} =
\frac{d}{d\tau}A $, for any $ A $.

If we integrate the above action over the Grassmann odd
coordinate $d\xi$, the action for a super-particle follows:
\begin{equation}
 \int \; d\tau \;\;( \frac{1}{N} \dot{x}^a \dot{x}_a + \varepsilon^2
\dot{\theta}^a \theta_a - \frac{2\varepsilon^2 M}{N} \dot{x}^a
\theta_a ). 
\label{af}
\end{equation}
Defining the two momenta 
\begin{equation}
p^{\theta }_a : = \frac{ \overrightarrow{\partial} L}
{ \partial {\dot{\theta}^a}} = \epsilon^2 \theta^a,\;\; p_a : =
\frac{\partial L}{ \partial \dot{x}^a} = \frac{2}{N}( 
\dot{x}_a - M p^{\theta a}), 
\label{ptf}
\end{equation}
the two Euler-Lagrange equations follow:
\begin{equation} \frac{dp^a}{d \tau} = 0,\;\;\; \frac{dp^{\theta a}}{d \tau} =
\varepsilon ^2 \frac{M}{2} p^a. 
\label{el}
\end{equation}
Variation of the action (Eq.(\ref{af})) with respect to $M $ and
$N$ gives 
the two constraints
\begin{equation}
\chi^1: = p^a a^{\theta}_a = 0, \quad \chi^2 = p^a p_a = 0, \;\;
a^{\theta}_a:= i p^{\theta}_a + \varepsilon^2 \theta_a,
\label{chi}
\end{equation} 
while  
$ \chi^3{ }_a: = - p^{\theta }_a + \epsilon^2 \theta_a = 0 $
(Eq.(\ref{ptf})) is the third type of constraints of the action(\ref{af}).
For $\varepsilon^2 = -i$ we find  that $
a^{\theta}{ }_a = \tilde{a} 
^a,\;\;$ which agrees with Eq.(\ref{at}), while $\chi^3{ }_a =
\tilde{\tilde a}_a = 0, $  which makes a choice between
$\tilde{a}^a$ and $\tilde{\tilde{a}}^a$. 

We find the generators  of the Lorentz transformations for the
action(\ref{af}) to be 
\begin{equation}
 M^{ a b} = L^{a b} + S^{a b} \;,\; L^{a b} = x^a p^b - x^b p^a
\;,\; S^{a b} = \theta^a p^{ \theta b} - \theta^b p^{ \theta a}
=  \tilde{S} ^{a b} + \tilde{\tilde{S}}{}^{a b},
\label{mabl}
\end{equation} 
which agree with definitions in Eq.(\ref{vecsp}) and 
show that parameters of the Lorentz transformations are the
same in both spaces.

We define the Hamilton function:
\begin{equation}
 H:= \dot{x}^a p_a + \dot{\theta}{ }^a p^{\theta}{ }_a - L =
\frac{1}{4} N p^a p_a + \frac{1}{2} M p^a (\tilde a_a +
i \tilde{\tilde a }{ }_a) 
\label{haml} 
\end{equation}
and the corresponding Poisson brackets
\begin{equation}
\{A,B\}_p=  
\frac{ \partial A}{ \partial x^a} \frac{ \partial B}{ \partial
p_a}  - \frac{ \partial A}{ \partial p_a} \frac{ \partial B}{ \partial
x^a} +  \frac{ \overrightarrow{ \partial A}}{\partial \theta
^a} \frac{ \overrightarrow{ \partial B}}{\partial p^\theta_a} +
  \frac{ \overrightarrow{ \partial A}}{\partial p^\theta_a} 
\frac{ \overrightarrow{ \partial B}}{\partial \theta^a}, 
\label{poil}
\end{equation} 
which fulfill  the algebra of the generalized
commutators\cite{mankoc1999} of Eq.\ref{gc}.

If we take into account the constraint $\chi^3{ }_a = \tilde{\tilde
a}{ }_a = 0\;$ in the Hamilton function (which just means that
instead of H the Hamilton function $ H + \sum_i \alpha^i
\chi^i + \sum_a \alpha^3{ }_a \chi^3{ }^a $ is taken, with parameters 
$\alpha^i, i=1,2 $ and 
$ \alpha^3{ }_a = -\frac{M}{2} p_a$, $a=0,1,2,3$, $5,..,d$ 
chosen in such a way that the Poisson brackets
of the three types of constraints with the new Hamilton function
are equal to zero) and in all dynamical quantities, we find:
\begin{equation}
H = \frac{1}{4} N p^a p_a + \frac{1}{2} M p^a \tilde a_a,\;\;
\chi^1 = p^a p_a = 0,\;\; \chi^2 = p^a \tilde a_a =
0,
\label{hamla}
\end{equation}
\begin{equation}
\dot{p}_a = \{ p_a, H \}_P = 0,\;\; \dot{\tilde a}{ }_a = \{ 
\tilde{a}_a, H \}_P = iM p_a,
\nonumber 
\end{equation}
which agrees with the Euler-Lagrange equations (\ref{el}).

We further find 
\begin{equation}
 \dot {\chi}^i = \{ H, \chi^i \}_P = 0,\;\;i =1,2,\;\;\; \dot
{\chi}^3{ }_a = \{ H, \chi^3{ }_a \}_P = 0,\;\;a =
0,1,2,3,5,..,d,
\nonumber 
\end{equation}
which guarantees that the three
constraints will not change with the time parameter $\tau$ and
that $\dot{\tilde M}{ }^{ab} = 0 $, with $ \tilde M { }^{ab} =
L^{ab} + \tilde{S}^{ab}$,  saying that $ \tilde M{ }^{ab} $
is the constant of motion.

The Dirac brackets, which can be obtained from the Poisson
brackets of Eq.(\ref{poil}) by adding to these brackets on the right
hand side a term $ - \{A, \tilde{\tilde a}^c \}_P \cdot$ $ ( -
\frac{1}{2i} \eta_{ce} ) \cdot $ $ \{ \tilde{\tilde a}{ }^e, B
\}_P $, give  for the dynamical quantities, 
which are observables, the same results as the Poisson brackets.
This is true also for $ \tilde a^a,$ ( $\{ \tilde
a^a, \tilde a^b \}_D = i\eta^{ab} = \{ 
\tilde a^a, \tilde a^b \}_P $),  which is the
dynamical quantity but not  an observable since its odd
Grassmann character causes  super-symmetric
transformations. We also find that $\{ \tilde a^a, \tilde{\tilde
a}{ }^b \}_D 
= 0 = \{ \tilde a^a, \tilde{\tilde a}{ }^b \}_P $ .
The Dirac brackets give  different results only for the quantities
$ \theta^a $ and $ p^{\theta a} $ and  for $\tilde {\tilde
a}{ }^a $ among themselves: $ \{ \theta^a, p^{\theta b}
\}_P = \eta^{ab}, \{ \theta^a, p^{\theta b}
\}_D = \frac{1}{2} \eta^{ab} $, $ \{ \tilde {\tilde a}{ }^a, \tilde
{\tilde a}{ }^b \}_P = 2i \eta^{ab}, \{ \tilde {\tilde a}{ }^a,
\tilde {\tilde a}{ }^b \}_D = 0 $. According to the above  properties
of the Poisson brackets, we suggested \cite{mankoc1995,mankoc1999} that 
in the quantization 
procedure the Poisson brackets (\ref{poil}) rather than the Dirac
brackets are used, so that variables $\tilde{\tilde a}^a $,
which are removed from all dynamical quantities, stay as
operators. Then $\tilde a^a $ and 
$\tilde{\tilde a}{ }^a $ are expressible with $\theta^a $ and
$p^{\theta a} $ (Eq.(\ref{at}))  and the algebra of linear operators
introduced in section \ref{dirac},  can be used. We
shall 
show, that  suggested quantization procedure leads to the
Dirac equation, which is the differential equation in ordinary
and Grassmann space and has all desired properties.

In the proposed quantization procedure $\; -i \{ A,B \}_p $ goes to
either a commutator or to an anticommutator, according to the
Poisson brackets (\ref{poil}). The operators $\theta ^a , p^{\theta a}
$ ( in the coordinate representation they become $ \theta^a 
\longrightarrow \theta^a , \; p^{\theta}_a \longrightarrow i 
\frac{\overrightarrow{\partial }}{\partial \theta^a} $) fulfill
the Grassmann odd Heisenberg algebra, while the operators 
$ \; \tilde{a}^a \; $ and $\; \tilde{\tilde{a}}{}^{a}\; $ fulfill
the Clifford algebra (Eq.(\ref{ateta})).

The constraints (Eqs.(\ref{chi})) lead to
the Weyl-like and  the Klein-Gordon equations
\begin{equation}
p^a \tilde{a} _a | \tilde{\Phi} > = 0 \;,\; p^a p_a |
\tilde{\Phi}> = 0 , \; {\mathrm with} \;  p^a \tilde{a}_a p^b \tilde{a}_b =
p^a p_a . 
\label{dkl}
\end{equation}
Trying to solve the eigenvalue problem $ \tilde{\tilde a}{ }^a 
| \tilde {\Phi} > = 0,\;\; a=(0,1,2,3,5,...,d), $ we find that no
solution of this eigenvalue problem exists, which means that
the third constraint $ \tilde{\tilde a}{ }^a = 0 $ can't be
fulfilled in the operator form (although we take it into account
in the operators for all dynamical variables in order that
operator equations would agree with classical equations). We can
only take it into account 
in the  expectation value form 
\begin{equation}
< \tilde{\Phi} | \tilde{\tilde a}{ }^a | \tilde{\Phi} > = 0.
\label{phittl}
\end{equation}
Since $ \tilde{\tilde a}{ }^a $ are Grassmann odd operators,
they change monomials (Eq.(\ref{phi})) of an Grassmann odd character
into monomials of an Grassmann even character and opposite,
which is the super-symmetry transformation.
It means that Eq.(\ref{phittl}) is fulfilled for monomials of either odd
or even Grassmann character and that superpositions of the
Grassmann odd and the Grassmann even monomials are not solutions
for this system. 

We  define the projectors
\begin{equation} 
 P_{\pm} = \frac{1}{2} ( 1 \pm  \sqrt{ (-)^{\tilde
\Upsilon \tilde{\tilde 
\Upsilon}}} \tilde{ \Upsilon} \tilde{\tilde \Upsilon}),\;\;\;\;
(P_{\pm})^2 = 
P_{\pm}, 
\label{proj} 
\end{equation}
where $\tilde \Upsilon$ and $ \tilde{\tilde \Upsilon}$ are the two
operators  defined  for any dimension d as follows
$ \tilde \Upsilon = \; i^{\alpha} $ $ \prod_{a=0,1,2,3,5,..,d} \tilde{
a}{ }^a $ $\sqrt{\eta^{aa}},$ $\;\; \tilde{\tilde \Upsilon} =
i^{\alpha} \prod_{a=0,1,2,3,5,..,d} 
$ $\tilde{ \tilde{a}}{ }^a \sqrt{\eta^{aa}},$
with $\alpha $ equal either to $ d/2 $ or to $ (d-1)/2 $ for
even and odd dimension $d$ of the space, respectively. 
It can be checked that $( \tilde \Upsilon )^2 = 1 = ( \tilde{
\tilde{\Upsilon}} )^2 $.

We can use the projector $P_{\pm}$ of Eq.(\ref{proj}) to project
out of monomials  either the Grassmann odd or the Grassmann even
part. Since this projector commutes with the Hamilton function $
 (\{ P_{\pm}, H \} = 0 ) $,  it means that eigenfunctions of $
H $, which fulfill the Eq.(\ref{phittl}), have either an odd or an even
Grassmann character. 
In order that in the second quantization procedure  fields
$ | \tilde{\Phi} > $ would describe fermions, it is meaningful
to accept  in the fermion case Grassmann  odd monomials only.

\section{Particles in gauge fields}
 \label{pgf}

The dynamics of a point particle in gauge fields, the
gravitational field in $d$-dimensions, which then, as we shall show,
manifests in the subspace $d=4$ as ordinary gravity and all the
Yang-Mills fields, can be obtained by 
transforming  vectors from a freely falling to an external 
coordinate system \cite{wessandbagger1983}.  
To do this, supervielbeins ${\mathbf e}^{a}{ } _{\mu} $ have to be 
introduced, which in our case depend on ordinary and on
Grassmann coordinates, as well as on 
two types of parameters $ \tau^i = ( \tau, \xi ) $.  The index a 
refers to a freely falling coordinate system ( a Lorentz index),
the index $\mu$ refers to an external coordinate system ( an
Einstein index). 
 
We write the transformation of vectors as follows
$ \partial_i X^a= {\mathbf e}^{ a} { }_{\mu} \partial_i X^{\mu} \;,\;
\partial_i X^{\mu} = {\mathbf f}^{ \mu} { }_a \partial_i X^a \;,\;
\partial_i $ $= ( \partial_{\tau} , \partial_{\xi} ).$
From here it follows that
$ {\mathbf e}^{ a} { }_{\mu} {\mathbf f}^{ \mu} { }_b = \delta^a { }_b \;,\;  
{\mathbf f}^{ \mu} { }_{a} {\mathbf e}^{ a} { }_{\nu} = \delta^{\mu} {
}_{\nu}.$ 

Again we make a Taylor expansion of vielbeins with respect to 
$ \xi,\; $ 
$ {\mathbf e}^{ a} { }_{\mu} = e^{a} { }_{\mu} + \varepsilon^2 \xi
\theta^b e^{a} { }_{ \mu b} \;,\; {\mathbf f}^{ \mu} { }_a $ $= f^{
\mu} { }_a - \varepsilon^2 \xi \theta^b
f^{\mu} { }_{a b}.$

Both expansion coefficients  again depend  on ordinary
and on Grassmann coordinates. Having an even Grassmann character
$ e^{a} { }_{\mu}$  will describe the spin 2 part of a
gravitational field. The coefficients $ 
e^{a} { }_{\mu b}$  define the
spin connections \cite{mankoc1992,mankoc1999}. 

It follows that
$   e^{ a} { }_{\mu} f^{\mu} { }_b = \delta^a { }_b \;,\;  
f^{\mu} { }_{a} e^{a} { }_{\nu} = \delta^{\mu} { }_{\nu}
\;,\; e^{a} { }_{\mu b} f^{\mu} { }_c = e^{a} { }_{\mu}
f^{ \mu} { }_{c b}.$

We find the metric tensor ${\mathbf g}_{\mu \nu} = {\mathbf e}^{a}
{ }_{\mu} {\mathbf e}^{ }_{a \nu} ,\;
{\mathbf g}^{ \mu \nu} ={\mathbf f}^{ \mu} { }_{a} {\mathbf f}^{\nu a}$. 
Rewriting the action from section \ref{lfmp} in terms of an external
coordinate system, using the Taylor expansion of 
super-coordinates $ X^{\mu}$ and super-fields $ {\mathbf{e}}^{a} {
}_{\mu}$ and
integrating the action over the Grassmann odd parameter $\xi$,
the action follows
\begin{eqnarray}
 I=\int d\tau \; \{ \frac{1}{N} g^{\mu \nu} \dot{x}^\mu
\dot{x}^\nu \; - \; \varepsilon^2 \frac{ 2 M}{N} \theta_a e^{a} {
}_{\mu} \dot{x}^\mu \; + \; \varepsilon^2 \frac{1}{2}(
\dot{\theta}^\mu \theta_a -\theta_a \dot{\theta}^\mu) e^{a} {
}_{\mu} \; + 
\nonumber\\ 
 + \;  \varepsilon^2 \frac{1}{2} (\theta^b \theta_a
-\theta_{a} \theta^b ) e^{a} { }_{  \mu b} \dot{x}^\mu \} ,
\label{acgr}
\end{eqnarray}
which defines the two momenta of the system
$ p_{\mu} = \frac{\partial L}{\partial \dot{x}^\mu} = p_{0 \mu} +
 \frac{1}{2} \tilde{S}^{ab} e_{a \mu b} , \;\;
 p^\theta_\mu = -i \theta_a e^{a} { }_{\mu}
$
( $\varepsilon^2 = -i $ ).
Here $ p_{0 \mu} $ are the canonical (covariant) momenta of a
particle. 
For $ p^{\theta}_{a} = p^{\theta}_{\mu} f^{\mu} { }_{a}$, it follows
that $ p^{\theta}_{a}$ is proportional to $\theta_{a}$. Then $
\tilde{a}_{a} = i 
(p^{\theta}_{a} - i \theta_{a}), 
$ while $ \tilde{\tilde{a}}_{a}= 0 $.  We may further write
\begin{equation}
 p_{ 0 \mu} = p_{ \mu} - \frac{1}{2} \tilde{S}^{a b} e_{a \mu b}
= p_{ \mu} - \frac{1}{2} \tilde{S}^{a b} \omega_{a b \mu} \;,\;
\omega_{a b \mu}=\frac{1}{2} (e_{a \mu b} - e_{b \mu a}),
\label{cmg}
\end{equation} 
which is the usual expression for the covariant momenta in
gauge gravitational fields \cite{wessandbagger1983}.
One can find  the two constraints
\begin{equation}
 p_0^\mu p_{0 \mu} = 0 = p_{0 \mu} f^{\mu} { }_a \tilde{a}^a .
\label{cong} 
\end{equation}

We shall comment on the breaking of symmetries which leads in ($d=4$)-
dimensional subspace as ordinary gravity and all the gauge fields
in section \ref{bltsm}.

\section{ Breaking 
 $SO(1,13)$ through $SO(1,7) \times SO(6)$ to 
$SO(1,3) \times SU(2) \times U(1) \times SU(3)$}
 \label{bltsm}

In this section, we shall first discuss a possible breaking of
symmetry, which leads from the unified theory of only spins and
gravity in d dimensions to spins and charges and to 
 the symmetries and assumptions of the Standard Model, on the
algebraic level (\ref{ac}).
We shall then comment on the breaking of symmetries on the level of
canonical momentum  
for the particle in the presence of the gravitational field
(\ref{dar}). 

We shall present as well the possible explanation for that
postulate of the Standard Model, which requires that only left
handed weak charged massless doublets and right handed weak
charged massless singlets exist, and accordingly   
 connect spins and charges of fermions.

\subsection{Algebraic considerations of symmetries}
 \label{ac}

 The algebra of the group $ SO(1,d-1) $ {\it or} $ SO(d) $ 
contains \cite{mankoc1995,borstnikandmankoc1999} $ n $ subalgebras
defined by 
operators $ \tau ^{A i}, A = 1,n ; i = 1,n_A $,
where $ n_A $ is the number of elements of each
subalgebra, with the properties 
\begin{equation}
 [ \tau ^{Ai} , \tau ^{B j} ] = i \delta ^{AB
} f^{A ijk } \tau ^{A k}, 
\label{taug}
\end{equation}
if operators $ \tau ^{A i} $  can be expressed as
linear superpositions  of operators $ M^{ab} $ 
\begin{equation}
 \tau^{A i} = {\mathit c} ^{A i} { }_{ab} M^{ab}, \;\;
{\mathit c} ^{A i}{ }_{ab} = - {\mathit c} ^{A i}{ }_{ba}, \;\;
A=1,n, \;\;
i=1,n_{A}, \;\;a,b=1,d. 
\label{tauga} 
\end{equation}

Here $ f^{A ijk} $ are structure constants
of the ($ A $) subgroup with $ n_{A} $ operators.
According to the three kinds of operators $ {\mathcal S}^{ab} $, two of
spinorial and one of vectorial character, there are  three kinds
of operators $ \tau^{A i} $ defining subalgebras of
spinorial and vectorial character, respectively, those of
spinorial types being expressed with either $ \tilde S^{ab} $ or
$ \tilde{ \tilde S}{ }^{ab} $ and those of vectorial type being
expressed by $ S^{ab} $. All three kinds
of operators are, according to Eq.(\ref{taug}), 
defined by the same coefficients $ {\mathit c}^{A i} { }_{ab} $
and the same structure constants $ f^{A i j k } $.
From Eq.(\ref{taug}) the following relations among constants ${\mathit
c}^{A i}{ }_{ab} $ follow
\begin{equation}
 -4 {\mathit c}^{A i}{ }_{ab} {\mathit c}^{B j b}{ }_c -
\delta^{A B} f^{A ijk} {\mathit c}^{A k}{ }_{ac}
= 0. 
\label{taugb}
\end{equation}

When we look for coefficients $ c^{A i}{ }_{ab} $ which
express operators $ \tau ^{A i} $, forming a subalgebra
$ SU(n) $ of an algebra $ SO(2n) $ in terms of $ M^{ab} $, the
procedure is rather simple\cite{georgi1982,mankoc1997}. We
find: 
\begin{equation} 
 \tau^{A m} = -\frac{i}{2} (\tilde \sigma^{A m})_{jk}
 \{ M^{(2j-1) (2k-1)} +
M^{(2j) (2k)} + i M^{(2j) (2k-1)} - i M^{(2j-1) (2k)}
\}.
\label{tausab} 
\end{equation}
Here $(\tilde \sigma^{A m})_{jk}$ are the traceless matrices
which form the algebra of $ SU(n) $.
One can easily prove that operators $ \tau^{A m} $ fulfill the
algebra of the group $ SU(n) $  for any of three
choices for operators $ M^{ab} : S^{ab}, \tilde S^{ab},
\tilde{\tilde S}{ }^{ab}$.

While the coefficients are the
same for all three kinds of operators, the representations depend on the
operators $M^{ab}$. After solving the
eigenvalue problem  for  invariants of 
subgroups, the representations can be presented as polynomials
of coordinates $\theta^a,$ or $ dx^a \wedge,$ $a =
0,1,2,3,5,..,14 $. The operators 
of spinorial character define the fundamental representations of
the group and the subgroups, while the operators of vectorial
character define the adjoint representations of the groups.
 We shall from now on, for the
sake of simplicity, refer to the polynomials of 
 Grassmann coordinates only. \\

We first analyze the space
of $2^d$ vectors for $d=14$ with respect to commuting operators
(Casimirs) of subgroups $SO(1,7)$ and $SO(6)$, so that 
 polynomials of $\theta^0, \theta^1, \theta^2, \theta^3,
\theta^5,\theta^6, \theta^7$ and $\theta^8$ are used to describe
states of the group 
SO(1,7) and then polynomials of $\theta^9, \theta^{10},
\theta^{11}, \theta^{12}, \theta^{13}$ and $\theta^{14}$
further to describe states of the group $SO(6)$. The group
$SO(1,13)$ has the rank equal to $r=7$, since it has $7 $
commuting operators (namely for example ${\mathcal S}^{01}, {\mathcal S}^{23},
{\mathcal S}^{56}, ...,{\mathcal S}^{13\;14} $), while the ranks of the
subgroups $SO(1,7)$ and 
$ SO(6)$ are accordingly $r=4$ and $r=3$, respectively. We may
further decide to arrange the basic states in the space of
polynomials of $\theta^0,...,\theta^8$ as eigenstates of $4 $
Casimirs of the subgroups $SO(1,3), SU(2),$ and $U(1)$ (the 
 first has  $r=2$, the second and the third have $r=1 $)  of the
group $SO(1,7)$, and the basic states in the space of polynomials of
$\theta^9,...,\theta^{14}$ as eigenstates of $r=3$ Casimirs of
subgroups $SU(3)$ and $U(1)$ ( with $r=2$ and $r=1$,
respectively) of the group $SO(6)$.  \\

We presented in Table I the eight Weyl spinors, two by two - 
one left ( $\tilde{\Gamma}^{(4)} = -1$) and one right (
$\tilde{\Gamma}^{(4)} = 1$) handed - 
connected by $\tilde{\gamma}^m, m=0,1,2,3$ into Weyl bi-spinors.
Half of vectors have Grassmann odd (odd products of
$\theta^m$ ) and half Grassmann even character.
The two four vectors of the same Grassmann character are
connected by the discrete time 
reversal operation $\theta^0 \rightarrow - \theta^0$ (
ref.\cite{mankocandnielsen1999}), while the
two four vectors, which 
differ in Grassmann character, are connected by the operation of
$\tilde{a}^a$.

According to Eqs.(\ref{taug}, \ref{tauga}, \ref{taugb}), one can
express the 
generators of the subgroups $SU(2)$ and $U(1)$ of the group
$SO(1,7)$ in terms of the generators ${\mathcal S}^{ab}$.

We find (since the indices $0,1,2,3$ are reserved for the
subgroup $SO(1,3)$) 

\begin{eqnarray}
\tau^{31}: = \frac{1}{2} ( {\mathcal S}^{58} - {\mathcal S}^{67} ),\quad
\tau^{32}: = \frac{1}{2} ( {\mathcal S}^{57} + {\mathcal S}^{68} ),\quad
\tau^{33}: = \frac{1}{2} ( {\mathcal S}^{56} - {\mathcal S}^{78} ).
\label{su2w}
\end{eqnarray}
One also finds
\begin{equation}
\tau^{41}: = \frac{1}{2} ( {\mathcal S}^{56} + {\mathcal S}^{78} ).
\label{u1w}
\end{equation}

The algebra of  Eq.(\ref{taug}) follows (since the
operators $\tau^{Ai}$ have an even Grassmann character, the
generalized commutation relations agree with the usual
commutators, denoted by $[\;,\;]$).
\begin{equation}
\{\tau^{3i}, \tau^{3j}\} = i \epsilon_{ijk} \tau^{3k}, \quad
\{\tau^{41}, \tau^{3i}\} = 0. 
\label{csu2u1w}
\end{equation}
One notices that $\tau^{51}: = \frac{1}{2} ( {\mathcal S}^{58} +
{\mathcal S}^{67} )$ and $\tau^{52}: = \frac{1}{2} ( {\mathcal S}^{57} -
{\mathcal S}^{68} )$ together with $ \tau^{41}$ form the algebra of
the group $SU(2)$ and that the generators of this group commute
with $\tau^{3i}$.

We present in Table II the eigenvectors of the operators
$\tilde{\tau}^{33}$ and $(\tilde{\tau}^3)^2 =
(\tilde{\tau}^{31})^2 + (\tilde{\tau}^{32})^2 
+(\tilde{\tau}^{33})^2 $, which are at the same time the
eigenvectors of 
$\tilde{\tau}^{41}$, for  spinors. We find, with respect to
the group $SU(2)$,   two doublets and four 
singlets  of an even and
another two doublets and four singlets of an odd Grassmann
character.

\begin{center}
\begin{tabular}{||c|c||c||c|c|c||}
\hline 
\hline
a & i &$<\theta|\Phi^a{ }_i>$& $\tilde{\tau}^{33}$& $
\tilde{\tau}^{41}$& Grassmann\\ 
&&&&&character\\
\hline
\hline
1 & 1 &$\frac{1}{2}(1-i\tilde{a}^5 \tilde{a}^6)(1+ i\tilde{a}^7
\tilde{a}^8)$ &$-\frac{1}{2}$ &$ 0$&\\
1 & 2 &$-\frac{1}{2}(\tilde{a}^5 + i\tilde{a}^6)(
\tilde{a}^7 - i \tilde{a}^8) $ &$\frac{1}{2} $&$ 0$&\\
2 & 1 &$\frac{1}{2}(1+i\tilde{a}^5 \tilde{a}^6)(1- i\tilde{a}^7
\tilde{a}^8)$ &$\frac{1}{2}$ &$ 0$&\\
2 & 2 &$-\frac{1}{2}(\tilde{a}^5 - i\tilde{a}^6)(
\tilde{a}^7 + i \tilde{a}^8) $ &$-\frac{1}{2} $&$ 0$&\\
&&&&&even\\
3 & 1 &$\frac{1}{2}(1+i\tilde{a}^5 \tilde{a}^6)(1+ i\tilde{a}^7
\tilde{a}^8)$ &0&$\frac{1}{2}$ &\\
4 & 1 &$\frac{1}{2}(\tilde{a}^5 + i\tilde{a}^6)(
\tilde{a}^7 + i \tilde{a}^8) $ &0&$\frac{1}{2} $&\\
5 & 1 &$\frac{1}{2}(1-i\tilde{a}^5 \tilde{a}^6)(1- i\tilde{a}^7
\tilde{a}^8)$ &0&$-\frac{1}{2}$ &\\
6 & 1 &$\frac{1}{2}(\tilde{a}^5 - i\tilde{a}^6)(
\tilde{a}^7 - i \tilde{a}^8) $ &0&$-\frac{1}{2} $&\\
\hline
\hline
7 & 1 &$\frac{1}{2}(1+i\tilde{a}^5 \tilde{a}^6)(\tilde{a}^7-i
\tilde{a}^8)$ &$\frac{1}{2}$ &$ 0$&\\
7 & 2 &$-\frac{1}{2}(\tilde{a}^5 - i\tilde{a}^6)(1+ i
\tilde{a}^7 \tilde{a}^8) $ &$-\frac{1}{2} $&$ 0$&\\
8 & 1 &$\frac{1}{2}(1-i\tilde{a}^5 \tilde{a}^6)(\tilde{a}^7+i
\tilde{a}^8)$ &$-\frac{1}{2}$ &$ 0$&\\
8 & 2 &$-\frac{1}{2}(\tilde{a}^5 + i\tilde{a}^6)(1-i
\tilde{a}^7 \tilde{a}^8) $ &$\frac{1}{2} $&$ 0$&\\
&&&&&odd\\
9 & 1 &$\frac{1}{2}(1-i\tilde{a}^5 \tilde{a}^6)(\tilde{a}^7-i
\tilde{a}^8)$ &0&$-\frac{1}{2}$ &\\
10 & 1 &$\frac{1}{2}(\tilde{a}^5 + i\tilde{a}^6)(1+ i
\tilde{a}^7 \tilde{a}^8) $ &0&$\frac{1}{2} $&\\
11 & 1 &$\frac{1}{2}(1+i\tilde{a}^5 \tilde{a}^6)(\tilde{a}^7+i
\tilde{a}^8)$ &0&$\frac{1}{2}$ &\\
12 & 1 &$\frac{1}{2}(\tilde{a}^5 - i\tilde{a}^6)(1- i
\tilde{a}^7 \tilde{a}^8) $ &0&$-\frac{1}{2} $&\\
\hline
\hline
\end{tabular}
\end{center}

\noindent
Table II: The eigenstates of the operators for spinors
$\tilde{\tau}^{33}, \tilde{\tau}^{41}$ are 
presented. We find two doublets and four singlets of an even
Grassmann character and two doublets and four singlets of an odd
Grassmann character. One sees that complex conjugation
transforms one doublet of either odd or even Grassmann character
into another of the same Grassmann character changing the sign
of  the
value of $\tilde{\tau}^{33}$, while it transforms one singlet into
another singlet
 of the same Grassmann character and of the opposite value of $
\tilde{\tau}^{41}$.  One can check that $\tilde{a}^h, \;\;
h \in (5,6,7,8)$, transforms the doublets of an even Grassmann
character into 
singlets of an odd Grassmann character.

One sees that $\tilde{\tau}^{5i},\;i = 1,2$, transform doublets into
singlets (which can easily be understood if taking into account
that $\tilde{\tau}^{5i}$ close together with $\tau^{41}$ the algebra 
of $SU(2)$ and that the two $SU(2)$ groups are isomorphic to the
group $SO(4)$).

One also sees the following very important property of 
representations of the 
group $SO(1,7) $: {\it If applying the
operators $\tilde{S}^{ab}$, 
$a,b = 0,1,2,3,5,6,7,8$ on the direct product of polynomials of
Table I and Table II, which forms the representations of the
group $SO(1,7)$, one finds that a 
multiplet of $SO(1,7)$ exists, which contains  left handed $SU(2)$ doublets
and right handed $SU(2)$ singlets.} It exists also another multiplet
which contains left handed $SU(2)$ singlets
and right handed $SU(2)$ doublets. It turns out that the operators
$\tilde{S}^{mh}$, with $m=0,1,2,3$ and $h=5,6,7,8$, although
having an even Grassmann 
character,  change the Grassmann character of that part of the
polynomials which belong to Table I and Table II,
respectively, keeping the 
Grassmann character of the products of the two types of
polynomials unchanged.  This can  be
understood if taking into account that  $\tilde{S}^{mh} =
-\frac{i}{2} \tilde{a}^m \tilde{a}^h$ and that the operator
$\tilde{a}^m$ changes the  polynomials  of an odd Grassmann character of
Table I, into an even polynomial, transforming a left handed
Weyl spinor of one family 
into a right handed Weyl spinor of another family, 
while $\tilde{a}^{h}$ changes simultaneously the $SU(2)$ 
 doublet of an even Grassmann character into a singlet of
an odd Grassmann character.

The symmetry, called the mirror symmetry, presented in this
approach, is not broken, as none of the symmetry is
broken. We only have arranged basic states to demonstrate 
 possible symmetries.

We  can  express   the
generators of  subgroups $SU(3)$ and $U(1)$ of the group
$SO(6)$ in terms of the generators ${\mathcal S}^{ab}$ (according to
Eq.(\ref{tauga})).

We find (since the indices $9,10,11,12,13,14$ are reserved for the
subgroup $SO(6)$) 
\begin{eqnarray}
\tau^{61}: = \frac{1}{2} ( {\mathcal S}^{9\;12} - {\mathcal S}^{10\;11} ),\quad
\tau^{62}: = \frac{1}{2} ( {\mathcal S}^{9\;11} + {\mathcal S}^{10\;12} ),\quad
\tau^{63}: = \frac{1}{2} ( {\mathcal S}^{9\;10} - {\mathcal S}^{11\;12} ),\quad
\nonumber
\\
\tau^{64}: = \frac{1}{2} ( {\mathcal S}^{9\;14} - {\mathcal S}^{10\;13} ),\quad
\tau^{65}: = \frac{1}{2} ( {\mathcal S}^{9\;13} + {\mathcal S}^{10\;14} ),\quad
\tau^{66}: = \frac{1}{2} ( {\mathcal S}^{11\;14} - {\mathcal S}^{12\;13}
),\quad 
\nonumber
\end{eqnarray}
\begin{eqnarray}
\tau^{67}: = \frac{1}{2} ( {\mathcal S}^{11\;13} + {\mathcal S}^{12\;14} ),\quad
\tau^{68}: = \frac{1}{2\sqrt{3}} ( {\mathcal S}^{9\;10} + {\mathcal
S}^{11\;12}  - 2{\mathcal S}^{13\;14}).\\
\label{su3c}
\end{eqnarray}

One  finds in addition 
\begin{equation}
\tau^{71}: = -\frac{1}{3}( {\mathcal S}^{9\;10} + {\mathcal S}^{11\;12}
+ {\mathcal S}^{13\;14} ).
\label{u1c}
\end{equation}

The algebra for the subgroups $SU(3)$ and $U(1)$ follows from
the algebra of the Lorentz group $SO(1,13)$
\begin{equation}
\{\tau^{6i}, \tau^{6j}\} = i f_{ijk} \tau^{6k}, \quad
\{\tau^{71}, \tau^{6i}\} = 0, {\mathrm \;\; for\;\; each \;\;i}. 
\label{csu3u1c}
\end{equation}
The coefficients $f_{ijk}$ are the structure constants of the group
$SU(3)$. 

We  can find the eigenvectors of the Casimirs of the groups
$SU(3)$ and $U(1)$  for spinors  as polynomials of
$\theta^h$, $h=9,...,14$. The eigenvectors, which are
polynomials of an 
even Grassmann character, can be found in
ref.\cite{mankoc1997}. We shall
present here only not yet 
published \cite{borstnikandmankoc1999} polynomials
of an odd Grassmann character.

\noindent
Table III: The eigenstates of the operators for spinors $\tilde{\tau}^{63},
\tilde{\tau}^{68}, \tilde{\tau}^{71}$ are presented for odd
Grassmann character polynomials. We find four triplets, four
anti-triplets and eight singlets.
 One sees that complex conjugation
transforms one triplet into anti-triplet, while
$\tilde{\tau}^{8i}$ transform triplets into anti-triplets or singlets.

\begin{center}
\begin{tabular}{||c|c||c||c|c|c||}
\hline
\hline 
a & i &$<\theta|\Phi^a{ }_i>$& $\tilde{\tau}^{63}$& $
\tilde{\tau}^{68}$& $ \tilde{\tau}^{71}$\\ 
\hline
\hline
1 & 1 & $\frac{1}{\sqrt{2^3}}(1+i\tilde{a}^{13} \tilde{a}^{14})(\tilde{a}^9
-i \tilde{a}^{10}) (1+i\tilde{a}^{11} \tilde{a}^{12}) $
& $\frac{1}{2}$ & $ \frac{1}{2\sqrt{3}} $ & $ \frac{1}{6}$ \\
1 & 2 &$\frac{1}{\sqrt{2^3}}(1+i\tilde{a}^{13}
\tilde{a}^{14})(1+i\tilde{a}^9 \tilde{a}^{10})
(\tilde{a}^{11}-i\tilde{a}^{12})$ 
&$-\frac{1}{2}$ &$ \frac{1}{2\sqrt{3}}$ & $\frac{1}{6}$ \\
1 & 3 &$-\frac{1}{\sqrt{2^3}}(\tilde{a}^{13}-i \tilde{a}^{14})(1+i\tilde{a}^9
 \tilde{a}^{10}) (1+i\tilde{a}^{11}\tilde{a}^{12})$
&$0$ &$ -\frac{1}{\sqrt{3}}$& $\frac{1}{6}$ \\
\hline
2 & 1 &$\frac{1}{\sqrt{2^3}}(1+i\tilde{a}^{13} \tilde{a}^{14})(1-i\tilde{a}^9
\tilde{a}^{10}) (\tilde{a}^{11}+i\tilde{a}^{12})$
&$\frac{1}{2}$ &$ \frac{1}{2\sqrt{3}}$ & $\frac{1}{6}$ \\
2 & 2 &$\frac{1}{\sqrt{2^3}}(1+i\tilde{a}^{13}
\tilde{a}^{14})(\tilde{a}^9+i \tilde{a}^{10})
(1-i\tilde{a}^{11}\tilde{a}^{12})$ 
&$-\frac{1}{2}$ &$ \frac{1}{2\sqrt{3}}$& $\frac{1}{6}$ \\
2 & 3 &$-\frac{1}{\sqrt{2^3}}(\tilde{a}^{13}-i \tilde{a}^{14})(\tilde{a}^9+i
 \tilde{a}^{10}) (\tilde{a}^{11}+i\tilde{a}^{12})$
&$0$ &$ -\frac{1}{\sqrt{3}}$& $\frac{1}{6}$ \\
\hline
3 & 1 &$\frac{1}{\sqrt{2^3}}(\tilde{a}^{13}+i\tilde{a}^{14})(\tilde{a}^9
-i \tilde{a}^{10}) (\tilde{a}^{11}+i\tilde{a}^{12})$
&$\frac{1}{2}$ &$ \frac{1}{2\sqrt{3}}$& $\frac{1}{6}$ \\
3 & 2 &$\frac{1}{\sqrt{2^3}}(\tilde{a}^{13}+i
\tilde{a}^{14})(1+i\tilde{a}^9 \tilde{a}^{10})
(1-i\tilde{a}^{11}\tilde{a}^{12})$ 
&$-\frac{1}{2}$ &$ \frac{1}{2\sqrt{3}}$& $\frac{1}{6}$ \\
3 & 3 &$-\frac{1}{\sqrt{2^3}}(1-i\tilde{a}^{13} \tilde{a}^{14})(1+i\tilde{a}^9
 \tilde{a}^{10}) (\tilde{a}^{11}+i\tilde{a}^{12})$
&$0$ &$ -\frac{1}{\sqrt{3}}$& $\frac{1}{6}$ \\
\hline
4 & 1 &$\frac{1}{\sqrt{2^3}}(\tilde{a}^{13}+i\tilde{a}^{14})(1-i\tilde{a}^9
\tilde{a}^{10}) (1+i\tilde{a}^{11}\tilde{a}^{12})$
&$\frac{1}{2}$ &$ \frac{1}{2\sqrt{3}}$& $\frac{1}{6}$ \\
4 & 2 &$\frac{1}{\sqrt{2^3}}(\tilde{a}^{13}+i
\tilde{a}^{14})(\tilde{a}^9+i \tilde{a}^{10})
(\tilde{a}^{11}-i\tilde{a}^{12})$ 
&$-\frac{1}{2}$ &$ \frac{1}{2\sqrt{3}}$ & $\frac{1}{6}$ \\
4 & 3 &$-\frac{1}{\sqrt{2^3}}(1-i\tilde{a}^{13} \tilde{a}^{14})(\tilde{a}^9
+i \tilde{a}^{10}) (1+i\tilde{a}^{11}\tilde{a}^{12})$
&$0$ &$ -\frac{1}{\sqrt{3}}$ & $\frac{1}{6}$ \\
\hline
\hline
5 & 1 & $\frac{1}{\sqrt{2^3}}(1-i\tilde{a}^{13} \tilde{a}^{14})(\tilde{a}^9
+i \tilde{a}^{10}) (1-i\tilde{a}^{11} \tilde{a}^{12}) $
& $-\frac{1}{2}$ & $ -\frac{1}{2\sqrt{3}} $ & $ -\frac{1}{6}$ \\
5 & 2 &$\frac{1}{\sqrt{2^3}}(1-i\tilde{a}^{13}
\tilde{a}^{14})(1-i\tilde{a}^9 \tilde{a}^{10})
(\tilde{a}^{11}+i\tilde{a}^{12})$ 
&$\frac{1}{2}$ &$- \frac{1}{2\sqrt{3}}$ & $-\frac{1}{6}$ \\
5 & 3 &$-\frac{1}{\sqrt{2^3}}(\tilde{a}^{13}+i \tilde{a}^{14})(1-i\tilde{a}^9
 \tilde{a}^{10}) (1-i\tilde{a}^{11}\tilde{a}^{12})$
&$0$ &$ \frac{1}{\sqrt{3}}$& $-\frac{1}{6}$ \\
\hline
6 & 1 &$\frac{1}{\sqrt{2^3}}(1-i\tilde{a}^{13} \tilde{a}^{14})(1+i\tilde{a}^9
\tilde{a}^{10}) (\tilde{a}^{11}-i\tilde{a}^{12})$
&$-\frac{1}{2}$ &$ -\frac{1}{2\sqrt{3}}$ & $-\frac{1}{6}$ \\
6 & 2 &$\frac{1}{\sqrt{2^3}}(1-i\tilde{a}^{13}
\tilde{a}^{14})(\tilde{a}^9-i \tilde{a}^{10})
(1+i\tilde{a}^{11}\tilde{a}^{12})$ 
&$\frac{1}{2}$ &$ -\frac{1}{2\sqrt{3}}$& $-\frac{1}{6}$ \\
6 & 3 &$-\frac{1}{\sqrt{2^3}}(\tilde{a}^{13}+i \tilde{a}^{14})(\tilde{a}^9-i
 \tilde{a}^{10}) (\tilde{a}^{11}-i\tilde{a}^{12})$
&$0$ &$ \frac{1}{\sqrt{3}}$& $-\frac{1}{6}$ \\
\hline
7 & 1 &$\frac{1}{\sqrt{2^3}}(\tilde{a}^{13}-i\tilde{a}^{14})(\tilde{a}^9
+i \tilde{a}^{10}) (\tilde{a}^{11}-i\tilde{a}^{12})$
&$-\frac{1}{2}$ &$ -\frac{1}{2\sqrt{3}}$& $-\frac{1}{6}$ \\
7 & 2 &$\frac{1}{\sqrt{2^3}}(\tilde{a}^{13}-i
\tilde{a}^{14})(1-i\tilde{a}^9 \tilde{a}^{10})
(1+i\tilde{a}^{11}\tilde{a}^{12})$ 
&$\frac{1}{2}$ &$ -\frac{1}{2\sqrt{3}}$& $-\frac{1}{6}$ \\
7 & 3 &$-\frac{1}{\sqrt{2^3}}(1+i\tilde{a}^{13} \tilde{a}^{14})(1-+i\tilde{a}^9
 \tilde{a}^{10}) (\tilde{a}^{11}-i\tilde{a}^{12})$
&$0$ &$ \frac{1}{\sqrt{3}}$& $-\frac{1}{6}$ \\
\hline
8 & 1 &$\frac{1}{\sqrt{2^3}}(\tilde{a}^{13}-i\tilde{a}^{14})(1+i\tilde{a}^9
\tilde{a}^{10}) (1-i\tilde{a}^{11}\tilde{a}^{12})$
&$-\frac{1}{2}$ &$ -\frac{1}{2\sqrt{3}}$& $-\frac{1}{6}$ \\
8 & 2 &$\frac{1}{\sqrt{2^3}}(\tilde{a}^{13}-i
\tilde{a}^{14})(\tilde{a}^9-i \tilde{a}^{10})
(\tilde{a}^{11}+i\tilde{a}^{12})$ 
&$\frac{1}{2}$ &$ -\frac{1}{2\sqrt{3}}$ & $-\frac{1}{6}$ \\
8 & 3 &$-\frac{1}{\sqrt{2^3}}(1+i\tilde{a}^{13} \tilde{a}^{14})(\tilde{a}^9
-i \tilde{a}^{10}) (1-i\tilde{a}^{11}\tilde{a}^{12})$
&$0$ &$ \frac{1}{\sqrt{3}}$ & $-\frac{1}{6}$ \\
\hline
\hline
9 & 1 &$\frac{1}{\sqrt{2^3}}(1+i\tilde{a}^{13} \tilde{a}^{14})(\tilde{a}^9
+i \tilde{a}^{10}) (1+i\tilde{a}^{11}\tilde{a}^{12})$
&$0$ &$ 0$ & $\frac{1}{2}$ \\
10 & 1 &$\frac{1}{\sqrt{2^3}}(1+i\tilde{a}^{13}
\tilde{a}^{14})(1+i \tilde{a}^9 \tilde{a}^{10})
(\tilde{a}^{11}+i\tilde{a}^{12})$ 
&$0$ &$ 0 $ & $\frac{1}{2}$ \\
11 & 1 &$\frac{1}{\sqrt{2^3}}(\tilde{a}^{13}+i \tilde{a}^{14})(1+i\tilde{a}^9
 \tilde{a}^{10}) (1+i\tilde{a}^{11}\tilde{a}^{12})$
&$0$ &$ 0$ & $\frac{1}{2}$ \\
12 & 1 &$\frac{1}{\sqrt{2^3}}(\tilde{a}^{13}+i \tilde{a}^{14})(\tilde{a}^9
+i \tilde{a}^{10}) (\tilde{a}^{11}+i\tilde{a}^{12})$
&$0$ &$ 0$ & $\frac{1}{2}$ \\
13 & 1 &$\frac{1}{\sqrt{2^3}}(1-i\tilde{a}^{13}
\tilde{a}^{14})(\tilde{a}^9-i \tilde{a}^{10})
(1-i\tilde{a}^{11}\tilde{a}^{12})$ 
&$0$ &$ 0$ & $-\frac{1}{2}$ \\
14 & 1 &$\frac{1}{\sqrt{2^3}}(1-i\tilde{a}^{13} \tilde{a}^{14})(1-i\tilde{a}^9
 \tilde{a}^{10}) (\tilde{a}^{11}-i\tilde{a}^{12})$
&$0$ &$ 0$ & $-\frac{1}{2}$ \\
15 & 1 &$\frac{1}{\sqrt{2^3}}(\tilde{a}^{13}-i
\tilde{a}^{14})(1-i\tilde{a}^9 \tilde{a}^{10})
(1-i\tilde{a}^{11}\tilde{a}^{12})$ 
&$0$ &$ 0$ & $-\frac{1}{2}$ \\
16 & 1 &$\frac{1}{\sqrt{2^3}}(\tilde{a}^{13}-i \tilde{a}^{14})(\tilde{a}^9-i
 \tilde{a}^{10}) (\tilde{a}^{11}-i\tilde{a}^{12})$
&$0$ &$ 0$ & $-\frac{1}{2}$ \\
\hline
\hline
\end{tabular}
\end{center}
 
One finds four triplets and four anti-triplets as well as eight
singlets. Besides the eigenvalues of the commuting operators
$\tilde{\tau}^{63}$ 
and $\tilde{\tau}^{68}$ of the group $SU(3)$ also the eigenvalue
of $\tilde{\tau}^{71}$ forming $U(1)$, is presented.
The operators $\tilde{\tau}^{81}: = \frac{1}{2} (
\tilde{S}^{9\;12} + \tilde{S}^{10\;11} ),\quad 
\tau^{82}: = \frac{1}{2} ( \tilde{S}^{9\;11} - \tilde{S}^{10\;12} ),\quad
\tau^{83}: = \frac{1}{2} ( \tilde{S}^{9\;14} + \tilde{S}^{10\;13} ),\quad
\tau^{84}: = \frac{1}{2} ( {\mathcal S}^{9\;13} - {\mathcal S}^{10\;14} ),\quad
\tau^{85}: = \frac{1}{2} ( \tilde{S}^{11\;14} + \tilde{S}^{12\;13}),\quad 
\tau^{86}: = \frac{1}{2} ( \tilde{S}^{11\;13} - \tilde{S}^{12\;14}),\quad 
 $ which  transform triplets of the group $SU(3)$ into
anti-triplets and singlets with respect to the group $SU(3)$.

The spinorial representations of the group $SO(1,13)$ are the 
direct product of polynomials of Table I, Table II and
Table III.

We can find all the members of a spinorial multiplet of the
group $SO(1,13)$ by 
applying $\tilde{S}^{ab}$ on any initial Grassmann odd product of 
polynomials, if one polynomial is taken from Table I, another from Table
II and the third from Table III. In the same multiplet there are
triplets, singlets and 
anti-triplets with respect to $SU(3)$, which are doublets or singlets
with respect to $SU(2)$, and are left and right handed with
respect to $SO(1,3)$. 

We can arrange in the same sense also eigenstates of operators
of vectorial character, with bosonic character. In this paper we
shall not do that.

\subsection{Dynamical arrangement of representations of $SO(1,13)$
with respect to subgroups $SO(1,7)$ and $SO(6)$}
 \label{dar}

To see  how  Yang-Mills fields enter into the theory,
we shall rewrite the Weyl-like equation in the presence of the
gravitational field (\ref{cong}) in terms of
components of fields which determine  
gravitation in the four dimensional subspace and of those 
which determine  gravitation in higher dimensions, assuming
that the coordinates of ordinary space with indices higher than
four stay compacted to unmeasurable small dimensions (or can not
at all be noticed for some other reason). 
Since  Grassmann space only manifests itself through  average
values of observables, compactification of a part of 
Grassmann space has no meaning.  However, since
parameters of  the Lorentz transformations in a freely falling
coordinate system for both spaces have
to be the same, no transformations to the fifth or
higher coordinates may occur 
at measurable energies. Therefore, at low energies, the four
dimensional subspace 
of  Grassmann space with the generators defining the Lorentz
group $ SO(1,3)$ is (almost) decomposed from the rest of the 
Grassmann space with the generators forming the (compact) group
$ SO(d-4) $, because of the decomposition of  ordinary
space. This is valid on the classical level only.

According to the previous subsection, the breaking of symmetry
of $SO(1,13)$
should, however, appears in steps,  first  through
$SO(1,7)\times SO(6)$ and later to the final symmetry, which is
needed in the Standard Model for massless particles.

We shall comment on possible ways of spontaneously broken
symmetries by studying the Weyl equation in the presence of
gravitational fields in d dimensions for massless particles
(Eqs.(\ref{cmg}, \ref{cong}))
\begin{equation}
\tilde{\gamma}^{a} p_{0a} = 0, \quad p_{0a} = f^{\mu}{ }_a 
p_{0\mu}, \quad p_{0\mu} = p_{\mu} - \frac{1}{2}
\tilde{S}^{ab}\omega_{ab\mu}.  
\label{dgebb}
\end{equation}
\subsubsection{Standard Model case}
 \label{smodc}

To make discussions more transparent we shall first comment on
 the well known case  of the Standard model. Before the breaking of the
symmetry $SU(3) \times SU(2) \times U(1)$ into $SU(3) \times
U(1)$, the canonical momentum $p_{0\alpha}$ ($\; \alpha = 
0,1,2,3$ and $d=4$)  includes the gauge fields, connected with the
groups $SU(3)$, $SU(2)$ and $U(1)$. We shall pay attention
on only 
the groups $SU(2)$ and $U(1)$, which are involved in the breaking
of symmetry 
\begin{equation}
p_{0\alpha} = p_{\alpha} - g \tau^{i}A^{i}{ }_{\alpha} - g'Y B_{\alpha},  
\label{smp}
\end{equation} 
where $g$ and $g'$ are the two coupling constants.
Introducing $\tau^{\pm} = \tau^1 \pm i \tau^2$, the
superposition follows $ A^{\pm}{
}_{\alpha} = A^1{ }_{\alpha} \mp i \; A^2{ }_{\alpha} $. If defining $ A^3{
}_{\alpha} = \frac{g/g'}{\sqrt{1 + (g/g')^2}} Z_{\alpha} + \frac{1}{\sqrt{1
+ (g/g')^2}} A_{\alpha}$ and $B_{\alpha} = -\frac{1}{\sqrt{1 +
(g/g')^2}} Z_{\alpha}  +
\frac{g/g'}{\sqrt{1 + (g/g')^2}} A_{\alpha}$, so that the
transformation is orthonormalized, one can easily rewrite 
Eq.(\ref{smp}) as follows

\begin{equation}
p_{0\alpha} = p_{\alpha} - \frac{g}{2} (\tau^{+}A^{+}{
}_{\alpha} + \tau^{-}A^{-}{ }_{\alpha}) + \frac{gg'}{\sqrt{g^2 +
g'^2}} Q A_{\alpha} + \frac{g^2}{\sqrt{g^2 + g'^2}} Q' Z_{\alpha}. 
\label{smpt}
\end{equation}
with
\begin{equation}
Q = \tau^3 + Y, \quad Q' = \tau^3 - (\frac{g'}{g})^2 Y.
\label{qq'}
\end{equation} 
 
In the Standard Model $<Q>$ is the conserved quantity and $<Q'>$ is
not, since $<Q>$ is zero for the Higgs
fields in the ground state, while $<Q'>$ is nonzero ( 
$<Q'> = - \frac{1}{2} ( 1 + (\frac{g'}{g})^2)$). 

  If no symmetry is spontaneously  broken, that is if no
Higgs breaks symmetry by making a choice for his ground state
symmetry, the only thing which has been done by introducing
linear superpositions of fields, is the 
rearrangement of fields, which  always can be done without any
consequence, except that it may help to better see the
symmetries. 

Spontaneously breaking of  symmetries causes the non-conservation of
quantum numbers, as well as massive clusters of fields.

\subsubsection{Spin connections and gauge fields leading to the
Standard Model}
 \label{scgf}

We shall rewrite the canonical momentum of Eq.(\ref{dgebb})
to manifest possible ways of breaking symmetries of $SO(1,13)$
down to the symmetries of the Standard model. We first write 

\begin{equation}
\tilde{\gamma}^{a} p_{0a} = 0 = \tilde{\gamma}^{a} f^{\mu}{ }_a
p_{0\mu} = (\tilde{\gamma}^{m} f^{\alpha}{ }_m  +
\tilde{\gamma}^{h} f^{\alpha}{ }_h )
p_{0\alpha}  + (\tilde{\gamma}^{m} f^{\sigma}{ }_m  +
\tilde{\gamma}^{h} f^{\sigma}{ }_h  ) p_{0\sigma},
\label{dgex}
\end{equation}
with $\alpha, m \in \{0,1,2,3 \} $ and $\sigma, h \in \{5,...,14
\} $ to separate the $d=4$ dimensional subspace out of $d = 14$
dimensional space. We may further rearrange the canonical
momentum $p_{0\mu}$
\begin{equation}
p_{0\mu} = p_{\mu} - \frac{1}{2} \tilde{S}^{h_1 h_2}\;
\omega_{h_1 h_2 \mu}- \frac{1}{2} \tilde{S}^{k_1 k_2}\;
\omega_{k_1 k_2 \mu} - \frac{1}{2} \tilde{S}^{h_1 k_1}\;
\omega_{h_1 k_1 \mu},
\label{p0a}
\end{equation}
with $h_i \in \{0,1,..,8 \}$ and $k_i \in \{9,...,14 \}$
so that $\tilde{S}^{h_1h_2}$ define the algebra of the subgroup
$SO(1,7)$, while $\tilde{S}^{h_1h_2}$ define the algebra of the
subgroup $SO(6)$. The generators $\tilde{S}^{h_1k_1}$ rotate 
states of a multiplet of the group $SO(1,13)$ into each other.  

Taking into account subsection \ref{ac}  we may
rewrite the generators $\tilde{S}^{ab}$ in terms of the
corresponding generators of subgroups $\tilde{\tau}^{Ai}$ and
accordingly, similarly to the Standard Model case, introduce
new fields (see subsection \ref{smodc}), which are 
superpositions of the old ones
\begin{eqnarray}
g \;A^{31}{ }_{\mu} = \frac{1}{2} (\omega_{58\mu} - \omega_{67\mu}), \quad
g \;A^{32}{ }_{\mu} &=& \frac{1}{2} (\omega_{57\mu} + \omega_{68\mu}), \quad\nonumber\\
g \;A^{33}{ }_{\mu} &=& \frac{1}{2} (\omega_{56\mu} - \omega_{78\mu}), \quad\quad\quad
\label{su2a}
\\ 
g \;A^{41}{ }_{\mu} = \frac{1}{2} (\omega_{56\mu} + \omega_{78\mu}), \quad\quad\quad\quad
\nonumber
\\
g \;A^{51}{ }_{\mu} = \frac{1}{2} (\omega_{58\mu} + \omega_{67\mu}), \quad
g \;A^{52}{ }_{\mu} &=& \frac{1}{2} (\omega_{57\mu} - \omega_{68\mu}). \quad
\label{su2u1af}
\end{eqnarray} 
It follows then 
\begin{equation}
\frac{1}{2} \tilde{S}^{h_1h_2}\; \omega_{h_1 h_2 \mu} =
g \; \tilde{\tau}^{Ai} A^{Ai}{ }_{\mu},
\label{su2u1a}
\end{equation}
where for $A=3$,  $i = 1,2,3$, for $A=4$,  $i = 1$ and for $A=5$
$i=1,2$. Accordingly, the fields $A^{Ai}_{}\mu$ are the gauge
fields of the group $SU(2)$, if $A=3$ and of $U(1)$ if $A=4$. 
Since $\tilde{\tau}^{41}$ and $\tilde{\tau}^{5i}$ form the group
$SU(2)$ as well, the corresponding fields could be the gauge
fields of this group. The breaking of symmetry should make a choice
between  the gauge groups $U(1)$ and $SU(2)$.

We leave the notation for spin connection fields in the case that
$h_i \in \{ 0,1,2,3\}$ unchanged. We also leave unchanged the 
spin connection fields for the case, that $h_1 = 0,1,2,3$ and
$h_2 = 5,6,7,8$ as well as for the case, that $h_1 \in
\{0,1...,8 \}$ and $k_1 \in \{9,..,14 \}$, while we  arrange 
terms with $k_i \in \{8,...,14\} $ to demonstrate the symmetry
$SU(3)$ and $U(1)$
\begin{eqnarray}
g \;A^{61}{ }_{\mu} &=& \frac{1}{2} (\omega_{9\;12\mu} - \omega_{10\;11\mu}), \quad
g \;A^{62}{ }_{\mu} = \frac{1}{2} (\omega_{9\;11\mu} + \omega_{10\;12\mu}), \quad 
\nonumber\\
g \;A^{63}{ }_{\mu} &=& \frac{1}{2} (\omega_{9\;10\mu} - \omega_{11\;12\mu}), \quad 
g \;A^{64}{ }_{\mu} = \frac{1}{2} (\omega_{9\;14\mu} - \omega_{10\;13\mu}), \quad
\nonumber\\
g \;A^{65}{ }_{\mu} &=& \frac{1}{2} (\omega_{9\;13\mu} + \omega_{10\;14\mu}), \quad
g \;A^{66}{ }_{\mu} = \frac{1}{2} (\omega_{11\;14\mu} - \omega_{12\;13\mu}), \quad 
\label{su3a} 
\\
g \;A^{67}{ }_{\mu} &=& \frac{1}{2} (\omega_{11\;13\mu} + \omega_{12\;14\mu}), \quad\nonumber\\
g \;A^{68}{ }_{\mu} &=& \frac{1}{2\sqrt{3}} (\omega_{9\;10\mu} + \omega_{11\;12\mu} - 
2  \omega_{13\;14\mu}),\quad\nonumber
\\
\sqrt{\frac{2}{3}}g \;A^{71}{ }_{\mu} &=& - \sqrt{\frac{2}{3}}\frac{1}{2} (\omega_{9\;10\mu} + 
\omega_{11\;12\mu} + 
\omega_{13\;14\mu})= g\sqrt{\frac{2}{3}} A'^{71}_{\mu} = g' A'^{71}_{\mu}.\quad
\label{su3u1a}
\end{eqnarray}
We may accordingly define fields $
g \;A^{81}{ }_{\mu} = \frac{1}{2} (\omega_{9\;12\mu} + \omega_{10\;11\mu}), \quad 
g \;A^{82}{ }_{\mu} = \frac{1}{2} (\omega_{9\;11\mu} -
\omega_{10\;12\mu}), \quad $ $
g \;A^{83}{ }_{\mu} = \frac{1}{2} (\omega_{9\;14\mu} + \omega_{10\;13\mu}), \quad 
g \;A^{84}{ }_{\mu} = \frac{1}{2} (\omega_{9\;13\mu} - \omega_{10\;14\mu}), \quad
g \;A^{85}{ }_{\mu} = \frac{1}{2} (\omega_{11\;14\mu} +
\omega_{12\;13\mu}), \quad $ $
g \;A^{86}{ }_{\mu} = \frac{1}{2} (\omega_{11\;13\mu} - \omega_{12\;14\mu}) \quad 
 $, so that it follows
\begin{equation}
\frac{1}{2} \tilde{S}^{k_1k_2}\; \omega_{k_1 k_2 \mu} =
g' \; \tilde{\tau}^{Ai} A'^{Ai}{ }_{\mu},
\label{su3u1gf}
\end{equation}
with $ A = 6,7,8$ and all $A'^{Ai}_{\mu}= A^{Ai}_{\mu} $, except for $A=7, i=1$, which 
is defined in Eq.(\ref{su3u1a}). 
While $A^{6i}{}_{\mu},\; i \in \{1,..,8 \}$, form the
gauge field of the group $SU(3)$ and $A^{71}{ }_{\mu}$
corresponds to the gauge group $U(1)$, terms $g
\tilde{\tau}^{7i}\; A^{7i}{ }_{\mu}$ transform $SU(3)$ triplets 
into singlets and anti-triplets. Again, without additional
requirements, all the coupling constants  $g$ are equal.
To be in agreement with what the Standard model needs as an
input, we further  rearrange the gauge fields belonging to
the two $U(1)$ fields, one coming from the subgroup $SO(1,7)$
the other from the subgroup $SO(6)$.
We therefore define
\begin{equation}
Y_1 =  \tau^{41} + \tau^{71},\quad 
Y_2 = - \tau^{41} + \tau^{71}\quad 
\label{y1y2}
\end{equation}
and accordingly similarly to the Standard Model case of
subsection \ref{smodc} we make the corresponding  superpositions of the fields
$A^{41}{ }_{\mu}$ and $
A'^{71}{ }_{\mu})$.

The rearrangement of fields demonstrates all the symmetries of the
massless particles of the Standard Model and more. For further comments on the coupling constants
of the fields before and after the break of symmetries see ref.\cite{mankocandnielsen2002}

Taking into account Tables I, II and III one finds for the
quantum numbers of spinors, which belong to a multiplet of
$SO(1,7)$ with left handed $SU(2)$ doublets and right handed
$SU(2)$ singlets and which are triplets or singlets with respect
to $SU(3)$, the ones, presented on Table IV. We use the
names of the Standard model to denote triplets and singlets with
respect to $SU(3)$ and $SU(2)$.

\noindent
Table IV: Expectation values for the generators $\tilde{\tau}^{63}$
and $\tilde{\tau}^{68}$ of the group $SU(3)$ and the generator
$\tilde{\tau}^{71}$ of the group $U(1)$, the two groups are
 subgroups of the group $SO(6)$, and of the generators 
$\tilde{\tau}^{33}$ of the group $SU(2)$ , $\tilde{\tau}^{41}$
of the group $U(1)$ and $\tilde{\Gamma}^{(4)}$ of the group
$SO(1,3)$, the three groups are subgroups of the group $SO(1,7)$,
for the multiplet (with respect to $SO(1,7)$), which contains
left handed ($<\Gamma^{(4)}> = -1$) $SU(2)$ doublets and right
handed ($<\Gamma^{(4)}> = 1$) $SU(2)$ singlets. In addition, 
values for $\tilde{Y}_1$ and $\tilde{Y}_2$ are also presented.
Index $i$ of $u_i, d_i, \nu_i $ and $e_i$ runs over four
families presented in Table I.
\begin{center}
\begin{tabular}{||c|c|cccccc|cccccc||}
\hline
\hline
  & & 
   \multicolumn{6}{c|}{SU(2) doublets} & 
   \multicolumn{6}{c|}{SU(2) singlets}\\ 
  & & $\tilde{\tau}^{33}$& $
\tilde{\tau}^{41} $& 
$\tilde{\tau}^{71} $&$\tilde{Y}_1$&$\tilde{Y}_2$ &
$\tilde{\Gamma}^{(4)}$ & $\tilde{\tau}^{33}$& 
$\tilde{\tau}^{41}$ & $\tilde{\tau}^{71}$ & $\tilde{Y}_1 $& $\tilde{Y}_2$&
$\tilde{\Gamma}^{(4)}$ \\ \hline
SU(3) triplets & & &&&&&&&&&&& \\
$\tilde{\tau}^{6\;3}\;\;$  =  & 
$u_i$& 
1/2 & 0 & 1/6 & 1/6 & 1/6 & - 1 & 0 & 1/2 & 1/6 & 2/3 & -1/3 & 1
\\ 
 ( $ \frac{1}{2},\;\;$  $ -\frac{1}{2},\;\;$
 $ 0\;$  )&&&&&&&&&&&&&\\
$\tilde{\tau}^{6\;8} $  = & 
$d_i$ & -1/2 & 0 & 1/6 & 1/6 & 1/6 & -1 & 0 & -1/2 & 1/6 & -1/3
& 2/3 & 1 \\
 ( $\frac{1}{2 \sqrt{3}},$
$\frac{1}{2 \sqrt{3}},$  $-\frac{1}{ \sqrt{3}}$  )&&&&&&&&&&&&&\\  \hline 
SU(3) singlets & & &&&&&&&&&&& \\
$\tilde{\tau}^{6\; 3} =  0$ & $\nu_i$ & 1/2 & 0 & -1/2 & 
-1/2 & -1/2 & -1 & 0 & 1/2 & -1/2 & 0 & -1 & 1\\
$\tilde{\tau}^{6\; 8} = 0$ & 
$e_i$ & -1/2 & 0 & -1/2 & 
-1/2 & -1/2 & -1 & 0 & -1/2 & -1/2 & -1 & 0 & 1\\ \hline
\hline
\end{tabular}
\end{center}

We see that, besides $\tilde{Y}_2$, these are just the quantum
numbers needed for massless fermions of the Standard Model. The
value for the additional hyper charge $\tilde{Y}_2$ is nonzero
for the right 
handed neutrinos, as well as for other states, except right
handed electrons.

Since no symmetry is broken yet, all the gauge fields are of the
same strength. To come to the symmetries of massless fields of
the Standard Model,  surplus symmetries
should be broken so that all the fields fields $\omega_{ab\mu}$ which do not determine the fields
$A^{Ai} { }_{\mu}$, $A=3,6$ (Eqs.(\ref{su2a},\ref{su3a})) and $A^{41}_{\mu}$ and
$A^{71}_{\mu}$  should be invisible at low energies. 

The mirror symmetry should also be broken so that multiplets
of $SO(1,7)$ with right handed $SU(2)$ doublets and left handed
$SU(2)$ singlets become very massive.  All the surplus multiplets,
either bosonic or fermionic should become of large enough masses
not to be measurable at low energies.
 
The proposed approach predicts four rather than three families
of fermions.

Although in this paper, we do not discuss possible ways of
appearance of spontaneously broken 
symmetries, bringing the symmetries of the group $SO(1,13)$
down to symmetries of the Standard model (for these discussions the reader  should 
look at refs. \cite{mankocandnielsen2002,borstnikandmankocandnielsen2002}, 
which will also appear at the proceedings), we still would like to
know, whether there are terms in the Weyl equation
(Eq.\ref{dgex}) which may behave like the Yukawa couplings. We
see that indeed the term $\tilde{\gamma}^{h}f^{\sigma}{ }_h p_{0
\sigma}$, with $h \in \{ 5,6,7,8 \}$ and $\sigma \in \{5,6,..
\}$ really may, if operating on a right handed $SU(2)$ singlet
transform it to a left handed $SU(2)$ doublet. We also can find
among scalars the terms with quantum numbers of Higgs bosons
(which are $SU(2)$ doublets with respect to operators of the
vectorial character.)
All this is in preparation and not yet finished or fully understood.

\section{CONCLUDING REMARKS}
 \label{cr}

In this paper, we demonstrated that if assuming that the space has $d$
commuting and $d$ anti-commuting coordinates, then, for $d\ge 14$, all
spins in $d$ dimensions, described in the vector space spanned
over the space of anti-commuting
coordinates, demonstrate in four dimensional subspace
as the spins and all the charges, unifying spins and charges of fermions
and bosons independently, although the super-symmetry, which
guarantees the same number of fermions and bosons, is a manifesting
symmetry.  The anti-commuting coordinates can be represented by
either Grassmann coordinates or by the K\" ahler differential
forms. 

We demonstrated that either our approach or the approach of
differential forms suggest four families of quarks and leptons,
rather than three.
We have shown that starting (in any of the two approaches) with
the Lorentz symmetry in the tangent space in $d\ge 14$, spins
degrees of freedom ( described by dynamics in the space of
anti-commuting coordinates) manifests in four dimensional subspace
as spins and color, weak and hyper charges, with one additional
hyper charge, in a way that only left handed weak charge
doublets together with right handed weak charge singlets appear,
if the symmetry is spontaneously broken from $SO(1,13)$ first to
$SO(1,7)$ and $SO(6)$, so that a multiplet of $SO(1,7)$ with
only left handed $SU(2)$ doublets and right handed $SU(2)$
singlets survive, while the mirror symmetry is broken, and then
to $SO(1,3), SU(2), SU(3) $ and $U(1).$

We have demonstrated that the gravity in d dimensions manifests
as ordinary gravity and all gauge fields in four-dimensional
subspace, after the breaking of symmetry and the accordingly
changed coupling constant.
We also have shown that there are terms in the Weyl equations,
which in four-dimensional subspace manifest as Yukawa couplings.

The two approaches, the K\" ahler one after the generalization,
which we have been suggested, and our, lead to the same results.

A lot of work and ideas are still needed to show that the approach, although 
a very promising one, is showing the right way behind the Standard Model.

\section*{Acknowledgements}   
The author would like to
acknowledge the  work, done together with Holger Bech
Nielsen, which is the generalization of the  approach-
proposed by the author- to
the K\" ahler differential forms as well as very fruitful
discussions, and  the work, done together
with Anamarija Bor\v stnik, which is the breaking of the SO(1,13)
symmetry.


\title*{Semitopological Q-Rings%
\thanks{Contributed to the 4th Workshop "What Comes %
Beyond The Standard Model", Bled,%
Slovenia, July 17-27 2001 in honor of Holger Bech Nielsen's 60th%
Birthday.} }
\author{M. Axenides\thanks{e-mail:axenides@mail.demokritos.gr}
}\institute{%
Institute of Nuclear Physics,
  N.C.S.R. Demokritos,  15310, Athens, Greece} 

\authorrunning{M. Axenides}
\titlerunning{Semitopological Q-Rings}

\maketitle

\begin{abstract}
Semitopological Vortices (Q-Rings) are identified to be classical 
soliton configurations whose stability is attributed to both 
topological and nontopological charges. We discuss some recent 
work on the simplest possible realization of such a configuration 
in a scalar field theory with an unbroken $U(1)$ global symmetry. 
We show that Q-Rings correspond to local minima of the energy, 
exhibit numerical solutions of their field configurations and 
derive virial theorems demonstrating their stability.  
\end{abstract}


As we celebrate the 60th birthday of Holger Bech Nielsen we can 
without doubt assess his contributions to the development of the 
theory of strings and vortices to  bear the strongest possible 
impact. Indeed  his early work on the development of multiparticle 
dual models\cite{KN1,KN2,KN3} was soon after followed by the 
introduction of the string picture in the study of strong 
interaction physics \cite{HBN}. At the time it improved 
tremedously our physical understanding of dual models\cite{PF}. 
The string concept, of course, was bound to become much more 
useful in the unification of particle interactions with gravity. 
Aside from Holger's contribution to the development of the "string 
idea" he much later provided the first covariant formulation of a 
vortex in a theory with spontaneously broken abelian gauge 
symmetry \cite{NO}. The stability of such a gauged vortex is due 
the presence of a topological charge. The cosmic role of such 
topological defects in the phase transitions of the early universe 
has been important. It is in the spirit of this line work of 
Holger's that we will present a novel class of vortex like 
configurations that share some of the properties of topological 
solitons as well as those that are non-topological in 
character.Hence their identification as semitopological. The work 
was done in collaboration with E.G.Floratos,S.Komineas and 
L.Perivolaropoulos\cite{AFKP}  

 Non-topological solitons (Q balls)  are localized time dependent field 
configurations with a rotating internal phase and their stability 
is due to the conservation of a Noether charge $Q$\cite{c85}. They 
have been studied extensively in the literature in one, two and 
three dimensions\cite{lp92}. In three dimensions, the only 
localized, stable configurations of this type have been assumed to 
be of spherical symmetry hence the name Q balls. The 
generalization of two dimensional (planar) Q balls to three 
dimensional Q strings leads to loops which are unstable due to 
tension. Closed strings of this type are naturally produced during 
the collisions of spherical Q balls and have been seen to be 
unstable towards collapse due to their tension\cite{bs00,lpwww}. 

There is a simple mechanism however that can stabilize these 
closed loops. It is based on the introduction of an additional 
phase on the scalar field that twists by $2\pi N$ as the length of 
the loop is scanned. This phase introduces additional pressure 
terms in the energy that can balance the tension and lead to a 
stabilized configuration, the {\it Q ring}. This type of pressure 
is analogous to the pressure of the superconducting string 
loops\cite{Witten:1985eb} (also called `springs'\cite{hht88}). In 
fact it will be shown that Q rings carry both Noether charge and 
Noether current and in that sense they are also superconducting. 
However they also differ in many ways from superconducting 
strings. Q rings do not carry two topological invariants like 
superconducting strings but only one: the winding $N$ of the phase 
along the Q ring. Their metastability is due not only to the 
topological twist conservation but also  due to the conservation 
of the Noether charge as in the case of ordinary Q balls. Due to 
this combination of topological with non-topological invariants Q 
rings may be viewed as semitopological defects. In what follows we 
demonstrate the existence and metastability of Q rings in the 
context of a simple model. We use the term 'metastability' instead 
of `stability' because {\it finite size} fluctuations can lead to 
violation of cylindrical symmetry and decay of a Q ring to a Q 
ball as demonstrated by our numerical simulations. 

Consider a complex scalar field whose dynamics is determined by 
the Lagrangian 
\begin{equation} \label{model} {\cal L}={1\over 2} 
\partial_\mu \Phi^* 
\partial^\mu \Phi - U(|\Phi |) 
\end{equation}
The model has a global $U(1)$ symmetry and the associated 
conserved Noether current is 
\begin{equation} \label{current} J_\mu = Im(\Phi^* 
\partial_\mu \Phi) 
\end{equation} 
with conserved Noether charge $ Q=\int d^3 x \; J_0 $.  
Provided that the potential of (\ref{model}) 
satisfies certain conditions \cite{c85,lp92} the model accepts 
stable Q ball solutions which are described by the ansatz 
$ \Phi = f(r) e^{i \omega t}$.  The energy density of this Q ball 
configuration is localized and spherically symmetric. The 
stability is due to the conserved charge $Q$. 

In addition to the $Q$ ball there are other similar stable 
configurations with cylindrical or planar symmetry but infinite, 
not localized energy in three dimensions. For example an infinite 
stable Q string that extends along the z axis is described by the 
ansatz 
\begin{equation} \label{stranz} 
\Phi = f(\rho) e^{i \omega t}
\end{equation} 
where $\rho$ is the azimouthal radius.  This configuration has also been 
called `planar' or 'two dimensional' Q ball\cite{lp92}. 

The energy of this configuration can be made finite and localized 
in three dimensions by considering closed Q strings. These 
configurations which have been shown to be produced during 
spherical Q ball collisions\cite{bs00,lpwww} are unstable towards 
collapse due to their tension. In order to stabilize them we need 
a pressure term that will balance the effects of tension. This 
term appears if we substitute the string ansatz (\ref{stranz}) by 
the ansatz of the form 
\begin{equation} \label{qringanz}
\Phi = f(\rho) e^{i 
\omega t} e^{i \alpha(z)} 
\end{equation} 
where $\alpha(z)$ is a phase that 
varies uniformly along the z axis. This phase introduces a new 
non-zero $J_z$ component to the conserved current density 
(\ref{current}). The corresponding current is of the form 
\begin{equation} \label{jzcons} 
I_z=  \int d z \;{{d \alpha}\over {d z}} \; 
2\pi\int d\rho \; \rho \; f^2 
\end{equation}  
Consider now closing the 
infinite Q string ansatz (\ref{qringanz}) to a finite (but large) 
loop of size $L$. The energy of this configuration may be 
approximated by 
\begin{eqnarray*} 
E &=& {{Q^2}\over {4 \pi L \int d \rho \; 
\rho \; f^2}} + \pi \; L \; \int d \rho \; \rho \; f'^2  
\\ &+& {{(2\pi N)^2 \pi}\over L}\int d \rho \; \rho \; f^2 + 2 \pi  L \int d 
\rho \; \rho U(f) \\ &\equiv & I_1 + I_2 + I_3 + I_4 
\end{eqnarray*} 
where we have assumed $\alpha (z) = {{2 \pi N}\over L} z $ and the terms 
$I_i$ are all positive. Also $Q$ is the charge conserved in $3D$ 
defined as 
\begin{equation}
Q=\omega 2\pi L\int d\rho \; \rho \;f^2 
\end{equation} 
The winding $2 \pi N=\int d z \;{{d \alpha}\over {d z}}$ is 
topologically conserved 
and therefore the current (\ref{jzcons}) is very similar to the 
current of superconducting strings.   

After a rescaling 
$ \rho \longrightarrow \sqrt{\lambda_1} \rho$, 
$ z\longrightarrow \lambda_2 z $ the rescaled energy may be written 
as 
\begin{equation} 
E={1 \over {\lambda_1 \lambda_2}} I_1 + \lambda_2 I_2 
+{\lambda_1 \over \lambda_2} I_3 + \lambda_1 \lambda_2 I_4 
\end{equation} 
This configuration can be metastable towards collapse since 
Derrick's theo\-rem\cite{d64} is evaded due to the time 
dependence\cite{k97,akpf00} of the configuration (\ref{qringanz}). 
Demanding metastability towards collapse in any direction we 
obtain the virial conditions  
\begin{eqnarray} I_3 + I_4 &=& I_1 
\label{virial1} 
\\ I_2 + I_4 &=& I_1 + I_3 \label{virial2} 
\end{eqnarray} 

In order to check the validity of these conditions numerically we 
must first solve the ode which $f$ obeys. This is of the form 
\begin{equation}
f'' +{1\over \rho} f' + (\omega^2 -(2\pi N)^2/L^2) f -U'(f) = 0 
\label{fode} \end{equation} 
with boundary conditions $ f(\infty)=0$ and  ${df\over d\rho}(0)= 0$. 
Equation (\ref{fode}) is identical with the corresponding 
equation for 2D Qballs\cite{akpf00} (see ansatz (\ref{stranz})) 
with the replacement of $\omega^2$ by 
\begin{equation} 
\omega^2 -{{(2\pi 
N)^2}\over {L^2}} \equiv \omega'^2
\end{equation} 
Solutions of (\ref{fode}) 
for various $\omega'$ and 
$U(f) = {1 \over 2}f^2 - {1 \over 3} f^3 + {B \over 4}f^4$ 
with $B=4/9$ were obtained in Ref. 
\cite{akpf00}. Now it is easy to see that the first virial 
condition (\ref{virial1}) may be written as 
\begin{equation}
\omega'^2 \int d\rho \; \rho \; f^2 = 2 \int d\rho \; \rho U(f) 
\label{virnew} 
\end{equation} 
This is exactly the virial theorem for 2D 
Qballs (infinite Q strings) with $N=0$ and field ansatz given by 
(\ref{stranz}) with $\omega$ replaced by $\omega'$. The validity 
of this virial condition has been verified in Ref. \cite{akpf00}. 
This therefore is an effective verification of our first virial 
condition (\ref{virial1}). 

The second virial condition (\ref{virial2}) can be written (using 
the first virial (\ref{virial1})) as
\begin{equation}
2 I_3 = I_2
\end{equation} 
which implies that
\begin{equation}
{{2\pi N^2} \over L^2} = {{\int d\rho \; \rho f'^2}\over {\int 
d\rho \; \rho f^2}} \label{vir2} 
\end{equation} 
This can be viewed as a 
relation determining the value of $L$ required for balancing the 
tension ie for metastability.   

These virial conditions can be used to lead to a determination of 
the energy  as 
\begin{equation} 
E=2(I_1+I_3)  
\end{equation} 
In the thin wall limit where 
$2 \pi \int d \rho \rho f^2 = A f_0^2$ ($A$ is the surface of a 
cross section of the Q ring) this may be written as 
\begin{equation} 
E \simeq 
{Q^2 \over {2 L A f_0^2}} +{{(2\pi N)^2 A f_0^2}\over {2 L}} 
\label{etwa} 
\end{equation}  
and can be minimized with respect to $f_0^2$. 
The value of $f_0$ that minimizes the energy in the thin wall 
approximation is 
\begin{equation}
f_0=\sqrt{Q\over {2\pi N A}} 
\end{equation} 
Substituting this value back on 
the expression (\ref{etwa}) for the energy we obtain 
\begin{equation} 
E={{2 \pi N Q}\over L} 
\end{equation} 
This is consistent with the corresponding 
relation for spherical Q balls which in the thin wall 
approximation lead to a linear increase of the energy with $Q$. 

The above virial conditions demonstrate the persistance of the Q 
ring configuration towards shrinking or expansion in the two 
periodic directions of the Q ring torus for large radius. In order 
to study the Q rings of any size and its stability properties 
towards any type of fluctuation we must  study the full evolution 
of a Q ring in 3D by performing energy minimization and  numerical 
simulation of  evolution. This is precisely what we did for a 
potential energy given by : 
 \begin{equation} 
\label{potential} U(\phi) = {1 \over 2}|\Phi|^2 - {1 \over 3} 
|\Phi|^3 + {B \over 4}|\Phi|^4 
\end{equation} 
The ansatz we used that 
captures the above mentioned properties of the Q ring is 
\begin{equation} 
\label{eq:ansatz} 
\Phi = f(\rho,z)\; e^{i [\omega t + n\phi]} 
\end{equation} 
where the center of the coordinate system now is in the center of 
the torus that describes the Q ring and the ansatz is valid for 
{\it any} radius of the Q ring. We have also replaced $N$ by $n$. 

The energy of this configuration is 
\begin{eqnarray}  
E  &=& {1 \over 2} {Q^2 
\over \int f^2 dV} 
  + {1 \over 2} \int \left[ \left({\partial f \over \partial \rho} \right)^2
       + {n^2 f^2 \over \rho^2} \right]\;dV \nonumber \\
  &+& {1 \over 2} \int 
       \left[ \left({\partial f \over \partial z} \right)^2 \right]\;dV 
       + \int U(f)\,dV  \label{eq:energy2}
\end{eqnarray} 
The field equation for $\Phi$ is 
\begin{equation}
\label{eq:equation} 
\ddot{\Phi} - \Delta \Phi + \Phi - |\Phi| \Phi 
+ B |\Phi|^2 \Phi = 0 
\end{equation} 
Substituting the ansatz (\ref{eq:ansatz}) we find that $f(\rho,z)$ 
should satisfy 
\begin{equation}
\label{eq:ansatzequation} 
 {\partial^2 f \over \partial \rho^2} 
  + {1 \over \rho}\,{\partial f \over \partial \rho}
  - {n^2 f \over \rho^2} + {\partial^2 f \over \partial z^2}
  + (\omega^2-1) f + f^2 - B f^3 = 0
\end{equation} 
In order to solve this equation we minimized the energy 
(\ref{eq:energy2}) at fixed $Q$ using the algorithm 
\begin{eqnarray} {\partial 
f \over \partial \tau} &=& - {\delta E \over \delta f} \Rightarrow  
\label{eq:dissipative}\\ {\partial f \over 
\partial \tau} &=& {\partial^2 f \over \partial \rho^2} 
  + {1 \over \rho}\,{\partial f \over \partial \rho} 
  - {n^2 f \over \rho^2} + {\partial^2 f \over \partial z^2}
  \nonumber \\ &+& (\omega^2-1) f  + f^2 - B f^3 \label{fevol}
\end{eqnarray} 
with boundary conditions 
$f(0,z)=0$, ${{\partial f(\rho,z)}\over  {\partial z}}|_{z=0}=0$. 
The validity of the 
algorithm is checked by  
\begin{equation} 
dG/d\tau= \delta 
E/\delta f\; df/d\tau = - (dE/d\tau)^2<0 
\end{equation} 
In (\ref{eq:dissipative}) $\omega$ is defined as 
\begin{equation} \label{omegdef}  
\omega={Q \over {\int{f^2\,dV}}} 
\end{equation} 
In the algorithm, we have used the initial ansatz: 
\begin{equation} 
f(\rho,z) = \hbox{const} \; \exp^{-{(\rho-\rho_0)^2+z^2 \over 
\hbox{const}}} 
\end{equation} 
where $\rho_0$ is a fixed initial radius. 
The energy minimization resulted to a non-trivial configuration 
$f(\rho,z)$ for a given set of parameters $B, n, Q$ in the 
expression for the energy. We then used (\ref{omegdef}) to 
calculate $\omega$ and constructed the full Q ring configuration 
using (\ref{eq:ansatz}).While the details of our numerical 
analysis can be found elsewhere we just report the main results.

It was verified that the Q ring configurations evolve with 
practically no distortion and are metastable despite their long 
evolution. Finite size nonsymmetric fluctuations were found to 
lead to a break up and eventual decay of the Q ring to one or more 
Q balls. Thus a Q ring is a metastable as opposed to a stable 
configuration. 

The Q ring configuration we have discovered is the simplest 
metastable ring-like defect known so far. Previous attempts to 
construct metastable ring-like configurations were based on pure 
topological arguments (Hopf maps) and required gauge fields to 
evade Derrick's theorem due to their static 
nature\cite{Faddeev:1997zj,Perivolaropoulos:2000gn}. This resulted 
in complicated models that were difficult to study analytically or 
even numerically. Q rings require only a single complex scalar 
field and they appear in all theories that admit stable Q balls 
including the minimal supersymmetric standard model (MSSM). The 
simplicity of the theory despite the non-trivial geometry of the 
field configuration is due to the combination of topological with 
non-topological charges that combine to secure metastability 
without added field complications. 

The derivation of metastability of this configuration opens up 
several interesting issues that deserve detailed investigation. 
They pertain to the various mechanisms of formation of Q Rings 
(Kibble and Affleck-Dine mechanisms, Q ball collisions etc.) as 
well as on the dependence of the winding N on Q. We hope to have 
something interesting to report in the forthcoming 70th birthday 
celebration of Holger.


\title*{Non-local Axial Anomalies  in the Standard Model}
\author{D. Melikhov and B. Stech}
\institute{%
Institut f\"ur Theoretische Physik, Universit\"at Heidelberg, 
Philosophenweg 16, D-69120, Heidelberg, Germany}

\authorrunning{D. Melikhov and B. Stech}
\titlerunning{Non-local Axial Anomalies  in the Standard Model}

\maketitle

\begin{abstract}
We demonstrate that the amplitude 
$\langle\rho\gamma|\partial^\nu (\bar q\gamma_\nu \gamma_5 q)|0\rangle$ 
does not vanish in the limit of zero quark masses. 
This represents a new kind of violation of the classical equation of motion 
for the axial current and should be interpreted as the axial anomaly for 
bound states. The anomaly emerges in spite of the fact that the one loop 
integrals are ultraviolet-finite as guaranteed by the finite-size of  
bound-state wave functions. As a result, the amplitude behaves like 
$\sim 1/p^2$ in the limit of a large momentum $p$ of the current. 
This is to be compared with the amplitude 
$\langle \gamma\gamma|\partial^\nu (\bar q\gamma_\nu \gamma_5 q)|0\rangle$ 
which remains finite in the limit $p^2\to\infty$. 

The observed effect requires the modification of the classical equation of 
motion of the axial-vector current by non-local operators. The non-local 
axial anomaly is a general phenomenon which is effective for axial-vector 
currents interacting with spin-1 bound states. 
\end{abstract}

\section{Introduction}
It is well-known that the classical equations of motion can be violated in 
quantum field theories. One of the best-known examples of such a violation 
is the axial vector current: in the limit of vanishing quark masses the 
divergence of the neutral hadronic axial vector current should be zero
but in fact is proportional to a local operator containing the photon fields. 
This is the famous Adler-Bell-Jackiw anomaly \cite{abj}. 
The standard model is free of these local anomalies: 
the anomalies of the hadronic (quark) part of the divergence is compensated by the 
leptonic part. Such cancelations are in fact required for any renormalizable field 
theory. 

Also the amplitude of the decay $\pi^0\to \gamma\gamma$, the process which led to the 
discovery of the axial anomaly, is free of anomalies! Only 
when one relates the $\pi^0\to \gamma\gamma$ decay amplitude via PCAC to matrix 
elements of the pure hadronic part of the axial-vector current, one is able to use 
the above mentioned local anomaly term for an estimate of the decay rate.

In this article we will show that the standard model has non-local anomalies and that
these anomalies are the only ones which do not cancel. They arise from the coupling of 
the of the local axial vector current to two vector particles where at least one of 
them is a bound state. Examples are the couplings of the axial vector current to a photon
and a vector meson, or to two vector bound states.  
For the leptonic axial current an example is the coupling to a photon and  
orthopositronium, or two orthopositronium states.

We will demonstrate the existence of such anomalies for the case where the isovector 
axial current emits a photon and a $\rho$-meson in the limit $m_q\to 0$. 
Since this process necessitates a quark loop, it is evident that there is no cancellation 
by the leptonic part of the axial current. This anomaly turns out to be non local, and thus
has no consequences for the renormalizability of the standard model. 
An example for a physical process for which such an anomaly is important is weak
annihilation in radiative $B\to\rho\gamma$ decays:  
The $B$ meson is annihilated by the $\bar u\gamma_5\gamma_\nu b$ axial  
current, and the $\bar d\gamma_5\gamma_\nu u$ part of the axial current 
for light quarks generates the photon and the $\rho$ meson. 

It is interesting that these anomalies had so far not been discovered even though their
derivation is not difficult. 
There are many examples in the literature in which - in the limit of massless quarks - 
the corresponding divergence of the axial current has been put equal to zero by relying on
the validity of the classical equation of motion. 

This article follows our recent analysis presented in Ref. \cite{ms}.

\section{The Adler-Bell-Jackiw anomaly}
The analysis of two-photon decays of pseudoscalar mesons in the late 60-s led to 
the discovery of the famous axial anomaly \cite{abj}: the divergence of the axial vector 
current violates the classical equation of motion and does not vanish in the limit of zero 
fermion masses. 
For a quark of mass $m_q$ and charge $eQ_q$ the properly modified equation of motion 
contains a local anomalous term and has the form\footnote{
We use the following notations: 
$e=\sqrt{4\pi\alpha_{\rm em}}$, 
$\gamma^5=i\gamma^0\gamma^1\gamma^2\gamma^3$,  
$\epsilon^{0123}=-1$, 
${\rm Sp}\left (\gamma^5\gamma^{\mu}\gamma^{\nu}\gamma^{\alpha}\gamma^{\beta}\right )
=4i\epsilon^{\mu\nu\alpha\beta}$, $F_{\mu\nu}={\partial_\mu A_\nu-\partial_\nu A_\mu}$.
$\epsilon_{abcd}=\epsilon_{\alpha\beta\mu\nu}a^\alpha b^\beta c^\mu d^\nu$ for any
4-vectors $a,b,c,d$.}
\begin{eqnarray}
\partial^\nu (\bar q\gamma_\nu\gamma_5 q)=2im_q \bar q\gamma_5 q+N_c\frac{(eQ_q)^2}{16\pi^2} 
F\tilde F, \qquad \tilde F_{\mu\nu}=\epsilon_{\mu\nu\alpha\beta}F^{\alpha\beta}.  
\end{eqnarray}
with $F_{\mu\nu}$ the electromagnetic field tensor. This modification of the 
classical equation of motion accounts for the fact that 
the form factor $G^\gamma$ defined by the 2-photon matrix element 
\begin{eqnarray}
\label{Ggamma}
\langle \gamma(q_1)\gamma(q_2)|\partial^\nu (\bar q\gamma_\nu \gamma_5 q)|0\rangle
=e^2\epsilon_{q_1 \epsilon^*_1 q_2 \epsilon^*_2}G^\gamma(p^2,q_1^2,q_2^2),  
\end{eqnarray}
does not vanish for $m_q=0$ but turns out to be a constant independent of the current momentum $p=q_1+q_2$: 
\begin{eqnarray}
\label{abj-anomaly}
G^\gamma(p^2,q_1^2=q_2^2=0)=-{2N_c(Q_q)^2}/{4\pi^2}. 
\end{eqnarray} 
In this letter we study the properties of axial currents when one of 
the photons, $\gamma(q_2)$, is replaced by a vector meson $V(q_2)$, e. g. a $\rho$-meson.   
We demonstrate that the form factor $G^V$ defined according to the relation
\begin{eqnarray}
\label{GV}
\langle \gamma(q_1)V(q_2)|\partial^\nu (\bar q'\gamma_\nu \gamma_5 q)|0\rangle
=e\epsilon_{q_1 \epsilon^*_1 q_2 \epsilon^*_2}G^V(p^2,q_1^2,q_2^2)   
\end{eqnarray}
has also an anomalous behavior and does not vanish for massless quarks. 
This occurs in spite of the fact that the vector meson is a bound $q\bar q$ state and 
the corresponding loop graph has no ultraviolet divergence. 
Moreover, the anomalous behavior is observed for both, the neutral and the charged 
axial-vector currents. 

The classical equation of motion reads 
\begin{eqnarray}
\label{class}
\partial^\nu (\bar q'\gamma_\nu\gamma_5 q)=i(m_q+m'_q)\bar q'\gamma_5 q
+e(Q_{q'}-Q_{q})A^\nu \bar q'\gamma_\nu\gamma_5 q,  
\end{eqnarray}
where $A^\nu$ is the electromagnetic field. The $\langle \gamma V|...|0\rangle$ matrix 
element of the second term on the r.h.s. of  (\ref{class}) 
vanishes to order $e$. Therefore, to this order, the classical equation of motion 
(\ref{class}) predicts $G^V=0$ for $m_{q'}=m_q=0$. 
We find however to order $e$ and for large $p^2$ 
\begin{eqnarray}
G^V\sim M_V f_V/p^2,    
\end{eqnarray}
where $M_V$ and $f_V$ denote the mass and the decay constant of the vector meson. 
Because of the dependence on $p^2$, the newly found deviation from the classical 
equation of motion for bound states corresponds to a non-local anomaly. 

We are interested in the region $|p^2|\gg m_q^2, \Lambda_{QCD}^2$, in which 
case the quarks in the triangle diagram 
have high momenta and their propagation can be treated perturbatively. 
We discuss in parallel the $\gamma\gamma$ and $\gamma V$ final states 
in order to show how the anomaly emerges in both cases. 
As in Ref. \cite{dz}, we consider the spectral representation for the axial-vector 
current itself before forming the divergence. In distinction to \cite{dz}, where the 
spectral representation in $p^2$ was considered, we use the spectral representation in the 
variable $q_2^2$. This allows us to take bound state properties into account. 

\section{The absorptive part of the triangle amplitude}
The amplitude of the single-flavor axial current between the vacuum and the 
two-photon and the photon-vector meson states,   
respectively, can be written in the form 
\begin{eqnarray}
\epsilon^{*\beta}(q_2)\epsilon^{*\alpha}(q_1)T_{\nu\alpha\beta}(q_1,q_2). 
\end{eqnarray}
The absorptive part $t_{\nu\alpha\beta}$ of $T_{\nu\alpha\beta}$ is calculated 
by setting the two quarks attached to the external particle with the momentum $q_2$ on the mass shell, 
see Fig 1. 
\begin{figure}
\centering
\includegraphics[width=12cm]{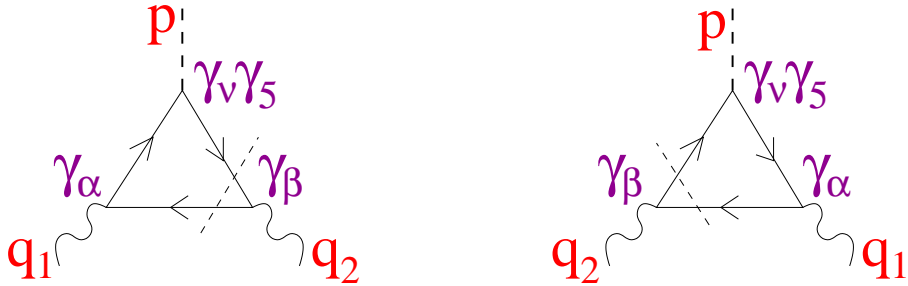}
\caption{%
Diagrams describing 
$\langle \gamma(q_1)\gamma(q_2)|\bar q \gamma_\nu\gamma_5 q|0\rangle$ 
and/or $\langle \gamma(q_1)V(q_2)|\bar q \gamma_\nu\gamma_5 q|0\rangle$, with 
$p=q_1+q_2$. 
The cut corresponds to the calculation of the absorptive part in the variable $q_2^2$.}
\end{figure}
$t_{\nu\alpha\beta}$ is our basis for the spectral representation of $T_{\nu\alpha\beta}$ in terms of the variable $q_2^2$. 
The coupling at the vertex $\beta$ is $\gamma_\beta e Q_q$ if the particle 2 is a photon, and 
$-\gamma_\beta g(q_2^2)/\sqrt{N_c}$ if it is a vector meson.\footnote{
The full vertex has the form \cite{m}
$\Gamma_\beta=-1/\sqrt{N_c}
[\gamma_\beta-\frac{1}{\sqrt{s}+2m}(k_1-k_2)_\beta/(\sqrt{s}+2m)]g((k_1+k_2)^2)$,  
but the term proportional to $(k_1-k_2)_\beta$ does not contribute to the trace. 
The overall $(-)$ sign is the standard choice of the phase of the vector meson wave
function which leads to a positive leptonic decay constant.}   
The coupling $g(q_2^2)$ will be further discussed below. By taking the trace and performing the 
integration over the internal momentum in the loop it is straighforward to obtain 
$t_{\nu\alpha\beta}$. The result is automatically gauge-invariant 
\begin{eqnarray}
q_1^\alpha t_{\nu\alpha\beta}(q_1,q_2)=0, \qquad q_2^\beta t_{\nu\alpha\beta}(q_1,q_2)=0. 
\end{eqnarray}
It is therefore possible to write the covariant decomposition of $t_{\nu\alpha\beta}(q_1,q_2)$ 
in terms of three invariant amplitudes 
\begin{eqnarray}
\label{imampl}
t_{\nu\alpha\beta}(q_1,q_2) &=&
-p_\nu \epsilon_{\alpha\beta q_1 q_2}ic_0
+(q_1^2\epsilon_{\nu\alpha\beta q_2}-q_{1\alpha}\epsilon_{\nu q_1\beta  q_2})ic_1 \\
 & & {}+(q_2^2\epsilon_{\nu\beta\alpha q_1}-q_{2\beta} \epsilon_{\nu q_2\alpha q_1})ic_2.    
\end{eqnarray}
This Lorentz structure is chosen in such a way that no kinematical singularities appear. 
We take $\gamma(q_1)$ to be a real photon, $q_1^2=0$. Hence, the term 
containing the invariant amplitude 
$c_1$ does not contribute to the divergence of the current.  
Setting in addition $m_q=0$ one obtains for $c_0$ and $c_2$ with 
$s=q_2^2$
\begin{eqnarray}
\label{im-inv}
c_0(p^2,s)=-\frac{\zeta(s)}{4\pi}\frac{s}{(s-p^2)^2}, \qquad
c_2(p^2,s)=-\frac{\zeta(s)}{4\pi}\frac{p^2}{(s-p^2)^2},   
\end{eqnarray} 
where $\zeta(s)=2N_c Q_q^2 \theta (s)$ for the $\gamma\gamma$ process and 
$\zeta(s)=-2\sqrt{N_c} Q_q g(s)\theta (s)$ for the  $\gamma V$ process.  

Clearly, the absorptive part $t_{\nu\alpha\beta}(q_1,q_2)$ of the axial-vector current 
matrix element respects the classical equation of motion, that is
\begin{eqnarray}
\label{eq1}
p^2\, c_0(p^2,s)-s\,c_2(p^2,s)=0.  
\end{eqnarray}
 
\section{The triangle amplitude and its divergence}
The full amplitude $T_{\nu\alpha\beta}(q_1,q_2)$ has the same Lorentz structure as its absorptive 
part 
\begin{eqnarray}
\label{ampl}
T_{\nu\alpha\beta}(q_1,q_2) &=&
-p_\nu \epsilon_{\alpha\beta q_1 q_2}iC_0
+(q_1^2\epsilon_{\nu\alpha\beta q_2}-q_{1\alpha}\epsilon_{\nu q_1\beta  q_2})iC_1 \\
 & & {}+(q_2^2\epsilon_{\nu\beta\alpha q_1}-q_{2\beta} \epsilon_{\nu q_2\alpha q_1})iC_2.    
\end{eqnarray}
The absence of any contact terms in $T_{\nu\alpha\beta}$ can be verified by 
reducing out one of the photons and using the conservation of the electromagnetic current. 
The invariant amplitudes $C_i$ can be represented by the following dispersion integrals 
\begin{eqnarray}
C_i(p^2,q_1^2=0,q_2^2)=\frac{1}{\pi}\int_{0}^{\infty}\frac{c_i(p^2,s)}{s-q_2^2-i0}ds. 
\end{eqnarray}
All the integrals converge and thus need no subtraction.

Taking the divergence of $T_{\nu\alpha\beta}$ we find 
\begin{eqnarray}
ip^\nu T_{\nu\alpha\beta}=
-\frac{1}{\pi}\left\{p^2 \int_{0}^{\infty}\frac{c_0(p^2,s)}{s-q_2^2}ds-
q_2^2 \int_{0}^{\infty}\frac{c_2(p^2,s)}{s-q_2^2}ds \right\}\epsilon_{q_1 \alpha q_2 \beta}. 
\end{eqnarray}
The form factor $G$ defined in Eqs. (\ref{Ggamma}) and (\ref{GV}) now reads  
\begin{eqnarray}
\label{Gint}
G(p^2,q_2^2)=\frac{p^2}{4\pi^2}\int_{0}^{\infty}\frac{\zeta(s)}{(s-p^2)^2}ds
\end{eqnarray}
In the case of the $\gamma\gamma$ process $\zeta(s)$ is a constant. The integral can be performed and 
gives the well-known value shown in Eq (\ref{abj-anomaly}). 
In the case of the $\gamma V$ matrix element the integrals converge even better since $g(s)$ which appears in $\zeta(s)$ descibes the spatial size of the vector meson. We conclude from 
Eq. (\ref{Gint}) that the divergence of the axial-vector current is nonzero for $m_q=0$ not only for the $\gamma \gamma $ but also for the 
$\gamma V$ final state! Namely, 
\begin{eqnarray}
\label{GVint}
G^V(p^2,q_2^2)=2\sqrt{N_c}e Q_q \frac{-p^2}{4\pi^2}\int_{0}^{\infty}\frac{g(s)}{(s-p^2)^2}ds. 
\end{eqnarray}
The behavior with respect to $p^2$ is however different from the $\gamma\gamma$ case and has the form 
$G^V(p^2)\sim 1/p^2$ for the large values of $p^2$ where our formula applies. 

For the transition to the $\gamma\rho$ (isospin-1) arising from the isovector axial current we obtain 
\begin{equation}
\label{GV1}
G^{\rho}=(Q_u+Q_d)\kappa \frac{f_\rho M_\rho}{p^2}, \qquad
\kappa=-\frac{\sqrt{N_c}}{4\pi^2}\frac{p^4}{f_\rho M_\rho}\int_{0}^{\infty}\frac{g(s)}{(s-p^2)^2}ds. 
\end{equation}
The parameter $\kappa$ in this equation is
non zero for $m_q=0$ and $|p^2|\to \infty$. $f_\rho$ is defined by the relation 
$\langle \rho^-|\bar d\gamma_\nu u|0\rangle=f_\rho M_\rho \epsilon^*_\nu$. 

Eq. (\ref{GV1}) takes into account the soft contribution to the form factor $G^{\rho}$.  
For large $|p^2|$ one should take care of the QCD evolution of the $\rho$-meson 
wave function 
from the soft scale $\mu^2\sim$ 1 GeV$^2$ to the scale $\mu^2 \sim |p^2|$.\footnote
{We want to  point here to the similarity of the form factor $G^\rho$ with the 
$\pi\gamma$ transition form factor $F_{\pi\gamma\gamma^*}(p^2)$. For a detailed analysis of 
the latter we refer to Ref. \cite{mr}. Likewise, the form factor $G^{\rho\rho}$ 
describing the amplitude $\langle \rho\rho|\partial^\nu (\bar q\gamma_\nu \gamma_5 q)|0\rangle$ has some common feature with 
the pion elastic form factor.}
This can be done most directly by expressing $\kappa$ in terms of the $\rho$-meson 
light-cone distribution amplitudes \cite{bb}   
\begin{eqnarray}
\nonumber
\langle\rho(q_2)|\bar d(x)\gamma_\lambda u(0)|0\rangle&=&
-iq_{2\lambda} (\epsilon^*x)f_\rho M_\rho\int\limits_0^1 du
e^{iuq_2x}\Phi(u) \\
 & & {}+\epsilon^*_\lambda f_\rho M_\rho \int\limits_0^1 du e^{iuq_2x}g_\perp^{(v)}(u),  
\\
\langle\rho(q_2)|\bar d(x)\gamma_\lambda \gamma_5u(0)|0\rangle&=&-\frac{1}{4}
\epsilon_{\lambda\eta\rho\sigma}\epsilon^{*\eta} q_2^\rho x^\sigma 
f_\rho M_\rho\int\limits_0^1 du e^{iuq_2x}g_\perp^{(a)}(u).   
\end{eqnarray} 
The diagrams of Fig 2 explain the procedure we follow: The quark propagator connects the 
axial current with the photon and the two distant space time points are bridged by the 
$\rho$ meson wave function. This leads to the following expression for $\kappa$  
\begin{figure}
\centering
\includegraphics[width=12cm]{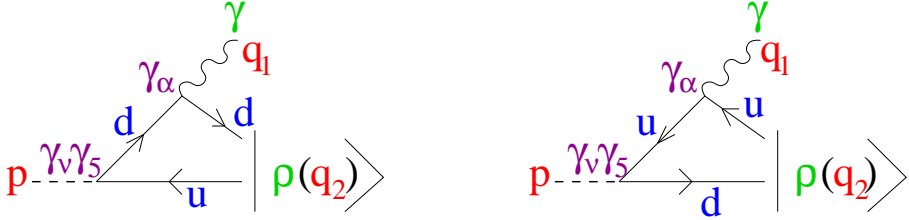}
\caption{%
The diagrams for the matrix element 
$\langle \gamma(q_1)\rho^-(q_2)|\bar d \gamma_\nu\gamma_5 u|0\rangle$, 
with two space time points bridged by the $\rho$ meson light-cone distribution amplitude.}
\end{figure}
\begin{eqnarray}
\kappa=-\int\limits_0^1 du \left[
 \frac{1+u}{4u^2}g_\perp^{(a)}
-\frac{1-u}{u}g_\perp^{(v)}
-\frac{1}{1-u}\Phi\right]. 
\end{eqnarray}
The leading-twist distribution amplitudes 
\begin{eqnarray}
g_\perp^{(a)}&=&6u(1-u), 
\nonumber\\
g_\perp^{(v)}&=&\frac{3}{4}\left(1+(2u-1)^2\right),
\nonumber\\
\Phi&=&\frac{3}{2}u(1-u)(2u-1) 
\end{eqnarray}
give the main contribution for large $p^2$ and lead 
to the value $\kappa=-3/2$. 
Corrections to this value are calculable 
in terms of the higher twist distribution amplitudes \cite{bb}. 


Summing up, our results are as follows: 

\begin{enumerate}
\item The divergence of the axial vector current and thus the form factor $G$ does not 
vanish in the limit $m_q\to 0$. This holds for the $\gamma\rho$ final state as well 
as for the $\gamma\gamma$ final state. The observed effect for vector mesons requires 
a proper modification of the equation of motion for the axial current. 
For large momenta $p$ of the axial-vector current the corresponding 'bound state anomaly' 
can be described in terms of a non-local operator appearing at order $e$ 
(there are no local operators of the appropriate dimension): 
\begin{eqnarray}
\nonumber
&&{\rm Single\; flavor\;current}:
\\
&&\partial^\nu(\bar q\gamma_\nu\gamma_5 q)=2im\;\bar q \gamma_5 q +
\frac{(eQ_q)^2N_c}{16\pi^2}F\tilde F \\
 & & {}+ e\kappa\,Q_q\,\Box^{-1}
\left\{\partial_\mu(\bar q\gamma_\nu q)\cdot\tilde F^{\mu\nu}\right\}+O(\Box^{-2}). 
\\
{}\nonumber
\\
&&{\rm Isovector\; charged\; current}:
\nonumber\\
&&\partial_\nu(\bar d\gamma_\nu\gamma_5 u)=
i(m_u+m_d)\bar u \gamma_5 d 
-i e (Q_u-Q_d)\bar u \gamma_\nu\gamma_5 d\cdot A^\nu
\nonumber\\
&& {}+ e\kappa\frac{(Q_u+Q_d)}{2}\Box^{-1}\left\{
\partial_\mu(\bar u\gamma_\nu d)\cdot\tilde F_{\mu\nu}\right\}+O(\Box^{-2}). 
\end{eqnarray}
$\kappa$ can be expanded in a power series of $\alpha_s$.  
In leading order one obtains the value $\kappa=-3/2$. 

As pointed out in \cite{bmns}, the $\rho\gamma$ anomaly is important in 
rare decays: for instance, in $B\to\rho\gamma$ decays, it substantially 
corrects the weak annihilation amplitude which carries the CP violating 
phase.  

\item The amplitude for the $\rho\rho$ final state also stays finite for $m_q=0$. The
corresponding non-local anomalous term for the divergence of the isovector axial current 
$\partial_\nu(\bar q\gamma_\nu\gamma_5 {\tau^a}q)$ appears already at order $O(e^0)$. 
The operator structure of the anomalous term is more complicated in this case. 
One of the possible lowest-dimension operators which has a nonvanishing 
$\langle\rho\rho|...|0\rangle$ matrix element and thus will contribute to the 
anomalous term (in leading order in $1/p^2$) is the product of the two isovector 
tensor currents 
\begin{eqnarray}
\epsilon^{abc}\epsilon^{\mu\nu\alpha\beta}
\Box^{-1}\left(
\bar q \sigma_{\mu\nu}      {\tau^b}q\cdot  
\bar q \sigma_{\alpha\beta} {\tau^c}q\right). 
\end{eqnarray} 
Accordingly, for large $|p^2|$ the $\langle\rho\rho|...|0\rangle$ amplitude of the 
divergence of the axial-vector current is given by the factorized matrix element of the anomalous term and 
has a $1/p^2$ suppression 
\begin{equation}
\begin{array}{rcl}
\langle \rho(q_1)\rho(q_2)|\partial^\nu(\bar d\gamma_\nu\gamma_5
u)|0\rangle
&=&
\epsilon_{q_1 \epsilon^*_1 q_2 \epsilon^*_2}G^{\rho\rho},\\ 
G^{\rho\rho} &\sim& f_\rho^2 /p^2+O(1/p^4).  
\end{array}
\end{equation}

\item We have illustrated the appearance of non-local anomaly due to 
vector mesons in QCD. This anomaly is of a general nature and should be present in 
any theory containing $J^P=1^-$ bound states. For example, the anomaly will also 
contribute to the generation of orthopositronium by the leptonic axial-vector current.
In contrast to the conventional local anomalies, there is no obvious cancellation 
of the non-local anomalies in the standard model. 
\end{enumerate}

\section*{Acknowledgments}
It is a pleasure to thank V. Braun, H. G. Dosch, A. Khodjamirian, 
O. Nachtmann, M. Neubert, 
O. P\`ene and V. I. Zakharov for discussions. D.M. is grateful to the Alexander 
von Humboldt-Stiftung for financial support.

\title*{Could there be a  Fourth Generation?\thanks{We would like to thank David Sutherland for discussions and useful remarks.}}
\author{%
C.D. Froggatt\thanks{E-mail: c.froggatt@physics.gla.ac.uk} 
and J.E. Dubicki\thanks{E-mail: j.dubicki@physics.gla.ac.uk}}
\institute{%
Department of Physics and Astronomy
Glasgow University, Glasgow G12 8QQ, Scotland}

\authorrunning{C.D. Froggatt and J.E. Dubicki}
\titlerunning{Could there be a  Fourth Generation?}
\maketitle

\begin{abstract}
We investigate the possibility of incorporating a chiral fourth-generation into a GUT model. We find that in order to do so, precision fits to electroweak
observables demand the introduction of light ($< M_Z$) supersymmetric particles. This also enables us to provide decay channels to the fourth-generation 
quarks. Perturbative consistency sets an upper bound on the coloured supersymmetric spectrum. The mass of the lightest Higgs boson is calculated and found 
to be above the present experimental lower limit.
\end{abstract}

\section{Introduction}
\label{sec1}
Despite the success of the Standard Model (SM), it is far from complete. For instance, the origin of the inter-generation mass hierarchy and issues 
such as baryon asymmetry are yet to be resolved. We would expect new physics to contribute to these areas. At the more fundamental level, nobody knows of 
any deep explanation as to why there should only be three generations of quarks \& leptons. For these reasons, and others, extensions of the 
three-generation SM are being investigated. We investigate the possibility of incorporating a fourth-generation into a GUT model.\medskip \\
As is well known, all fourth-generation models must adhere to certain experimental constraints, the first of which stems from precise measurements of the 
decay characteristics of the $Z$-boson performed at LEP. This has set a lower bound of $M_F\ge\frac{M_Z}{2}$ on any non-SM particles that couple to 
the $Z$-boson. Ignoring for the moment the unnatural hierarchy emerging within the neutrino sector, we assume a Dirac mass 
$\sim\left(\frac{M_Z}{2}\right)$ for the heavy neutral lepton.
We label the fourth-generation doublets explicitly as,
\begin{equation*}
        Q_4 = {\left( \begin{array}{c} T \\ B \end{array} \right)}_4 
                        \quad T_4^c \quad B_4^c; \qquad
        L_4 = {\left( \begin{array}{c} N \\ E \end{array} \right)}_4 
                        \quad N_4^c \quad E_4^c 
                        \nonumber 
\end{equation*}
\medskip \\
Physics beyond the Standard Model is severely constrained by precision electroweak data\footnote{We assume ${\left|V_{tb}\right|}^2$, 
${\left|V_{TB}\right|}^2 \sim 1$ and ${\left|V_{Tb}\right|}^2$, ${\left|V_{tB}\right|}^2 \ll 1$ so that contributions to the
$Z \rightarrow b\overline{b}$ decay can be ignored.}.
Assuming SM contributions, fits to LEP data give the radiative correction parameter\footnote{S is the well known radiative correction parameter (weak 
isospin symmetric), normalized to zero for the SM with $M_{Higgs} =$ 100 GeV.} $S=-0.04 \pm 0.11$. For a heavy ($\gg M_Z$) degenerate fourth-generation we 
obtain $\Delta S = \frac{2}{3\pi}$ and so is ruled out at $99.8\%$ $C.L$. However, Maltoni {\it et al.}~\cite{maltoni} have shown that particles with mass 
$M \sim \frac{M_Z}{2}$ give drastically different contributions to $S$. It is possible for a heavy neutrino $N$ with mass $M_N \sim \frac{M_Z}{2}$ to 
cancel the contributions from the heavy $T,B,E$ with the SM solution
\begin{equation*}  M_E > M_N ; \quad M_T > M_B \end{equation*}
\begin{equation*}  M_N \sim \frac{M_Z}{2} ; \quad M_B = M_B^{min} \end{equation*}
\begin{equation}\label{fit}  
        (M_E-M_N) \sim 3(M_T-M_B) \sim 60 GeV
\end{equation}
and the $N$ must be relatively stable to avoid detection ({\it i.e.} mixing matrix elements $V_{Ne,\mu,\tau} < 10^{-6}$). An extra generation can be 
accomodated below the $1\sigma$ level (or even two generations at $1.5\sigma$)\footnote{Updated fits to recent LEP data show that an extra generation 
with $M_N = M_U \sim$ 180 GeV; $M_D = M_E \sim$ 130 GeV gives a $\chi^2$ fit for the fourth-generation as good as that of the three-generation 
case~\cite{okun}. Perturbative unification is not consistent with such large masses.}.
However, it was shown by Gunion {\it et al.}~\cite{gunion} that the fourth-generation charged lepton must be relatively light ($<$ 60 GeV) in 
order to stay in the perturbative regime below the GUT scale. Although they worked within a supersymmetric framework, similar results are expected to 
hold in the SM~\cite{pirogov}. Decreasing $M_E$ below $\sim$ 60 GeV to achieve perturbative unification is not consistent with the above SM fits 
(Eq.(\ref{fit})) to the precision data. Therefore, we require the cancellations to arise from another sector and so we consider a supersymmetric theory.  
Indeed, it is shown in ~\cite{maltoni} that light supersymmetric particles can also effect the fit to precision data, allowing for a fourth-generation 
below the $2\sigma$ level. In particular, neutralinos (${\tilde{\chi}}^0_1$) and charginos (${\tilde{\chi}}^{\pm}_1$) with masses $<M_Z$ can provide the 
correct sign contributions whilst being consistent with current experimental limits if nearly degenerate~\cite{L3}.\medskip \\
In this paper we investigate the possibility of consistently incorporating a fourth-generation into a $N=1$ R-parity conserving supergravity model. 
We assume a structure akin to that of the minimal supersymmetric standard model (MSSM3), adding a complete chiral fourth-generation and its associated 
SUSY partners (the so called MSSM4). Specifically we require $(i)$ perturbative values for all Yukawa couplings at energies up to the GUT scale and 
$(ii)$ gauge coupling unification. These two constraints will be termed collectively as perturbative unification. \medskip\\
Having satisfied precision data fits, it remains for us to provide solutions that $(i)$ unify perturbatively
at the GUT scale and $(ii)$ evade the direct experimental searches performed at $CDF$. 
In section~\ref{sec2} we investigate the specific decay channels of the fourth-generation quarks and ensure we can provide consistency
with the experimental direct searches. The solution we present requires the introduction of light SUSY particles 
($M_{{\tilde{\chi}}^{0}},M_{{\tilde{\chi}}^{\pm}},M_{\tilde{B}}$), so the two-body decays $T \rightarrow \tilde{B} {\tilde{\chi}}^{+}$
and $B \rightarrow \tilde{B} {\tilde{\chi}}^{0}$ are kinematically allowed. These will always dominate over the one-loop FCNC
decays $B \rightarrow b Z^0$ and two-generation decays $B \rightarrow c W^{-}$ that traditional searches have looked for, whilst at the 
same time suppress the decay rate $T \rightarrow b W^+$. We note a light, degenerate
chargino/neutralino pair is just what is needed to provide the necessary cancellations in precision data. \medskip \\
In section~\ref{sec3} we discuss the influence of specific fourth-generation/SUSY masses on precision data fits in more detail and point out new
features that appear when trying to satisfy all constraints. In section~\ref{sec4} we proceed with a renormalisation group study of the four generation 
MSSM. Under the assumption of a common mass scale for the coloured/weak sparticle spectrum ($M_{col}$/$M_{wk}$ respectively) we will derive upper bounds 
on $M_{col}$. This bound stems from the fact that if we are to achieve perturbative values for the top quark Yukawa coupling constant $h_t$ up to the 
GUT scale, then the effects of the coloured sparticle spectrum must be included in the running of the strong coupling $\alpha_3$ at an early stage.
Finally, in section~\ref{sec5} we investigate the lightest Higgs mass in the MSSM4. \medskip \\
\section{Experiment and a Fourth Generation}
\label{sec2}
In this study we assume the hierarchy within the fourth generation is such that $M_T > M_B$ and $M_E > M_N$. We begin our discussion by considering
the leptonic sector. Under the assumption that the mixing between ($E,N$) and the first three generations is negligible, the decay 
$E \rightarrow N W^*$ will be dominant. In order to evade experimental bounds, the mass difference $\Delta M_L = M_E - M_N$ must be small enough to 
result in a virtual $W^*$ whose decay products are too soft to be triggered. Regarding the heavy neutrino, LEPII has set the bound $M_N >$ 70 - 80 GeV 
based on the search for $N \rightarrow l W^*$ ($l=e, \mu$ or $\tau$) where the mixing matrix elements $V_{Ne,\mu,\tau} > 10^{-6}$~\cite{lepton}. A heavy 
neutrino with this mass is inconsistent with perturbative unification. However,  we have previously assumed that the mixing of the fourth 
generation leptonic sector is in fact negligible ($V_{Ne,\mu,\tau} < 10^{-6}$) so the neutrino is stable enough to leave the detector and in this case 
only the DELPHI bound of $M_N >$ 45 GeV from measurements of the $Z$-width applies. From figure (4) in~\cite{lepton} we can see that 
$M_E \sim M_N \sim$ 50 GeV is allowed.  \medskip \\
Turning to the quark sector, the requirement of perturbative unification places strict upper limits on the masses of the $T$ and $B$ and they certainly
must be below the top quark whose mass is $M_t = 174 \pm$ 5.1 GeV. Experimental searches for the fourth-generation quarks
have mainly concentrated on the $B$-quark where CDF have set a bound of $M_B >$ 199 GeV assuming the branching ratio
$BR(B \rightarrow b Z^0) \sim 1$. The search is also sensitive to other decay modes; for instance the decay $B \rightarrow b h^0$ or 
$B \rightarrow cW^-$ is triggered as long as $BR(R \rightarrow bZ)$ is not negligible, since the hadronic decay of the $h^0$ or $W$ are kinematically
similar to those of the $Z$. One possible escape might come about if we note that $h^0 \rightarrow N \overline{N}$ would compete with 
$h^0 \rightarrow b \overline{b}$ for $M_h^0 \sim M_Z^0$ (dominating for larger $M_h^0$) and would provide an invisible signature. For $M_B \sim$ 100 GeV 
and $M_h^0 \sim M_Z^0$ we would also expect $BR(B \rightarrow b h^0) \sim BR (B \rightarrow b Z^0)$~\cite{hou}. However, CDF still exclude a $B$-mass 
in the range 104 GeV $\rightarrow$ 152 GeV assuming $BR(B \rightarrow bZ^0) \ge 50 \%$ and no sensitivity to the other decay modes~\cite{quark}. Taking 
$M_{h^0} \sim$ 100 GeV (which places $M_B \sim$ 105 GeV for $B \rightarrow b h^0$ to be kinematically allowed) we obtain 
$BR(h^0 \rightarrow invisible) \sim 80\% - 90\%$ and so we can expect to rule this possibility out. Moreover, we have yet to account for the $T$ decays 
which turn out to be highly constrained if we assume $SM$-like processes. The channel $T \rightarrow b W^+$ is prohibited for the obvious reason that 
the $T$-quark would have been picked up in the $CDF$ search for the top quark. Although we might assume that $T \rightarrow B W^*$ is dominant by 
suppressing $V_{Tb}$ we must notice that, since the $T\overline{T}$ production cross-section is similar to that of $B\overline{B}$, we would effectively 
expect double the event rate on the $B$-quark search. Taking this into account would further strengthen existing bounds on the $B$-quark mass. \medskip\\ 
From the ideas presented so far we might conclude that the fourth-generation with perturbative unification is not consistent with experimental bounds
on the ($T,B$) quarks. We have not, however, considered the possibility of light SUSY particles providing decay channels for ($T,B$). In this situation
one can constrain the light ({\it i.e.} $< M_Z$) neutralino/chargino pair, which is already required by fits to precision data, in 
order to provide the following two-body decays 
\begin{equation}  
        T \rightarrow \tilde{B} {\tilde{\chi}}^{\pm}_1 ; \quad B \rightarrow \tilde{B} {\tilde{\chi}}^0_1
\label{chan}
\end{equation}
Ensuring both decay channels are kinematically accessible, combined with the fact that perturbative unification requires\footnote{As we increase $M_t$ 
within the allowed experimental range, the solution becomes harder to maintain. Low values of $M_t$ are consistent with the fit from the jet $+$ lepton 
channel which is thought to be more precise.}(for $M_t =$ 170 GeV) $M_Q <$ 110 GeV, where $M_Q = M_B \simeq  M_T$, places severe restrictions on the 
allowed spectrum. To be definite we choose
\newline
\parbox{5.0cm}{\begin{eqnarray*}
                M_{\tilde{\chi}} & \simeq & 50 \; GeV  \nonumber \\
                M_{\tilde{B}} & \simeq & 55 \; GeV \nonumber
\end{eqnarray*}}
\hfill \parbox{8.0cm}{\begin{eqnarray*}
                M_Q & \simeq & 105 - 107 \; GeV \nonumber \\
                M_t & = & 170 \; GeV \nonumber
\end{eqnarray*}}
\newline
where under the assumption of a nearly degenerate neutralino/chargino pair we define the notation
$\Delta M_{\tilde{\chi}} =  M_{{\tilde{\chi}}^{\pm}_1} - M_{{\tilde{\chi}}^0_1}$;
$M_{\tilde{\chi}} =  M_{{\tilde{\chi}}^0_1} \sim M_{{\tilde{\chi}}^{\pm}_1}$.
Although these masses seem to be contrived, we note that they can be obtained from reasonable assumptions about the supersymmetric sector. In order to 
obtain $\Delta M_{\tilde{\chi}} \sim M_{{\pi}^+}$ we require the hierarchy $|\mu| \gg M_1 \ge M_2$ where $|\mu|$ is the Higgs mixing parameter and is 
fixed by the requirement of radiative electroweak breaking. $M_1$ and $M_2$ are the electroweak
gaugino masses (for a review of supersymmetry see~\cite{susy}). This structure can occur naturally when the gaugino masses are dominated by 
loop corrections, originating from superstring models ({\it i.e.} the $O-II$ model~\cite{string}). We also require substantial 
${\tilde{B}}_L/{\tilde{B}}_R$ mixing to induce the light $\tilde{B}$ mass\footnote{The most recent $LEP$ search for $\tilde{B}$ squarks from 
$\tilde{B} \rightarrow b {\tilde{\chi}}^0$ decays is insensitive to $\Delta M_{{\tilde{B}}{\tilde{\chi}}} = M_{\tilde{B}} - M_{\tilde{\chi}} <$ 8 GeV, 
especially in the case of large mixing~\cite{pdg}.}.
\section{Precision Measurements and a Fourth Generation}
\label{sec3}
It is difficult to provide bounds from precision data without a fully consistent study taking into account exact particle masses, any light SUSY 
spectra present and mixings between different flavours. However, it is pointed out in~\cite{maltoni} that a highly degenerate neutralino/chargino pair 
($\Delta M_{\tilde{\chi}} \sim M_{{\pi}^+}$; $M_{\tilde{\chi}} \sim$ 60 GeV) can provide the necessary contributions needed to cancel that of the fourth-
generation whilst at the same time being consistent with LEP bounds. Looking at their results, we see that the magnitude of the contribution to the 
fitted parameters from this sector is highly dependent on $M_{\tilde{\chi}}$. Indeed, demanding $M_{\tilde{\chi}} \simeq$ 50 GeV in our model might be 
too restrictive. One possible solution presents itself if we assume that the lightest supersymmetric particle (LSP) is in fact the fourth-generation 
sneutrino $\tilde{N}$. We have no phenomenological bounds on this scenario, the ${\tilde{\chi}}^0_1$ would decay invisibly via 
${\tilde{\chi}}^0_1 \rightarrow {\upsilon}_{\tau} \tilde{N}$ assuming the $3-4$ mixing angle is non-zero~\cite{gunion}. This does, however, allow us to 
increase $M_{\tilde{\chi}}$ whilst decreasing $M_{\tilde{B}}$. The following masses are possible
\newline
\parbox{5.0cm}{\begin{eqnarray*}
                M_{\tilde{\chi}} & \simeq & 55 \; GeV \nonumber \\
                M_{\tilde{B}} & \simeq & 50 \; GeV \nonumber \\
                M_{\tilde{N}} & \simeq & 46 \; GeV \nonumber 
\end{eqnarray*}}
\hfill \parbox{8.0cm}{\begin{eqnarray*}
                M_Q& \simeq & 105 - 107 \; GeV \nonumber \\
                M_t & = & 170 \; GeV \nonumber
\end{eqnarray*}}
\newline
The sneutrino $\tilde{N}$ is now the LSP and is stable due to $R$-parity. The bottom squark will decay via 
$\tilde{B} \rightarrow ij \tilde{N}$ (where $(i,j) = (c,\tau)$ or $(b,\upsilon_{\tau})$) depending on which decay is kinematically allowed. Such decays 
involve the factor $V_{Bi} V_{Nj}$, leading to a long lifetime\footnote{Current bounds looking at stable/long-lived squarks exclude the range 
5 GeV $\le M_{\tilde{B}} \le$ 38 GeV if the mixing in the squark mass matrix is large~\cite{long}.}.
If the coupling of $\tilde{N}$ to the $Z$-boson is small then it is possible to have $M_{\tilde{N}} < \frac{M_Z}{2}$. This allows us to decrease
$M_{\tilde{B}}$ even further, hence increase $M_{\tilde{\chi}}$ and still retain the decay channels in Eq.(\ref{chan}). The maximum value obtained in this 
case is $M_{\tilde{\chi}}^{max} \le$ 60 GeV, consistent with precision fits as shown in~\cite{maltoni}. \medskip \\
The fits to precision data performed in~\cite{maltoni} were for a different set of fourth-generation masses. Although the non-universal contributions 
(see Maltoni, Ph.D. thesis~\cite{maltoni} for terminology) remain the same in the limit $|M_i-M_j| \ll M_Z$ (where $i,j$ label the two 
fermions within the same $SU(2)_L$ doublet), the universal contributions will differ. However, we could arrange to have significant $SU(2)_L$ breaking in 
the ${(\tilde{T},\tilde{B})}_L$ doublet if $M_{\tilde{T}}$ was found to be sufficiently large. This would provide universal contributions to the precision 
parameters~\cite{maltoni}, thus compensating for the difference between our model with highly degenerate fourth-generation fermion doublets and the fits 
as performed in~\cite{maltoni}. 
\section{Renormalization Group Study of the MSSM4}
\label{sec4}
Here we investigate the effect of the fourth-generation on the evolution of couplings to the GUT scale, where we require gauge coupling unification.
This places upper limits on the masses of the extra particles to ensure their Yukawa couplings run perturbatively to the unification scale $M_U$
($h^2(\mu) \le 4\pi$, $M_Z \le \mu \le M_U$). Starting at the low-energy scale $M_Z$, the electroweak gauge couplings $\alpha_1(M_Z)$, $\alpha_2(M_Z)$
are fixed through the relations $\frac{1}{\alpha_i(M_Z)} = \frac{3}{5} \frac{\left(1-{\sin^2 \theta_W}\right)}{\alpha_{em}(M_Z)} ;
\frac{{\sin^2 \theta_W}}{\alpha_{em}(M_Z)}$
for $i=1,2$ respectively. The strong coupling $\alpha_3(M_Z)$ is taken from the Particle Data Group (PDG)~\cite{pdg} to be $0.1181 \pm 0.002$. We 
take the best fit values for $\frac{\alpha_{em}(M_Z)}{\sin\theta_W}$ from the PDG. In principle, one should extract the $Z$-pole couplings assuming the full 
MSSM4, thereby accounting for the fourth-generation fermions and light SUSY spectra in a fully consistent way\footnote{We account for the one-loop 
leading logarithmic corrections from the SUSY sector when running the RGE by employing the step-function approach~\cite{boer}. This procedure is 
accurate in the limit of heavy sparticles but fails for masses $\tilde{M} < M_Z$ where both logarithmic and finite corrections will influence the 
extraction of the couplings.}. 
We have also performed our study in the $\overline{MS}$ scheme although it is the $\overline{DR}$ scheme that is consistent with supersymmetry. However, 
differences between the $\overline{DR}$ and $\overline{MS}$ schemes are not significant at the low-energy scale as compared to other uncertainties.

\begin{figure}
\centering
\includegraphics[height=7.385cm,width=12cm]{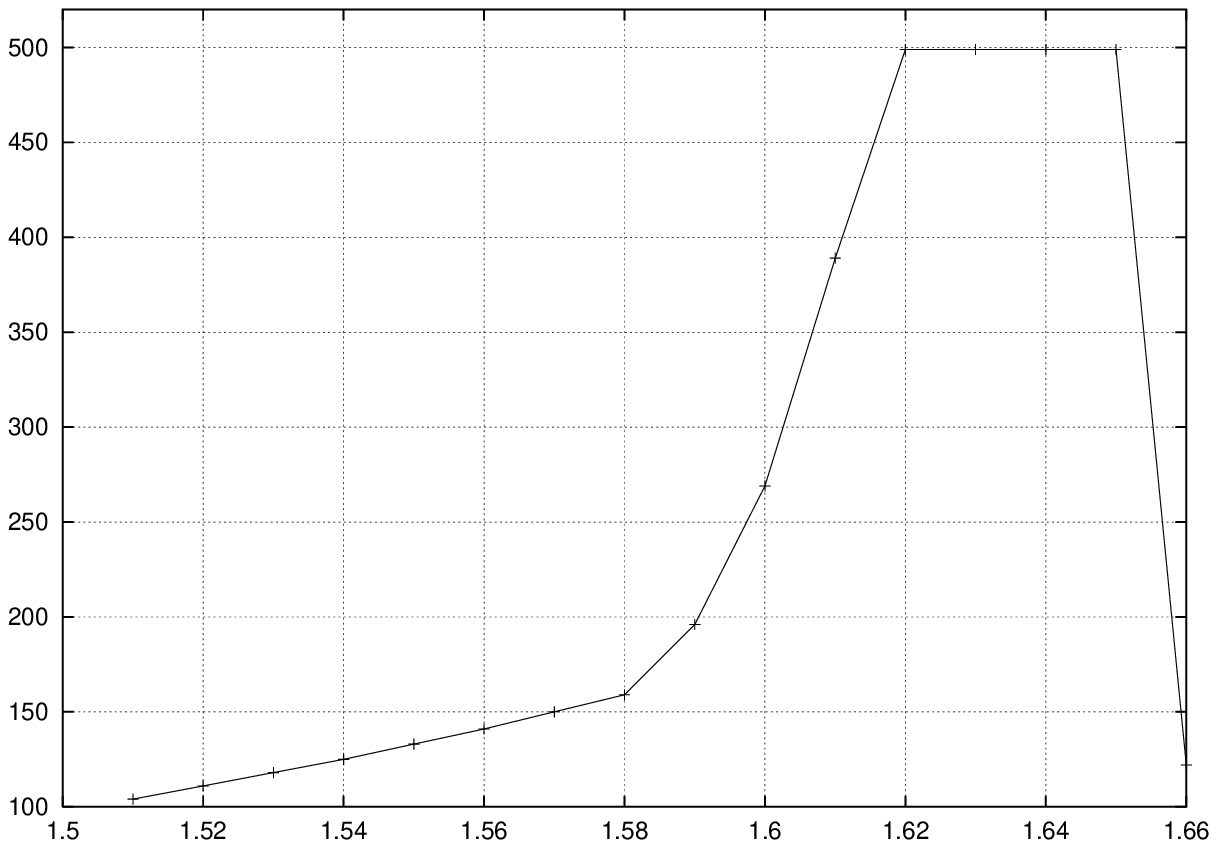}
\caption{Plot of $M_{col}^{max}$ vs. $\tan\beta$ for $M_T =$ 107 GeV; $M_B =$ 105 GeV; $M_N =$ 46 GeV; $M_E =$ 50 GeV; $M_t =$ 170 GeV.}
\label{susyr}
\end{figure}

In our analysis we neglect all Yukawa couplings from the first three generations except that of the $t$-quark whose mass we take to be 
$M_t^{pole} =$ 170 GeV. As is typical with four-generation models, we require small values of $\tan\beta$ (the ratio of the Higgs vev's)
so as to avoid $h_B (M_Z)\ge$$\cal O$($\sqrt{4\pi}$). Once all couplings at $M_Z$ have been fixed, we integrate up in energy scale using the two-loop 
renormalization group equations (RGE). The one-loop leading logarithmic threshold corrections from the SUSY sector are accounted for in the numerical 
procedure. We select the point where $\alpha_1(\mu) = \alpha_2(\mu)$ as the unification scale $M_U$ with coupling $\alpha_U(M_U)$. Any deviation in 
$\alpha_3(M_U) = \alpha_U(M_U)$, which we parameterize as $\delta = \frac{\alpha_3(M_U) - \alpha_U(M_U)}{\alpha_U(M_U)} $, can arise from either of two 
sectors. On the one hand, we have the $\overline{MS}$ vs. $\overline{DR}$ mismatch, experimental errors in  $\alpha_{em}(M_Z)$/$\sin\theta_W$ and the 
variations in the best fit values of $\alpha_{em}(M_Z)$ and $\sin\theta_W$ as the fourth-generation and light SUSY particles ($\tilde{M} < M_Z$) are 
included. However, more importantly, assuming no intermediate scales, high-energy threshold corrections from specific GUT/string models can provide 
contributions to $\delta$. Following~\cite{ross}, we note that these corrections (for particular models)can be large. Considering all the uncertainties 
together, we conservatively require unification to within $\pm 5 \%$ ($\delta = \pm 0.05$). \medskip \\
The SUSY threshold corrections in the MSSM4 are important as they can influence whether or not a particular set of masses ($M_t,M_T,M_B,M_N,M_E$)
will retain perturbative consistency to the GUT scale. This can be observed analytically if we write the one-loop leading logarithmic correction to the 
strong coupling from the SUSY sector
\begin{equation}  \frac{1}{\alpha_3^+(M_Z)} - \frac{1}{\alpha_3^-(M_Z)} = \frac{b_3^{MSSM4}-b_3^{SM4}}{2\pi} \ln \left(
        \frac{M_{col}}{M_Z} \right)
\label{match}
\end{equation}
where $M_{col}$ represents an effective threshold scale and $b_3^{MSSM4/SM4}$ represents
the one-loop beta function contribution to the strong coupling in the MSSM4/SM4 respectively. This correction is implemented at the scale $M_Z$
and accounts for the coloured sparticles with masses $M_{col} > M_Z$ in the running of the strong coupling. $\alpha^+$/$\alpha^-$ represents
the renormalized gauge coupling just above/below the scale $M_Z$. Eq.(\ref{match}) holds if the masses of the sparticles are $\le$ 2 TeV~\cite{polonsky}. 
It is obvious that the higher the scale $M_{col}$, the lower $\alpha_3^+(M_Z)$. Looking at the structure of the RGE it is evident that the strong coupling 
plays an important role in keeping the Yukawa couplings perturbative. We obtain an upper bound $M_{col}^{max}$, above which the initial $\alpha_3$ is too 
small to counteract the effect of the fourth-generation couplings. \medskip \\ 
In practice, we perform a full numerical study, accounting for the threshold corrections from $t,T,B,{\tilde{M}}_i$ (where i runs over all sparticles with
$\tilde{M} > M_Z$) by changing the $\beta$-functions and using the step-function approach in the running of the gauge couplings~\cite{boer}. We assume 
separate degeneracies amongst the coloured and weak SUSY spectrum 
\begin{eqnarray}
 M_{wk} & = & |\mu| =  M_{\tilde{L}} = M_{\tilde{H}} = M_{\tilde{W}}= M_H \label{limit} \\
 M_{col} & = & M_{\tilde{Q}} = M_{\tilde{g}} \nonumber
\end{eqnarray}
We take the fixed value $M_{wk} =$ 500 GeV though in principle it could be anywhere up to $\sim$ 1 TeV. Of course, considering separate degeneracies 
amongst the coloured and weak sparticle spectra is an approximation. In fact, our model demands the introduction of light sparticles to ensure consistency 
with experimental searches for the fourth-generation and precision data bounds, thus providing significant deviations from degeneracy.  Nevertheless, the 
study does serve to show that the combined effect of the coloured spectrum is bounded from above in the process of retaining perturbative consistency of 
the model. See figure~\ref{susyr} for a plot of $M_{col}^{max}$ vs. $\tan\beta$. In particular we conclude that perturbative unification is only
possible for the small range $1.5 < \tan\beta < 1.7$. Further restrictions on the allowed $\tan\beta$ range arise from the mass of the lightest Higgs 
boson that must exceed its experimental lower bound (see section~\ref{sec5}).\medskip \\ 

\section{ The Higgs Sector of the MSSM4}
\label{sec5}
The effect of the fourth-generation on the lightest Higgs mass ($M_h$) is investigated. For our particular solution of 
($M_t,M_T,M_B,M_N,M_E,{\tilde{M}}_i$) that is consistent with perturbative unification, we see the upper bound $M_h^{max}$ increases through loop 
corrections from the extra particles. Normally, the Higgs sector of minimal supersymmetry models is fixed by two parameters $(i) \tan\beta$ and 
$(ii) M_A$ (the pseudoscalar Higgs mass). We fix $\tan\beta \sim 1.63$ from the requirement of perturbative unification (see figure~\ref{susyr}) and so 
we are left with the one free parameter $M_A$. Assuming this is to be included in the heavy Higgs sector ($M_H$) we vary 
500 GeV $\le |\mu| = M_A \le$ 1 TeV to be consistent with Eq.(\ref{limit}). \medskip \\
In the MSSM4, for $M_h \ge$ 100 GeV, the channel ($h \rightarrow \mbox{invisible}$), where invisible represents ($N\overline{N},E\overline{E},\tilde{N}
\overline{\tilde{N}}, \tilde{\chi}\overline{\tilde{\chi}}$), will open and would dominate over conventional $h \rightarrow b\overline{b}$ rates.
Exclusion limits will now come from the missing energy search ($e^+e^- \rightarrow Zh \rightarrow Z + \mbox{missing energy}$) that currently sets the 
lower bound at 114.4 GeV~\cite{invisible}. We therefore need to check, for our constrained set of parameters 
($M_T,M_B,M_N,M_E,{\tilde{M}}_i,\tan\beta$), that $M_h^{max}$ falls above this experimental lower bound. We shall employ the one-loop effective potential 
with contributions from the top/stop and fourth-generation fermions/sfermions. Since we are at low $\tan\beta$ we can ignore the contributions from the 
third generation bottom/sbottom masses.
\medskip \\
In general, the sfermion mass matrix can be written as
\begin{equation*}
        M_{\tilde{i}} = \left(\begin{array}{cc}
                M_{Si}^2 + M_i^2+\Delta_{Di} & M_i \left(A_i - \mu \cot \beta \right) \\
                M_i \left(A_i - \mu \cot \beta \right)  & M_{{\overline{S}}i}^2 + M_i^2+\Delta_{{\overline{D}}i} 
        \end{array} \right)
\end{equation*}
where $\Delta_D$ and $\Delta_{\overline{D}}$ represents the $D$-term contributions. We set the soft supersymmetry breaking parameters for the 
coloured squarks $M_{Si} = M_{{\overline{S}}i} = M_{col}$ ($i=t,T,B$) and, for a given $\tan\beta$, we vary $M_{col}$ within the range as plotted in 
figure~\ref{susyr}. For the sleptons $M_{Si} = M_{{\overline{S}}i} = |\mu|$ ($i=E,N$). We randomly vary the mixing parameters ($A_t, A_T, A_B, A_N, A_E$) 
and retain the maximum value returned for the lightest Higgs mass. As shown in previous studies of the Higgs sector, this will occur for 
$A \sim \sqrt{6} M_{col}$. Large mixings will induce light sparticles ($< M_Z$) which are required in the MSSM4 to provide the 
decay channels to the fourth-generation quarks. Since we are at low $\tan\beta$ ($\sim 1.63$) we require relatively heavy squarks in order to increase the 
Higgs mass above its experimental lower bound. We find that $M_h^{max} >$ 114 GeV for $M_{col} \ge$ 350 GeV. This is consistent with results from 
section~\ref{sec4}. The absolute upper bound (that occurs when $M_{col} =$ 500 GeV) is $M_h^{abs}\sim$ 120 GeV. 

\section{Conclusions}
\label{sec6}
We have seen that it is possible to incorporate a fourth-generation into a GUT model, requiring the existence of supersymmetric particles in order
to provide the necessary cancellations in precision fits. We constrain the fourth-generation masses to be $M_T =$ 107 Gev; $M_B =$ 105 GeV; $M_N =$ 46 GeV;
$M_E =$ 50 GeV. The MSSM4 SUSY spectrum needs to be relatively light in order to retain perturbative consistency to the unification scale 
($M_{col} \le$ 500 GeV). On the other hand ($M_{col} \ge$ 350 GeV) is set from the requirement that the lightest Higgs mass must be above its present 
experimental lower bound. In order to provide decay channels to the fourth-generation quarks, it might be that the LSP is the sneutrino $\tilde{N}$. This 
has implications for dark matter constraints. The SUSY spectrum needed to satisfy all constraints can not be obtained from MSUGRA scenarios with universal 
parameters at the unification scale.

\newcommand{\pd}{\partial}

\title*{Another Complex Bateman  Equation}
\author{%
D. B. Fairlie\thanks{E-mail:david.fairlie@durham.ac.uk}}
\institute{%
Department of Mathematical Sciences
University of Durham, Durham DH1 3LE}

\authorrunning{D.B. Fairlie}
\titlerunning{Another Complex Bateman  Equation}
\maketitle

\begin{abstract}
A further class of complex covariant field equations is investigated. 
These equations possess several common features: they may be solved,
or partially solved in terms of implicit functional relations, they 
possess an infinite number of inequivalent 
Lagrangians which vanish on the space of solutions of the equations 
of motion, they are invariant under linear transformations of the 
independent variables, and thus are signature-blind and are
consequences of first order equations of hydrodynamic type. 
\end{abstract}

\section{Introduction}
This paper is another in a series devoted to an investigation of simple equations exhibiting covariance of solutions. These equations have arisen in the study of generalisations of the Bateman equation \cite{gov}, in the equations arising from continuations of the String and Brane Lagrangians to the situation where the target space has fewer dimensions than the base space \cite{baker} and a complex form of these equations
\cite{fai}. The simplest example of these is a complexification of the Bateman equation.
What we have called the Complex Bateman equation is the following equation for a real function $\phi$ defined over the space of variables $(x_1,\ x_2;\ \bar x_1,\ \bar x_2)$;
\begin{equation}
\phi_{x_1}\phi_{\bar x_1}\phi_{x_2\bar x_2} + \phi_{x_2}\phi_{\bar x_2}\phi_{x_1\bar x_1}-\phi_{x_1}\phi_{\bar x_2}\phi_{\bar x_1 x_2}-\phi_{x_2}\phi_{\bar x_1}\phi_{x_1\bar x_2}\,=\,0.\label{complex1}
\end{equation}
(Here subscripts denote differentiation). This equation was shown to be
completely integrable \cite{chaundy}\cite{leznov}, with solution given by
solving for $\phi$ the following constraint upon two arbitrary functions of three variables, $F,\ G$;
\[ F(\phi;\, x_1,x_2)\,=\,G(\phi;\,\bar x_1,\bar x_2).\]
From the form of the solution, or from the equation itself, it is manifest that if $\phi$ is a solution, any function of $\phi$ will also be a solution and thus that the equation exhibits covariance. It is also invariant under separate diffeomorphisms of the pairs of variables $(x_1,\ x_2)$ and $(\bar x_1,\ \bar x_2)$. In fact a subclass of solutions is given by the sum of `holomorphic'  and `antiholomorphic' functions
\[\phi\,=\,f(x_1,x_2) +g(\bar x_1,\bar x_2).\]  
A general characteristic of such equations is that they possess an infinite number of inequivalent Lagrangians. The equations of motion are partially, or sometimes fully solveable in implicit form, as in the examples cited. The fully integrable eqautions arise from kinematical first order equations of hydrodynamic type.

\section{ Another Complex Bateman Equation}

Now there is another possibility for complexification; we could take instead
\begin{equation}
\bar\phi_x \phi_x\phi_{tt}-\bar\phi_x \phi_t\phi_{tx}-\bar\phi_t \phi_x\phi_{tx}+\bar\phi_t\phi_t\phi_{xx}\,=\,0,\label{complex2}
\end{equation}
together with its complex conjugate. These equations also exhibit covariance; $\phi$ may be replaced by any function of itself, and the same for $\bar\psi$ and the equations remain invariant.
Where do these equations come from?
Take the hydrodynamic equations
\begin{eqnarray}
\frac{\partial u}{\partial t} &=&v\frac{\partial u}{\partial x},\label{int1}\\
\frac{\partial v}{\partial t} &=&u\frac{\partial v}{\partial x},\label{int2}
\end{eqnarray}
and set $\displaystyle{u\,=\, \frac{\bar\phi_t}{\bar\phi_x},\ v\,=\, \frac{\phi_t}{\phi_x}}$, and the equation (\ref{complex2}), together with its complex conjugate are reproduced. Indeed, all that is necessary is to set
in an alternative reduction, $u=\bar\phi$ and $v=\phi$ and the same equations arise in consequence. 
These equations admit an infinite number of conserved quantities \cite{mulvey};
If $S_n$ denotes the symmetric polynomial of degree $n$ in $u,\ v$, then
\begin{equation}
\frac{\pd }{\pd t}S_n \,=\,\frac{\pd }{\pd x}(uv S_{n-1})\label{sym}
\end{equation} is a conservation law. This is easily proved by induction and from the
iterative definition: $S_n\,=\, u^n+vS_{n-1}$.                  
These equations can be integrated by the usual hodographic method of interchanging dependent and independent variables, where they become
\begin{eqnarray}
\frac{\partial x}{\partial v} &+&v\frac{\partial t}{\partial v}\,=\,0,\label{inv1}\\
\frac{\partial x}{\partial u} &+&u\frac{\partial t}{\partial u}\,=\,0,\label{inv2}
\end{eqnarray}
which can be solved in terms of two arbitrary functions $f,\ g$ to give
\[ t\,=\,f'(u)+g'(v);\ \ \ x\,=\,f(u)-uf'(u)+g(v)-vg'(v)\]
with primes denoting differentiation with respect to the argument. If $u=\bar\phi$ and $v=\phi$ this parametrisation is a  solution to the alternative complexification. Note that the requirements that $(t,\ x)$ be real imposes a further constraint upon the functions $(f,\ g)$. Of course, if $(\phi,\ \bar\phi)$ are treated as independent real functions, no such restriction exists. Now the second order equations (\ref{complex2}) are Poincar\'e invariant; indeed are covariant under general inhomogeneous linear transformations of the independent variables.
 This must be true also for the first order equations (\ref{inv1}),
(\ref{inv2}). They are clearly translation invariant; if $(t,\ x)$ transform as
\[t'\,=\,at+bx,\\\ x'\,=\,ct+dx,\]
then invariance will be maintained if 
\[u'\, =\, \frac{du-c}{a-bu},\ \ \ \ v'\, =\, \frac{dv-c}{a-bv}.\]
\section{2-dimensional Born-Infeld equation}
We may remark parenthetically that the same equations(\ref{int1}),(\ref{int2}) also yield the general solution to one form of the so-called Born-Infeld equation in two dimensions in light-cone co-ordinates \cite{mulvey}\cite{arik};
\[
\phi_x^2 \phi_{tt}+\phi_x^2 \phi_{tt}-(\lambda+2\phi_x\phi_t)\phi_{xt}\,=\,0.
\]
This is achieved by setting
\[ \frac{\pd\phi}{\pd x}\,=\,\frac{\sqrt{\lambda}}{\sqrt{u}-\sqrt{v}},\ \ \ \frac{\pd\phi}{\pd t}\,=\,\frac{\sqrt{\lambda uv}}{\sqrt{u}-\sqrt{v}}.\]
The integrability constraints upon these equations is just the Born-Infeld equation itself. 
Thus the primacy of the first order hydrodynamic equations is again manifest. This is a phenomenon which has been noticed before; that the same first order equations yield different second order ones depending upon the assumptions made about the dependency of the unknown functions in the first order equations upon the functions which enter into the second order equations \cite{intrev}\cite{fai}.

\section{Lagrangian}
The construction of a Lagrangian for (\ref{complex2}) follows along the lines of \cite{fai}. Introduce an auxiliary field $\psi$ and consider the singular Lagrangian
\begin{equation}
{\cal L} \,=\, \left(\frac{\partial\bar\phi}{\partial t}\frac{\partial\psi}{\partial x}-\frac{\partial\psi}{\partial t}\frac{\partial\bar\phi}{\partial x}\right)\frac{\frac{\partial\phi}{\partial t}}{\frac{\partial\phi}{\partial x}}.\label{lag}
\end{equation}
The equation of motion corresponding to variations in the field $\psi$ is simply
equation (\ref{complex2}). Similarly for the variations with respect to $\bar\phi$ we obtain
\begin{equation}
\psi_x \phi_x\phi_{tt}-\psi_x \phi_t\phi_{tx}-\psi_t \phi_x\phi_{tx}+\psi_t\phi_t\phi_{xx}\,=\,0,\label{complex3}
\end{equation}
i.e. a similar equation with $\bar\phi$ replaced with $\psi$. But the third equation, corresponding to variations with respect to $\phi$ is just
\begin{equation}
\frac{\partial}{\partial t}\left[(\bar\phi_t\psi_x-\bar\phi_x\psi_t)\left(\frac{1}{\phi_x}\right)\right]
-\frac{\partial}{\partial x}\left[(\bar\phi_t\psi_x-\bar\phi_x\psi_t)\left(\frac{\phi_t}{\phi_x^2}\right)\right]\,=\,0.\label{complex4}
\end{equation}
This is satisfied if $\psi$ is a function of $\bar\phi$; then equation (\ref{complex3}) is the same as equation (\ref{complex2}). Incidentally, we see here a situation which has been remarked upon before in the context of free field equations \cite{dbf}, and equations arising from Born-Infeld Lagrangians, namely that the Lagrangian itself is a constant, or else a divergence on the space of solutions of the equations of motion. It is also evident that 
the factor $\displaystyle{\frac{\frac{\partial\phi}{\partial t}}{\frac{\partial\phi}{\partial x}}}$ may be replaced by any homogeneous function of $\displaystyle{\left(\frac{\partial\phi}{\partial t},\ \frac{\partial\phi}{\partial x}\right)}$ of weight zero, without affecting the equations of motion.
 
\section{Multi-field Lagrangian}
The Lagrangian can be constructed along similar lines to that for the single field;
one choice is
\begin{equation}
{\cal L}=\frac{\pd(\bar\phi^1,\ \bar\phi^2,\ \theta)}{\pd(x_1, x_2, x_3)}\left(\frac{\frac{\pd(\phi^1,\ \phi^2)}{\pd(x_1,\ x_2)}}{\frac{\pd(\phi^1,\ \phi^2)}{\pd(x_1,\ x_3)}}\right)\\,+\, {\rm cc}.\label{newlag}
\end{equation}
Variation with respect to $ \theta$ gives a combination of the equations of motion for $\phi^1$ and $\phi^2$ and their complex conjugates; variations with respect to the  fields
$\bar\phi^1$ and $\bar\phi^2$  yields other linear combinations which together imply the following equations,where $j$ takes the values (1,\ 2);

\begin{equation}
\det\left|\begin{array}{ccccc}
0&0&\bar\phi^1_{x_1}&\bar\phi^1_{x_2}& \bar\phi^1_{x_3} \\
0&0&\bar\phi^2_{x_1}&\bar\phi^2_{x_2}& \bar\phi^2_{x_3} \\
\phi^1_{x_1}&\phi^2_{x_1}&\phi^j_{x_1x_1}&\phi^j_{x_1x_2}&\phi^j_{x_1x_3}\\
\phi^1_{x_2}&\phi^2_{x_2}&\phi^j_{x_2x_1}&\phi^j_{x_2x_2}&\phi^j_{x_2x_3}\\
\phi^1_{x_3}&\phi^2_{x_3}&\phi^j_{x_3x_1}&\phi^j_{x_3x_2}&\phi^j_{x_3x_3}\end{array}
\right|\,=\,0.\label{result}
\end{equation}
and that $\theta$ is a function of $\bar\phi^1$ and $\bar\phi^2$, in much the same manner as $\psi$ is a function of $\bar\phi$ in the single field case.
As in the case of a single pair of complex fields, these equations follow from
a set of hydrodynamic equations.
\begin{eqnarray}
\frac{\pd u^i}{\pd x_1}&+& v^1\frac{\pd u^i}{\pd x_3}\,+\, v^2\frac{\pd u^i}{\pd x_2}\,=\,0\ \ i=1,2\label{hydro1}\\
\frac{\pd v^i}{\pd x_1}&+& u^1\frac{\pd v^i}{\pd x_3}\,+\, u^2\frac{\pd v^i}{\pd x_2}\,=\,0\ \ i=1,2.\label{hydro4}
\end{eqnarray} 
Once again, these equations remain the same up to a constant factor under a general linear transformation of co-ordinates; this may be seen most easily if they are written in a 
homogeneous notation by introducing vectors $\xi^\mu,\ \eta^\mu;\ \mu\,=\,0,1,2$ such that
$\displaystyle{u^i\, =\, \frac{\xi^i}{\xi^0},\ \ v^i\, =\, \frac{\eta^i}{\eta^0}}$ so that the equations may be written as
\[\sum_0^2 \xi^\mu\frac{\pd v^i}{\pd x_\mu}\,=\, 0;\ \ \ \sum_0^2 \eta^\mu\frac{\pd u^i}{\pd x_\mu}\,=\,0,\]
making the invariance up to a factor  of the hydrodynamic equations under linear transformations of the co-ordinates and the vectors $\vec\xi,\ \vec\eta$ manifest.

Set $u^i\,=\,\phi_i$ and choose the following set of these equations and their derivatives:
\begin{eqnarray}
&&\phi^1_{x_1}+v^1\phi^1_{x_2}+v^2\phi^1_{x_3}\,=\,0\nonumber\\
&&\phi^2_{x_1}+v^1\phi^2_{x_2}+v^2\phi^2_{x_3}\,=\,0\nonumber\\
&&\phi^1_{x_1x_1}+v^1\phi^1_{x_1x_2}+v^2\phi^1_{x_1x_3}+v^1_{x_1}\phi^1_{x_2}+v^2_{x_1}\phi^1_{x_3}\,=\,0\nonumber\\
&&\phi^1_{x_1x_2}+v^1\phi^1_{x_2x_2}+v^2\phi^1_{x_2x_3}+v^1_{x_2}\phi^1_{x_2}+v^2_{x_2}\phi^1_{x_3}\,=\,0\nonumber\\
&&\phi^1_{x_1x_3}+v^1\phi^1_{x_2x_3}+v^2\phi^1_{x_3x_3}+v^1_{x_3}\phi^1_{x_2}+v^2_{x_3}\phi^1_{x_3}\,=\,0.\nonumber
\end{eqnarray}
Eliminate the first derivatives ($\phi^1_{x_2},\ \phi^1_{x_3}$) from the final three equations ,
solve the first pair of equations for $(v^1,\ v^2)$ and subsitute in the undifferentiated terms of the result, setting $v^1 =\bar\phi^1,\ v^2=\bar\phi^2$. The ensuing equation is just  one member of (\ref{result}). 
\section{The fundamental hydrodynamic equations}
All second order integrable equations of the type discussed here and in earlier work
\cite{leznov},\cite{fai} are consequences of the general first order equations, for which an implicit solution may be constructed  following Leznov \cite{leznov2}.
Consider a $2n$ dimensional Euclidean space with indepenent co-ordinates $(x_i,\ \bar x_i,\ i=1\dots n)$ and construct the differential operators
\begin{equation}
D\,=\,\frac{\pd}{\pd x_n} +\sum^{n-1}_{j=1}u^j\frac{\pd}{\pd x_j},\ \ \ \
\bar D\,=\,\frac{\pd}{\pd\bar x_n} +\sum^{n-1}_{j=1}v^j\frac{\pd}{\pd\bar x_j}
\label{hoo}
\end{equation}
Since $D\bar x_i\,=\,0,\ \bar D x_i\,=\,0,$  $D$ may be considered a holomorphic differential operator and $\bar D$  an antiholomorphic operator.. Now imposing the zero curvature condition, $[D,\ \bar D]\,=\,0$ requires that
\begin{equation}
Dv^i\,\equiv\, v^i_{x_n}+\sum u^jv^i_{x_j}\,=\,0,\ \ \ \bar Du^i\,\equiv\, u^i_{\bar x_n}+\sum v^ju^i_{\bar x_j}\,=\,0.\label{lezimp}
\end{equation}
These are the general first order equations mentioned above. Since $D,\ \bar D$ commute, these equations imply that $\bar D v^i$ is a solution to the same equation as $v^i$ satisfies. The integration of the equation $D\bar D v^i\,=\,0$ requires that $\bar D v^i$
is a general anti-holomorphic function, hence,
\begin{equation}
\bar D v^i\,=\, v^i_{\bar x_n}+\sum v^jv^i_{\bar x_j}\,=\,V^i(v^k;\bar x_l),\ \ \  Du^i\,=\, u^i_{ x_n}+\sum u^ju^i_{\bar x_j}\,=\,U^i(u^k;x_l).\label{lezimp2}
\end{equation}
Indeed $f(\bar D)v^i$, for arbitrary differentiable $f$ is also a solution to the equation
for $v^i$.
Suppose now we take $(n-1)$ functions $\phi^i$ constrained by the $(n-1)$ relations
\begin{equation}
Q^i(\phi^j;x_k)\,=\,P^i(\phi^j;\bar x_k),\  i=1\dots n-1.\label{relate}
\end{equation}
The arbitrary functions $Q^i,\ P^i$ depend upon $(2n-1)$ co-ordinates.
They imply straightforwardly
\begin{equation}\phi^j_{x_k}\,=\, (P^i_{\phi^j}-Q^i_{\phi^j})^{-1}Q^i_{x_k},\
\phi^j_{\bar x_k}\,=\, -(P^i_{\phi^j}-Q^i_{\phi^j})^{-1}P^i_{\bar x_k}.\label{cons}
\end{equation}
Suppressing the vector indices, suppose $u$ is a function  $u(\phi, x)$  and
$v$ is a function  $u(\phi,\bar x)$.  Then the equations (\ref{lezimp}) imply that
\begin{equation}
D\phi^j\,=\,\phi^j_{x_n}+\sum_1^{n-1}v^k\phi_{x_k}\,=\,0,\\\
\bar D \phi^j\,=\,\phi^j_{\bar x_n}+\sum_1^{n-1}u^k\phi_{\bar x_k}\,=\,0,\label{lezimp3}
\end{equation}
In other words this requires that $\phi^j$ be both holomorphic and antiholomorphic in this definition of holomorphicity.
Subsituting the derivatives from (\ref{cons}) and multiplying on the left by the matrix $(P_\phi-Q_\phi)$ the equations become
\begin{equation}
Q^j_{x_n}+\sum_1^{n-1}v^kQ^j_{x_k}\,=\,0,\\\
P^j_{\bar x_n}+\sum_1^{n-1}u^kP^j_{\bar x_k}\,=\,0,\label{lezimp4}
\end{equation}
which leads to the identifications
\begin{equation}
v\,=\, -(Q_x)^{-1}Q_{x_n}, \ \ \ u\,=\, -(P_{\bar x})^{-1}P_{\bar x_n}.\label{defu}
\end{equation}
In consequence any function of $\phi,\bar x$ ($\phi, x$) will be annihilated by
$D\  (\bar D).$ In particular
\begin{equation}
D\phi\,=\,\bar D\phi \,=\,DQ\,=\,\bar DP\,=\,DQ\,=\,\bar D P\,=\,0.\label{holo}
\end{equation}
The last two results follow a forteriori from the equality $P\,=\,Q$.
The functions $\phi^j$ satisfy the multi-field complex Bateman equation
\cite{fai}.
\section{Partially integrable covariant equations.}
It appears likely that in the case where the difference between the number of dimensions of the base space exceeds that of the target space by more than one, the equations of motion are only partially integrable, though this is by no means a definitive conclusion.
In the case of one field dependent on three co-ordinates, the equation which results from the  Euclidean Lagrangian
\[ {\cal L} =\sqrt{\phi_{t}^2+ \phi_{x}^2+ \phi_{y}^2}\] 
is
\[
\phi_{tt}(\phi_{x}^2+ \phi_{y}^2)+\phi_{xx}(\phi_{y}^2+
\phi_{t}^2)+\phi_{yy}(\phi_{t}^2+ \phi_{x}^2)
=2\phi_{tx}\phi_t\phi_x+2\phi_{yt}\phi_y\phi_t+2\phi_{xy}\phi_x\phi_y.
\]

This equation possesses a large class of solutions given implicitly by solving
\[tF(\phi)+xG(\phi)+yK(\phi)\,=\,{\rm constant}\]
for $\phi$.
It comes from  the following first order system;
\begin{eqnarray}
uu_x+vv_x&=&u_t+v_y+v^2u_t-uv(u_y+v_t)+u^2v_y\nonumber\\
uv_x-vu_x &=& v_t-u_y,\nonumber
\end{eqnarray}
where $\displaystyle{u\,=\, \frac{\phi_t}{\phi_x},\ v\,=\, \frac{\phi_y}{\phi_x}}$.
This construction suggests further analysis to try to determine whether the system is fully integrable or not. It is also not known whether there exist other Lagrangian formulations of these equations. 

\section*{Acknowledgement}
The author is indebted to the Leverhulme Trust for the award of an 
Emeritus Fellowship and to the Clay Mathematics Institute for
employment during the investigations reported here. 

\title*{Non-Associative Loops for Holger Bech Nielsen}
\author{%
P.H. Frampton$^{(a)}$, S.L. Glashow$^{(b)}$,
T.W. Kephart$^{(c)}$ and R.M. Rohm$^{(a)}$}
\institute{%
{}$^{(a)}$ Department of Physics and Astronomy,
University of North Carolina,
Chapel Hill, NC 27599-3255.
{}$^{(b)}$ Department of Physics, Boston University,%
Boston, MA 02215.
{}$^{(c)}$ Department of Physics and Astronomy, Vanderbilt University,\\
Nashville, TN 37235.}

\authorrunning{P.H. Frampton}
\titlerunning{Non-Associative Loops for Holger Bech Nielsen}
\maketitle

\begin{abstract}
Finite groups are of the  greatest importance in science.
Loops are a simple generalization of finite groups: they
share all the group axioms except for the requirement
that the binary operation be associative.
The least loops that are not themselves groups are
those of order five. We offer a brief discussion of these loops
and challenge the reader (especially Holger)
to find  useful applications
for them  in physics.
\end{abstract}


\section{Introduction}


Many physical systems have symmetries, and groups are the natural
mathematical objects to describe those symmetries (finite groups for
discrete symmetries and infinite continuous groups for continuous
symmetries). If the elements of a group act independently, then the
group is
abelian; if not, it is non-abelian and commutativity amongst the group
elements is lost. For discrete groups, this corresponds to an asymmetry
of the group multiplication table about its principal diagonal, {\it i.
e.,} $ab\neq ba$ for all $a$ and $b\in G$. However, group multiplication
is associative by definition,
\begin{equation}
(ab)c=a(bc)\,,
\end{equation}
and the concept of nonassociative operations  \cite{Schafer} has played a limited role
in science. Nevertheless, it has not been totally absent. Its main
point of entry into physics has been through octonions. Also called
octaves or Cayley numbers, they define
the only division algebra aside from
the real, complex and quarternionic numbers.
An early, but seemingly fruitless, application
of non-associativity in physics is
an octonionic version of quantum mechanics formulated
by Jordan, von Neumann, and Wigner\cite{J,JVW,A}. Attempts have been made
to use
octonions in particle physics to describe quark structure and other aspects of
internal structure.  For reviews see \cite{GunaydinGursey,Costa,Lohmus}. There
are also  an eight-dimensional
octonionic instantons \cite{Grossman,Fubini:1985jm} and applications to
superstrings \cite{Duff:1990wu,Harvey:1991eg}.
Here we observe that the minimal
non-associative  structures are not octonions, but objects called loops.
Let us first define them.


\section{Loops}


A {\bf loop} of order $n$
is a set $L$ of $n$ elements
with a binary operation \cite{Bruck} such that for $a$ and
$b$ elements of $L$, the equations
\begin{equation}
ax=b ~~~~~ and ~~~~~ ya=b
\label{binary}
\end{equation}
each has a unique  solution in $L$.  Furthermore, a loop possesses
an identity element $e$ which satisfies:
\begin{equation}
ex=xe=x\;\;\forall \;x\in L\,
\label{identity}
\end{equation}
The conditions Eqs.(\ref{binary}) and (\ref{identity})
imply that the multiplication table is a Latin square \cite{DK,LM,Heydayat}.
The multiplication table of a finite group is such
a Latin square, which was defined by Euler as a
square matrix with $n^{2}$ entries of $n$ different elements, none
occuring twice in the same row or column.

Any Latin square whose first row and column are identical defines a {\it
loop} whose upper-left entry is the identity element.  It follows that
any Latin square uniquely defines a loop, although different Latin squares
may define isomorphic loops.  This is because a Latin square remains a Latin
square under any permutation of its columns.  Thus, one can rearrange
any Latin square so that one row is identical to one column.  Once this is
done, that row and column label the elements of the loop and their common
element is the identity element.


A system whose multiplication table has non-identical first
row and column is a quasi-group which
is like a loop but which lacks the identity element
of Eq. (\ref{identity}). We do not consider these structures here.


In contrast to a group multiplication table, the binary opearation
defined by a  Latin square need not
be associative. However, all loops corresponding to
Latin squares with
$n\le 4$ satisfy equation (1). They yield the  groups
$I$, $Z_{2}$ and
Z$_{3}$
at orders 1, 2, and 3, and either
$Z_{2}\times Z_{2}$ or  $Z_{4}$ at order 4.


The situation becomes more
interesting at $n=5$, for which there are five distinct loops. One of
these is
the group $Z_5$. The remaining four are non-associative loops.
For  $n = 6$, there are two
groups, $Z_{2}\times Z_{3}$ and $D_{3}$, and  107 non-associative loops.


The number of non-associative loops rises {\it very} rapidly with $n$ and is
known only for small values.
The number of reduced Latin squares (those in the form with identical
first row and first
column as in all the examples below) is known to be 
9,408;\  16,942,080;\  535,281,401,856;\ 
 377,597,570,964,258,816\  and 
$$7,580,721,483,160,132,811,489,280$$
 at orders
n = 6; 7; 8; 9 and 10 respectively. For
n=11 the number of reduced Latin squares, and
hence the (smaller) number of non-associative loops which corresponds to the
number of isomorphism classes of Latin squares which contain at least
one reduced Latin square per class, is not yet known (see {\it e.g}
\cite{DK,Heydayat}).


Loops
are known to arise in the geometry of projective planes \cite{Pickert},
in combinatorics,
in knot theory \cite{Bar-Natan} and in non-associative algebras, but
have yet to play a role  in physics.
Thus  we present all the $n=5$ cases and some (not all!)
of the n=6 non-associative loops as a challenge to Holger and
others, who may find them to be interesting and useful for reasons
too subtle to have been revealed  to us.


We begin by presenting
all of the five $n=5$ multiplication tables (see p. 129 of\cite{DK})
in a form familiar from group theory. Case (1a) is the group $Z_5$
(the fifth roots of unity)
whilst  the other four are inequivalent non-associative n=5 loops.
 Case (1b) is special in that the square of any
element is the identity element.
As we discuss is \S 3, all 5-loops define commutation algebras that
satisfy the Jacobi identity.
\bigskip
\bigskip

\bigskip
\bigskip
\bigskip
\bigskip
\bigskip

$
\begin{array}{llllll}
\times  & \vline ~~ 1 & 2 & 3 & 4 & 5 \\
\hline 1  & \vline ~~ 1 & 2 & 3 & 4 & 5 \\
2 & \vline ~~ 2 &3 & 4 & 5 & 1 \\
3 & \vline ~~ 3 & 4 & 5 & 1 & 2 \\
4 & \vline ~~ 4 & 5 & 1 & 2 & 3 \\
5 & \vline ~~ 5 & 1 & 2 & 3 & 4
\end{array}
$ $\; \;
\begin{array}{llllll}
\times_{1} & \vline ~~ 1 & 2 & 3 & 4 & 5 \\
\hline 1 & \vline ~~ 1 & 2 & 3 & 4 & 5 \\
2 & \vline ~~ 2 & 1 & 4 & 5 & 3 \\
3 & \vline ~~ 3 & 5 & 1 & 2 & 4 \\
4 & \vline ~~ 4 & 3 & 5 & 1 & 2 \\
5 & \vline ~~ 5 & 4 & 2 & 3 & 1
\end{array}
$ $\;\;
\begin{array}{llllll}
\times_{2} & \vline ~~ 1 & 2 & 3 & 4 & 5 \\
\hline 1 & \vline ~~ 1 & 2 & 3 & 4 & 5 \\
2 & \vline ~~ 2 & 1 & 5 & 3 & 4 \\
3 & \vline ~~ 3 & 4 & 2 & 5 & 1 \\
4 & \vline ~~ 4 & 5 & 1 & 2 & 3 \\
5 & \vline ~~ 5 & 3 & 4 & 1 & 2
\end{array}
$ $\;\;$

(1a)\ \ \ \ \ \ \ \ \ \ \ \ \ \ \ \ \ \ \ \ \ \ \ \ \ \ \ (1b)\ \ \ \ \
\ \
\ \ \ \ \ \ \ \ \ \ \ \ \ \ \ \ \ \ \ \ \ \ \ (1c)

\bigskip
\bigskip
\bigskip
\bigskip
\bigskip
\bigskip
\bigskip
\bigskip
\bigskip
\bigskip

$
\begin{array}{llllll}
\times _{3} & \vline ~~ 1 & 2 & 3 & 4 & 5 \\
\hline
1 & \vline ~~ 1 & 2 & 3 & 4 & 5 \\
2 & \vline ~~ 2 & 1 & 4 & 5 & 3 \\
3 & \vline ~~ 3 & 4 & 5 & 1 & 2 \\
4 & \vline ~~ 4 & 5 & 2 & 3 & 1 \\
5 & \vline ~~ 5 & 3 & 1 & 2 & 4
\end{array}
$ $\;\;
\begin{array}{llllll}
\times _{4} & \vline ~~ 1 & 2 & 3 & 4 & 5 \\
\hline
1 & \vline ~~ 1 & 2 & 3 & 4 & 5 \\
2 & \vline ~~ 2 & 3 & 4 & 5 & 1 \\
3 & \vline ~~ 3 & 5 & 2 & 1 & 4 \\
4 & \vline ~~ 4 & 1 & 5 & 3 & 2 \\
5 & \vline ~~ 5 & 4 & 1 &
2 & 3
\end{array}
$ $\;\;$

(1d)\ \ \ \ \ \ \ \ \ \ \ \ \ \ \ \ \ \ \ \ \ \ \ \ \ \ \ \ \ \ (1e)

\vfill

\newpage

\bigskip
\bigskip

\noindent  Here we present three illustrative examples of the 107
distinct non-associative 6-loops. Each of these defines a commutation
algebra that satisfies the Jacobi identity:

\bigskip
\bigskip
\bigskip
\bigskip

$
\begin{array}{lllllll}
\times _{1}^{6} & \vline ~~ 1 & 2 & 3 & 4 & 5 & 6 \\
\hline
1 & \vline ~~ 1 & 2 & 3 & 4 & 5 & 6 \\
2 & \vline ~~ 2 & 1 & 4 & 3 & 6 & 5 \\
3 & \vline ~~ 3 & 5 & 1 & 6 & 4 & 2 \\
4 & \vline ~~ 4 & 6 & 5 & 1 & 2 & 3 \\
5 & \vline ~~ 5 & 3 & 6 & 2 & 1 & 4 \\
6 & \vline ~~ 6 & 4 & 2 & 5 & 3 & 1
\end{array}
$\qquad
$
\begin{array}{lllllll}
\times _{2}^{6} & \vline ~~ 1 & 2 & 3 & 4 & 5 & 6 \\
\hline 1 & \vline ~~ 1 & 2 & 3 & 4 & 5 & 6 \\
2 & \vline ~~ 2 & 1 & 6 & 5 & 3 & 4 \\
3 & \vline ~~ 3 & 6 & 1 & 2 & 4 & 5 \\
4 & \vline ~~ 4 & 5 & 2 & 1 & 6 & 3 \\
5 & \vline ~~ 5 & 3 & 4 & 6 & 1 & 2 \\
6 & \vline ~~ 6 & 4 & 5 & 3 & 2 & 1
\end{array}
$

\bigskip
\bigskip
\bigskip
\bigskip

$
\begin{array}{lllllll}
\times _{3}^{6} & \vline ~~ 1 & 2 & 3 & 4 & 5 & 6 \\
\hline
1 & \vline ~~ 1 & 2 & 3 & 4 & 5 & 6 \\
2 & \vline ~~ 2 & 1 & 5 & 6 & 4 & 3 \\
3 & \vline ~~ 3 & 4 & 1 & 5 & 6 & 2 \\
4 & \vline ~~ 4 & 3 & 6 & 1 & 2 & 5 \\
5 & \vline ~~ 5 & 6 & 2 & 3 & 1 & 4 \\
6 & \vline ~~ 6 & 5 & 4 & 2 & 3 & 1
\end{array}
$  $\;\;$ $ \\
$

\newpage

The following two examples of non-associative n=6 loops
define commutator algebras that fail to satisfy the Jacobi identity:

\bigskip
\bigskip
\bigskip
\bigskip

$
\begin{array}{lllllll}
\times _{4}^{6} & \vline ~~ 1 & 2 & 3 & 4 & 5 & 6 \\
\hline
1 & \vline ~~ 1 & 2 & 3 & 4 & 5 & 6 \\
2 & \vline ~~ 2 & 1 & 4 & 5 & 6 & 3 \\
3 & \vline ~~ 3 & 6 & 1 & 2 & 4 & 5 \\
4 & \vline ~~ 4 & 5 & 6 & 1 & 3 & 2 \\
5 & \vline ~~ 5 & 3 & 2 & 6 & 1 & 4 \\
6 & \vline ~~ 6 & 4 & 5 & 3 & 2 & 1
\end{array}
$\qquad
$
\begin{array}{lllllll}
\times _{5}^{6} & \vline 1 & 2 & 3 & 4 & 5 & 6 \\
\hline
1 & \vline 1 & 2 & 3 & 4 & 5 & 6 \\
2 & \vline 2 & 6 & 5 & 1 & 3 & 4 \\
3 & \vline 3 & 1 & 4 & 2 & 6 & 5 \\
4 & \vline 4 & 3 & 6 & 5 & 1 & 2 \\
5 & \vline 5 & 4 & 1 & 6 & 2 & 3 \\
6 & \vline 6 & 5 & 2 & 3 & 4 & 1
\end{array}
$


\bigskip
\bigskip

\section{Physics Challenge}


In this section  we  suggest a few possible applications of loops
to physics. We challenge the reader
to develop  a useful application to physics from these notions or any
others.
First, it may be useful to point out that the condition of associativity
which is required of groups is a natural condition for symmetry
transformations, since it is an automatic consequence of the composition
of mappings.  Such mappings between particle states, or between states
in a Hilbert space, give rise to the familiar symmetry groups.  Groups
themselves act as transformation groups on themselves, and this action
is consistent with the group action because of associativity.  For a
finite group, for instance, the multiplication table of the group gives a
representation of the group as a set of n permutations
\begin{equation}
g_i(g_j) = \pi_i(g_j) = g_i \times g_j
\end{equation}
and clearly
\begin{equation}
(g_i \times g_j) (g) = g_i (g_j (g) )
\end{equation}
is a consequence of associativity.  For a loop multiplication table,
we again get a set of permutations, but the multiplication by
composition of the permutations is not consistent with the loop multiplication, for
the same reason.  Thus our intuition about groups as transformations may
be a hindrance in interpreting loops in physical applications.


I. One could imagine defining a group product in a way similar to the
definition of a $q$-deformed bosonic commutator algebra where a
fermionic
anticommutator piece is added, $i.e$., here we would consider an
associative
group algebra product $a\cdot b$ deformed by a non-associative loop
algebra
piece $a*b$ to generate an algebra with product
\begin{equation}
a\otimes b=(1-\epsilon )a\cdot b+ \epsilon a*b.
\end{equation}
This may be a way of introducing dissipation or decoherence into a
system.


II. We could try to start with a space $S$ and factor out a loop $L$
similar to an orbifold construction, where a finite group is factored
out.
Such an  $S/L$ loopifold could have application in string theory
although its implementation is made non-trivial by the absence of
matrix representations of the loop.


III.  It is also a consequence of nonassociativity that a representation
of
a loop in terms of linear transformations is never faithful; since
matrix multiplication associates, the nonassociativity must be
annihilated
in any map from the loop to operators on a vector space.  In order
to bypass this obstacle, it is useful to construct  an object familiar
to finite group representation theory, a loop (or group) algebra.  We
take formal linear combinations of the elements of the loop (with
coefficients in ${\bf R}$ or ${\bf C}$), with
multiplication carried out termwise according to the loop multiplication
table.   This procedure defines a vector space whose basis elements are
the loop elements and a natural (but non-associative) multiplication
operation between vectors. We denote the non-associative algebra
corresponding to a group $L$ as $A(L)$.

In particular, the loop elements themselves act as linear
transformations
on $A$ via either left- or right-multiplication.
If $L$ is a group, this action admits  the decomposition
of $A$ into subspaces corresponding to
the irreducible representations
of the group.  For non-associative loops,
the situation is less clear because   matrix
multiplication does not follow the loop multiplication.

However, the algebra associated with a loop has another interesting
property.  To any $A(L)$ (associative or not), we may define the bracket
of two elementa $a\,, \in A$ as
\begin{equation}
[a,b] = a \times b - b \times a\;.
\end{equation}
It is evident that this operation yields an element of $A$, and
furthermore
that it is antisymmetric: $[a,b]=-[b,a]$. However, for non-associative
loops it is far from evident that the bracket operation satisfies the Jacobi
identity,
\begin{equation}
[a,[b,c]] + [b,[c,a]] + [c,[a,b]]=0\;.
\label{jacobi}
\end{equation}
Eq. (\ref{jacobi})  is always satisfied if $L$ is a group.  Every finite
group, through the commutator algebra thus defined, corresponds uniquely to
a Lie algebra.

What we find fascinating is that some (but not all) non-associative
loops do yield bracket operations that  satisfy the Jacobi identity,
thereby defining
commutator algebras that are Lie algebras.
Curiously (and as indicated above),
 all of the non-associative loops with $n=5$
are of this class, but only  some of those with n=6.


One could imagine using loops as objects to replace flavor
or horizontal symmetries in particle physics, or using them as
``pregroups."
For example, let us rewrite the $\times _{1}$ loop of Table (1b) in the
form


$
\begin{array}{llllll}
\times _{1} & 1 & a & b & c & d \\
1 & 1 & a & b & c & d \\
a & a & 1 & c & d & b \\
b & b & d & 1 & a & c \\
c & c & b & d & 1 & a \\
d & d & c & a & b & 1
\end{array}
$
\bigskip
\bigskip

\noindent For this case, the bracket operation of the loop
 algebra satisfies the Jacobi identity.
The structure  of the algebra is revealed  in terms of the
linear combinations

\[
K=(a+b+c+d)/2
\]

\[
u_1=(a+b-c-d)/2
\]

\[
u_2=(a-b+c-d)/2
\]

and

\[
u_3=(a-b-c+d)/2
\]
The bracket operation reveals that $K$ (and the identity element)
commute with the other
operations and the $u_i$ \ satisfy the $su_2 $\ algebra
$ [u_i,u_j] = -2\epsilon_{ijk}u_k$.
The nonassociativity lurks
still in the products of these elements, resembling a twisted version of
the Pauli matrices; in this basis they are
given by $K\times u_i=u_i\times K=-u_i/2$, $u_1\times u_2 =3u_3/2$,
$u_2\times u_1 =-u_3/2$, and cyclic permutations of these.
We also have the relations $K^{2}=1+\frac{3K}{2}$ and
$ u_i^{2}=1-K/2$.  It is interesting to note that the combination
$1-K/2$ commutes and associates with the other elements, and the
relation $\Sigma_i u_i^{2}=3(1-K/2) $\ suggests an interpretation
as a Casimir operator for the  $su_2 $; we leave this and other
details for the interested reader to interpret and, hopefully,
apply to physics.



\section*{Acknowledgements}

PHF and RMR acknowledge support by the US Department of Energy under
grant number DE-FG02-97ER-41036. The work of
SLG was supported in part by the National Science Foundation under grant
number NSF-PHY-0099529. TWK was supported by DOE grant number
DE-FG05-85ER-40226.


\newcommand{\holhs}[1]{\hspace{#1 em}}
\newcommand{\dd}[2]{\frac{\partial{#1}}{\partial{#2}}}
\newcommand{\ag}{\alpha}
\newcommand{\bg}{\beta}
\newcommand{\gam}{\gamma}
\newcommand{\del}{\delta}
\newcommand{\eps}{\epsilon}
\newcommand{\ve}{\varepsilon}
\newcommand{\zg}{\zeta}
\newcommand{\thg}{\theta}
\newcommand{\kg}{\kappa}
\newcommand{\lb}{\lambda}
\newcommand{\sg}{\sigma}
\newcommand{\rg}{\rho}
\newcommand{\fg}{\phi}
\newcommand{\vf}{\varphi}
\newcommand{\og}{\omega}
\newcommand{\Gam}{\Gamma}
\newcommand{\Del}{\Delta}
\newcommand{\Fg}{\Phi}
\newcommand{\Sg}{\Sigma}
\newcommand{\Og}{\Omega}
\newcommand{\Lb}{\Lambda}
\newcommand{\bxi}{\bar{\xi}}
\newcommand{\tu}{\tilde{u}}
\newcommand{\tx}{\tilde{x}}
\newcommand{\tf}{\tilde{f}}
\newcommand{\tG}{\tilde{G}}
\newcommand{\tT}{\tilde{T}}
\newcommand{\tpr}{t^{\prime}}
\newcommand{\lh}{\left(}
\newcommand{\rh}{\right)}
\newcommand{\ld}{\left.}
\newcommand{\rd}{\right.}
\newcommand{\pl}{\partial}
\newcommand{\nb}{\nabla}
\newcommand{\tg}{\mbox{\,tg\,}}
\newcommand{\ctg}{\mbox{\,ctg\,}}
\newcommand{\der}{\partial}
\newcommand{\Der}{D}
\newcommand{\sDer}{\slashed{D}}
\newcommand{\sder}{\slashed{\partial}}

\title*{Particles, Fluids and Vortices}
\author{%
J.W.\ van Holten\thanks{E-mail:t32@nikhef.nl}}
\institute{%
NIKHEF, Amsterdam NL}

\authorrunning{J.W.\ van Holten}
\titlerunning{Particles, Fluids and Vortices}
\maketitle

\begin{abstract}
Classical particle mechanics on curved spaces is related
to the flow of ideal fluids, by a dual interpretation of the Hamilton-Jacobi
equation. As in second quantization, the procedure relates the description
of a system with a finite number of degrees of freedom to one with infinitely
many degrees of freedom. In some two-dimensional fluid mechanics models a
duality transformation between the velocity potential and the stream function
can be performed relating sources and sinks in one model to vortices in the
other. The particle mechanics counterpart of the dual theory is reconstructed.
In the quantum theory the strength of sources and sinks, as well as vorticity
are quantized; for the duality between theories to be preserved these
quantization conditions must be related. 
\end{abstract}

\section{Particles \label{s1}}

The free motion of a classical particle with unit mass, moving in a smooth
space with metric $g_{ij}(x)$ is described by the Lagrangean
\begin{equation}
L = \frac{1}{2}\, g_{ij}(x) \dot{x}^i \dot{x}^j,
\label{1.1}
\end{equation}
where as usual the overdot represents a time-derivative. The Euler-Lagrange
equations imply that the particle moves on a geodesic:
\begin{equation}
\frac{D^2 x^i}{Dt^2} =
 \ddot{x}^i + \Gam_{jk}^{\;\;\;\;i} \dot{x}^j \dot{x}^k = 0.
\label{1.2}
\end{equation}
The canonical formulation of this theory is constructed in terms of the momenta
\begin{equation}
p_i = \frac{\der L}{\der \dot{x}^i} = g_{ij} \dot{x}^j,
\label{1.3}
\end{equation}
and the hamiltonian
\begin{equation}
H = \frac{1}{2}\, g^{ij} p_i p_j.
\label{1.4}
\end{equation}
The time-development of any scalar function $F(x,p)$ of the phase-space
co-ordin\-ates is then determined by the Poisson brackets
\begin{equation}
\frac{dF}{dt} = \left\{ F, H \right\} = \frac{\der F}{\der x^i}
 \frac{\der H}{\der p_i} - \frac{\der F}{\der p_i} \frac{\der H}{\der x^i}.
\label{1.5}
\end{equation}
In particular the Hamilton equations themselves read
\begin{equation}
\dot{x}^i = \frac{\der H}{\der p_i}, \holhs{2}
\dot{p}_i = - \frac{\der H}{\der x^i}.
\label{1.6}
\end{equation}
A third formulation of the classical theory is provided by Hamilton's
principal function\footnote{The terminology follows ref.\cite{goldstein}.}
$S(x,t)$, which is the solution of the partial differential equation
\begin{equation}
\frac{\der S}{\der t}\, = - H(x, p= \nabla S).
\label{1.7}
\end{equation}
For the case at hand this Hamilton-Jacobi equation takes the form
\begin{equation}
\frac{\der S}{\der t}\, = - \frac{1}{2}\, g^{ij}\, \nabla_i S\, \nabla_j S.
\label{1.8}
\end{equation}
Particular solutions $S$ are provided by the action for classical paths
$x^i(\tau)$ obeying the Euler-Lagrange equation (\ref{1.2}), starting at
time $\tau = 0$ at an initial point $x^i(0)$, and reaching the point
$x^i(t)=x^i$ at time $\tau = t$:
\begin{equation}
S(x,t) = \left. \int_0^t d\tau L(x,\dot{x})\, \right|_{x^i(\tau)}.
\label{1.9}
\end{equation}
An example of the class of theories of this type is that of a particle
moving on the surface of the unit sphere, $S^2$. A convenient co-ordinate
system is provided by the polar angles $(\thg, \vf)$, in terms of which
\begin{equation}
L(\thg, \vf) = \frac{1}{2}\, \lh \dot{\thg}^2 + \sin^2 \thg\, \dot{\vf}^2
 \rh,
\label{1.10}
\end{equation}
for a particle of unit mass. The corresponding hamiltonian is
\begin{equation}
H = \frac{1}{2}\, \lh p_{\thg}^2 + \frac{p_{\vf}^2}{\sin^2 \thg} \rh =
 \frac{{\bf J}^2}{2},
\label{1.11}
\end{equation}
with the momenta and velocities related by
\begin{equation}
p_{\thg} = \dot{\thg}, \holhs{2} p_{\vf} = \sin^2 \thg\, \dot{\vf}.
\label{1.12}
\end{equation}
The second equality (\ref{1.11}) relates the hamiltonian to the Casimir invariant
of angular momentum, the components of which are constants of motion given by
\begin{equation}
J_x = - \sin \vf\, p_{\thg} - \cos \vf \ctg \thg\, p_{\vf}, \holhs{1.5}
J_y = \cos \vf\, p_{\thg} - \sin \vf \ctg \thg\, p_{\vf}, \holhs{1.5}
J_z = p_{\vf}.
\label{1.13}
\end{equation}
The geodesics on the sphere are the great circles; they can be parametrized by
\begin{equation}
\cos \thg(\tau) = \sin \ag \sin \og (\tau - \tau_*), \holhs{2}
 \tg(\vf(\tau) - \vf_*) = \cos \ag\, \tg \og (\tau - \tau_*),
\label{1.14}
\end{equation}
where $\ag$ is a constant, and $\tau_*$ and $\vf_*$ are the time and
longitude at which the orbit crosses the equator: $\thg_* = \pi/2$.
On these orbits the angular frequency is related to the total angular
momentum by
\begin{equation}
\og^2 =  2H = {\bf J}^2,
\label{1.14.0}
\end{equation}
Observe that, for an orbit reaching the point with co-ordinates $(\thg, \vf)$
at time $\tau_* + t$, the following relations hold:
\begin{equation}
\begin{array}{l}
\cos \og = \sin \thg\, \cos (\vf - \vf_*), \holhs{2}
\sin \og t = \sqrt{ 1 - \sin^2 \thg\, \cos^2 (\vf - \vf_*) }, \\
 \\
\sin \ag = \displaystyle{ \frac{\cos \thg}{\sqrt{ 1 - \sin^2 \thg \cos^2 (\vf - \vf_*) }}.}
\end{array}
\label{1.14.1}
\end{equation}
The last equation implicitly describes the orbit $\thg(\vf)$, defining a great
circle which cuts the equator at $\thg = \thg_* = \pi/2$ and $\vf = \vf_*$,
at an angle $\ag$ defined by the direction of the angular momentum:
\begin{equation}
\frac{J_z}{\sqrt{ {\bf J}^2 }} = \cos \ag, \holhs{2} \frac{J_{\perp}}{\sqrt{
 {\bf J}^2 }} = \sin \ag, \holhs{1} J_{\perp} = \sqrt{J_x^2 + J_y^2}.
\label{1.14.3}
\end{equation}
The Hamilton-Jacobi equation for this system reads
\begin{equation}
\frac{\der S}{\der t} = - \frac{1}{2}\, \left[ \lh \frac{\der S}{\der \thg}
 \rh^2 + \frac{1}{\sin^2 \thg} \lh \frac{\der S}{\der \vf}\rh^2 \right].
\label{1.15}
\end{equation}
The solution corresponding to the orbit (\ref{1.14}) is
\begin{equation}
S(\thg,\vf,t) = \frac{1}{2t}\,
 \arccos^2 \left[ \sin \thg\, \cos (\vf - \vf_*) \right],
\label{1.16}
\end{equation}
which satisfies the equations
\begin{equation}
\begin{array}{lll}
\displaystyle{ \frac{\der S}{\der \thg} }& = & \displaystyle{ p_{\thg} = -
 \frac{\og \cos \thg\, \cos(\vf - \vf_*)}{\sqrt{ 1 - \sin^2 \thg
 \cos^2 (\vf - \vf_*) }} , }\\
 & & \\
\displaystyle{ \frac{\der S}{\der \vf} }& = & \displaystyle{ p_{\vf} =
 \frac{\og \sin \thg\, \sin(\vf - \vf_*)}{\sqrt{ 1 - \sin^2 \thg
 \cos^2 (\vf - \vf_*) }}, }\\
 & & \\
\displaystyle{ \frac{\der S}{\der t} }& = & \displaystyle{ - H = - \frac{\og^2}{2}.}
\end{array}
\label{1.17}
\end{equation}
In this approach, the expressions on the right-hand side are obtained by
{\em defining} $\og$ via the last expression, in agreement with (\ref{1.14.1}).
The same principle of energy conservation/time-translation invariance implies
that $S$ does not depend on $\tau_*$.

\section{Fluids \label{s2}}

The Hamilton-Jacobi equation (\ref{1.8}) can itself be obtained in a
straightforward way from a variational principle: introduce a Lagrange
multiplier field $\rg(x)$ and construct the
action functional
\begin{equation}
A(\rg,S) = \int dt \int d^n x\, \sqrt{g}\, \rg \lh \dd{S}{t} + \frac{1}{2}\, g^{ij}\,
 \nabla_i S\, \nabla_j S \rh.
\label{2.1}
\end{equation}
The square root of the (time-independent) background metric has been included
to make the integration measure invariant under reparametrizations. Of course,
we could absorb it in the definition of Lagrange multiplier field, but then
$\rg$ would transform as a density rather than as scalar.

The Hamilton-Jacobi equation follows by requiring the action to be
stationary w.r.t.\ variations of $\rg$:
\begin{equation}
\frac{1}{\sqrt{g}}\, \frac{\del A}{\del \rg} = \dd{S}{t} +
 \frac{1}{2}\, g^{ij}\, \nabla_i S\, \nabla_j S = 0.
\label{2.2}
\end{equation}
On the other hand, the stationarity of $A(\rg,S)$ w.r.t.\ $S$ gives
\begin{equation}
- \frac{1}{\sqrt{g}}\, \frac{\del A}{\del S} = \dd{\rg}{t} +
  \nabla_i \lh g^{ij} \rg \nabla_j S \rh = 0.
\label{2.3}
\end{equation}
This equation can be interpreted as the covariant equation of continuity for
a fluid\footnote{For background, see e.g.\ ref.\cite{ll}.} with density $\rg$
and local velocity
\begin{equation}
v_i = \nabla_i S \holhs{1} \Rightarrow \holhs{1}
 \dd{\rg}{t} + \nabla_i \lh \rg v^i \rh = 0.
\label{2.4}
\end{equation}
In this interpretation the gradient of the Hamilton-Jacobi equation gives the
covariant Euler equation
\begin{equation}
\dd{v_i}{t} + v^j \nabla_j v_i = 0, \holhs{2}
\nabla_j v_i = \dd{v_i}{x^j} - \Gam_{ji}^{\;\;\;\;k} v_k.
\label{2.5}
\end{equation}
Eq.(\ref{2.4}) states that the fluid flow is of the potential type.
Indeed, in the absence of torsion the Riemann-Christoffel connection
$\Gam_{ij}^{\;\;\;\;k}$ is symmetric and  the local vorticity vanishes:
\begin{equation}
\nabla_i v_j - \nabla_j v_i = 0.
\label{2.7}
\end{equation}
For the fluid flow to be incompressible, the velocity field must be divergence
free:
\begin{equation}
\nabla \cdot v = \Del S = 0,
\label{2.8}
\end{equation}
where $\Del = g^{ij}\, \nabla_i \nabla_j$ is the covariant laplacean on
scalar functions over the space. It follows that the number of incompressible
modes of flow on the manifold equals the number of zero-modes of the scalar
laplacean. For example, for flow on the sphere $S^2$ (or any other compact
Riemann surface) there is only one incompressible mode, the trivial one with
$v^i = 0$ everywhere.

For a given geometry $g_{ij}(x)$, the solution of the Hamilton-Jacobi equation
(\ref{1.8}), (\ref{2.2}) provides a special solution of the Euler equation
(\ref{2.5}); for a conservative system: $\der S/\der t = -H =$ constant, it
implies $\der v_i / \der t = 0$ and $v^j \nabla_j v_i = 0$. Accordingly, this
solution describes geodesic flow starting from the point $(\thg_*, \vf_*)$.

To turn this into a complete solution of the fluid-dynamical equations
(\ref{2.4}), (\ref{2.5}) it remains to solve for the density $\rg$. The
equation of continuity takes the form
\begin{equation}
\dd{\rg}{t} + \nabla_i (\rg \nabla^i S) = 0.
\label{2.10}
\end{equation}
It follows that a stationary flow, with $\rg$ not explicitly depending on
time $t$, is possible if
\begin{equation}
\nabla \cdot (\rg \nabla S) = 0.
\label{2.11}
\end{equation}
In addition to the trivial solution $\rg = \rg_0 =$ constant, $v = \nabla
S/m = 0$, it is possible to find non-trivial solutions of equation
(\ref{2.11}) for spatially varying density $\rg$. As an example, we consider
flow in a 2-dimensional space; in this case one can introduce a generalized
stream function $T(x,t)$, dual to the fluid momentum, and write
\begin{equation}
\rg \nabla^i S = \frac{1}{\sqrt{g}}\, \ve^{ij} \nabla_j T.
\label{2.11.1}
\end{equation}
Then for theories of the type (\ref{1.8}):
\begin{equation}
\rg = \frac{\ve^{ij} \nabla_i S \nabla_j T}{\sqrt{g} (\nabla S)^2}\,
 =\, \frac{\ve^{ij} \nabla_i S \nabla_j T}{2 H \sqrt{g}}.
\label{2.11.2}
\end{equation}
With $H$ constant, the factor $2H$ in the denominator can be absorbed
into the definition of $\tT = T/2H$, and hence the density is given by
\begin{equation}
\rg = \frac{1}{\sqrt{g}}\, \ve^{ij} \nabla_i S \nabla_j \tT
    = \frac{1}{\sqrt{g}}\, \ve^{ij} v_i \nabla_j \tT ,
\label{2.11.3}
\end{equation}
for the pseudo-scalar function $T$ the gradient of which is dual to
$\rg \nabla S$. Note also, that eq.(\ref{2.11.1}) implies $\nabla S \cdot
\nabla T = v \cdot \nabla T = 0$.

As an illustration, we again consider the unit sphere $S^2$. The velocity
field is given by the momenta (\ref{1.17}) per unit mass:
\begin{equation}
v_{\thg} = - \frac{\og \cos \thg\, \cos(\vf - \vf_*)}{
 \sqrt{ 1 - \sin^2 \thg \cos^2 (\vf - \vf_*) }}, \holhs{2}
v_{\vf} = \frac{\og \sin \thg\, \sin(\vf - \vf_*)}{
 \sqrt{ 1 - \sin^2 \thg \cos^2 (\vf - \vf_*) }}.
\label{2.11.4}
\end{equation}
Taking into account that on the sphere the non-vanishing components
of the connection are
\begin{equation}
\Gam_{\thg \vf}^{\;\;\;\;\vf} = \frac{\cos \thg}{\sin \thg}, \holhs{2}
\Gam_{\vf\vf}^{\;\;\;\;\thg} = - \sin \thg \cos \thg,
\label{2.13}
\end{equation}
a straightforward calculation shows that indeed
\begin{equation}
v_j v^j = \og^2, \holhs{2} v^j \nabla_j v_i = 0, \holhs{2} \dd{v_i}{t} = 0.
\label{2.14}
\end{equation}
The first two equations actually imply $v^j (\nabla_i v_j - \nabla_j v_i) = 0$,
in agreement with the absence of local circulation. From these results
it follows, that the flowlines are geodesics (great circles) given by
eq.(\ref{1.14.1}), and stationary.

For the gradient of the stream function $T$ to be orthogonal to the velocity
field (\ref{2.11.4}), it must satisfy the linear differential equation
\begin{equation}
v \cdot \nabla T = 0 \holhs{1} \Leftrightarrow \holhs{1} \tg (\vf - \vf_*)\,
 \nabla_{\vf} T = \sin \thg\, \cos \thg\, \nabla_{\thg} T.
\label{2.14.1}
\end{equation}
The general solution can be obtained by separation of variables, and is a
function of the single variable: $T(\thg,\vf) = f\lh y \rh$, with $y = \tg \thg\,
\sin (\vf - \vf_*) = \ctg \ag$. For such a scalar field
\begin{equation}
\nabla_{\thg} T = \frac{\sin (\vf - \vf_*)}{\cos^2 \thg}\, \left.
 f^{\prime}(y)\right|_{y = \tiny{\ctg} \ag}, \holhs{2} \nabla_{\vf} T = \tg \thg\,
 \cos(\vf - \vf_*)\, \left. f^{\prime}(y)\right|_{y = \tiny{\ctg} \ag}.
\label{2.16}
\end{equation}
The corresponding density $\rg$ is then
\begin{equation}
\rg(\thg,\vf) = \frac{\bar{\rg}(\ag)}{\cos \thg}\, =\, - \frac{1}{\og \sin \ag
 \cos \thg}\, \left. f^{\prime}(y)\right|_{y = \tiny{\ctg} \ag}.
\label{2.17}
\end{equation}
The simplest, most regular solution is obtained for $\bar{\rg}(\ag) = \rg_*
\sin \ag$:
\begin{equation}
\rg(\thg,\vf) = \frac{\rg_* \sin \ag}{\cos \thg} =
 \frac{\rg_*}{\sqrt{ 1 - \sin^2 \thg \cos^2 (\vf -\vf_*)}}.
\label{2.18}
\end{equation}
This solution corresponds to
\begin{equation}
T(\thg,\vf) = \og \rg_*\, \ag(\thg,\vf) \holhs{1} \Leftrightarrow \holhs{1}
 f(y) = \og \rg_* \mbox{\,arcctg\,} y.
\label{2.18.1}
\end{equation}
Observe, that in this case $T$, like $\ag$, is an angular variable; indeed,
$\ag$ increases by $2 \pi n$ on any loop winding around the point $(\thg =
\pi/2; \vf = \vf_*)$ $n$ times.

The solution (\ref{2.18}) possesses singular points at $\thg = \pi/2$,
$\vf = \vf_* + n \pi$, corresponding to a source for $n = 0$, and a sink
for $n = 1$. This can be established from the expression for $\nabla \cdot v$:
\begin{equation}
\nabla \cdot v = \frac{\og \sin \thg \cos (\vf - \vf_*)}{
\sqrt{1 - \sin^2 \thg \cos^2 (\vf - \vf_*)}},
\label{2.19}
\end{equation}
which becomes $(+\infty, -\infty)$ at the singular points. However, a more
elegant way to establish the result, is to make use of the stream function
(\ref{2.18.1}) and consider the flux integral
\begin{equation}
\Fg(\Gam) = \oint_{\Gam} dl\, \rg v_n,
\label{2.20}
\end{equation}
representing the total flow of material across the closed curve $\Gam$ per
unit of time. Consider a contour $\Gam$ winding once around the singularity
at $(\thg = \pi/2; \vf = \vf_*)$; on such a curve $\ag$ increases from $0$
to $2\pi$. Then
\begin{equation}
\Fg(\Gam) = \oint_{\Gam} \sqrt{g} \ve_{ij} \rg v^i dx^j =
 \oint_{\Gam} \nabla_i T dx^i = 2\pi \og \rg_*.
\label{2.21}
\end{equation}
This represents the total flow of matter from the hemisphere centered on the
source at $(\thg = \pi/2; \vf = \vf_*)$ to the hemisphere centered on its
antipodal point, the sink at $(\thg = \pi/2; \vf = \vf_* + \pi)$.

\section{Vortices \label{s3}}

The dual relationship between the velocity potential $S$ and the stream
function $T$ suggests to study the dynamics of a fluid for which $T$ is the
velocity potential:
\begin{equation}
v_i = \frac{1}{\rg_*}\, \nabla_i T.
\label{3.1}
\end{equation}
The constant $\rg_*$ has been included for dimensional reasons.
Like before, this velocity field is stationary: $\der v_i/\der t = 0$, but
it is not geodesic. Indeed, the velocity field describes motion under the
influence of an external potential; specifically:
\begin{equation}
v \cdot \nabla v_i = \frac{1}{2}\, \nabla_i\, v^2 =
 \frac{1}{2\rg_*^2}\, \nabla_i (\nabla T)^2 =
 \frac{1}{2 \rg_*^2}\, \nabla_i (\rg \nabla S)^2.
\label{3.2}
\end{equation}
Here $\rg(x)$ and $S(x)$ denote the previously defined functions mapping
the manifold to the real numbers ---e.g.\ (\ref{1.16}) and (\ref{2.18})
for fluid motion on a sphere--- irrespective of their physical
interpretation. Now again, as $(\nabla S)^2 = 2 H = \og^2 =$ constant,
it follows that
\begin{equation}
v \cdot \nabla v_i = \frac{\og^2}{2\rg_*^2}\, \nabla_i\, \rg^2 \equiv -
 \nabla_i h.
\label{3.3}
\end{equation}
Combining eqs.(\ref{3.2}) and (\ref{3.3}):
\begin{equation}
\frac{1}{2}\, v^2 = - (h - h_0) = \frac{\og^2 \rg^2}{2 \rg_*^2},
\label{3.4}
\end{equation}
where $h$ represents the external potential. Because of the potential nature
of the flow, eq.\ (\ref{3.1}), the local vorticity again vanishes: $\nabla_i
v_j - \nabla_j v_i = 0$,  but as eq.(\ref{2.21}) shows, this is not necessarily
true globally. Indeed, in singular points of the original geodesic fluid flow
(with sources/sinks), the dual flow generally has vortices/ anti-vortices.

Continuing our example from the previous sections, we can illustrate these
results in terms of flow on the unit sphere, for which $T/\rg_* =
\og \ag$ and $v_i = \og \nabla_i \ag$:
\begin{equation}
v_{\thg} = - \frac{\og \sin (\vf - \vf_*)}{1 - \sin^2 \thg \cos^2 (\vf - \vf_*)},
 \holhs{2}
v_{\vf} = - \frac{\og \sin \thg \cos \thg \cos(\vf - \vf_*)}{1 - \sin^2 \thg
 \cos^2 (\vf - \vf_*)}.
\label{3.5}
\end{equation}
It follows, as expected, that
\begin{equation}
v^2 = \og^2 (\nabla \ag)^2 = \frac{\og^2 \rg^2}{\rg_*^2}\, =
 \frac{\og^2}{1 - \sin^2 \thg \cos^2 (\vf - \vf_*)}.
\label{3.6}
\end{equation}
A further remarkable property, is that the dual flow is divergence free:
\begin{equation}
\nabla \cdot v = 0 \holhs{1} \Leftrightarrow \holhs{1} \Del \ag = 0,
\label{3.7}
\end{equation}
where-ever $v$ is well-defined; obviously, the result can only be
true because of the two singular points $(\thg = \pi/2; \vf = \vf_*)$
and $(\thg = \pi/2; \vf = \vf_* + \pi)$, where $v_i$ and its divergence
are not well-defined, i.e.\ topologically the velocity field is defined
on a cylinder, rather than a sphere. These two points are centers of
vorticity, as follows directly from eq.(\ref{2.21}), which in the present
context can be rewritten as
\begin{equation}
\oint_{\Gam} v_i dx^i = 2 \pi \og,
\label{3.8}
\end{equation}
for any closed curve $\Gam$ winding once around the singular point
$(\pi/2,\vf_*)$; as this curve also winds once around the other singular
point in the opposite direction, they clearly define a pair of vortices of
equal but opposite magnitude.

As the flow is divergence free, it follows that in this case there can be
non-trivial incompressible and stationary flow modes: for constant
density $\rg_1$ one has
\begin{equation}
\dd{\rg_1}{t} = 0, \holhs{2} \nabla \rg_1 = 0 \holhs{1} \Rightarrow \holhs{1}
 \nabla \cdot (\rg_1 v) = 0,
\label{3.9}
\end{equation}
and the equation of continuity is satisfied.

The nature of the flow lines defined by eq.(\ref{3.5}) is clear: they are
parallel circles of equidistant points around the centers of vorticity.
On these circles the velocity is constant in magnitude, implying by
(\ref{3.6}) that $\sin \thg \cos(\vf - \vf_*) \equiv \cos \bg =$ constant.
For example, for $\vf_* = 0$ we get $x = \cos \bg =$ constant; the flow
line then is the circle where this plane of constant $x$ cuts the unit
sphere. On these flow lines
\begin{equation}
v_{\thg} = - \og_1 \sin (\vf - \vf_*), \holhs{2}
v_{\vf} = - \og_1 \cos \bg \cos \thg,
\label{3.10}
\end{equation}
with
\begin{equation}
\og_1 = \frac{v^2}{\og}\, = \frac{\og}{1 - \sin^2 \thg \cos^2 (\vf - \vf_*)}\,
 = \frac{\og}{\sin^2 \bg}.
\label{3.11}
\end{equation}

\section{The dual particle model \label{s4}}

Having clarified the nature of the (incompressible) flow described by
the dual velocity potential $T/\rg_*$, we now reconstruct the corresponding
particle-mechanics model for which $T/\rg_*$ is Hamilton's principal
function. From eqs.(\ref{3.4}), (\ref{3.6}) we observe that the hamiltonian
is of the form $H_1 = K + h$, with for the specific case at hand a
kinetic-energy term:
\begin{equation}
K = \frac{1}{2}\, g^{ij} p_i p_j\, \rightarrow\,
 \frac{1}{2}\, \lh p_{\thg}^2 + \frac{p_{\vf}^2}{\sin^2 \thg} \rh,
\label{4.2}
\end{equation}
and the potential (normalized for later convenience such that $2H = \og\og_1$):
\begin{equation}
h(\thg, \vf) = h_0 - \frac{\og^2 \rg^2}{2 \rg_*^2}\, \rightarrow\,
 \frac{\og\og_1}{2} \lh 1 - \frac{\og/\og_1}{1 - \sin^2 \thg \cos^2
 (\vf - \vf_*)} \rh.
\label{4.3}
\end{equation}
The corresponding lagrangean $L_1 = K - h$ produces the Euler-Lagrange equations
\begin{equation}
\begin{array}{lll}
\dot{p}_{\thg} & = & \displaystyle{\ddot{\thg}\, =\, \sin \thg \cos \thg\, \dot{\vf}^2
 + \frac{\og^2 \sin \thg \cos \thg\, \cos^2 (\vf - \vf_*)}{(1 - \sin^2 \thg
 \cos^2 (\vf - \vf_*) )^2}, }\\
 & & \\
\dot{p}_{\vf} & = & \displaystyle{ \frac{d}{dt} \lh \sin^2 \thg\, \dot{\vf} \rh\, =\,
 - \frac{\og^2 \sin^2 \thg \sin (\vf - \vf_*) \cos (\vf - \vf_*)}{(1 -
 \sin^2 \thg \cos^2 (\vf - \vf_*) )^2}. }
\end{array}
\label{4.5}
\end{equation}
These equations have solutions
\begin{equation}
\cos \thg = \sin \bg\, \sin \og_1 t, \holhs{2}
\tg (\vf - \vf_*) = \tg \bg\, \cos \og_1 t,
\label{4.6}
\end{equation}
with $\bg$ a constant, implying the relation
\begin{equation}
\sin \thg\, \cos (\vf - \vf_*) = \cos \bg.
\label{4.7}
\end{equation}
Solving for the velocity (and taking into account the unit mass)
\begin{equation}
p_{\thg} = v^{\thg} = - \og_1 \sin (\vf - \vf_*), \holhs{2}
p_{\vf} = \sin^2 \thg\, v^{\vf} = - \og_1 \cos \bg\, \cos \thg,
\label{4.8}
\end{equation}
in agreement with (\ref{3.10}). From these results we can compute Hamilton's
principal function
\begin{equation}
S_1(\thg,\vf,t) = \int_0^t d\tau\, L_1[\thg(\tau),\vf(\tau)] = \frac{1}{2t}\,
 \mbox{arcctg}^2 \lh \tg \thg \sin (\vf - \vf_*) \rh.
\label{4.9}
\end{equation}
This function indeed satisfies the Hamilton-Jacobi equations
\begin{equation}
\dd{S_1}{\thg} = p_{\thg}, \holhs{2} \dd{S_1}{\vf} = p_{\vf},
\label{4.11}
\end{equation}
with $(p_{\thg}, p_{\vf})$ as given by eq.(\ref{4.8}), and
\begin{equation}
\dd{S_1}{t} =\, -\frac{\og\og_1}{2} = -\frac{1}{2}\, \left[ \lh \dd{S_1}{\thg}
 \rh^2 + \frac{1}{\sin^2 \thg} \lh \dd{S_1}{\vf} \rh^2 \right] - h(\thg,\vf).
\label{4.10}
\end{equation}
Using the relation $\ctg \ag = \tg \thg \sin (\vf - \vf_*) = \ctg \og_1 t$,
the equations (\ref{4.11}) can be recast in the form
\begin{equation}
p_i = \og \nabla_i \ag = \frac{1}{\rg_*}\, \nabla_i T.
\label{4.12}
\end{equation}
Hence $T/\rg_*$ can indeed be identified with Hamilton's principal function
of this system.

Repeating the arguments of sect.\ \ref{s2}, the action (\ref{2.1}) for the
Hamilton-Jacobi theory is now generalized to:
\begin{equation}
A(\rg,S_1;h) = \int dt \int d^n x\, \sqrt{g}\, \rg \lh \dd{S_1}{t} + \frac{1}{2}\,
 g^{ij} \nabla_i S_1 \nabla_j S_1 + h \rh.
\label{4.13}
\end{equation}
Reinterpretation of $S_1$ as a velocity potential for fluid flow: $v = \nabla
S_1$, leads back directly to the inhomogeneous Euler equation
\begin{equation}
\dd{v_i}{t} + v \cdot \nabla v_i = - \nabla_i h,
\label{4.14}
\end{equation}
which for stationary flow becomes eq.(\ref{3.3}). Variation of this
action w.r.t.\ $S_1$ gives the equation of continuity for $\rg$, as before;
note that in this action $h$ plays the role of an external source for the
density $\rg$.

\section{Quantum theory \label{s5}}

The quantum theory of a particle on a curved manifold is well-established.
For the wave function to be well-defined and single-valued, the momenta must
satisfy the Bohr-Sommerfeld quantization conditions
\begin{equation}
\oint_{\Gam} p_i dx^i = 2 \pi n \hbar,
\label{5.1}
\end{equation}
for any closed classical orbit $\Gam$. For the free particle of unit mass
on the unit sphere the left-hand side is
\begin{equation}
\int_0^T v^2 d\tau = \og^2 T = 2 \pi \og,
\label{5.2}
\end{equation}
where $T = 2\pi/\og$ is the period of the orbit. Hence the quantization
rule amounts to quantization of the rotation frequency (the angular momentum):
$\og = n \hbar$.

For the dual model, the same quantity takes the value
\begin{equation}
\oint_{\Gam} v_i dx^i = \int_0^{T_1} v^2 d\tau = \frac{\og^2 T_1}{\sin^2 \bg}\,
 = \og\, \og_1 T_1 = 2 \pi \og,
\label{5.4}
\end{equation}
and again $\og = n \hbar$. As the quantization conditions in the two dual
models are the same, the duality can be preserved in the quantum theory.

If this is to be true also in the fluid interpretation, the quantization
conditions must be respected at that level as well. Now the first quantization
condition for the integral (\ref{5.2}) is interpreted in the fluid dynamical
context as a quantization of the fluid momentum, cf.\ eq.(\ref{2.11.4}). The
second quantization condition (\ref{5.4}) has a twofold interpretation: first,
according to eqs.(\ref{2.20}), (\ref{2.21}) it quantizes the strength of the
fluid sources and sinks in the model of free geodesic flow; the agreement
between the two quantization conditions is then obvious: in order for the
strength of the source/sink to satisfy a quantization condition, the amount
of fluid transfered from one to the other must be quantized as well.

In the context of the dual model however, the condition imposes the
quantization of vorticity in the quantum fluid \cite{feynman}. In the more
general context of quantum models of fluids in geodesic flow on a compact
two-dimensional surface and their duals described by the stream functions,
this observation shows that duality at the quantum level requires the
quantization of sources in one model to be directly related to
the quantization of vorticity in the dual one. This situation
closely parallels the relation between the quantization of monopole charge
\cite{dirac} and the quantization of the magnetic flux of fluxlines
\cite{abrikosov} in three dimensions.

\newcommand{\jlpba}{\beta_\alpha}
\newcommand{\bb}{\beta_\beta}
\newcommand{\ga}{\gamma^\alpha}
\newcommand{\gb}{\gamma^\beta}
\newcommand{\kvt}{\sqrt{t}}
\newcommand{\hn}{h^\vee}
\newcommand{\kn}{k^\vee}
\newcommand{\dab}{{\delta_\alpha}^\beta}
\newcommand{\pa}{\partial}
\newcommand{\nn}{\nonumber \\ }
\newcommand{\hf}{\frac{1}{2}}         
\newcommand{\paj}{P_{-\alpha}^j}
\newcommand{\vmab}{V_{-\alpha}^\beta}
\newcommand{\vab}{V_\alpha^\beta}
\newcommand{\vib}{V_i^\beta}
\newcommand{\db}{\pa_\beta}
\newcommand{\dtb}{\delta_\theta^\beta}
\newcommand{\fabc}{{f_{ab}}^c}
\newcommand{\rton}{R_2^{(n)}}
\newcommand{\rjn}{R^j_{(n)}}
\newcommand{\binomial}[2]{\left (\begin{array}{c} {#1}\\ {#2} \end{array}
\right )}
\newcommand{\zb}{\overline{z}}
\newcommand{\xb}{\overline{x}}
\newcommand{\cpp}{{C_+}^+}
\newcommand{\cnp}{{C_0}^+}
\newcommand{\cmp}{{C_-}^+}
\newcommand{\cmn}{{C_-}^0}
\newcommand{\cmm}{{C_-}^-}
\newcommand{\bra}[1]{\langle {#1}|}
\newcommand{\ket}[1]{|{#1}\rangle}
\newcommand{\C}{\mbox{\hspace{1.24mm}\rule{0.2mm}{2.5mm}\hspace{-2.7mm} C}}
\newcommand{\Q}{\mbox{\hspace{1.24mm}\rule{0.2mm}{2.7mm}\hspace{-2.7mm} Q}}
\newcommand{\Z}{\mbox{$Z\hspace{-2mm}Z$}}
\newcommand{\Nat}{\mbox{\hspace{.04mm}\rule{0.2mm}{2.8mm}\hspace{-1.5mm} N}}
\newcommand{\spa}{\hspace{1 cm},\hspace{1 cm}}
\renewcommand{\Im}{{\rm Im}\,}
\newcommand{\eq}[1]{eq.(\ref{#1})}
\newcommand{\Eq}[1]{Eq.(\ref{#1})}

\newcommand{\NP}[1]{Nucl.\ Phys.\ {\bf #1}}
\newcommand{\PL}[1]{Phys.\ Lett.\ {\bf #1}}
\newcommand{\CMP}[1]{Commun.\ Math.\ Phys.\ {\bf #1}}
\newcommand{\PR}[1]{Phys.\ Rev.\ {\bf #1}}
\newcommand{\PRL}[1]{Phys.\ Rev.\ Lett.\ {\bf #1}}
\newcommand{\PTP}[1]{Prog.\ Theor.\ Phys.\ {\bf #1}}
\newcommand{\PTPS}[1]{Prog.\ Theor.\ Phys.\ Suppl.\ {\bf #1}}
\newcommand{\MPL}[1]{Mod.\ Phys.\ Lett.\ {\bf #1}}
\newcommand{\IJMP}[1]{Int.\ Jour.\ Mod.\ Phys.\ {\bf #1}}
\newcommand{\IM}[1]{Invent.\ Math.\ {\bf #1}}
\newcommand{\SJNP}[1]{Sov. J. Nucl. Phys.\ {\bf #1}}

\title*{Results on 2D Current Algebras}

\author{%
J.L. Petersen\thanks{E-mail:jenslyng@nbi.dk}}
\institute{%
The Niels Bohr Institute, Copenhagen, Denmark}

\authorrunning{J.L. Petersen}
\titlerunning{Results on 2D Current Algebras}
\maketitle

\section{Introduction}
A brief account i presented on a series of results on free field realizations
of 2-d affine current algebras and techniques for working out correlators,
in particular in non-integrable, admissible representations. A more
detailed account of these results, obtained in collaboration with Yu Ming and 
J\o rgen Rasmussen, may be found i refs. 
\cite{PRY1,PRY2,PRY3,PRY4,PRY5,PRY6,PRY7,JR}.The use of these may be used in 
inderect ways in formulations of perturbative string vacua and 2-D quantum 
gravity.

\section{$SL(2)$ current algebra and admissible representations}
We first describe how to obtain conformal blocks for $SL(2)$
WZW theories in the case of non-integrable representations, in particular for
admissible representations \cite{KK,MFF}. 
For previous approaches to the same problem, see 
\cite{BF,ATY,FGPP,Dot,JLP-A,SV,FIM,FV,FF,MFF,An,AY,FM,FM95}. 
Our goal here is to obtain a formulation based on the 
Wakimoto \cite{Wak} free field realization. The problem is complicated by the
need 
for introducing a second screening charge in the case of admissible 
representations. This screening operator involves a fractional power of a free 
(antighost) field \cite{BO}. Also the formalism requires the introduction of
several other fractional powers of free fields.
We have found that everything may
be treated rather neatly by means of fractional calculus.

As an explicit verification that our formalism works we have managed to prove 
that the conformal blocks we obtain satisfy the Knizhnik-Zamolodchikov equations
\cite{KZ,CF}. As an additional bonus we have been able to provide a proof of an 
interesting suggestion by Furlan, Ganchev, Paunov and Petkova \cite{FGPP} for
how conformal blocks of the $SL(2)$ WZW theory reduce to the conformal blocks
of minimal models. 

When the level of the affine $\widehat{SL(2)}_k$ algebra is $k$, we define
\begin{equation}
k+2\equiv t=p/q
\end{equation} 
where $p,q$ are positive coprime integers 
for admissible representations. For these
there are degenerate representations whenever the spin is given by
\begin{equation}
2j_{r,s}+1=r-st,\ \ \ 
r=1,...,p-1,\ \ \ 
s=0,...,q-1
\end{equation}
The Wakimoto free field realization is in terms of a scalar field, $\varphi$
and a pair of bosonic dimension $(1,0)$ ghosts, $(\beta,\gamma)$:
\begin{eqnarray}
\varphi(z)\varphi(w)&\sim&\log(z-w), \ \ \ 
\beta(z)\gamma(w) \sim \frac{1}{z-w}\nn
J^+(z)&=&\beta(z)\nn
J^3(z)&=&-\gamma\beta(z)-\sqrt{\frac{t}{2}}\pa\varphi(z)\nn
J^-(z)&=&-\gamma^2\beta(z)+k\pa\gamma(z)-\sqrt{2t}\gamma\pa\varphi(z)
\end{eqnarray}
They satisfy
\begin{eqnarray}
J^+(z)J^-(w)&\sim&\frac{2}{z-w}J^3(w)+\frac{k}{(z-w)^2}\nn
J^3(z)J^\pm(w)&\sim&\pm\frac{1}{z-w}J^\pm(w)\nn
J^3(z)J^3(w)&\sim&\frac{k/2}{(z-w)^2}\nn
\end{eqnarray}
The central charge of the corresponding virasoro algebra is $c=3k/t$.
Fateev and Zamolodchikov \cite{FZ} introduced a very useful formalism for 
primary fields, which we shall adopt. In general there is a multiplet of primary
fields $\phi_j^m(z)$. One combines these introducing an extra variable, $x$ as
follows
\begin{equation}
\phi_j(z,x)=\sum_m\phi_j^m(z)x^{j-m}
\end{equation}
For integrable representations $2j$ is an integer and we simply get a polynomial
in $x$. However, for fractional spins we have highest weight, lowest weight or
continuous representations, and the $x$-dependence can be arbitrarily 
complicated. The new primary field satisfies the following OPE
\begin{eqnarray}
J^a(z)\phi_j(w,x)&\sim&\frac{1}{z-w}D_x^a\phi_j(w,x)\nn
D_x^+=-x^2\pa_x+2xj,\ \ \ 
D_x^3&=&-x\pa_x+j,\ \ \ 
D_x^-=\pa_x
\end{eqnarray}
One easily verifies that
\begin{equation} 
\phi_j(z,x)={[}1+x\gamma(z){]}^{2j}e^{-j\sqrt{\frac{2}{t}}\varphi(z)}
\label{primary}
\end{equation}
Finally there are the two screening charge currents \cite{BO}
\begin{equation}
S_1(z)=\beta(z)e^{\sqrt{\frac{2}{t}}\varphi(z)}, \ \ \
S_{-t}(z)=\beta(z)^{-t}e^{-t\sqrt{\frac{2}{t}}\varphi(z)}=(S_1(z))^{-t}
\label{screening}
\end{equation}
We see in these last equations the need for being able to treat fractional 
powers of free fields. 

Our treatment of Wick contractions in such situations is based on the
following identity, trivially valid for $-t$ a positive integer, but non-trivial
for general $t$:
\begin{eqnarray}
\beta(z)^{-t}F(\gamma(w))
&=&:{[}\beta(z)+\frac{1}{z-w}\pa_{\gamma(w)}{]}^{-t}
F(\gamma(w)):\nn
&=&\sum_{n\in\Z}\binomial{-t}{n}:\beta^{n}(z)(z-w)^{t+n}
\pa^{-t-n}_{\gamma(w)}F(\gamma(w)):
\end{eqnarray}
Examples of the use of fractional calculus \cite{MR} are the Riemann-Liouville
operator
\begin{equation}
\pa^{-a}f(z)=\frac{1}{\Gamma(a)}\int_0^z(z-t)^{a-1}f(t)dt, \ a>0
\end{equation}
and
\begin{equation}
\pa_x^a x^b=\frac{\Gamma(b+1)}{\Gamma(b-a+1)}x^{b-a}
\end{equation}
In addition we shall need unconventional (asymptotic) expansions like
\begin{equation}
e^x=\pa_x^ae^x=\sum_{n\in\Z}\frac{1}{\Gamma(n-a+1)}x^{n-a}
\end{equation}
for $x$ an operator.

According to the above, we may treat the free field representation of an 
$N$-point conformal block in terms of the following integral of a free field
correlator:
\begin{eqnarray}
&&\bra{j_N}\prod_{n=2}^{N-1}{[}1+x_n\gamma(z_n){]}^{2j_n}
e^{-j_n\sqrt{\frac{2}{t}}\varphi(z_n)}\nn
&\times&\oint\prod_{k=1}^{s}\frac{dv_k}{2\pi i}\beta^{-t}(v_k)
e^{-t\sqrt{\frac{2}{t}}\varphi(v_k)}
\times\oint\prod_{l=1}^{r}\frac{dw_l}{2\pi i}\beta(w_l)
e^{\sqrt{\frac{2}{t}}\varphi(w_l)}\ket{j_1}
\label{npoint}
\end{eqnarray}
Here $r$ and $s$ are the number of screening charges of the first and
second kind respectively. There has to be a precise
choice of the bra and ket as described in \cite{PRY1,PRY2}. 

By carefully analyzing the three-point function it is possible to verify fusion
rules previously obtained using different techniques.
For a three point function, $j_1+j_2-j_3=r-st$ with $r,s$ 
being the number of screening
operators.  Using $2j_i+1=r_i-s_it$, we find the following fusion rule, 
already written down in \cite{AY,FM}  
\begin{eqnarray}
1+|r_1-r_2|\leq&r_3&\leq p-1-|r_1+r_2-p|\nn
|s_1-s_2|\leq&s_3&\leq q-1-|s_1+s_2-q+1|\nn
\end{eqnarray}
referred to as their rule I. There is in addition a rule II:
\begin{eqnarray}
1+|p-r_1-r_2|\leq&r_3&\leq p-1-|r_1-r_2|\nn
1+|q-s_1-s_2-1|\leq&s_3&\leq q-2-|s_1-s_2|\nn
\end{eqnarray}
It is less trivial to obtain, but it follows both by a proper analysis of the
three point function and from our analysis of the 4-point functions (see below).

It is now possible to write down the general 
$N$-point function with $M=r+s$ screening charges as follows
\begin{eqnarray}
W_N&=&\int\prod_{i=1}^M\frac{dw_i}{2\pi i}W_N^\varphi W_N^{\beta\gamma}\nn
W_N^\varphi&=&\prod_{m<n}(z_m-z_n)^{2j_mj_n/t}
\prod_{i=1}^M\prod_{m=1}^{N-1}(w_i-z_m)^{2k_ij_m/t}
\prod_{i<j<M}(w_i-w_j)^{2k_ik_j/t}\nn
W^{\beta\gamma}&=&\int\prod_{m=2}^{N-1}\frac{du_m}{2\pi i}\Gamma(2j_m+1)
u^{2j_m-1}e^{1/u_m}\prod_{i=1}^M(\sum_{l=1}^{N-1}\frac{x_l/u_l}{w_i-z_l})^{-k_i}
\end{eqnarray}
where $z_1=x_1=0$.
Here the powers $k_i$ are $-1$ and $t$ respectively for screening charges of
the first and second kind. 

The $N$-point conformal block, $W(z_N,x_N,...,z_1,x_1)$, is a function of $N$
pairs of variables, $(z_i,x_i)$. An interesting proposal of Furlan, Ganchev, 
Paunov and  Petkova \cite{FGPP} is that when we put $x_i=z_i$
this block agrees up to normalisation with a corresponding block in the 
minimal conformal theory with the same $p,q$ as for the $\widehat{SL(2)}_k$ 
theory with $t=k+2=p/q$. We have proved this result and clarified how
it is related to hamiltonian reduction based on $J^+(z)\sim 1$.

We have used the above formalism to carry the computation of 4-point 
conformal blocks generalizing the techniques developed by Dotsenko and 
Fatteev \cite{DF}. The calculation is very technical and we refer to 
ref.\cite{PRY5} for details. Here we merely give the highly non trvial result
for the operator algebra. Expressing the OPE of two primary fields as
\begin{equation}
\phi_{j_2}(z,\zb;x,\xb)\phi_{j_1}(0,0;0,0)=
\sum_j\frac{(x\xb)^{j_1+j_2-j}}{(z\zb)^{h(j_1)+h(j_2)-h(j)}}
C^{\lambda}_{\lambda_1\lambda_2}\phi_j(0,0;0,0)
\end{equation}
We use the following notation (Fusin rule I)
\begin{eqnarray}
\lambda&=&2j_i+1\nn
h(j)&=&j(j+1)/t\nn
j&=&j_1+j_2-r+st\nn
G(x)&\equiv&\Gamma(x)/\Gamma(1-x)=1/G(1-x)
\end{eqnarray}
Then we have found
\begin{eqnarray}
C^{\lambda}_{\lambda_1\lambda_2}&=&t^{-2rs}\prod_{i=1}^r G(i/t)\prod_{i=1}^s G(it-r)
\prod_{i=0}^{s-1}\frac{G(\lambda_1+it)G(\lambda_2+it)}
{G(1+\lambda-(1+i)t)}\nn
&\times&\prod_{i=0}^{r-1}\frac{G(1-s+(1-\lambda_1+i)/t)G(1-s+(1-\lambda_2+i)/t)}
{G(1+s-(1+\lambda+i)/t)}
\end{eqnarray}

\section{Generalization to algebras other than $SL(2)$}
Much of the above formalism may be generalized to arbitrary algebras. Whereas
a large amount of work has already been carried out on free field realizations,
\cite{FF,GMMOS,BF,BMP,KOS,Ito0,Ito,Dot,Ku,ATY,Tay,deBF},
we have provided several proofs and new results, not least concerning 
extremely compact general
expressions and other missing elements, see ref.\cite{PRY7} for more details.

Consider some simple Lie algebra of dimension $d$ and rank $r$.
Let $\{j_a\}$ denote a set of Lie algebra generators with $\{ e_\alpha,f_\alpha\}$
being raising and lowering operators and $\{h_i\}$ being Cartan algebra
generators. The set of (positive) roots is denoted $(\Delta_+)$ $\Delta$. Simple
roots are $\{\alpha_i\}, i=1,...,r$, and $\theta$ denotes
the highest root. The root dual to $\alpha$ is $\alpha^\vee=2\alpha/\alpha^2$.
Commutation relations are
\begin{equation}
{[}j_a,j_b{]}={f_{ab}}^cj_c
\end{equation}
or in the Cartan-Weyl basis
\begin{equation}
{[}h_i,e_\alpha{]}=(\alpha_i^\vee,\alpha)e_\alpha\ \ ,
\ \ {[}h_i,f_\alpha{]}=-(\alpha_i^\vee,\alpha)f_\alpha
\end{equation}
for which
\begin{equation}
{[}e_\alpha,f_\alpha{]}=h_\alpha=G^{ij}(\alpha_i^\vee,\alpha^\vee)h_j,\ \ 
{[}h_i,e_j{]}=A_{ij}e_j
\label{cartanmatrix}
\end{equation}
with $e_j\equiv e_{\alpha_j}$ for $\alpha_j$ a simple root, and where 
$A_{ij}=\alpha_i^\vee\cdot\alpha_j=G_{ij}\alpha_j/2$ is
the Cartan matrix. The Cartan-Killing form is $tr(j_aj_b)=\kappa_{ab}$.
Dynkin labels $\Lambda_k$ of weight $\Lambda$ are defined by
$$\Lambda=\Lambda_k\Lambda^{(k)}\ \ , \ \ \Lambda_k=(\alpha_k^\vee,\Lambda)$$
with $\{\Lambda^{(k)}\}$ the fundamental weights. We introduce a coordinate 
$x^\alpha$ (generalizing the $x$ for $SL(2)$)  for every positive root.
Further, define Lie algebra elements
\begin{equation}
e(x)=x^\alpha e_\alpha\ \ ,\ \ f(x)=x^\alpha f_\alpha 
\label{e(x)f(x)}
\end{equation}
and group elements
$$g_+(x)=e^{e(x)}\ \ .\ \ g_-(x)=e^{f(x)}$$
and thus on a representation space corresponding to highest weight $\Lambda$
$$\ket{\Lambda,x}=g_-(x)\ket{\Lambda}$$ 
It proves extremely convenient to introduce the adjoint matrix representation
of $e(x)$
\begin{equation}
C_a^b(x)={C(x)_a}^b={(x^\beta C_\beta)_a}^b=-x^\beta {f_{\beta a}}^b
\label{adje}
\end{equation}
In a Cartan-Weyl basis, $a=(\alpha,i,-\beta), \alpha, \beta \in \Delta_+$ 
and $C_a^b=0$ unless either 
$a=-\alpha$ or $b=+\alpha$ with $\alpha\in\Delta_+$.
We then obtain differential operator realizations from
\begin{equation}
j_a\ket{\Lambda,x}=J_{-a}(x,\pa,\Lambda)\ket{\Lambda,x}
\end{equation}
Let 
\begin{equation} 
B(u)=\frac{u}{e^u-1}=\sum_{n=0}^\infty \frac{B_n}{n!}u^n
\end{equation}
be the generating function of Bernulli numbers. Using the power series 
expansion, we may replace $u$ by any matrix.
Define the $d\times (d-r)/2$ matrix
\begin{equation}
{V(x)_a}^\beta={(e^{-C(x)})_a}^\gamma{B(-C(x))_\gamma}^\beta
\end{equation}
with $\alpha,\beta,\gamma$ all denoting positive roots. It is then possible to express 
the result as
\begin{equation}
J_a(x,\pa,\Lambda) ={V(x)_a}^\beta\pa_\beta + (e^{-C(x)})_a^i\Lambda_i
\end{equation}
Notice that $\left(e^{-C}\right)_\alpha^i=0$ ($\alpha\in\Delta_+$) and 
$\left(e^{-C}\right)_j^i=\delta_j^i$.
For any given algrebra, such expressions truncate at finite order.

The free field realization for the current algebra is in terms of bosonic 
ghost fields, $(\jlpba(z), \gb(z))$ and Liouville type scalars 
$\phi_i(z), i=1,...,r$ satisfying the OPEs
\begin{equation}
\jlpba(z)\gb(w)=\frac{\delta_\alpha^\beta}{z-w}\ \ ,\ \ \phi_i(z)\phi_j(w)=G_{ij}
\ln (z-w)
\end{equation}
with $G_{ij}$ the metric of \eq{cartanmatrix}.
For central charge, $k$ and $k^\vee=2k/\theta^2$ and dual Coxeter number
$h^\vee$, we introduce
$$t=\frac{\theta^2}{2}(k^\vee +h^\vee)$$
We seek a set of currents satisfying
\begin{equation}
J_a(z)J_b(w)=\frac{\kappa_{ab}k}{(z-w)^2}+\frac{{f_{ab}}^c J_c(w)}{z-w}
\end{equation}
The free field realization may then be obtained from the differential  operator
realization simply be the replacements
$$\pa_\alpha\rightarrow \jlpba(z)\ \ ,\ \ x^\alpha\rightarrow \ga(z)\ \ ,\ \ 
\Lambda_i\rightarrow \sqrt{t}\pa\phi_i(z)$$
followed by normal ordering.
In addition, for negative roots we must add to $J_{-\alpha}(z)$ the ``anomalous" 
term
\begin{eqnarray}
F_\alpha^{\mbox{anom.}}(\gamma(z),\pa\gamma(z))&=&\pa\ga(z)F_{\alpha\beta}(\gamma(z))\nn
F_{\alpha\beta}(\gamma)&=&\frac{2k}{\alpha^2}(V^{-1}(\gamma))_\beta^\alpha + (V^{-1}(\gamma))_\beta^\mu
\pa_\sigma V_\mu^\delta(\gamma)\pa_\delta V_{-\alpha}^\sigma(\gamma)
\end{eqnarray}
where $V^{-1}$ is to be understood as obtained from inverting the sqare matrix
with positive roots only in rows and columns. All roots indicated,
$\alpha, \beta, \delta, \mu$, are positive.

For every simple root, there are two kinds of screening operators, the ones of
the ``second kind" with fractional powers of free fields. We find that
they are given by
(see also \cite{FF,BMP,Ito0,Ku,ATY,deBF} for many results in special cases and 
different mostly more cumbersome forms)
\begin{eqnarray}
s^{(1)}_j(z)&=&:E_{\alpha_j}(-\gamma(z),-\beta(z)))::e^{-\alpha_j^2\phi_j(z)/2\sqrt{t}}:\nn
s^{(2)}_j(z)&=& (s^{(1)}_j(z))^{-2t/\alpha_j^2}
\end{eqnarray}
We have provided proofs previously missing, in particular in the last case,
that these satisfy the required conditions.

Finally we have provided explicit expressions for free field realizations of
primary fields. For a general representation with highest weight 
$\Lambda=\Lambda_k\Lambda^{(k)}$ the primary field may be simply expressed in terms of those
for the fundamental repesentations by
\begin{equation}
\phi_\Lambda(\gamma(w),x)=\prod_{k=1}^r {[}\phi_{\Lambda^{(k)}}(\gamma(w),x){]}^{\Lambda_k}
\end{equation}
For the fundamentals we here restrict to the case of $SL(N)$, where roots are 
conveniently labelled by double indices $(ij), i,j =1,...,N$. Introduce the 
matrix representation $F^{(N)}(x)$ of $f(x)$ \eq{e(x)f(x)} by 
${(F^{(N)})_i}^j=x^{ij}$ for $i<j$ and $0$ otherwise, and let
$$e^{F^{(N)}(x)}(I(k))$$
denote the $k\times k$ matrix obtained from the $N\times N$ matrix 
$e^{F^{(N)}(x)}$ by using the first $k$ rows, and the columns given by the
set $I(k)=\{i_1,...,i_k\}$. Then we have obtained
\begin{equation}
\varphi_{\Lambda^{(k)}}(\gamma(z),x)=\sum_{I(k)}\det\left( e^{F^{(N)}(x)}(I(k))\right)
\det\left( e^{F^{(N)}(\gamma(z))}(I(k))\right) 
\end{equation}
Using the screening charges, we may then obtain correlators for non-integer
Dynkin labels of the form
\begin{equation}
\Lambda_k=A_{ki}r^i-G_{ki}s^it=\hat{r}_k-\hat{s}_k\frac{\theta^2}{\alpha_k^2}\hat{t}
\end{equation}
with $\hat{t}=\frac{2}{\theta^2}t=k^\vee + h^\vee$. These are degenerate
representations for $r^i,s^i,\hat{r}_k,\hat{s}_k$ integer, \cite{KK}, and 
admissible representations for $\hat{t}$ rational, \cite{KW}.
 
\section*{Acknowledgement}
I want to thank the organizers of the 
4th Workshop 'What comes beyond the Standard model', Bled 2001, 
for allowing me to contribute
this note in honour of Holger Bech Nielsen's 60 year birthday. I want to thank
Holger for his powerful impact on me as a scientist and as a person,
and for the (admittedly few) times
we have collaborated. Finally I want to thank my friends, Jørgen Rasmussen
and Ming Yu for the happy collaboration on the work briefly presented here.

\def\la{\mathrel{\mathpalette\fun <}}
\def\ga{\mathrel{\mathpalette\fun >}}
\def\fun#1#2{\lower3.6pt\vbox{\baselineskip0pt\lineskip.9pt
\ialign{$\mathsurround=0pt#1\hfil##\hfil$\crcr#2\crcr\sim\crcr}}}

\newcommand{\plaqr}{$\biggl(\quad\raisebox{-2pt}{\mbox{\framebox(0,12){\phantom{a}}\hspace{-5.6mm}
\dashbox{3}(12,12)[b]{\phantom{a}}}}\biggr)$}
\newcommand{\plaql}{$\biggl(\raisebox{-2pt}{\mbox{\framebox(0,12){\phantom{a}}\hspace{-1.3mm}
\dashbox{3}(12,12)[b]{\phantom{a}}}}\biggr)$}
\newcommand{\plaqt}{$\biggl(\raisebox{-2pt}{\mbox{\raisebox{12pt}{\framebox(11,0){\phantom{a}}}\hspace{-5.5mm}
\dashbox{3}(12,12)[b]{\phantom{a}}}}\biggr)$}
\newcommand{\plaqb}{$\biggl(\raisebox{-2pt}{\mbox{\framebox(11,0)
{\phantom{a}}\hspace{-5.5mm}
\dashbox{2}(12,12)[b]{\phantom{a}}}}\biggr)$}

\newcommand{\link}{\begin{array}{l}\begin{picture}(22,4)
    \put(0,2.5){\circle*{4}}
    \put(20,2.5){\circle*{4}}
    \put(0,2.5){\line(1,0){20}}
\end{picture}\end{array}}

\title*{Phase Transition in Gauge Theories and the Planck Scale Physics}
\author{%
L.V. Laperashvili and D.A. Ryzhikh}
\institute{%
Institute of Theoretical and Experimental Physics,
Moscow, Russia }

\authorrunning{L.V. Laperashvili and D.A. Ryzhikh}
\titlerunning{Phase Transition in Gauge Theories and the Planck Scale Physics}
\maketitle

\section{Introduction}

The modern physics of electroweak and strong interactions is described by
the Standard Model (SM), unifying the Glashow--Salam--Weinberg electroweak
theory and QCD -- theory of strong interactions.

The gauge group of the SM is:
\begin{equation}
   SMG = SU(3)_c\times SU(2)_L\times U(1)_Y,            \label{2}
\end{equation}
which describes the present elementary particle physics up to the
scale $\sim 100$ GeV.

Considering the physical processes at very small (Planck scale) distances,
physicists can make attempts to explain the well--known laws of physics
as a consequence of the more fundamental laws of Nature. Random Dynamics (RD)
was suggested and developed in Refs.[\cite{2a}--\cite{2k}] as a
theory of physical
processes proceeding at small distances of order of the Planck length
$\lambda_P$:
\begin{equation}
\lambda_P=M_{Pl}^{-1}, \quad{\mbox{where}} \quad
                M_{Pl}=1.22\cdot 10^{19}\,{\mbox{GeV}}.     \label{1}
\end{equation}
Having an interest in fundamental laws of physics leading to the description
of the low--energy SM phenomena, observed by the contemporary experiment,
we can consider two possibilities:

\vspace{0.1cm}

1. At very small (Planck scale) distances
{\underline{\it our space--time is continuous}}, and there exists a theory with
a very high symmetry.

\vspace{0.1cm}

2. At very small distances
{\underline{\it our space--time is discrete}}, and this discreteness
influences on the Planck scale physics.

The item 2 is a base of the RD theory.

The theory of Scale Relativity (SR) \cite{10na} is also related with item 2
and has a lot in common with RD. In the SR the resolution of experimental
measurements plays in quantum physics a completely new role with respect
to the classical one and there exists a minimal scale of the space-time
resolution: $\epsilon_{min}=\lambda_P$, which can be considered as a
fundamental scale of our Nature. In this case, our (3+1)--dimensional
space is discrete on the fundamental level. This is an initial point of
view of the present theory, but not an approximation.

The lattice model of gauge theories is the most convenient formalism
for the realization of the RD ideas. In the simplest case we can imagine
our space--time as a regular hypercubic (3+1)--lattice with the parameter
$a$ equal to the fundamental scale:
\begin{equation}
 a = \lambda_P = 1/M_{Pl} \sim 10^{-33}\, cm.                   \label{3}
\end{equation}
But, in general, we do not know (at least on the level of our today
knowledge) what lattice--like structure plays role in the description
of the physical processes at very small distances.

\section{G-theory, or Anti-Grand Unification Theory\\ (AGUT)}

Most efforts to explain the Standard Model (SM) describing well all
experimental results known today are devoted to Grand Unification
Theories (GUTs). The supersymmetric extension of the SM consists of taking the
SM and adding the corresponding supersymmetric partners \cite{32a}.  The Minimal
Supersymmetric Standard Model (MSSM) shows \cite{33a} the possibility of the
existence of the grand unification point at
$\mu_{GUT}\sim 10^{16}$ GeV.
Unfortunately, at present time experiment does not indicate any manifestation
of the supersymmetry. In this connection, the Anti--Grand Unification
Theory (AGUT) was developed in Refs.[\cite{2a}--\cite{2k}] and [\cite{17p}--\cite{38}]
as a realistic alternative to SUSY GUTs. According to this theory, supersymmetry does
not come into the existence up to the Planck energy scale (\ref{1}).
The Standard Model (SM) is based on the group SMG described by Eq.(\ref{2}).
AGUT suggests that at the scale $\mu_G\sim \mu_{Pl}=M_{Pl}$
there exists the more fundamental group $G$ containing $N_{gen}$
copies of the Standard Model Group SMG:
\begin{equation}
G = SMG_1\times SMG_2\times...\times SMG_{N_{gen}}\equiv (SMG)^{N_{gen}},
                                                  \label{76y}
\end{equation}
where $N_{gen}$ designates the number of quark and lepton generations.

If $N_{gen}=3$ (as AGUT predicts), then the fundamental gauge group G is:
\begin{equation}
    G = (SMG)^3 = SMG_{1st gen.}\times SMG_{2nd gen.}\times SMG_{3rd gen.},
                                        \label{77y}
\end{equation}
or the generalized one:
\begin{equation}
         G_f = (SMG)^3\times U(1)_f,           \label{78y}
\end{equation}
which was suggested by the fitting of fermion masses of the SM
(see Refs.\cite{35}).

Recently a new generalization of AGUT was suggested in Refs.\cite{37}:
\begin{equation}
           G_{\mbox{ext}} = (SMG\times U(1)_{B-L})^3,    \label{79y}
\end{equation}
which takes into account the see--saw mechanism with right-handed neutrinos,
also gives the reasonable fitting of the SM fermion masses and describes
all neutrino experiments known today.

By reasons considered in this paper, we prefer
not to use the terminology "Anti-grand unification theory, i.e. AGUT",
but call the theory with the group of symmetry $G$, or $G_f$,
or $G_{ext}$, given by Eqs.(\ref{76y})-(\ref{79y}), as "G--theory",
because, as it will be shown below, we have a possibility of the
Grand Unification near the Planck scale using just this theory.

The group $G_f$ contains the following gauge fields:
$3\times 8 = 24$ gluons, $3\times 3 = 9$ W-bosons and $3\times 1 + 1 = 4$
Abelian gauge bosons. The group $G_{ext}$ contains:
$3\times 8 = 24$ gluons, $3\times 3 = 9$ W-bosons and $3\times 1 +
3\times 1 = 6$ Abelian gauge bosons.

There are five Higgs fields in AGUT, extended by Froggatt and Nielsen \cite{35}
with the group of symmetry $G_f$ given by Eq.(\ref{78y}).
These fields break AGUT to the SM what means that their vacuum expectation
values (VEV) are active.
The authors of Refs.\cite{35} used three parameters -- three independent VEVs with
aim to find the best fit to conventional experimental data
for all fermion masses and mixing angles in the SM.
The result was encouraging.

The extended AGUT by Nielsen and Takanishi \cite{37}, having
the group of symmetry $G_{ext}$ (see Eq.(\ref{79y})),
was suggested with aim to explain the neutrino oscillations.
Introducing the right--handed neutrino in the model, the authors replaced the
assumption 1 and considered U(48) group instead of U(45), so that
$G_{ext}$ is a subgroup of U(48): $G_{ext}\subseteq U(48)$. This group
ends up having 7 Higgs fields.

In contrast to the "old" extended AGUT by Froggatt--Nielsen (called here
as $G_f$--theory), the new results of $G_{ext}$--theory are more encouraging,
and it is possible to conclude
that the $G$--theory, in general, is successful in describing of
the SM experiment.

\section{Multiple Point Principle}

AGUT approach is used in conjunction with the Multiple Point
Principle proposed in Ref.\cite{17p}.
According to this principle Nature seeks a special point --- the Multiple
Critical Point (MCP) --- which is a point on the phase diagram of the
fundamental regulirized gauge theory G (or $G_f$, or $G_{ext}$), where
the vacua of all fields existing in Nature are degenerate having the same
vacuum energy density.
Such a phase diagram has axes given by all coupling constants
considered in theory. Then all (or just many) numbers of phases
meet at the MCP.

MPM assumes the existence of MCP at the Planck scale,
insofar as gravity may be "critical" at the Planck scale.

The usual definition of the SM coupling constants:
\begin{equation}
  \alpha_1 = \frac{5}{3}\frac{\alpha}{\cos^2\theta_{\overline{MS}}},\quad
  \alpha_2 = \frac{\alpha}{\sin^2\theta_{\overline{MS}}},\quad
  \alpha_3 \equiv \alpha_s = \frac {g^2_s}{4\pi},     \label{81y}
\end{equation}
where $\alpha$ and $\alpha_s$ are the electromagnetic and SU(3)
fine structure constants, respectively, is given in the Modified
minimal subtraction scheme ($\overline{MS}$).
Here $\theta_{\overline{MS}}$ is the Weinberg weak angle in $\overline{MS}$ scheme.
Using RGE with experimentally
established parameters, it is possible to extrapolate the experimental
values of three inverse running constants $\alpha_i^{-1}(\mu)$
(here $\mu$ is an energy scale and i=1,2,3 correspond to U(1),
SU(2) and SU(3) groups of the SM) from the Electroweak scale to the Planck
scale. The precision of the LEP data allows to make this extrapolation
with small errors (see \cite{33a}). Assuming that these RGEs for
$\alpha_i^{-1}(\mu)$ contain only the contributions of the SM particles
up to $\mu\approx \mu_{Pl}$ and doing the extrapolation with one
Higgs doublet under the assumption of a "desert", the following results
for the inverses $\alpha_{Y,2,3}^{-1}$ (here $\alpha_Y\equiv \frac{3}{5}
\alpha_1$) were obtained in Ref.\cite{17p} (compare with \cite{33a}):
\begin{equation}
   \alpha_Y^{-1}(\mu_{Pl})\approx 55.5; \quad
   \alpha_2^{-1}(\mu_{Pl})\approx 49.5; \quad
   \alpha_3^{-1}(\mu_{Pl})\approx 54.0.
                                                        \label{82y}
\end{equation}
The extrapolation of $\alpha_{Y,2,3}^{-1}(\mu)$ up to the point
$\mu=\mu_{Pl}$ is shown in Fig.1.

\begin{figure*}
\begin{center}
\noindent\includegraphics[width=100mm, height=85mm]{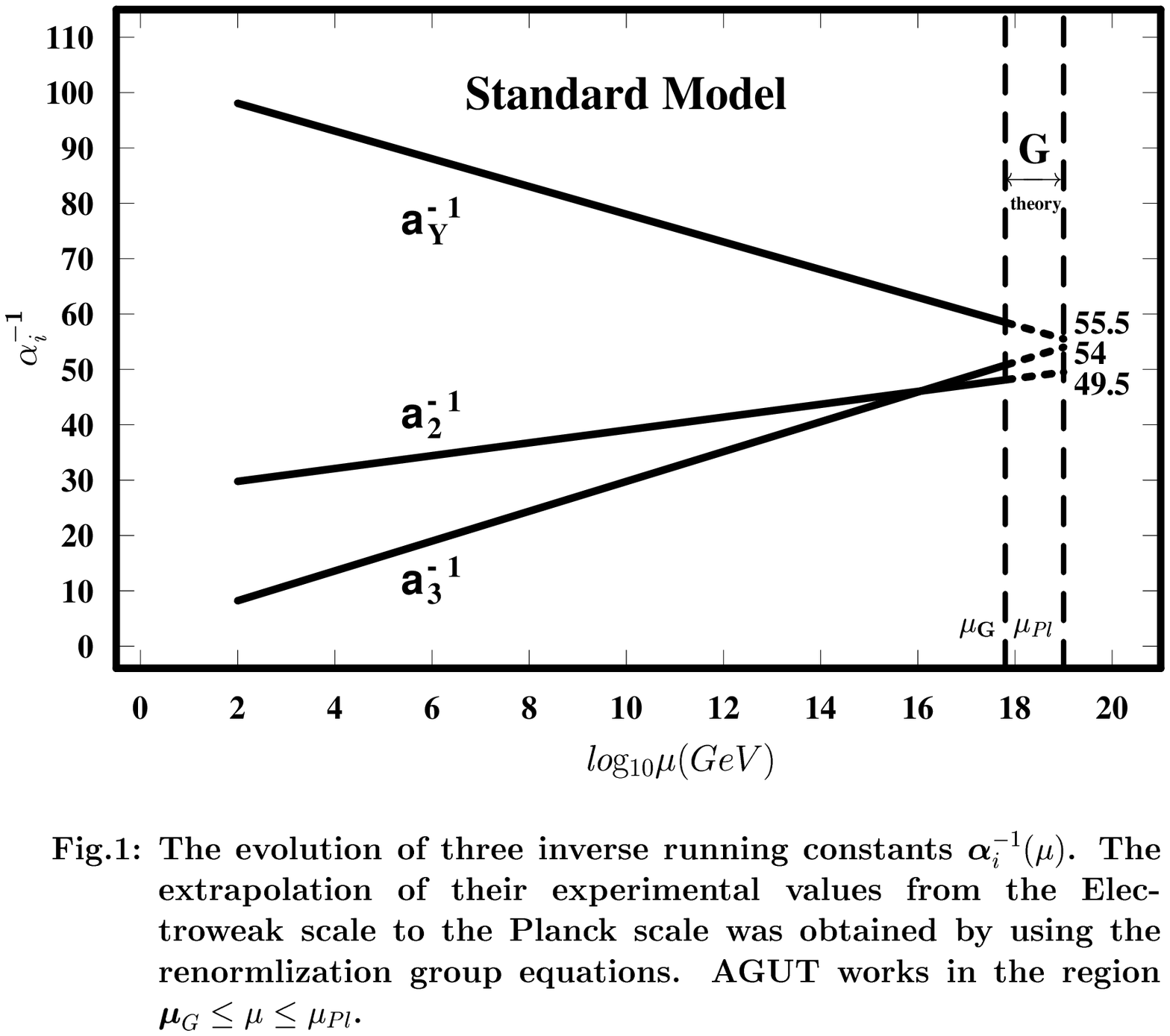}
\end{center}
\end{figure*}

According to AGUT, at some point $\mu=\mu_G < \mu_{Pl}$ the fundamental
group $G$ (or $G_f$, or $G_{\mbox{ext}}$)
undergoes spontaneous breakdown to its diagonal subgroup:
\begin{equation}
      G \longrightarrow G_{diag.subgr.} = \{g,g,g || g\in SMG\},
                                                          \label{83y}
\end{equation}
which is identified with the usual (low--energy) group SMG.
The point $\mu_G\sim 10^{18}$ GeV also is shown in Fig.1, together with
a region of G--theory, where AGUT works.

The AGUT prediction of the values of $\alpha_i(\mu)$ at $\mu=\mu_{Pl}$
is based on the MPM assumption about the existence of phase
transition boundary point MCP at the Planck scale, and gives these values
in terms of the corresponding critical couplings $\alpha_{i,crit}$
[\cite{2g},\;\cite{2k},\;\cite{17p}]:
\begin{equation}
            \alpha_i(\mu_{Pl}) = \frac {\alpha_{i,crit}}{N_{gen}}
                       = \frac{\alpha_{i,crit}}{3}
                \quad{\mbox{for}}\quad i=2,3, \,{\mbox{(also for $i>3$)}},
                                   \label{84y}
\end{equation}
\begin{equation}
\alpha_1(\mu_{Pl}) = \frac{\alpha_{1,crit}}{\frac{1}{2}N_{gen}(N_{gen} + 1)}
                   = \frac{\alpha_{1,crit}}{6} \quad{\mbox{for}}\quad U(1).
                                      \label{85y}
\end{equation}

\section{Lattice Theories}

The philosophy of MPM leads to the necessity to investigate the phase
transition in different gauge theories.
A lattice model of gauge theories is the most convenient formalism for the
realization of the MPM ideas. As it was mentioned above, in the simplest
case we can imagine our space--time as a regular hypercubic
(3+1)--lattice with the parameter $a$ equal to the fundamental
(Planck) scale: $a = \lambda_P$.

The lattice SU(N) gauge theories was first introduced by K.Wilson \cite{1s}
for studying the problem of confinement. He suggested the following
simplest action:
\begin{equation}
         S = - \frac{\beta}{N}\sum_p Re(Tr{\cal U}_p),     \label{36}
\end{equation}
where the sum runs over all plaquettes of a hypercubic lattice
and ${\cal U}_p$ belongs to the fundamental representation of SU(N).
The simplest Wilson lattice action for $U(1)$ gauge theory has the form:
\begin{equation}
     S_W = \beta \sum_p \cos\Theta_p, \quad {\mbox{where}}\quad
                                   {\cal U}_p = e^{i\Theta_p}.   \label{35a}
\end{equation}
For the compact lattice QED: $\beta = 1/e_0^2$, where $e_0$ is the bare
electric charge.

The Villain lattice action for the $U(1)$ gauge theory is:
\begin{equation}
         S_V = (\beta/2)\sum_p(\Theta_p - 2\pi k)^2, \quad k\in Z.   \label{39}
\end{equation}

The critical value of the effective electric fine structure
constant $\alpha$
was obtained in Ref.\cite{10s} in the compact QED described by the Wilson and
Villain actions (\ref{35a}) and (\ref{39}), respectively:
\begin{equation}
\alpha_{crit}^{lat}\approx 0.20\pm 0.015\quad
{\mbox{and}} \quad {\tilde \alpha}_{crit}^{lat}\approx 1.25\pm 0.10
\quad{\mbox{at}}\quad
\beta_T\equiv\beta_{crit}\approx{1.011}.
\label{47}
\end{equation}
Here

\vspace{-12mm}
\begin{equation}
\alpha = \frac{e^2}{4\pi}\quad{\mbox{and}}\quad
\tilde \alpha = \frac{g^2}{4\pi}
\label{47*}
\end{equation}
are the electric and magnetic fine structure constants, containing
the electric charge $e$ and magnetic charge $g$.

\begin{figure*}
\begin{center}
\noindent\includegraphics[width=100mm, height=142mm]{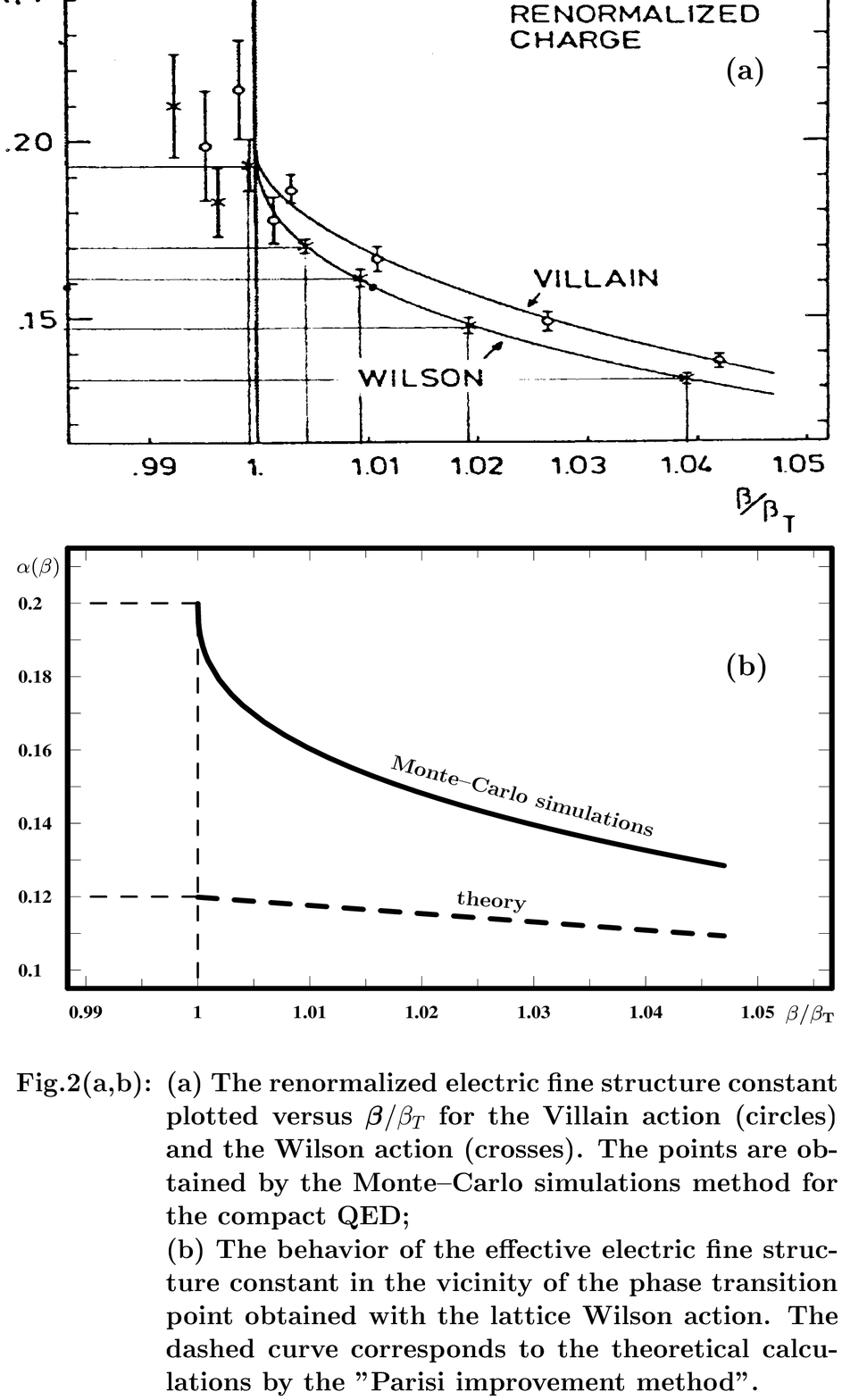}
\end{center}
\end{figure*}

The result of Ref.\cite{10s} for the behavior of $\alpha(\beta)$ in 
the vicinity
of the phase transition point $\beta_T$ is shown in Fig.2(a) for the Wilson
and Villain lattice actions. Fig.2(b) demonstrates the comparison of the
function $\alpha(\beta)$ obtained by Monte Carlo method for the Wilson
lattice action and by theoretical calculation of the same quantity.
The theoretical (dashed) curve was calculated by so-called "Parisi improvement
formula" \cite{13p}:
\begin{equation}
    \alpha (\beta )=[4\pi \beta W_p]^{-1}.     \label{48}
\end{equation}
Here $W_p=<\cos \Theta_p >$ is a mean value of the plaquette energy.
The corresponding values of $W_p$ are taken from Ref.\cite{9s}.

\begin{figure*}
\begin{center}
\noindent\hspace*{-5mm}\includegraphics[width=130mm, height=105mm]{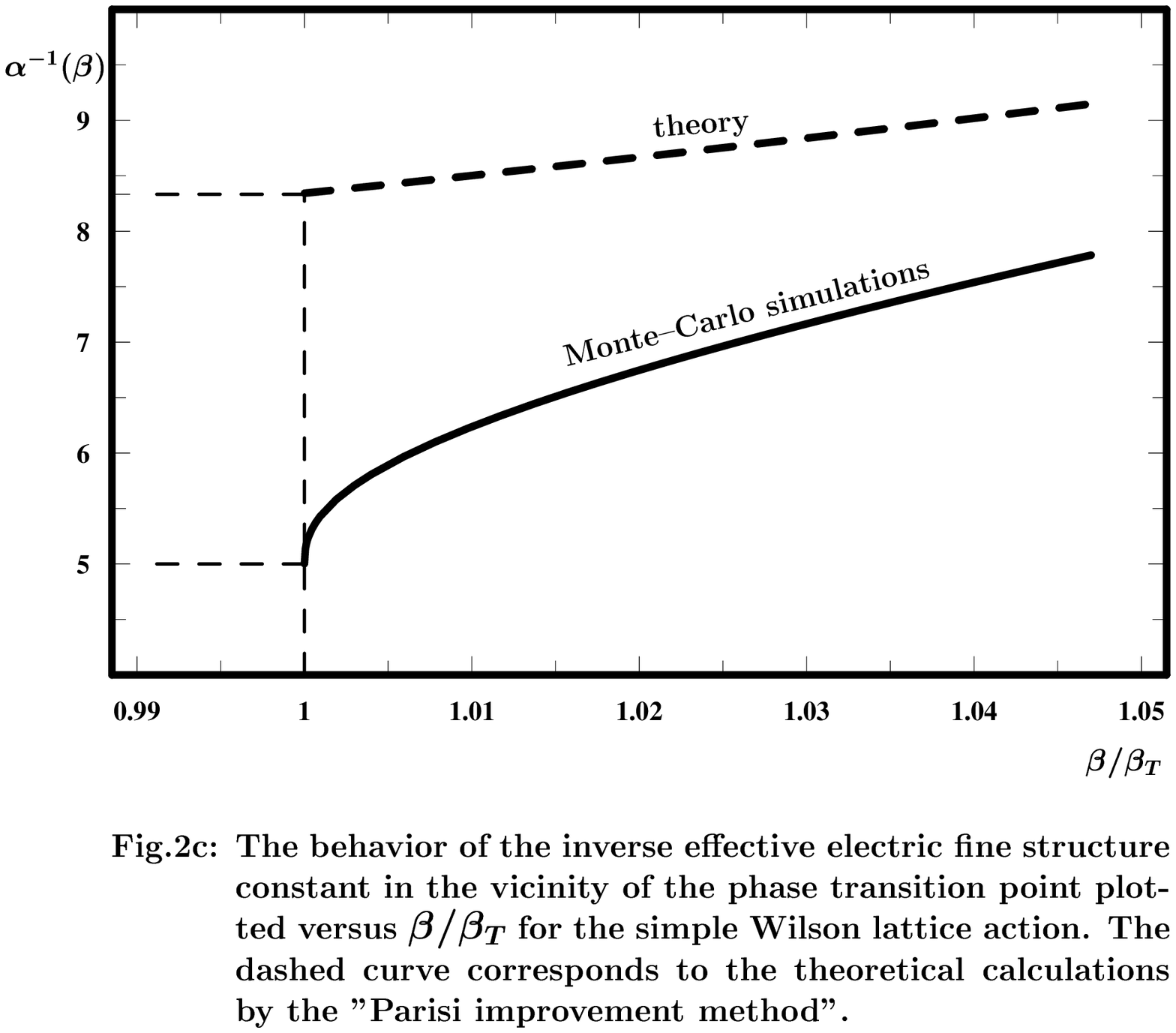}
\end{center}
\end{figure*}

The theoretical value of $\alpha_{crit}$ is less than the "experimental"
(Monte Carlo) value (\ref{47}):
\begin{equation}
      \alpha_{crit}\mbox{(in\,\,lattice\,\,theory)}\approx{0.12}.
                                                    \label{49}
\end{equation}
According to Fig.2(c):
\begin{equation}
   \alpha_{crit.,theor.}^{-1}\approx 8.                   \label{50a}
\end{equation}
This result does not coincide with the lattice result (\ref{47}),
which gives the following value:
\begin{equation}
   \alpha_{crit.,lat.}^{-1}\approx 5.                   \label{50b}
\end{equation}
The deviation of theoretical calculations of $\alpha(\beta )$ from the
lattice ones, which is shown in Fig.2(b,c),
has the following explanation: Parisi improvement formula (\ref{48})
is valid in Coulomb--like phase where the mass of artifact monopoles is infinitely
large and photon is massless. But in the vicinity of the phase
transition (critical) point the monopole mass $m\to 0$ and photon
acquires the non--zero mass $m_0\neq 0$ (on side of the confinement). This
phenomenon leads to the "freezing" of $\alpha$: the effective electric fine
structure constant is almost unchanged in the confinement phase and approaches
to its maximal value $\alpha=\alpha_{max}$. The authors of Ref.\cite{14p}
predicted that in the confinement phase, where we have the formation of
strings, the fine structure constant $\alpha$ cannot be infinitely large, but
has the maximal value: $\alpha_{max} \approx \frac{\pi}{12}\approx 0.26$
due to the Casimir effect for strings.

\section{Lattice Artifact Monopoles and Higgs Monopole Model}

Lattice monopoles are responsible for the confinement in lattice
gauge theories what was confirmed by many numerical and theoretical
investigations.

In the previous papers [\cite{17p}--\cite{19p}] the calculations of the U(1)
phase transition (critical) coupling constant were connected with the
existence of artifact monopo\-les in the lattice gauge theory and also
in the Wilson loop action model \cite{19p}.

In Ref.\cite{19p} we (L.V.L. and H.B.Nielsen) have put forward the speculations
of Refs.[\cite{17p},\cite{18p}] suggesting that the modifications of the form of
the lattice action might not change too much the phase transition value of the
effective continuum coupling constant. The purpose was to investigate this
approximate stability of the critical coupling with respect to a somewhat
new regularization being used instead of the lattice, rather than just
modifying the lattice in various ways.
In \cite{19p} the Wilson loop action was considered in the
approximation of circular loops of radii $R\ge a$. It was shown that the
phase transition coupling constant is indeed approximately independent
of the regularization method: ${\alpha}_{crit}\approx{0.204}$,
in correspondence with the Monte Carlo simulation result on lattice:
${\alpha}_{crit}\approx{0.20\pm 0.015}$ (see Eq.(\ref{47})).

But in Refs.[\cite{20p}--\cite{22p}], instead of using the lattice or Wilson loop
cut--off, we have considered the Higgs Monopole Model (HMM) approximating
the lattice artifact monopoles as fundamental pointlike particles described
by the Higgs scalar field.
The simplest effective dynamics describing the
confinement mechanism in the pure gauge lattice U(1) theory
is the dual Abelian Higgs model of scalar monopoles \cite{13s}, [\cite{20p}--\cite{22p}] (shortly HMM).
This model, first suggested in Refs.\cite{13s}, considers the
following Lagrangian:
\begin{equation}
    L = - \frac{1}{4g^2} F_{\mu\nu}^2(B) + \frac{1}{2} |(\partial_{\mu} -
           iB_{\mu})\Phi|^2 - U(\Phi),\quad              \label{5y}
{\mbox{where}}\quad
 U(\Phi) = \frac{1}{2}\mu^2 {|\Phi|}^2 + \frac{\lambda}{4}{|\Phi|}^4
\end{equation}
is the Higgs potential of scalar monopoles with magnetic charge $g$, and
$B_{\mu}$ is the dual gauge (photon) field interacting with the scalar
monopole field $\Phi$.  In this model $\lambda$ is the self--interaction
constant of scalar fields, and the mass parameter $\mu^2$ is negative.
In Eq.(\ref{5y}) the complex scalar field $\Phi$ contains
the Higgs ($\phi$) and Goldstone ($\chi$) boson fields:
\begin{equation}
          \Phi = \phi + i\chi.             \label{7y}
\end{equation}
The effective potential in the Higgs Scalar ElectroDynamics (HSED)
was first calculated by Coleman and Weinberg \cite{20s} in the one--loop
approximation. The general method of its calculation is given in the
review \cite{21s}. Using this method, we can construct the effective potential
for HMM. In this case the total field system of the gauge ($B_{\mu}$)
and magnetically charged ($\Phi$) fields is described by
the partition function which has the following form in Euclidean space:
\begin{equation}
      Z = \int [DB][D\Phi][D\Phi^{+}]\,e^{-S},     \label{8y}
\end{equation}
where the action $S = \int d^4x L(x) + S_{gf}$ contains the Lagrangian
(\ref{5y}) written in Euclidean space and gauge fixing action $S_{gf}$.
Let us consider now a shift:
\begin{equation}
 \Phi (x) = \Phi_b + {\hat \Phi}(x)                \label{9y}
\end{equation}
with $\Phi_b$ as a background field and calculate the
following expression for the partition function in the one-loop
approximation:
$$
  Z = \int [DB][D\hat \Phi][D{\hat \Phi}^{+}]
   \exp\{ - S(B,\Phi_b)
   - \int d^4x [\frac{\delta S(\Phi)}{\delta \Phi(x)}|_{\Phi=
   \Phi_b}{\hat \Phi}(x) + h.c. ]\}\\
$$
\begin{equation}
    =\exp\{ - F(\Phi_b, g^2, \mu^2, \lambda)\}.      \label{10y}
\end{equation}
Using the representation (\ref{7y}), we obtain the effective potential:
\begin{equation}
  V_{eff} = F(\phi_b, g^2, \mu^2, \lambda)        \label{11y}
\end{equation}
given by the function $F$ of Eq.(\ref{10y}) for the constant background
field $ \Phi_b = \phi_b = \mbox{const}$. In this case the one--loop
effective potential for monopoles coincides with the expression of the
effective potential calculated by the authors of Ref.\cite{20s} for scalar
electrodynamics and extended to the massive theory (see review \cite{21s}).

Considering the renormalization group improvement
of the effective Coleman--Weinberg potential \cite{20s},\cite{21s}, written
in Ref.\cite{22p} for the dual sector of scalar electrodynamics in the
two--loop approximation, we have calculated the U(1) critical values of
the magnetic fine structure constant:
\begin{equation}
{\tilde\alpha}_{crit} = g^2_{crit}/4\pi\approx 1.20,      \label{1cr}
\end{equation}
and electric fine structure constant:
\begin{equation}
 \alpha_{crit} = \pi/g^2_{crit}\approx 0.208             \label{2cr}
\end{equation}
by the Dirac relation:
\begin{equation}
          eg= 2\pi, \quad{\mbox{or}}\quad \alpha\tilde \alpha = \frac{1}{4}.
                                                       \label{3dr}
\end{equation}
The values (\ref{1cr}),(\ref{2cr}) coincide with the lattice result
(\ref{47}).

\section{Monopoles strength group dependence}

As it was shown in a number of investigations, the confinement in the SU(N) lattice gauge
theories effectively comes to the same U(1) formalism. The reason is the
Abelian dominance in their monopole vacuum: monopoles of the Yang--Mills
theory are the solutions of the U(1)--subgroups, arbitrary embedded into
the SU(N) group. After a partial gauge fixing (Abelian projection by
't Hooft \cite{24p}) SU(N) gauge theory is reduced to an Abelian
$U(1)^{N-1}$ theory with $N-1$ different types of Abelian monopoles.
Choosing the Abelian gauge for dual gluons, it is possible to describe
the confinement in the lattice SU(N) gauge theories by the analogous
dual Abelian Higgs model of scalar monopoles.

Considering the Abelian gauge and taking into account that
the direction in the Lie algebra of monopole fields are gauge
independent, we have found in Ref.\cite{22p} an average over these directions
and obtained \underline{the group dependence} \underline{relation} between the
phase transition
fine structure constants for the groups $U(1)$ and $SU(N)/Z_N$:
\begin{equation}
      \alpha_{N,crit}^{-1}
           = \frac{N}{2}\sqrt{\frac{N+1}{N-1}}
                          \alpha_{U(1),crit}^{-1}.
                                            \label{25z}
\end{equation}

\section{AGUT-MPM prediction of the Planck scale values of the
U(1), SU(2) and SU(3) fine structure constants}

As it was assumed in Ref.\cite{17p}, the MCP values
$\alpha_{i,crit}$ in Eqs.(\ref{84y}) and (\ref{85y}) coincide with
the critical values of the effective fine structure
constants given by the generalized lattice SU(3)--, SU(2)-- and U(1)--gauge
theories.

Now let us consider $\alpha_Y^{-1}\,(\approx \alpha^{-1})$ at the point
$\mu=\mu_G\sim 10^{18}$ GeV shown in Fig.1.
If the point $\mu=\mu_G$ is very close to the Planck scale
$\mu=\mu_{Pl}$, then according to Eqs.(\ref{82y}) and (\ref{85y}), we have:
\begin{equation}
         \alpha_{1st\, gen.}^{-1}\approx
    \alpha_{2nd\, gen.}^{-1}\approx \alpha_{3rd\, gen.}^{-1}\approx
    \frac{\alpha_Y^{-1}(\mu_G)}{6}\approx 9,        \label{88y}
\end{equation}
what is almost equal to the value (\ref{50a}):
$$
            \alpha_{crit.,theor}^{-1}\approx 8
$$
obtained by the Parisi improvement method (see Fig.2(c)).
This means that in the U(1) sector of AGUT we have $\alpha $ near
the critical point. Therefore, we can expect the existence of MCP
at the Planck scale.

It is necessary to mention that the lattice investigators were not able
to obtain the lattice triple point values $\alpha_{i,crit}$
(i=1,2,3 correspond to U(1),SU(2) and SU(3) groups) by Monte Carlo
simulation methods.
These values were calculated theoretically by Bennett and Nielsen
in Ref.\cite{17p} using the Parisi improvement method \cite{13p}:
\begin{equation}
    \alpha_{Y,crit}^{-1}\approx 9.2\pm 1,
    \quad \alpha_{2,crit}^{-1}\approx 16.5\pm 1, \quad
    \alpha_{3,crit}^{-1}\approx 18.9\pm 1.                 \label{89y}
\end{equation}
Assuming the existence of MCP at $\mu=\mu_{Pl}$
and substituting the last results in Eqs.(\ref{84y}) and (\ref{85y}),
we have the following prediction of AGUT \cite{17p}:
\begin{equation}
   \alpha_Y^{-1}(\mu_{Pl})\approx 55\pm 6; \quad
   \alpha_2^{-1}(\mu_{Pl})\approx 49.5\pm 3; \quad
   \alpha_3^{-1}(\mu_{Pl})\approx 57.0\pm 3.
                                                          \label{90y}
\end{equation}
These results coincide with the results (\ref{82y}) obtained by the
extrapolation of experimental data to the Planck scale
in the framework of pure SM (without any new particles) \cite{33a}, \cite{17p}.

Using the relation (\ref{25z}), we obtained the following relations:
\begin{equation}
    \alpha_{Y,crit}^{-1} : \alpha_{2,crit}^{-1} : \alpha_{3,crit}^{-1}
           = 1 : \sqrt{3} : 3/\sqrt{2} = 1 : 1.73 : 2.12.
                                                     \label{91y}
\end{equation}
Let us compare now these relations with the MPM prediction.

For $\alpha_{Y,crit}^{-1}\approx 9.2$  given by the first equation of
(\ref{89y}), we have:
\begin{equation}
 \alpha_{Y,crit}^{-1} : \alpha_{2,crit}^{-1} : \alpha_{3,crit}^{-1}
    = 9.2 : 15.9 : 19.5.                                     \label{92y}
\end{equation}
In the framework of errors the last result coincides with the
AGUT--MPM prediction (\ref{89y}).
Of course, it is necessary to take into account an approximate description
of confinement dynamics in the SU(N) gauge theories, which was
used in our investigations.

\section{The possibility of the Grand Unification Near the Planck Scale}

We can see new consequences of the extension of $G$--theory, if
$G$--group is broken down to its diagonal subgroup $G_{diag}$, i.e. SM,
not at $\mu_G\sim 10^{18}$ {GeV}, but at $\mu_G\sim 10^{15}$ {GeV}.
In this connection, it is very attractive to consider the gravitational
interaction.

\subsection{"Gravitational finestructure constant" evolution}

The gravitational interaction between two particles
of equal masses M is given by the usual classical Newtonian potential:
\begin{equation}
   V_g = - G \frac{M^2}{r} =
           - \left(\frac{M}{M_{Pl}}\right)^2\frac{1}{r}
                   = - \frac{\alpha_g(M)}{r},              \label{1x}
\end{equation}
which always can be imagined as a tree--level approximation of quantum
gravity.

\begin{figure*}
\begin{center}
\noindent\hspace*{-5mm}\includegraphics[width=130mm, height=110mm]{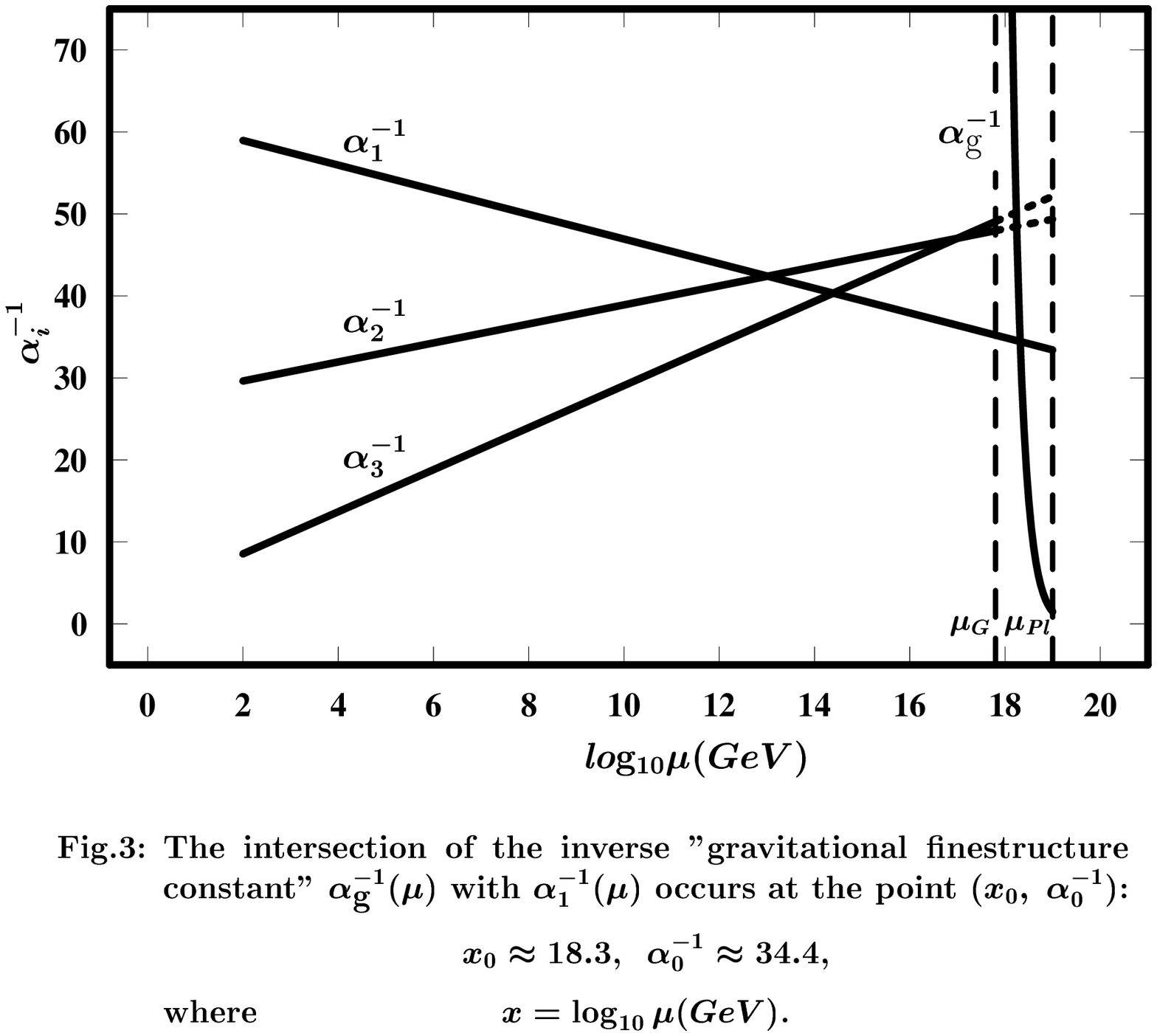}
\end{center}
\end{figure*}

Then the quantity:
\begin{equation}
      \alpha_g = \left(\frac{\mu}{\mu_{Pl}}\right)^2     \label{2x}
\end{equation}
plays a role of the "gravitational finestructure constant" and the
evolution of its inverse quantity is presented in Fig.3 together with the
evolutions of $\alpha_{1,2,3}^{-1}(\mu)$ (here we have returned to the
consideration of $\alpha_1$ instead of $\alpha_Y$).

Then we see the intersection of $\alpha_g^{-1}(\mu)$
with $\alpha_1^{-1}(\mu)$ in the region of $G$--theory at the point:
$$
               (x_0, \alpha_0^{-1}),
$$
where
\begin{equation}
      x_0 \approx 18.3,  \quad {\mbox{and}}\quad
       \alpha_0^{-1} \approx 34.4,                   \label{3x}
\end{equation}
and $\,x = \log_{10}\mu$.

\subsection{The consequences of the breakdown of $G$-theory
at $\mu_G\sim 10^{15}$ or $10^{16}$ GeV}

Let us assume now that the group of symmetry $G$ undergoes the breakdown
to its diagonal subgroup not at $\mu_G\sim 10^{18}$ GeV, but at
$\mu_G\sim 10^{15}$ GeV, i.e. before the intersection of
$\alpha_{2}^{-1}(\mu)$ and $\alpha_{3}^{-1}(\mu)$ at
$\mu\sim 10^{16}$
GeV.

As a consequence of behavior of the function $\alpha^{-1}(\beta)$
near the phase transition point, shown in Fig.2c, we have to expect the
change of the evolution of $\alpha_i^{-1}(\mu)$ in the region $\mu > \mu_G$
shown in Fig.1 by dashed lines. Instead of these dashed lines,
we must see the decreasing of $\alpha_i^{-1}(\mu)$, when they
approach MCP, if this MCP really exists at the Planck scale.

According to Fig.2c, in the very vicinity of the phase transition point
(i.e. also near the MCP at $\mu=\mu_{Pl}$), we cannot
describe the behavior of $\alpha_i^{-1}(\mu)$ by the one---loop
approximation RGE.

It is well known, that the one--loop approximation RGEs
for $\alpha_i^{-1}(\mu)$ can be
described in our case by the following expression~\cite{1a}:
\begin{equation}
  \alpha_i^{-1}(\mu) =
  \alpha_i^{-1}(\mu_{Pl}) + \frac{b_i}{4\pi}\log(\frac{\mu^2}{\mu^2_{Pl}}),
                                                \label{4x}
\end{equation}
where $b_i$ are given by the following values:
$$
   b_i = (b_1, b_2, b_3) =
$$
\begin{equation}
( - \frac{4N_{gen}}{3} -\frac{1}{10}N_S,\,\,
      \frac{22}{3}N_V - \frac{4N_{gen}}{3} -\frac{1}{6}N_S,\,\,
      11 N_V - \frac{4N_{gen}}{3} ).                   \label{5x}
\end{equation}
The integers $N_{gen},\,N_S,\,N_V\,$ are respectively the numbers
of generations,
Higgs bosons and different vector gauge fields of given "colors".

For the SM we have:
\begin{equation}
       N_{gen} = 3, \quad N_S = N_V =1,                    \label{6x}
\end{equation}
and the corresponding slopes describe the evolutions of
$\alpha_i^{-1}(\mu)$ up to $\mu = \mu_G$ presented in Fig.3.

\begin{figure*}
\begin{center}
\noindent\hspace*{-5mm}\includegraphics[width=85mm, height=113mm]{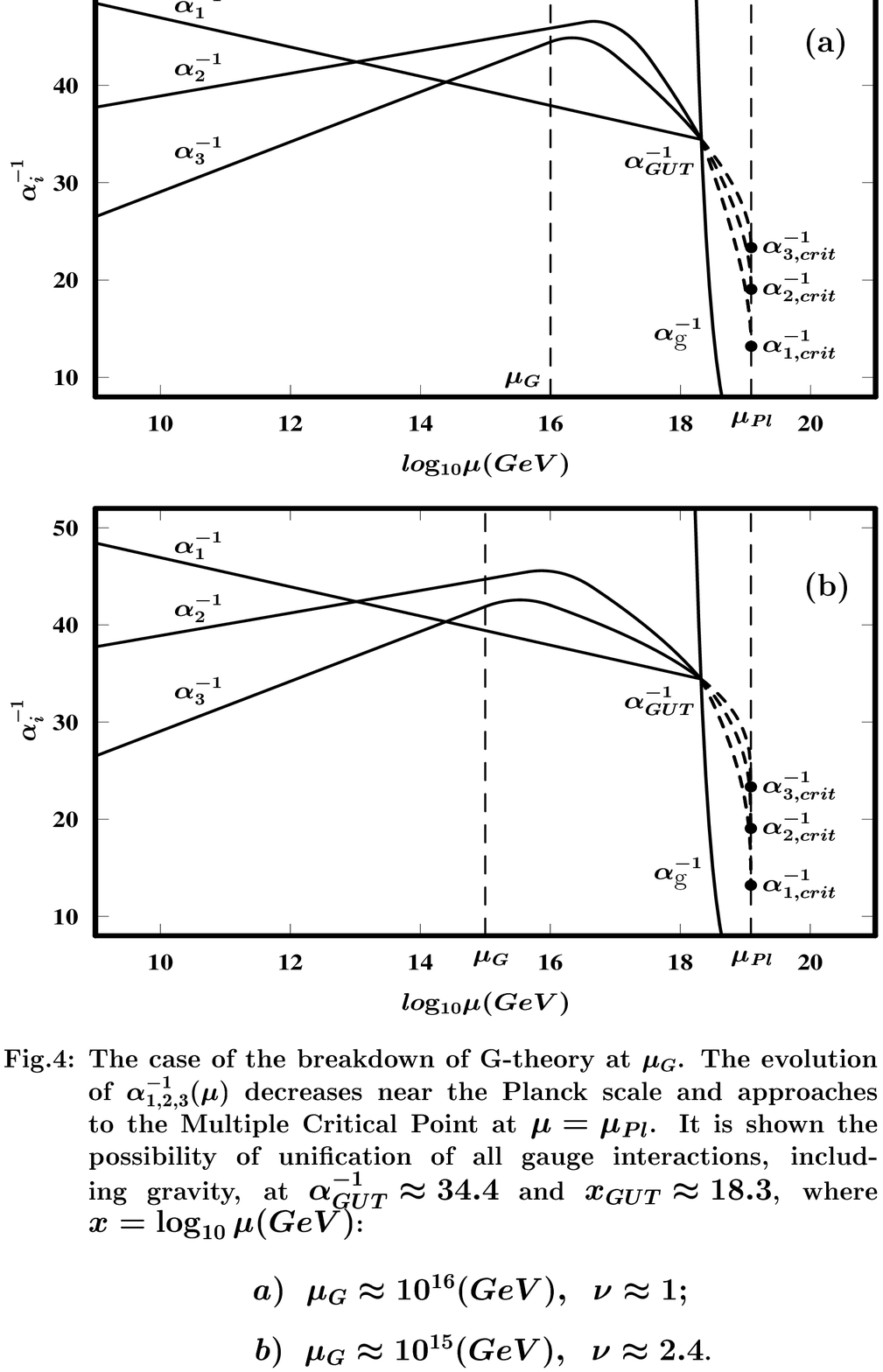}
\end{center}
\end{figure*}

But in the region $\mu_G\le \mu \le \mu_{Pl}$, when $G$--theory works,
we have $N_V = 3$ (here we didn't take into account the additional
Higgs fields which can change the number $N_S$), and the one--loop
approximation slopes are almost 3 times larger than the same ones for the SM.
In this case, it is difficult to understand that such evolutions give the
MCP values of $\alpha_i^{-1}(\mu_{Pl})$, which are
shown in Fig.4. These values were obtained by the following way:
$$
  \alpha_1^{-1}(\mu_{Pl}) \approx
  6\cdot \frac{3}{5}\alpha_{U(1),crit}^{-1}\approx 13, \quad
$$
$$
  \alpha_2^{-1}(\mu_{Pl}) \approx
  3\cdot \sqrt{3}\alpha_{U(1),crit}^{-1}\approx 19,\quad
$$
\begin{equation}
  \alpha_3^{-1}(\mu_{Pl}) \approx
  3\cdot \frac{3}{\sqrt 2}\alpha_{U(1),crit}^{-1}\approx 24,      \label{7x}
\end{equation}
where we have used the relation (\ref{25z}) with
\begin{equation}
  \alpha_{U(1),crit} =
  \frac{\alpha_{crit}}{\cos^2\theta_{\overline{MS}}}\approx 0.77\alpha_{crit},
                      \label{8x}
\end{equation}
taking into account our HMM result (\ref{2cr}):
$\alpha_{crit}\approx 0.208$, which coincides with the lattice result
(\ref{47}) and gives:
\begin{equation}
  \alpha_{U(1),crit}^{-1} \approx 3.7.
                                            \label{9x}
\end{equation}
In the case when $G$--group undergoes the breakdown to the SM not
at $\mu_G\sim 10^{18}$ GeV, but at $\mu_G\sim 10^{15}$ GeV, the artifact
monopoles of non-Abelian  SU(2) and SU(3) sectors of $G$--theory begin
to act more essentially.

According to the group dependence relation (\ref{25z})
(although now it is necessary to expect that it is very approximate)
we have, for example, the following estimation
at $\mu_G\sim 10^{15}$ GeV:
\begin{equation}
   \alpha_{U(1)}^{-1}(\mu_G) \sim 7\quad -\quad for\quad
       SU(3)_{1st\,\,gen.}, \,etc.                         \label{10x}
\end{equation}
which is closer to MCP than the previous value $\alpha_Y^{-1}\sim 9$,
obtained for the AGUT breakdown at $\mu_G\sim 10^{18}$ GeV.

It is possible to assume that $\beta$--functions of SU(2) and SU(3)
sectors of $G$--theory change their one--loop approximation behavior
in the region $\mu > 10^{16}$ GeV and $\alpha_{2,3}^{-1}(\mu)$
begin to decrease, approaching the phase transition (multiple critical)
point at $\mu = \mu_{Pl}$. This means that the asymptotic freedom
of non--Abelian theories becomes weaker near the Planck scale, what can
be explained by the influence of artifact monopoles.
It looks as if these $\beta$--functions have
singularity at the phase transition point and, for example,
can be approximated by the following expression:
\begin{equation}
   \frac{d\alpha^{-1}}{dt} = \frac{\beta(\alpha)}{\alpha}\approx
             A(1 - \frac{\alpha}{\alpha_{crit}})^{-\nu}\quad
              {\mbox{near the phase transition point}}.       \label{11x}
\end{equation}
This possibility is shown in Fig.4 for $\nu\approx 1$ and $\nu \approx 2.4$.

Here it is worth-while to comment that such a tendency was revealed
in the vicinity of the confinement phase by the forth--loop
approximation of $\beta$--function in QCD (see Ref.\cite{40a}).

\subsection{Does the [SU(5)]$^{\bf 3}$ SUSY unification exist near the Planck
scale?}

Approaching the MCP in the region of $G$--theory ($\mu_G\le \mu \le \mu_{Pl}$),
$\alpha_{2,3}^{-1}(\mu)$ show the necessity
of intersection of $\alpha_{2}^{-1}(\mu)$ with $\alpha_{3}^{-1}(\mu)$
at some point of this region if $\mu_G\sim 10^{15}$ or $10^{16}$ GeV
(see Fig.4).
If this intersection takes place at the point $(x_0,\,\alpha_0^{-1})$
given by Eq.(\ref{3x}), then we have the unification of all gauge
interactions (including the gravity) at the point:
\begin{equation}
  (x_{GUT};\,\alpha_{GUT}^{-1})\approx (18.3;\,34.4),     \label{12x}
\end{equation}
where $x = \log_{10}\mu$(GeV).
Here we assume the existence of [SU(5)]$^3$ SUSY unification having superparticles of masses
\begin{equation}
           M\approx 10^{18.3}\, {\mbox{GeV}}.        \label{13x}
\end{equation}
The scale $\mu_{GUT}=M$, given by Eq.(\ref{13x}), can be considered
as a SUSY breaking scale.

\begin{figure*}
\begin{center}
\noindent\hspace*{-5mm}\includegraphics[width=85mm, height=113mm]{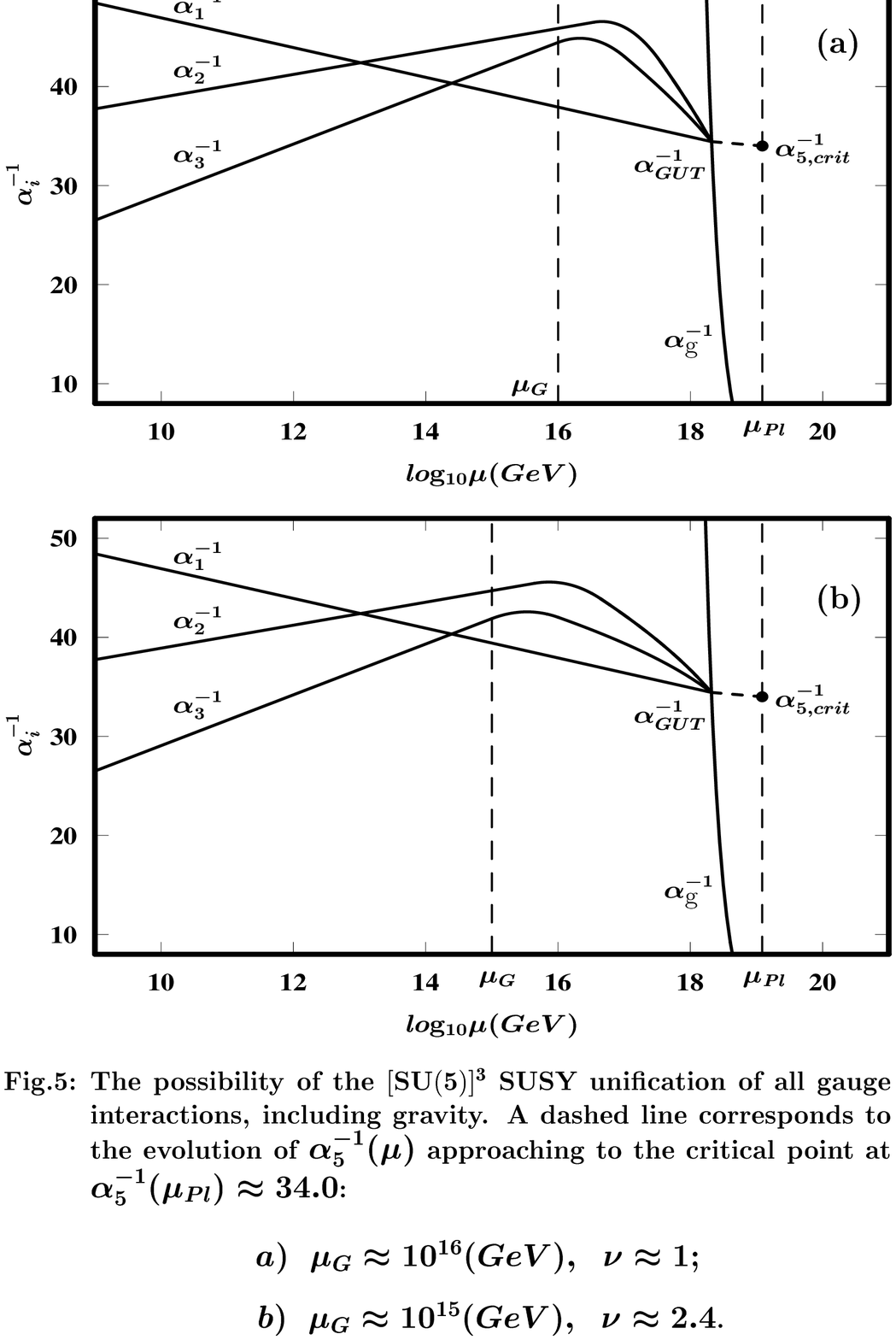}
\end{center}
\end{figure*}

Figures 5(a,b) demonstrates such a possibility of unification.
We have investigated the solutions of joint intersections of
$\alpha_g^{-1}(\mu)$ and all
$\alpha_i^{-1}(\mu)$ at different $x_{GUT}$ with different $\nu$
in Eq.(\ref{11x}). These solutions exist from
$\nu \approx 0.5$ to $\nu \approx 2.5$.

The unification theory with [SU(5)]$^3$--symmetry was suggested first
by S. Raj\-poot \cite{40}.

It is essential that the critical point in this theory,
obtained by means of Eqs.(\ref{84y}), (\ref{25z}) and (\ref{9x}),
is given by the following value:
\begin{equation}
      \alpha_{5,crit}^{-1}\approx 3\cdot \frac{5}{2}\sqrt
                              {\frac{3}{2}}\approx 34.0.      \label{14x}
\end{equation}
The point (\ref{14x}) is shown in Fig.5(a,b) presented for the cases:

1. $\nu \approx 1$, $\alpha^{-1}_{GUT}\approx 34.4$, $x_{GUT}\approx 18.3$
and $\mu_G\approx 10^{16}$ GeV shown in Fig.5(a);

2. $\nu \approx 2.4$, $\alpha^{-1}_{GUT}\approx 34.4$, $x_{GUT}\approx 18.3$
and $\mu_G\approx 10^{15}$ GeV shown in Fig.5(b).

We see that the point (\ref{14x}) is very close to the
unification point $\alpha_{GUT}^{-1}\approx 34.4$ given by Eq.(\ref{12x}).
This means that the unified theory, suggested here as the [SU(5)]$^3$ SUSY
unification, approaches the confinement phase at the Planck scale.
But the confinement of all SM particles is impossible in our world,
because then they have to be confined also at low energies what is not
observed in the Nature.

It is worth--while to mention that using the Zwanziger formalism for the
Abelian gauge theory with electric and magnetic charges
(see Refs.[\cite{41}--\cite{43}] and \cite{21p}), the possibility of unification
of all gauge interactions at the Planck scale was considered in Ref.\cite{38}
in the case when unconfined monopoles come to the existence near the Planck
scale. They can appear only in G--theory, because RGEs for monopoles
strongly forbid their deconfinement in the SM up to the Planck scale.
But it is not obvious that they really exist in the G--theory. This problem
needs more careful investigations, because our today knowledge about
monopoles is still very poor.

The unified theory, suggested in this paper, essentially differs in its
origin from the case considered in Ref.\cite{38}, because this theory
does not assume the existence of deconfining monopoles up to the
Planck scale, but assumes the influence of our space--time lattice
artifact monopoles near the phase transition (critical) point.

Considering the predictions of this unified theory for the low--energy
physics and cosmology, maybe in future we shall be able to answer
the question: "Does the [SU(5)]$^3$ SUSY unification
theory really exist near the Planck scale?"

\section{Reminiscences on a collaboration with Holger}

I got acquainted with Holger Bech Nielsen in the year 1970,
during the Roches\-ter Conference in Kiev (Ukraineq, USSR).
His brilliant achievements and childish charm conquered me.

When I returned to Tbilisi after this conference, I received
from Holger a very kind letter, a pretty present and 9 papers
of his activity. Then I understood that the destiny gave me a chance to meet
a genius.

That time my homeland of Georgia entered the USSR, and KGB immediately
called me to their office and began to cross-examine me, and shout:
``Why `Dear Lara' in his letter, but not `Dr. Laperashvili'?!
You will be thrown out of the Institute immediately! Be glad that you will
be not imprisoned!" Then I did not inform Holger about this story,
because I was afraid that he would stop writing to me. The director
of my Institute of Physics in Tbilisi saved me: he told to KGB, that
I am spy and I must find out a very important information (it was lie,
of course). And Holger continued to write his letters and to send his papers.
Six times he invited me to NBI, but without success: I was forbidden to go
abroad and to collaborate with him. And only after ``perestroyka",
in 1987, I began to collaborate with him and came to Copenhagen
first in 1990. After 20 years! But our friendship is alive up to now.

Dual strings, analogue model, Koba--Nielsen representation, KNO--scaling,
ANO--strings, spaggheti vacuum, Cheshire cat bag model, Nielsen--Ninomiya
theorem, Random dynamics, Multiple Point Principle, Anti--grand unification,
Higgs--Top masses prediction, etc.,etc.,etc, - Holger's name is well--known
in the World of Physics. He is in the first row of the World physicists.
There was time when all leading (high energy) physical journals were full
of his name in the titles of different papers.

I am sure that Holger made for Denmark so much that his name and activity
will add a glorious golden page to history of Denmark.
And of course, Russia, Georgia and the World Science will not forget
his outstanding services to physics. I am sure that he will produce
excellent physics in the nearest future, because he is fresh, active and full
of non--trivial ideas.

The collaboration with Holger Bech Nielsen was the best time of my life.
Always busy with physics, he has unimaginable flux of energy, fantastic
intuition and a great volume of knowledges. He operates by complicated
mathematical structures.

Holger is a poet in physics.

Simultaneously, he has the severity in physical investigations and
childish perception of human life. He is a bright person, full of
kindness and progressive ideas. Many people in different countries love him.

Dear Holger, I congratulate you for your 60th jubilee and wish you
many, many years activity in future, health, wealth and a Great Success.
Thank you very much for collaboration with me. I send you my Great Love.

\title*{Some Remarks on the ``Classical'' Large $N$ Limit}
\author{%
P. Olesen\thanks{To Holger Bech Nielsen on his 60th birthday}}
\institute{%
The Niels Bohr Institute, Blegdamsvej 17
DK-2100 Copenhagen, Denmark}

\authorrunning{P. Olesen}
\titlerunning{Some Remarks on the ``Classical'' Large $N$ Limit}
\maketitle

\begin{abstract} 
It has been proposed some time ago that the large $N-$limit can be understood
as a ``classical limit'', where commutators in some sense approach the
corresponding Poisson brackets. We discuss this in the light of 
some recent numerical
results for an SU($N$) gauge model, which do not agree with this
``classicality'' of the large $N-$limit. The world sheet becomes very
crumpled. We speculate that this effect would disappear in supersymmetric
models. 
\end{abstract}

In the 1970ies Holger and I discussed the confinement problem quite a lot.
In those days the picture of confinement by means of a string connecting
the quarks became more and more accepted.
It was known that the Nambu-Goto action could be written as a square root of
the Poisson bracket squared. One of the subjects we discussed was 
whether somehow the Poison bracket was related to the commutator like
in quantum mechanics, in some limit. However,
we did not gain any insight in this. After the work by Eguchi and Kawai
it became of course clear that the relevant limit was the large $N$
limit, where the action could be expressed in terms of commutators. 
Later on Hoppe \cite{hoppe} constructed a very suggestive relation between
Poisson brackets and commutators. This was then used in ingenious ways
by a number of authors \cite{floratos,fairlie,bars}, in connection
with the Eguchi-Kawai formulation of the large $N$ theory.

 
To give a brief review of this approach,
let us consider SU($N$) with the gauge field $A_\mu$ and the generators
$l_{\bf k},{\bf k}=(k_1,k_2)$,
\begin{equation}
(A_\mu)^j_i=\sum_{\bf k}a^{\bf k}_\mu~ (l_{\bf k})^j_i,
\end{equation}
where $a^{\bf k}_\mu$ are expansion coefficients to be integrated in functional
integrals. This expansion can be compared to the one for a string variable 
$X_\mu (\sigma,\tau)$,
\begin{equation}
X_\mu (\sigma,\tau)=\sum_{\bf k}a^{\bf k}_\mu~e^{i{\bf k\sigma}},
\end{equation}
with $\sigma=(\sigma,\tau)$. The variables $X_\mu$ are the Weyl
transform of the matrices $A_\mu$.
The generators can be constructed explicitly
in terms of the Weyl matrices, and have the commutation relation
\begin{equation}
[l_{{\bf k}_1},l_{{\bf k}_2}]=i\frac{N}{2\pi}~\sin \left(\frac{2\pi}{N} 
{\bf k}_1\times {\bf k_2}\right)~l_{{\bf k}_1+{\bf k}_2}\rightarrow i
\left({\bf k}_1\times {\bf k_2}\right)~l_{{\bf k}_1+{\bf k}_2}
\label{sine}
\end{equation}
for $N\rightarrow \infty$, where ${\bf k}_1\times {\bf k_2}=
(k_1)_1(k_2)_2-(k_2)_1(k_1)_2$. This leads to a comparison
of the commutator with the Poisson bracket. For example, one has
\begin{equation}
[A_\mu,A_\nu]=i\frac{N}{2\pi}\sum_{\bf k,p}a^{\bf k}_\mu~ a^{\bf p}_\nu~
\sin \left(\frac{2\pi}{N}{\bf k}\times {\bf p}\right)~
l_{{\bf k}_1+{\bf k}_2},
\label{modes}
\end{equation}
and
\begin{equation}
\{X_\mu (\sigma),X_\nu (\sigma)\}=i\sum_{\bf k,p}a^{\bf k}_\mu~ a^{\bf p}_\nu~
({\bf k}\times{\bf p})~e^{i({\bf k}+{\bf p})}.
\end{equation}
Using 
\begin{equation}
{\rm tr}~ l_{\bf k}l_{\bf p}\propto \delta_{\bf k+p,0},~~\int d^2\sigma~
e^{i({\bf k}+{\bf p}){\bf \sigma}}\propto \delta_{\bf k+p,0},
\end{equation}
one can derive that
\begin{equation}
{\rm tr}~[A_\mu,A_\nu]^2\rightarrow {\rm const.}~\int d^2\sigma 
\{X_\mu,X_\nu\}^2,
\end{equation}
where the constant (depending on $N$) can be removed by a suitable
normalization, and where it was used that
\begin{equation}
\frac{N}{2\pi}~\sin \left(\frac{2\pi}{N} {\bf k}_1\times {\bf k_2}\right)~
\rightarrow {\bf k}_1\times {\bf k_2}
\end{equation}
for $N\rightarrow\infty$, even {\it inside} the relevant sum over modes. In 
this sense the commutator approaches the  Poisson 
bracket \cite{floratos,fairlie,bars}. 


Unfortunately this argument, leading to a nice connection between strings and
fields in the large $N-$limit, depends on performing the limit $N\rightarrow
\infty$ inside sums over modes like in eq. (\ref{modes}). This is
evidently valid if the low modes dominate, since the difference
between sin $x$ and the linear function $x$ is considerable when $x$ is not 
close to 0.


Recently Anagnostopoulos, Nishimura and the author \cite{us} started to
investigate numerically the mode-distribution for large $N$ in a
four-dimensional SU($N$) model with the partition function
\begin{equation}
\int dA_\mu~\exp\left(\frac{1}{4g^2}{\rm tr}[A_\mu,A_\nu]^2\right).
\label{action}
\end{equation}
This model has been shown to exist \cite{krauth,hotta,austing}.
The expectation value of the commutator squared 
\begin{equation}
M=<{\rm tr}[A_\mu,A_\nu]^2>,
\label{olesen1}
\end{equation}
can be found from a scaling argument \cite{hotta} to be $N^2-1$, and we 
checked our
numerical method by seeing that it leads very precisely to this
result. 

Then we computed the corresponding expectation value of the Poisson
bracket by introducing the Weyl-transform\footnote{The 
Weyl transform $X_\mu(\sigma)$ of the matrix $A_\mu$ is given by $X_\mu(\sigma)
\propto\sum_k \exp (ik\sigma)~$tr$(l_kA_\mu)$}$X_\mu (\sigma)$ (defined
on a torus) of $A_\mu$, and we therefore found the behavior of
\begin{equation}
<P>=<\int d^2\sigma~\{X_\mu(\sigma),X_\nu (\sigma)\}^2>.
\end{equation}
Here the normalization is such that if only the very lowest modes are kept,
one has the result $<M>=<P>$. For the details of this construction, we
refer to the paper \cite{us}. One important point discussed 
in this paper is that the Poisson bracket is gauge dependent, and hence
$<P>$ is in general gauge dependent, in contrast to the gauge invariant
$<M>$. We have selected a gauge fixing analogous to the Landau gauge.
Of course, one can anyhow argue that if the commutator really approaches
the Poisson bracket, gauge invariance of the former implies gauge
invariance of the latter, at least approximatively.


In our numerical calculations we took $N=15,25,$ and 35. The results were 
quite embarrasing. Keeping only a few modes, we got $<M>\approx <P>$, but 
including all modes these two quantities differ considerably: Whereas very 
precisely we find $<M>=N^2-1$,
it turns out that $<P>$ rather grows like $O(N^4)$. Therefore the two
expectation values do not at all agree, and the discrepancy actually
increases with $N$. We can therefore say that in the model with the
action (\ref{action}), the Poisson bracket is not approximated by
the commutator. 

We also computed the average of the area of the world sheet,
\begin{equation}
<A>=<\int d^2\sigma \sqrt{\{X_\mu (\sigma),X_\nu (\sigma)\}^2}~>.
\end{equation}
It turns out that 
$<A>\approx <M>$. Thus, the expectation value of the Nambu-Goto action
does indeed behave approximately as the commutator. This does, however, not
show that the action (\ref{action}) is approximately equal to the Nambu-Goto
action, since we have computed only one moment of $A$.


The negative result given above shows that the question of
whether the large $N$ limit is a classical limit, depends on the dynamics, and
a simple model like (\ref{action}) is not able to suppress the higher
modes, and hence one cannot use the limit (\ref{sine}) inside sums over
modes. The string picture resulting from the behavior of $X_\mu (\sigma)$
turns out to be an extremely crumpled string, with an almost infinite
($\approx 33$) Hausdorff dimension. This is another way of seeing that 
higher modes are important, since dominance of lower modes would lead to 
a smooth string. 


As pointed out in \cite{us} the model (\ref{action}) does exist in a
world-sheet version, since the commutator can be replaced by star products
of the $X'$s, so the action can be rewritten as
\begin{equation}
\int dX_\mu~\exp\left[\frac{-1}{4g^2N}\int d^2\sigma \left(X_\mu (\sigma)
\star X_\nu (\sigma)-X_\nu (\sigma)\star X_\mu (\sigma)\right)^2\right],
\end{equation}
which exists because the corresponding matrix model exists. The world sheet
defined by this theory (with an infinite number of derivatives in the
star product) is the same as before, and is thus extremely crumpled.


Thus the dynamical question of the existence of a ``classical'' large 
$N-$limit boils down to finding a rather smooth string, since any
``violent'' string would require higher modes. Presumably the QCD
string is smooth. In this case the action (\ref{action}) is replaced by
a quenched action, which hopefully can manage to suppress the high
modes. Another possibility is to take a supersymmetric version of
(\ref{action}), since supersymmetry in general makes strings more smooth:
For example, in the vacuum energy the tachyon is removed by supersymmetry.
For the simple bosonic string it was shown long time ago by Alvarez
\cite{alvarez} that the bosonic string is very crumpled,
and by the author \cite{po} that the tachyon produces this crumpling. 
Alvarez' formula for the free energy \cite{alvarez},
\begin{equation}
F=\frac{1}{2\pi \alpha '}T\sqrt{R^2-R_c^2},
\end{equation}
can actually be interpreted \cite{po} as a formula for the lowest tachyonic 
energy, with the tachyon mass squared being proportional to $-R_c^2$.
The phenomenon of crumpling should thus not occur in the supersymmetric case. 
We hope that these questions will be investigated numerically for
supersymmetric versions of (\ref{action}) in the future.

Finally, I would like to take this opportunity to thank Holger for
pleasant years of discussions of very many aspects of physics.

\newcommand{\N}{{\cal N}}
\newcommand{\<}{\langle}
\newcommand{\oc}{{\cal O}_C}
\renewcommand{\a}{\alpha}
\renewcommand{\b}{\beta}
\newcommand{\half}{{1\over 2}}
\renewcommand{\>}{\rangle}
\def\tro{\tilde{\rho}}
\def\dt{\partial_{\tau}}
\def\ds{\partial_{\sigma}}
\def\R{\bar{R}}

\title*{String Theory and the Size of Hadrons}
\author{%
L. Susskind\thanks{SU-ITP 00-25}}
\institute{%
Department of Physics
Stanford University, Stanford, CA 94305-4060}


\authorrunning{L. Susskind}
\titlerunning{String Theory and the Size of Hadrons}
\maketitle

\begin{abstract}
We begin by outlining the ancient puzzle of off shell currents and
the infinite size particles
in a string theory of hadrons. We then consider the problem from
the modern AdS/CFT perspective. We argue that although hadrons
should be thought of as  ideal thin strings from the 5-dimensional
bulk
point of view, the 4-dimensional strings are a superposition of
``fat" strings of different thickness.

We also find that the warped nature of the target geometry
provides a mechanism
for taming the infinite zero point fluctuations which apparently
produce a
divergent result for hadronic radii.
\end{abstract}

\setcounter{footnote}{0}
\setcounter{equation}{0}
\section{Meeting Holger }

When I was a school kid  during the early 1950's
we used to have to read a magazine
called  ``The Reader's Digest".  It was full of corny articles about
patriotic platitudes which were very boring but it always had an
interesting section called ``My Most Unforgettable Character ".
 It was usually about a somewhat eccentric but admirable
 character that the writer had especially fond memories of.
Well, for me (and I suspect anyone else who knows
him), Holger will always be one of the most unforgettable characters
I've ever met.

Holger and I first met through the mail in 1970.
He had seen a paper that
I wrote claiming that the Veneziano amplitude described the
scattering of some kind of elastic strings. Unfortunately I no
longer have the hand written letter  but I can still see his
distinctive curly handwriting and the signature - Holger Bech
Nielsen. Most of all I remember his almost child-like enthusiam and
simplicity. He
too had been working on a similar idea \footnote{Nambu had also
been working on the same ideas but I don't believe that
 Holger or
I were aware of it. }. Unlike so many messages
that I've received over the years, this one had nothing to do with
staking a claim or as we say, pissing on territory. The letter
straightforwardly  expressed his excitement and joyously shared his own
ideas.
It was completely clear to me that I had met a larger than life,
most unforgettable character.

That year I invited Holger to spend a month visiting me in New
York and what a month it was. We ate too much, drank too much and yelled
too
loud but the physics excitement was palpable. I have never had more fun
doing physics than during that time.
At the time, string theory was of
course a theory of hadrons. Mesons were strings with quarks at
their ends. Both of us were disturbed by something that we thought
was a very serious shortcoming of the theory. At about that time
the electromagnetic properties of hadrons were under intensive
investigation at SLAC and other places. Electromagnetic form
factors of nucleons were already well measured. SLAC had measured
deep inelastic electroproduction and Feynman and Bjorken had
explained the data with their parton ideas. The problem that
puzzled Holger and me was that we could see no way to define the
local electromagnetic current of a hadron using string theory.
Every time we tried we got nonsense. Hadrons came out infinitely
big and in a sense, infinitely soft.  Holger and I had a wonderful
time thinking about the problem. I think it is fair to say that
many of the themes of my future work trace back to that brief
month and to Holger's profound influence on me .

What was our solution to the problem? Both of us were inclined to
think of the string as an idealized limit of a discrete system. In
my case I viewed it as a system of partons in the light cone
frame. Holger had a more covariant view which he had been
discussing with Aage Kraemmer in Copenhagen. According to this
view, the Koba Nielsen disc \cite{holger} was really the continuum
limit of an infinitely dense planar Feynman diagram or more
precisely, a sum over such diagrams. At that time we had no idea
why planar diagrams should dominate. That had to wait for 't
Hooft \cite{tooft}.

One of the ideas in the paper was that the geometry of a large
planar diagram defined a kind of metric which could be gauge fixed
to what would now be called the conformal gauge. We realized that
a proper treatment should include a sum over diagrams which could
be represented as a path integral over a diagram density. Holger
wanted to treat this degree of freedom as an additional dimension
which a decade  later, following the work of Polyakov, became the
Liouville field. I thought it was a bad idea since I could not see
how an additional infinite direction could fit into hadron
physics. For this reason we decided that the integration should be
dominated by some specific density that didn't fluctuate too much.

In fact the form factor problem forced us to conclude that the
continuum limit was just too extreme. Hadrons might be described
by fairly dense systems of partons but not a continuum. There had
to be a cutoff  which limited the zero-point fluctuations that blew
the string up to infinite size \cite{aage,karliner} and also removed the
hard effects
of discrete partons.
Together with Kraemmer we wrote a paper \cite{aage} formulating what we
called
the Dual Parton Model which tried to keep the good features of
strings without passing to the extreme limit. It is a great
pleasure to
 come back to this problem which so occupied
our thoughts during that month thirty one  years ago and to contribute
some new thoughts for Holger's Festschrift.

\setcounter{equation}{0}
\section{  The Puzzle of Infinite Size }

The obvious difficulties with hadronic string theory
involved the spectrum which invariably included massless vectors,
scalars and tensor particles. There were also  the subtle problems
of local currents
that Holger and I had wrestled with. Technically
speaking, there was no possibility of continuing string theory
 off the mass shell to construct the matrix elements
needed to describe the interaction of hadrons with
electromagnetism and gravitation \cite{aage}.
The natural candidates, vertex
operators like $\exp{ikX}$ can not be sensibly continued away from
specific discrete ``on shell" values of $k^2$. Closely connected
with this was the divergence encountered in attempting to compute
the hadronic electromagnetic or gravitational radius
\cite{aage} \cite{karliner}.
Thus string
theory was abandoned as a theory of hadrons and replaced by QCD.
The  success of string theory in understanding Regge Trajectories
and quark confinement was understood in terms of an approximate
string-like behavior of chromo-electric flux tubes. According to
this view, hadronic strings are not the infinitely thin idealized
objects of mathematical string theory but are thick tubes similar
to the quantized flux lines in superconductors\cite{olesen}.
The ideal  string theory
was relegated to the world of quantum gravity.

However more recent developments have strongly suggested that an
idealized form of string theory may exactly describe certain
gauge theories which are quite similar to
QCD \cite{confinejoe} \cite{hardjoe}.  We have returned
full circle to the suspicion that hadrons may be precisely
described by an idealized string theory, especially in the 't
Hooft limit \cite{tooft}.
The new string theories are certainly more
complicated than the original versions and it seems very plausible
that the problems with the massless spectrum of particles will be
overcome. Less however has been studied about the problems
connected with local currents. In this contribution I will show that the
new insights from the AdS/CFT correspondence provide a solution to
the form factor problem.

I begin by reviewing the problem. For definiteness we work in the
light cone frame in which string theory has the form of a
conventional Galilean-invariant Hamiltonian quantum mechanics. The
degrees of freedom of the first-quantized string include
 $D-2$ transverse coordinates  $X^m(\sigma)$ and the
Lagrangian for these variables is
\begin{equation}\label{sec2:eq1}
L=\frac{1}{4 \pi}\int_0^{2\pi P_-} d\sigma (\dot{X}\dot{X}-
(\alpha')^{-2}X'X')
\end{equation}
where $\dot{X}$
and $X'$ mean derivative with respect to light-cone time $\tau$
and string parameter $\sigma$. The light-cone momentum $P_-$ is
conjugate to the light like coordinate $x^-$. All irrelevant constants
have been
set to unity.

An important feature of the light-cone theory involves the local
distribution of $P_-$ on the string. The rule is that the
distribution of $P_-$ is uniform with respect to $\sigma$. In
other words the longitudinal momentum $dP_-$ carried on a segment
of string $d \sigma $ is exactly $d \sigma /{2 \pi}$.

Let us now consider the transverse density of $P_-$. In a space-time
field
theory this would be given by
\begin{equation}\label{sec2:eq2}
\rho(X) = \int dx^- T_{--}(X,x^-)
\end{equation}
where $T$ is the energy momentum tensor of the field theory.
Matrix elements of $\rho$ between strings of equal $P_-$ define
form factors for gravitational interactions of the string and are
entirely analogous to electromagnetic form factors.

The natural object in string theory to identify with $\rho(X)$ is
\begin{equation}\label{sec2:eq3}
\frac{1}{2\pi}\int d\sigma [\delta (X-X(\sigma))]
\end{equation}
In other words  $\rho(X)$ receives contributions from every element
of string localized at $X$. The Fourier transform of $\rho(X)$
\begin{equation}\label{sec2:eq4}
\tro(k)= \int d \sigma \exp{i k X(\sigma)}
\end{equation}
defines a system of form factors by its matrix elements between
string states.

The mean square radius of the distribution function is given by
\begin{equation}\label{sec2:eq5}
\R^2=\< \int X^2 \rho(X) \>
\end{equation}
and can be rewritten in terms of $\tro$.
\begin{equation}\label{sec2:eq6}
\R^2 = -\partial_k \partial_k \<\tro \> |_{k=0}.
\end{equation}
Eq.(\ref{sec2:eq6}) is the standard definition of the mean-square radius
in
terms of the momentum space form factor.

The squared radius is also given by
\begin{equation}\label{sec2:eq7}
\< X(\sigma)^2 \>.
\end{equation}
where the value of $\sigma$ is arbitrary.

For a field theory with a mass gap, such as pure QCD it is
possible to prove that $\R^2$ is finite. This follows from the
standard analytic properties of form factors.
The problem arises when we attempt to apply the world sheet field
theory to compute $\< X(\sigma)^2 \>$. An elementary calculation
based on the oscillator representation of $X$ gives a sum over
modes
\begin{equation}\label{sec2:eq8}
\<X^2\> \sim  \alpha'\sum_0^{\infty} \frac{1}{n} =\alpha' \log{\infty}.
\end{equation}
A related disaster occurs when we compute the form factor which is
easily seen to have the form
\begin{equation}\label{sec2:eq9}
\< \tro(k)\> = \exp{-k^2 \<X^2 \>}.
\end{equation}
Evidently it is only non-zero  at
$k^2=0$.

In a covariant description of string theory the problem has its
roots in the fact that the graviton vertex operator is
only well defined on
the mass shell of the graviton, $k^2=0$. Vertex operators to be
well defined must correspond to perturbations with vanishing world
sheet $\beta$ function. This implies that they should correspond
to on shell solutions of the appropriate space-time gravitational
theory. For the kinematical situation in which the graviton
carries vanishing $k_{\pm}$ the transverse momentum must vanish.
Thus no well defined off shell continuation of the form factor
exists.

One might wonder if the divergence of $X^2$  is special to the case of a
free world sheet field theory. The answer is that the divergence
can only be made worse by interactions. The 2-point function of a
unitary quantum field theory is at least as divergent as the
corresponding free field theory. This follows from the spectral
representation for the two point function and the positivity of
the spectral function. Thus it is hard to see how an ideal string
theory can ever describe hadrons.

\setcounter{equation}{0}
\section{ Light Cone Strings in AdS }

There are good reasons to believe that certain confining
deformations of maximally supersymmetric Yang Mills theory are
string theories albeit in higher dimensions. The strings move in a
5 dimensional space
\footnote{Strictly speaking the target space is $10$
dimensional with the form $AdS_5$ times a compact space such as $S_5$.
In this paper the compact factor plays no role.}
that is asymptotically AdS. In the 't Hooft
limit these theories are believed to be free string theories.
Evidently if this is so there must exist a well defined string
prescription for form factors in the 4-D theory.

What we will see is that although the theory in bulk of AdS is an
ideal thin-string theory the 4-D boundary field theory is not
described by thin strings. That may seem surprising. Suppose that
in the light-cone frame the thin 5-D string has the form
\begin{equation}\label{sec3:eq1}
X(\sigma), Y(\sigma)
\end{equation}
where $X$ are the transverse coordinates of 4-D Minkowski space
and $Y$ is the additional coordinate perpendicular to the boundary
of AdS. Then it would seem natural to consider the projection of
the string onto the $X$ plane to define a thin string. According to this
view the mean-squared radius would again be $\<X^2\>$ and we would
be no better off than before. Before discussing the resolution of
this problem let us work out the bosonic part of the light-cone string
Lagrangian in
AdS. I will make no attempt to derive the full supersymmetric
form of the theory in this paper. I believe the resolution of the
form factor problem does not require this. On the geometric side
I will also ignore the 5-sphere component of the geometry implied
by the usual R-symmetry of the $N=4$ supersymmetry.

The metric of AdS is given by
\begin{equation}\label{sec3:eq2}
ds^2 = R^2\frac{dx^+ dx^- -dX^2 -dY^2}{Y^2}
\end{equation}
I have defined the overall scale of the AdS (radius of curvature)
to be $R$.

In order to pass to the light cone frame we must also introduce
the world sheet metric $h_{ij}$. In the usual flat space theory it
is possible to fix the world sheet metric to be in both the light
cone gauge $\sigma_0 = \tau = x^+$ and also the conformal gauge
$h_{00} = -h_{11}, \ \ h_{01}=0$. However this is not generally
possible since it entails 3 gauge conditions which is one too
many. The special feature of flat space which permit the
over-fixing of the gauge is not shared by AdS. Thus we must give
up the conformal gauge if we wish to work in light-cone gauge.

Let us fix the gauge by choosing 2 conditions
\begin{eqnarray}\label{sec3:eq3}
\sigma_0 &=& x^+ \cr
h_{01} &=& 0.
\end{eqnarray}

Let us further define
\begin{equation}\label{sec3:eq4}
\sqrt{\frac{-h_{11}}{h_{00}}} =E.
\end{equation}
Setting $\alpha'=1$, an elementary calculation gives the Hamiltonian
\begin{equation}\label{sec3:eq5}
H=
\int d\sigma \left(P_X P_X +P_Y P_Y +\frac{R^4}{Y^4}
(\ds X \ds X + \ds Y \ds Y) \right).
\end{equation}
The precise version of the supersymmetric Hamiltonian was given in
\cite{thorn}.
This is a more or less conventional string action with the unusual
feature that the effective string tension scales like $1/Y^4$.
Thus the tension blows up at the AdS boundary $Y=0$ and tends to
zero at the horizon $Y= \infty$. This of course is a manifestation
of the usual UV/IR connection .

The  Hamiltonian could be obtained from an action
\begin{equation}\label{sec3:eq6}
S=\int d \sigma d\tau \left(\dot{X}\dot{X} +\dot{Y}\dot{Y}
-\frac{R^4}{Y^4}
(\ds X \ds X + \ds Y \ds Y) \right).
\end{equation}
This action thought of as a $1+1$ dimensional field theory is not
Lorentz invariant in the world sheet sense. However it is
classically scale invariant if we assume $X$ and $Y$ are dimension
zero. The Hamiltonian has dimension $1$ and therefore scales as
the inverse length of the $\sigma$ circle which is  $\sim P_-$.
We recognize this scale symmetry as space-time longitudinal boost
invariance
under which $H$ and $P_-$ scale oppositely and $X,Y$ are
invariant. No doubt the actual Lagrangian when properly
super-symmetrized retains this symmetry when quantized.

Let us consider the equal time correlation function
$\<X(0)X(\sigma)\>$
in the field theory defined by (\ref{sec3:eq6}) or more precisely in its

supersymmetrized version. By inserting a complete set of
eigenstates of the energy and (world sheet) momentum we obtain
\begin{eqnarray}\label{sec3:eq7}
\<X(0)X(\sigma)\>&=& \sum \int
e^{ip\sigma } |\<X(0)|E,p\>|^2 \frac{1}{p^2}dE dp \cr
&=& \int F(E,p)e^{ip\sigma }\frac{1}{p^2}dE dp
\end{eqnarray}
with $F\ge 0$

The measure of integration $dE dp/p^2$ follows from the fact that
$X$ has ``engineering" dimension zero under the longitudinal boost
rescaling. Furthermore the assumption that the scale invariance is
preserved in the quantum theory requires $F(E,p) = F(E/p)$ for
large $p,E$. It follows that as long as $F$ does not go to zero
in this limit that the correlation function diverges as $\sigma \to
0$. This would imply $X^2 = \infty$

\setcounter{equation}{0}
\section{Dressing the Vertex with $Y$  Dependence }

Let us consider the problem from the point of view of the vertex
operator $\exp{ikX}$. One problem that I have emphasized is that it
is not a solution of the on-shell condition. We can try to fix
this by replacing it with a solution of the wave equation for a
graviton in AdS space. The relevant equation is
\begin{equation}
\label{sec4:eq1}
\left(
\partial_{\mu}\partial^{\mu} +
Y^3 \partial_Y Y^{-3}\partial_Y
\right)\Phi =0
\end{equation}
where $\mu$ runs over the four dimensions of flat Minkowski space.

The particular solutions we are looking for are independent of the
$x^{\pm}$ and have the form
\begin{equation}\label{sec4:eq2}
\Phi = \exp{ikX} F(k,Y)
\end{equation}
where $F$ satisfies
\begin{equation}\label{sec4:eq3}
Y^3 \partial_Y Y^{-3} \partial_Y F(k,Y) = k^2 F
\end{equation}
Thus the on shell vertex operator has the form
\begin{equation}\label{sec4:eq4}
\int d\sigma \exp{ikX(\sigma)} F(k,Y(\sigma))
\end{equation}
The factor $F $ is a  dressing of the vertex, necessary to
make its matrix elements well defined for $k \neq 0$.

Let us consider the mean square radius of the hadron defined by
eq.(\ref{sec2:eq6}).
\begin{equation}
\label{sec4:eq5}
\R^2= -\partial_k \partial_k \<  F(k,Y) \exp{ikX}\> |_{k=0}
\end{equation}
or
\begin{equation}
\label{sec4:eq6}
\R^2=\< X^2 F(0,Y) -2iX\cdot F'(0,Y) -F''(0,Y) \>
\end{equation}
where $F''\equiv \partial_k\partial_k F$.

For a state of zero angular momentum in the $X$ plane the term
linear in $X$ vanishes and we have
\begin{equation}
\label{sec4:eq7}
\R^2=\< X^2 F(0,Y) -F''(0,Y) \>.
\end{equation}

One possibility for resolving the infinite radius problem is a
cancellation of the two terms in eq.(\ref{sec4:eq7}).
To compute $F$ and $F''$ we Taylor expand $F(k,Y)$ in powers of
$k$ and substitute into  eq.(\ref{sec4:eq3}). There are two linearly
independent
solutions.
\begin{equation}
\label{sec4:eq8}
F(k,Y) = Y^4 + \frac{1}{12}k^2 Y^6 +\cdots
\end{equation}
and
\begin{equation}
\label{sec4:eq9}
F(k,Y) =1 - \frac{1}{4}k^2 Y^2 +\cdots
\end{equation}

Only the second of these is relevant to the problem of vertex
operators. To see this we need only note that the vertex at $k=0$
is just the operator that measures $P_-$. For states with
$P_-=1$ this operator in just the identity. This implies that
$F(0,Y) =1$.

 Thus we find
\begin{equation}
\label{sec4:eq10}
\R^2=\<[ X^2  +Y^2]  \>
\end{equation}
and $\R^2$ is the sum  of a divergent term and a positive term.
The
mean radius continues to be divergent. Evidently cancellation is
not the answer.

The dressing of the vertex by the factor $F(k,Y)$ obviously
modifies the expression (\ref{sec2:eq3})
for the transverse density $\rho$. If we define the Fourier
transform of $F$ with respect to $k$ to be $\tilde{F}(X,Y)$
eq.(\ref{sec2:eq3}) is
replaced by
\begin{equation}
\label{sec4:eq11}
\rho \sim \int d \sigma \tilde{F}(X-X(\sigma),Y).
\end{equation}
This means that an ideal thin string in the AdS bulk space is
smeared out by the holographic projection onto the boundary. This
is of course the familiar UV/IR correspondence at work. Bulk
strings near the boundary are projected as very thin strings in
the 4-D theory but those far from the boundary are fat. The extra
term $\<Y^2\>$ in eq.(\ref{sec4:eq10}) represents this fattening.
Evidently I
have only made things worse by including the dressing.

Before discussing the solution to the problem let us make some
remarks about confining deformations in the context of AdS/CFT.
Bulk descriptions of  confining deformations of super Yang Mills theory
have an effective infrared ``wall" at
a value of $Y$ which represents the confinement  scale. In these
cases
the metric (\ref{sec3:eq2}) is modified in the infrared region.
\begin{equation}
\label{sec4:eq12}
ds^2= h(y) \left( dx^+dx^-  - dX^2 -dY^2 \right)
\end{equation}
where, as in the conformal case, $h\sim 1/Y^2$ for $Y \to 0$.
Assume that $h$ has a minimum at the confinement scale, $Y=Y^*$.

The light-cone hamiltonian is easily worked out,
\begin{equation}
\label{sec4:eq13}
H= \int d\sigma
\left(P_X P_X +P_Y P_Y +h(Y)^2
(\ds X \ds X + \ds Y \ds Y) \right)
\end{equation}

Consider a string stretched along the $X$ direction and choose $\sigma$
so that $\partial_{\sigma}X=1$. The potential energy of the string is
then given by
\begin{equation}
\label{sec4:eq14}
V(Y)=h(Y)^2
\end{equation}
which has a minimum at $Y=Y^*$. Thus  a classical long
straight string will be in equilibrium at this value of $Y$.
This classical bulk string
corresponds to a field theory configuration which, according to
the UV/IR connection, is thickened to a size $\sim Y^*$, that is,
the QCD scale.

Quantum fluctuations will cause the wave
function of the string to fluctuate away from $Y^*$.
The implication is that the
QCD string is a superposition of different thickness values
extending from infinitely thin to QCD scale. Indeed different
parts of the string can fluctuate in thickness over this range.
The portions of the string near $Y=0$ will be very thin and will
determine the large momentum behavior of the form factor.

\setcounter{equation}{0}
\section{Finiteness of $\< X^2 \>$}

I believe that despite the argument given at the end of Section 3
the value of $\< X^2 \> $ is finite. This can only be if the
function $F=\sum |\<X(0)|E,p\>|^2$ vanishes in the scaling limit
of large $E,p$. I will first give an intuitive argument and
follow it with a more technical renormalization group analysis
that is due to Joe Polchinski.

First suppose the string is ``stuck" at some value of $Y$. In that
case the action  for $X$ in eq.(\ref{sec3:eq6}) is a conventional string
action except that the string tension is replaced by $1/Y^4$. The
divergence in $X^2$ would then be given by
\begin{equation}\label{sec5:eq1}
\<X^2 \> = Y^2 |\log{\epsilon}|.
\end{equation}
If we ignore quantum fluctuations of $Y $ we could replace $Y$ by
$Y^*$. But $Y$ fluctuates as well as $X$ and can be expected to
fluctuate toward the boundary as $\epsilon$
tends to zero. This is just the usual UV/IR connection in AdS.
Therefore as we remove the cutoff the fluctuations of $X$ are
diminished because the string moves into a region of increasing
effective
stiffness. If for example the average value of $Y^2$ tends to zero
as $|1/\log{\epsilon}|$ or faster then the fluctuations of $X$ would
remain
bounded.  To see that this happens we consider the renormalization
running of the operator $X^2$.

Begin with the bare theory defined with a cutoff length $\epsilon$
on the world sheet. We can then ask how a given operator in this
bare theory is described in a renormalized version of the theory
with a cutoff at some longer distance $l$. A general operator
$\phi(X,Y)$ runs to lower momentum scales
according to the renormalization group equation

\begin{equation}\label{sec5:eq2}
(l \partial/\partial l) \phi(X,Y,l) =
(\alpha'/2) \nabla^2 \phi(X,Y,l).
\end{equation}
For example, consider flat space and the operator $X^2$. We look
for a solution of eq.(\ref{sec5:eq2}) with
\begin{equation}\label{sec5:eq3}
\phi(X, \epsilon) =X^2.
\end{equation}
The solution is
\begin{equation}\label{sec5:eq4}
\phi(X, l) = X^2 +\alpha' \log{l/{\epsilon}}.
\end{equation}
Thus if we regulate the theory at some fixed scale,
for example $l\sim 1$, the matrix elements of $X^2$ blow up as
send $\epsilon \to 0$.

By contrast, consider the the case of AdS space where
\begin{equation}\label{sec5:eq5}
\nabla^2 = R^{-2} ( Y^2 \partial_X^2 + Y^5 \partial_Y Y^{-3}
\partial_Y)\ .
\end{equation}
For a solution of the form $X^2 + f(l) Y^2$ this becomes
\begin{equation}\label{sec5:eq6}
(l \partial/\partial l) f = (2\alpha'/R^2) (1 - f)\ .
\end{equation}
With $f(\epsilon) = 0$ the solution is
\begin{equation}\label{sec5:eq7}
 f(l) =  1 - (\epsilon/l)^{2\alpha'/R^2}\ .
\end{equation}
So if we fix the scale $l$ and take the cutoff length $\epsilon$
to zero the matrix elements tend to finite
limits and the problem of infinite
radii is resolved . If, however,
 we expand in powers of
$\alpha'$ there are logarithmic divergences.

Note that the operator $X^2$ runs to a fixed point $X^2+Y^2$ which
is just the operator in eq.(\ref{sec4:eq10}) which represents the mean
squared
radius $\R^2.$

The reader may wonder how the finiteness of $X^2$ can be explained
in covariant gauges such as the conformal gauge in which the world
sheet theory has the form of a relativistic field theory. A
standard argument
insures that the singularity in a two point function can not be
less singular than a free field; in this case logarithmic.
The argument is based on the positivity of spectral functions
which in turn assumes the metric in the space of states is
positive. In general this is not  the case in covariant gauges.

\section{Discussion }

The original attempt to describe hadrons as idealized strings
was frustrated by the infinite zero point oscillations in the size
of strings. Early ideas for modifying string theory such as
replacing the idealized strings by fat  flux tubes or as
collections of partons which approximate strings fit well with
QCD but seemed to preclude an idealized mathematical string
description.

More recent evidence from AdS/CFT type dualities suggest that
idealized
string theory in higher dimensions may provide an exact
description of the 't Hooft limit of QCD-like theories.
I have argued that an ideal bulk string theory in five
dimensions is fully compatible with a fat non-ideal string in four
dimensions.

The fifth dimension can be divided into two regions.  The ``wall"
region near  $Y=Y^*$ corresponds to
the confinement scale $\Lambda$. If we ignore high frequency
fluctuations,
the string spends most of its time in this region. The usual UV/IR
spreading gives the string a thickness of order $\Lambda$. High
frequency fluctuations of small sections of string can occur
which cause it to  fluctuate toward $Y=0$, the region
corresponding to short distance behavior in space-time.
These fluctuations  will control the large momentum
behavior of form factors as well as deep inelastic matrix
elements. Such fluctuations  give the string a parton-like makeup.
We have also seen that these fluctuations stiffen the effective
string tension so much that the infinite zero point size that
Holger and I worried about so long ago is now
eliminated.

\section{Acknowledgements }
I am very grateful to Joe Polchinski for contributing the
renormalization group argument in Section 5. I also benefited from
discussions with Juan Maldacena. Finally I must thank Holger Bech
Nielsen for the inspiration which has stayed with me for so many
years.

\def\ave#1{\langle #1 \rangle}
\def\pd{\partial}

\title*{Relativity  and $c/\sqrt{3}$}
\author{%
S.I. Blinnikov, L.B. Okun and M.I. Vysotsky} 
\institute{%
ITEP, 117218 Moscow, Russia }

\authorrunning{S.I. Blinnikov, L.B. Okun and M.I. Vysotsky}
\titlerunning{Relativity  and $c/\sqrt{3}$}
\maketitle

\begin{abstract}
We define the critical coordinate velocity ${\rm v}_c$. A particle
moving radially in the Schwarz\-schild background with this velocity,
${\rm v}_c= c/\sqrt 3$, is neither accelerated, nor decelerated if
gravitational field is weak, $r_g \ll r$, where $r_g$ is the
gravitational radius, while $r$ is the current one.
We find that the numerical coincidence of ${\rm v}_c$  with velocity of sound
in ultrarelativistic plasma, $u_s$, is accidental, since two velocities
are different if the number of spatial dimensions is not equal to 3. 
\end{abstract}

{\bf To Holger Nielsen}

  We dedicate this note to our friend Holger Nielsen on the occasion 
of his 60th birthday. It  has been a great pleasure to discuss with
him exiting physical ideas at ITEP, CERN, the Niels Bohr Institute
and in other places around the world. We wish to Holger many new
discoveries and a long happy life.

\section{Motivation}

According to General Relativity (GR) clocks run slowly in the
presence of gravitational field, as a result, the coordinate
velocity of photons decreases. This is the reason for the delay of
radar echo from inner planets predicted and measured by I. Shapiro
\cite{vys1}. Propagation of ultrarelativistic particles is described
similarly to that of photons. That is why the retardation must
take place not only for photons but also for ultrarelativistic
particles. In this respect the latter drastically differ \cite{vys2}
from nonrelativistic bodies, velocity of which evidently increases
when they are falling radially onto a gravitating body (e.g., onto
the Sun). Obviously, there should be some intermediate velocity
${\rm v}_c$ which remains constant for a particle falling in
gravitational field of the Sun (or another star). The numerical
value ${\rm v}_c = c/\sqrt 3$ will be found in Sect. 2. When a
particle moves radially with this velocity in weak field it
``ignores'' gravity: it is neither accelerated, nor decelerated.
For nonradial trajectories gravity is never ignored: the
trajectories are bent for any velocity.

It is well known that $u_s = c/\sqrt 3$ is the speed of sound in
ultrarelativistic plasma and the question arises whether the
equality $u_s = {\rm v}_c$ has some physical reason, or it is a
numerical coincidence. To answer this question we find in Sect. 3
expressions for ${\rm v}_c$ and $u_s$ in spaces with number of
dimensions $n$ different from 3. Since for $n\neq 3$ we get ${\rm v}_c
\neq u_s$ we come to the conclusion that their coincidence at
$n=3$ does not have deep physical reason.

\section{Derivation of $\mbox{\boldmath${\rm v}_c = c/\sqrt 3$}$}

To simplify formulas, we put light velocity $c=1$, restoring it
when it is necessary. In what follows $G$ is gravitational constant;
gravitational radius $r_g$ of an object with mass $M$ equals
\begin{equation}
  r_g=2GM \; .
\label{rg}
\end{equation}

Let us start from definitions used in GR. The expression for
interval in the case of radial motion ($d\theta = d\varphi = 0$)
has the well known  Schwarzschild form:
\begin{equation}
ds^2 = g_{00}dt^2 -g_{rr}dr^2 \equiv d\tau^2 -dl^2 \;\; ,
\label{vyso1}
\end{equation}
where $g_{00} =(g_{rr})^{-1} =1-\frac{r_g}{r}$.
The local velocity $v$ of a particle measured by a local observer
at rest is:
\begin{equation}
v = \frac{dl}{d\tau} =\left(\frac{g_{rr}}{g_{00}}\right)^{1/2}
\frac{dr}{dt}= \frac{1}{g_{00}} \frac{dr}{dt} \;\; , \label{vyso2}
\end{equation}
while observer at infinity, where $g_{00}(\infty) =
g_{rr}(\infty) =1$, measures the so-called coordinate velocity at $r$:
\begin{equation}
{\rm v} = \frac{dr}{dt} =
v\left(\frac{g_{00}}{g_{rr}}\right)^{1/2} = g_{00}v \;\; .
\label{vyso3}
\end{equation}
In order to determine the time of radial motion from $a$ to $b$,
the infinitely distant observer should calculate the integral
\begin{equation}
t =\int_a^b\frac{dr}{{\rm v}} \;\; , \label{vyso4}
\end{equation}
that is why the coordinate velocity is relevant for radar echo.

For a particle moving in static gravitational field one can
introduce conserved energy (see ref. \cite{vys3}, eq. 88.9):
\begin{equation}
E =\frac{m\sqrt{g_{00}}}{\sqrt{1- v^2}} \;\; .
\label{vyso5}
\end{equation}
The expression for $E$ through ${\rm v}$:
\begin{equation}
E
=\frac{m\sqrt{g_{00}}}{\sqrt{1-({\rm v}/g_{00})^2}}
\label{vyso6}
\end{equation}
allows us to determine ${\rm v}(r)$ from the energy conservation:
\begin{equation}
E(r = \infty) =E(r) \;\; , \label{vyso7}
\end{equation}
\begin{equation}
{\rm v}^2 = g_{00}^2 -g_{00}^3 +g_{00}^3 {\rm v}_\infty^2 =
g_{00}^2 [1-g_{00}(1-{\rm v}_\infty^2)] \;\; . \label{vyso8}
\end{equation}
For the local velocity $v$ measured by a local observer
we obtain:
\begin{equation}
v^2 = 1-g_{00}(1-{\rm v}_\infty^2) \;\; , \label{vyso9}
\end{equation}
so, while $v$ always increases for a falling massive particle, reaching $c$ at
$r = r_g$, the behaviour of ${\rm v}$ is more complicated. Substituting
$g_{00} =1-\frac{r_g}{r}$ into (\ref{vyso8}), we get for weak gravitational field
($r\gg r_g$):
\begin{equation}
{\rm v}^2 = {\rm v}_\infty^2 +\frac{r_g}{r}(1-3{\rm v}_\infty^2)
\;\; . \label{vyso10}
\end{equation}
For the motion of a nonrelativistic particle (${\rm v}_\infty \ll 1$) the well-known
expression is reproduced:
\begin{equation}
{\rm v}^2 = {\rm v}_\infty^2 +\frac{2MG}{r} \;\; .
\label{vyso11}
\end{equation}
For ${\rm v}_\infty = {\rm v}_c = 1/\sqrt 3$ the coordinate velocity
of particle does not change, while it grows for ${\rm v}_\infty <
{\rm v}_c$ and diminishes  for ${\rm v}_\infty > {\rm v}_c$.
At $r = 3r_g$ according to eq.(\ref{vyso10}) the coordinate velocity becomes equal to ${\rm v}_c$.
However, for $r=3r_g$ our weak field approximation fails.

Let us dispose of the assumption of the weak field.
Coming back to expression (\ref{vyso8}) and substituting there $g_{00}
=1-\frac{r_g}{r}$, we observe that for ${\rm v}_\infty > {\rm v}_c$
the coordinate velocity always diminishes and becomes zero at $r=r_g$,
while in the case ${\rm v}_\infty < {\rm v}_c$ it grows up to the
value
\begin{equation}
{\rm v}_{\rm max}^2 =4/(27(1-{\rm v}_\infty^2)^2) \; ,
\label{vmax}
\end{equation}
which
is reached at
\begin{equation}
r_0 = \frac{3(1-{\rm v}_\infty^2)}{(1-3{\rm v}_\infty^2)}r_g \; ,
\label{r0}
\end{equation}
and after that diminishes to zero at $r=r_g$.
It is interesting to note that the velocity $v$ measured by local
observer equals ${\rm v}_c$ at the point where
${\rm v} = {\rm v}_{\rm max}$.

Thus, if the coordinate velocity is only mildly relativistic,
${\rm v}_\infty > c/\sqrt 3$, then ${\rm v}$ already decreases at
the free fall.

As an example of a non-radial motion let us consider the deflection
of light from a star by the Sun and compare it with the deflection of a
massive particle.
It is well known that the angle of deflection $\theta$ of photons
grazing the Sun is given by
\begin{equation}
 \theta_\gamma=\frac{2r_g}{R_\odot}
\label{thetagam}
\end{equation}
where $R_\odot$ is the radius of the Sun.
In the case of massive particles the deflection angle is larger:
\begin{equation}
  \theta = \theta_\gamma(1 + \beta^{-2}) \; ,
\label{theta}
\end{equation}
where  $\beta \equiv v_\infty/c < 1$.(See ref.\cite{MTW}, eq.
25.49, and ref.\cite{lightm}, problem 15.9, eq.13.)

\section{Speed of sound $u_s$ and critical speed
  v$_c$  in $n$ dimensions}

For ultrarelativistic plasma with equation of state $P=e/3$, where
$P$ is pressure and $e$ is energy density (including mass density),
we have for the speed of sound $u_s$:
\begin{equation}
 u_s^2 =c^2 \left.\frac{\partial P}{\partial e}\right|_{\rm ad} =
\frac{c^2}{3} \;\; , \label{vyso12}
\end{equation}
We use eq. 134.14 of ref.  \cite{vys4}, and correct misprint there, or
eq. 126.9 from \cite{lanliffm}; ``ad'' means adiabatic, i.e. for constant specific
entropy.
In order to obtain the expression for $u_s$ in the case when $n\neq 3$,
where $n$ is the number of spatial dimensions, let
us start with equation of state.

One can use a virial theorem to connect pressure $P$ and thermal
energy $\mathcal{E}$ of an ideal gas using classical equations of
particle motion (cf. \cite{vys6}). We have for a particle with
momentum ${\rm\bf p}$ and a Hamiltonian $H$:
\begin{equation}
 \dot{\rm\bf p} = - \frac{\pd H}{\pd {\rm\bf q}} \; ,
\label{vyso13}
\end{equation}
hence,
\begin{equation}
 {\rm\bf q} \dot{\rm\bf p} = - {\rm\bf q} \frac{\pd H}{\pd {\rm\bf q}} =
 {\rm\bf q} {\rm\bf F} \;\; ,
\label{vyso14}
\end{equation}
where ${\rm\bf F}$ is the force acting on the particle.
Let us average over time $t$:
\begin{equation}
 \ave{\dots} \equiv \frac{1}{t} \int_0^t \dots d\bar t \;\; .
\label{vyso15}
\end{equation}
Integrating  by parts we get:
\begin{equation}
   \ave{ {\rm\bf q} \dot{\rm\bf p}} = -  \ave{\dot{\rm\bf q} {\rm\bf p}} =
   \ave{ {\rm\bf q} {\rm\bf F}} \;\; .
\label{vyso16}
\end{equation}
For non-relativistic (NR) particles
\begin{equation}
  \dot {\rm\bf q} {\rm\bf p} = 2 E_{\rm kin} = {\rm\bf p}^2/m \; .
\label{vyso17}
\end{equation}
For extremely relativistic (ER) particles
\begin{equation}
  \dot{\rm\bf q} {\rm\bf p} =  E_{\rm kin} = c |{\rm\bf p}| \; .
\label{vyso18}
\end{equation}
Now for $N$ particles in a gas
\begin{equation}
   - \sum_{i=1}^N   \ave{   \dot{\rm\bf q}_i {\rm\bf p}_i } =
       \sum_{i=1}^N   \ave{ {\rm\bf q}_i {\rm\bf F}_i }.
\label{vyso19}
\end{equation}
(By the way, $-\frac{1}{2}\sum\limits_i \ave{ {\rm\bf q}_i {\rm\bf
F}_i }$ is called the {\em virial}.) If the gas is ideal (i.e.
non-interacting particles), then the force ${\rm\bf F}$ is
non-zero only at the collision of a particle with the wall, and
the virial reduces to an integral involving pressure:
\begin{equation}
 - \sum_{i=1}^N   \ave{   \dot{\rm\bf q} {\rm\bf p}}
      = - \int P \mathbf{n} {\rm\bf q} \, dS =
   -P\int \mbox{div}\, {\rm\bf q} \, dV = -3PV \; ,
\label{vyso20}
\end{equation}
where $\mathbf{n}$ is a unit vector normal to the wall area element $dS$ and
the Gauss theorem is used for transforming the surface integral to the volume one.
So, since the thermal energy $\mathcal{E}$  (not including mass) is just
the total kinetic energy of molecules,
\begin{equation}
  \mbox{NR} : \quad 2\mathcal{E} = 3PV, \quad P = 2\mathcal{E}/(3V)  \; ,
\label{vyso21}
\end{equation}
\begin{equation}
  \mbox{ER} : \quad \mathcal{E} = 3PV,  \quad P =  \mathcal{E}/(3V) \equiv e/3 \; .
\label{vyso22}
\end{equation}

The last equality holds since in extremely relativistic case
$E_{\rm kin} \gg m$. We see that $3$ here is due to $\mbox{div}\,
{\rm\bf q}=3$, i.e. the dimension of our space.

In a space of $n$ dimensions, following the same lines, we get
$\mbox{div}\, {\rm\bf q}=n$, so $P=e/n$ and for ER gas we obtain:
\begin{equation}
u_s = c/\sqrt n \;\; .
\label{vyso23}
\end{equation}
 Here we should use
$n$-volume $V_n$ instead of $V\equiv V_3$ and postulate the first
law of thermodynamics for adiabatic processes to be $d\mathcal{E}+PdV_n=0$,
so pressure would be the force per unit $V_{n-1}$ -- the boundary
of $V_n$.

The same equation of state follows from consideration of the
stress tensor $T_{ik}$ of ultrarelativistic plasma, which is diagonal and
traceless in a rest frame of plasma: $T_{00} = e$, $T_{ii} \equiv P = e/n$.

In order to find ${\rm v}_c$ in the case $n\neq 3$ we need an
$n+1$-dimensional spherically-symmetric static generalization of the
$3+1$-dimensional Schwarz\-schild metric which was found by
Tangherlini \cite{tang}. (See refs.\cite{MP} for the details
of aspherical and time-dependent black holes metrics in higher
dimensional spacetimes). The line element of the $n+1$-dimensional
Schwarz\-schild metric is
\begin{equation}
 ds^2 = \left(1 - \frac{r_{gn}^{n-2}}{r^{n-2}}\right) dt^2 -
 \left( 1 - \frac{r_{gn}^{n-2}}{r^{n-2}}\right)^{\!-1} dr^2 - r^2 d\Omega_{n-1}^2,
\label{Nsch}
\end{equation}
where $d\Omega_{(n-1)}$ is the line element on the unit
$(n-1)$-sphere
and the gravitational radius $r_{gn}$ is related to the black hole mass $M$:
\begin{equation}
r_{gn}^{n-2} = \frac{16 \pi G_n M}{(n - 1) \ A_{n-1}}.
\label{rgn}
\end{equation}
Here $A_{n-1}$ denotes the area of a unit $n-1$ sphere, which
is ${{2 \pi^\frac{n}{2}} / \Gamma(\frac{n}{2})}$ (for $n=3$:
$\Gamma(3/2)=\sqrt{\pi}/2$, $A_2=4\pi$).
We consider the spaces with $n \geq 3$.
The factor in the definition of $r_{gn}$
is taken from refs.\cite{MP}. 
The form of the metric (\ref{Nsch}) is very easy to guess.
In weak field  limit, when $g_{00} \to 1+2\varphi $
we have
\begin{equation}
   \varphi = - \frac{r_{gn}^{n-2}}{2r^{n-2}}
  \quad \mbox{for} \quad r \to \infty \; .
\label{pot}
\end{equation}
This leads in a natural way to the gravitational acceleration $\mathbf{g}$
with the radial component
\begin{equation}
   g_n = -\frac{\pd \varphi}{\pd r}= - \frac{(n-2) r_{gn}^{n-2}}{2r^{n-1}}
  \quad \mbox{for} \quad  r \to \infty \; ,
\label{gacc}
\end{equation}
which implies the constant flux of the acceleration $\mathbf{g}$
equal to
\begin{equation}
 A_{n-1}r^{n-1}g_n = 8\pi \frac{n-2}{n-1}  G_n M
\label{fluxg}
\end{equation}
through a sphere of area $A_{n-1}r^{n-1}$ at large $r$.
It is not hard to verify that the Ricci tensor $R_{ik}$ is zero for the
metric (\ref{Nsch}), that is the metric (\ref{Nsch}) satisfies Einstein
equations in vacuum and describes a spherically symmetric spacetime outside
a spherical gravitating body.

One should remember that the dimension of $G_n$ depends on $n$.
It is clear in the weak field limit from eqs. (\ref{pot}) and (\ref{rgn}),
since
the dimension       
of $[\varphi]$ is the square of velocity, i.e. zero for $c=1$, and hence
\begin{equation}
   [G_n] = \frac{L^{n-2}}{M} \; ,
\label{vyso27}
\end{equation}
or $[G_n] = L^{n-1}$, if $[M]=L^{-1}$.

For the coordinate velocity of a radially falling  particle we get from eq.
(\ref{vyso8}) for weak gravitational field:
\begin{equation}
{\rm v}^2 = {\rm v}_\infty^2 +\left(\frac{r_{gn}}{r}\right)^{n-2}
                     (1-3{\rm v}_\infty^2) \label{vyso28}
\end{equation}
instead of eq. (\ref{vyso10}). We see that in the case of
$n$-dimensional  space the expression for ${\rm v}_c$ remains the
same, ${\rm v}_c =c/\sqrt 3$. Number ``3'' here is not due to the
dimension of space, it is  simply due to cubic polynomial in
(\ref{vyso8}).

\section{Conclusions and Acknowledgements}

The speed of sound in relativistic, radiation dominated plasma
depends on the dimension  of space, while the critical velocity
${\rm v}_c=c/\sqrt{3}$ in the Schwarzschild metric is the same for
any dimension.

\bigskip
We are grateful to Ilya Tipunin for encouraging our search of
literature on black hole metrics in higher dimensions and to Yakov
Granowsky and Valentine Telegdi for their interest to the
discussed problem and stimulating questions. SB is partially
supported by RFBR grant 99-02-16205, and by a NSF grant at UCSC,
he is especially grateful to Stan Woosley for his hospitality. LO
is supported by A.v.Humboldt award and together with MV partly
supported by RFBR grant No. 00-15-96562.

PS. After the first version of this paper appeared on the web we
recieved an e-mail from Stanley Deser and Bayram Tekin in which
it was pointed out that critical velocity in weak field has been
considered by M.Carmeli: Lettere al Nuovo Cimento 3 (1972)379
and in his book ``Classical Fields'', 1982. We thank them for this 
bitter remark.

\renewcommand{\d}{\delta}
\renewcommand{\l}{\lambda}
\renewcommand{\L}{\Lambda}
\renewcommand{\b}{\beta}
\renewcommand{\a}{\alpha}
\renewcommand{\ni}{\noindent}
\newcommand{\n}{\nu}
\newcommand{\m}{\mu}
\renewcommand{\r}{\rho}
\newcommand{\q}{{\pi\over 5}}
\newcommand{\s}{\sigma}
\newcommand{\jgD}{{\cal D}}
\renewcommand{\S}{{\cal S}}
\newcommand{\vx}{\vec{x}}
\newcommand{\vL}{\vec{L}}
\newcommand{\va}{\vec{\alpha}}
\newcommand{\vH}{\vec{H}}
\newcommand{\tS}{\tilde{S}}
\newcommand{\V}{{\cal V}}
\renewcommand{\th}{\theta}
\newcommand{\tm}{\theta^M}
\newcommand{\tp}{\theta^{ph}}
\newcommand{\tpo}{\overline{\theta}^{ph}}
\newcommand{\tpom}{\overline{\theta}^{ph}(\tm)}
\newcommand{\intpi}{\int^\pi_{-\pi}}
\newcommand{\tU}{\tilde{U}}
\newcommand{\e}{\epsilon}
\newcommand{\ep}{\varepsilon}
\newcommand{\vph}{\varphi}
\newcommand{\oh}{\frac{1}{2}}
\newcommand{\oq}{\frac{1}{4}}
\newcommand{\ot}{\frac{3}{2}}
\newcommand{\dg}{\dagger}
\newcommand{\non}{\nonumber}
\renewcommand{\t}{\tau}
\newcommand{\rf}[1]{(\ref{#1})}
\newcommand{\ra}{\rightarrow}
\newcommand{\ph}{\phi}
\newcommand{\phd}{\phi^\dagger}
\newcommand{\CR}{\nonumber \\}
\newcommand{\ww}{w}
\newcommand{\rra}{\right\rangle}
\newcommand{\lla}{\left\langle}
\newcommand{\tr}{{\rm Tr}\,}

\title*{Finding Center Vortices}
\author{J. Greensite\thanks{E-mail: greensit@quark.sfsu.edu}}
\institute{%
Physics and Astronomy Department,
San Francisco State University,
 San Francisco, CA 94117 USA}

\authorrunning{J. Greensite}
\titlerunning{Finding Center Vortices}
\maketitle

\begin{abstract}
I report on recent progress in locating center vortex
configurations on thermalized lattices, generated by lattice Monte
Carlo simulations of SU(2) gauge theory.  A very promising method,
which appears to have some important advantages over previous techniques,
is center projection in the direct Laplacian center gauge.
\end{abstract}

\section{Introduction}

   It is a great pleasure for me to contribute this article in
celebration of Holger Bech Nielsen's 60th birthday.  As it happens, my
contribution to these proceedings is quite closely related to ideas
developed some twenty years ago by Holger, Poul Olesen and other
members of the Niels Bohr Institute \cite{CV}.  Holger and his
co-workers put forward a model of the QCD ground state known as the
``Copenhagen Vacuum,'' which is an explicit realization of the center
vortex theory of confinement, propounded at about the same time by 't
Hooft, Mack, and others \cite{tHooft}.  The Copenhagen Vacuum is based
on a calculation of the QCD effective action by techniques of
continuum perturbation theory.  The finding of Holger Nielsen and Poul
Olesen, based on an analysis of this effective action, was that the
QCD vacuum is dominated by a tangle (or ``spaghetti'') of center
vortices.  Unfortunately, this finding did not settle the issue of
confinement, since there are always doubts about the validity of
perturbation theory in the context of infrared physics.  After a brief
rush of interest, very little work was done on either the center
vortex theory in general, or the Copenhagen vacuum in particular,
after the early 1980's.

   Interest in the center vortex theory revived almost twenty years later,
in 1997, when methods were developed for locating vortex positions in
lattice gauge configurations, generated numerically by the Monte Carlo
technique \cite{Us}.  This advance led to another burst of activity,
this time mainly numerical, which accumulated a great deal of evidence
in favor of the vortex theory.  However, the methods which
are used to locate vortices in lattice configurations have recently 
come under critical scrutiny.  In particular,
what are these methods really doing, and can they fail?  A
number of authors have pointed out certain weaknesses, and these are 
related to the problem of Gribov copies.  In this article I would like
to describe the original method for vortex location and 
the difficulties that have been encountered, and to discuss an improved 
procedure which overcomes those difficulties.  The work I will report
was done in collaboration with Manfried Faber and  
{\v S}tefan Olejn\'{\i}k.

\section{Thin Center Vortices}

   A center vortex is an object whose field strength is concentrated
in a point-like region in D=2 dimensions, a line-like region in D=3
dimensions, and a surface-like region in D=4 dimensions.  When a vortex
is created on an arbitrary background, such that it is topologically
linked to some closed curve $C$, then a Wilson loop around that curve 
is changed
by a center element of the gauge group: 
\begin{equation}
      P\exp[i\oint A_\m dx^m] \ra z P\exp[i\oint A_\m dx^m]
\end{equation}
where $z\in Z_N$ is an element of the center subgoup of the gauge
group SU(N).
 
\begin{figure}[h!]
\centerline{\scalebox{0.40}{\includegraphics{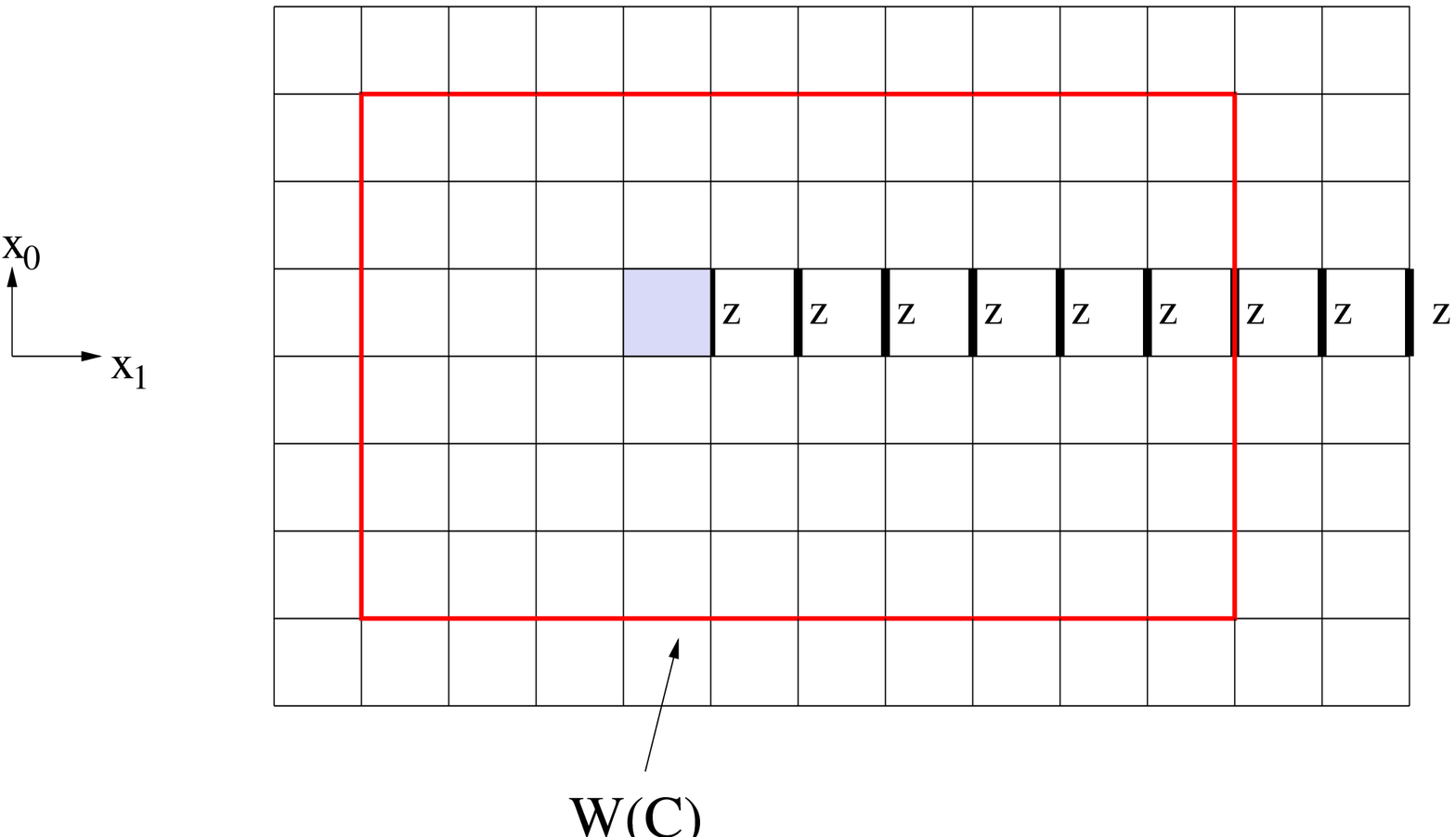}}}
\caption{%
Creation of a center vortex (shaded plaquette) in a plane.}
\label{makevort}
\end{figure}

   To create a thin center vortex on a two-dimensional lattice, 
make the following transformation on the links
\begin{equation}
       U_0(x) \ra z U_0(x) ~~~~~~~~ x_0=0, ~ x_1 > 0
\end{equation}
as illustrated in Fig.\ \ref{makevort}.  The action
changes only locally, at the shaded plaquette $P$, but Wilson
loops $U(C)$ change globally by a center element, if the loop
$C$ encloses the plaquette $P$.  The create a thin vortex in
higher dimensions, make the transformation shown on every $x_0-x_1$
plane.  The stack of shaded plaquettes forms a line-like object,
in the $x_2$ direction, in D=3 dimensions, and a surface-like
object in D=4 dimensions.  

\ni Since creation of a thin vortex changes plaquettes by
\begin{equation}
       \mbox{Tr}[UUU^\dg U^\dg] \rightarrow z  \mbox{Tr}[UUU^\dg U^\dg]
\end{equation}
on the vortex sheet (D=4), the action of a thin vortex is singular in
the continuum limit.  If center vortices are found in the QCD vacuum, 
then we expect these to be extended objects, in which the action of
of the vortex sheet is spread out into a surface of finite thickness
in physical units.  It is not hard to show that vortices percolating
through the lattice will disorder Wilson loops, leading to an 
area law.

\section{Direct Maximal Center Gauge}

   Although it is easy to describe how to create center vortices on
an arbitrary background, it is not obvious how to identify such
objects in any given configuration.  The method which our group proposed
in ref.\ \cite{Jan98} was to fix the lattice to a certain gauge,
known as ``direct maximal center gauge,'' and then map the links
of the gauge-fixed lattice onto center elements, a procedure known
as ``center projection.'' An important insight
into what this gauge is doing $-$ and how it can fail $-$ is due to
Engelhardt and Reinhardt \cite{ER}. 

   A typical thermalized lattice $U_\m(x)$, if printed out, looks like a
set of random numbers.  But it is not random, and in fact { locally},
at $\b \gg 1$, the link variables approximate a (classical) vacuum
configuration
\begin{equation}
       U_\m(x) \approx g(x) g^\dg (x+\hat{\m})
\end{equation}
Let us ask the question: What is the best fit, to a given lattice
$U_\m(x)$, by a pure gauge $g(x) g^\dg (x+\hat{\m})$?
We work with the SU(2) gauge group for simplicity.  The best fit
will minimize
\begin{eqnarray}
     d^2_F & =&  {1\over 4\V} \sum_{x,\m} \mbox{Tr} \left[
     \left( U_\m(x) - g(x)g^\dg(x+\hat{\m}) \right) 
     \times \Bigl(\mbox{h.c.}\Bigr) \right]
\non \\
           & =&  {1\over 4\V} \sum_{x,\m} 2 \mbox{Tr} \left[
                I - g^\dg(x) U_\m(x) g(x+\hat{\m}) \right]
\end{eqnarray}
where $\V$ is the lattice volume. Define
\begin{equation}
       {}^gU_\m(x) = g^\dg(x) U_\m(x) g(x+\hat{\m})
\end{equation}
Minimizing $d_F^2$ is the same as maximizing
\begin{equation}
      R_{L} =  \sum_{x,\m} \mbox{Tr}[{}^gU_\m(x)]
\end{equation}
which is the same as the lattice Landau gauge condition.  What this
demonstrates is that lattice Landau gauge is 
is equivalent to a best fit to a pure gauge.

   Next, consider trying to make a best fit to a thin vortex
configuration on a classical vacuum background
\begin{equation}
       V_\m(x) = g(x) Z_\m(x) g^\dg(x+\hat{\m})
\end{equation}
where $Z_\m(x)=\pm 1$.
This can be done in two steps.  First, 
since the adjoint representation is blind to $Z_\m$, we start by
finding the best fit
in the adjoint representation, minimizing
 
\begin{eqnarray}
     d^2_A  & =&  {1\over 4\V} \sum_{x,\m} \mbox{Tr}_A \left[
     \left( U_\m(x) - g(x)g^\dg(x+\hat{\m}) \right) 
     \times \Bigl( \mbox{h.c.}\Bigr) \right]
\non \\
         & =&  {1\over 4\V} \sum_{x,\m} 2 \mbox{Tr}_A \left[
                I - g^\dg(x) U_\m(x) g(x+\hat{\m}) \right]
\end{eqnarray}
which is the same as maximizing 
\begin{equation}
          R_{dmc} = \sum_{x,\m} \mbox{Tr}_A[{}^gU_\m(x)]
\label{dmc}
\end{equation}
where $\mbox{Tr}_A$ indicates the trace in the adjoint representation.
This is the direct maximal center gauge-fixing condition, which is 
obviously just lattice Landau gauge in the adjoint representation.
Having found $g(x)$ maximizing $R_{dmc}$, we then find the $Z_\m(x)$ 
which minimizes
\begin{equation}
     d^2 = {1\over 4\V} \sum_{x,\m} \mbox{Tr} \left[
     \left( U_\m(x) - g(x)Z_\m(x)g^\dg(x+\hat{\m}) \right) 
      \times \Bigl( \mbox{h.c.}\Bigr) \right]
\end{equation}
in the fundamental representation.  For fixed $g(x)$, this 
minimization is achieved by setting
\begin{equation}
     Z_\m(x) = \mbox{signTr}[{}^gU_\m(x)]
\end{equation}
This step is is known as ``center projection.''  It maps an SU(2) lattice
to a $Z_2$ lattice.

  The original and gauge-transformed lattices can be expressed,
respectively
\begin{eqnarray}
       U_\m(x) & =&  g(x)Z_\m(x) e^{i\d A_\m(x)} g^\dg(x+\hat{\m})
\non \\ 
   {}^gU_\m(x) & =&  Z_\m(x) e^{i\d A_\m(x)} 
\end{eqnarray}
with $Z_\m(x)$ the vortex background,  and $\d A_\m(x)$
the fluctuation around the background.  Direct maximal center gauge finds the
optimal  $Z_\m(x)$ minimizing $ \d A_\m(x)$.

   Unfortunately, as also pointed out in ref.\ \cite{ER}, there are
going to be problems with this approach to vortex finding, particularly
in the continuum limit.  It is clear that the action density of a thin 
vortex is singular at ``P-plaquettes'', which are plaquettes $p$ where
the product of $Z_\m(x)$ around the plaquette is negative, i.e.
$Z(p)=-1$.  P-plaquettes are dual to the thin vortex surface.
Since a thermalized lattice configuration
is locally close to a pure gauge at large $\b$, i.e.
$\oh \mbox{Tr}[U(p)] \approx 1 $,
it follows that the fit to a thin vortex is very bad at P-plaquette
locations.  As $\b\ra \infty$, the bad fit near P-plaquettes may 
overwhelm the good fit in the exterior region, and the best fit in the 
continuum limit may be no thin vortices at all. 

   Computer simulations tend to support this analysis.  In practice it
is impossible to find the global maximum of $R_L$ or $R_{dmc}$;
instead there are various methods of finding local maxima, known as
Gribov copies.  Using a sophisticated gauge-fixing technique which employs
simulated annealing, Bornyakov et al.\ \cite{BKP} were able to generate Gribov
copies in direct maximal center gauge with consistently higher values
of $R_{dmc}$ than those achieved by previous approaches based on the
over-relaxation method.  However, these ``better''
Gribov copies had much worse center dominance properties in the
scaling regime.  Center dominance is the agreement between string
tensions on the full and center projected lattices.  Using the
improved gauge fixing techniques, Bornyakov et al.\ found that
the projected string tension is lower than the full asymptotic
string tension by about 30\%.

\section{Direct Laplacian Center Gauge}

   In confronting the difficulties raised in refs.\ \cite{ER,BKP}, we
would like to retain the strategy of finding the best fit
to a vortex configuration, but avoid over-emphasizing the inevitable
bad fit at P-plaquettes, which seems to be the source of those
difficulties (cf. ref.\ \cite{Remarks}).  We therefore adopt a strategy 
inspired in part by related work of de Forcrand et al.\ 
\cite{Alex,Pepe}, which in turn draws on the earlier Laplacian Landau
gauge fixing of Vink and Weise \cite{Vink}.  The idea is to replace 
the SO(3) gauge transformation
$g(x)$ in \rf{dmc} by  a real $3\times 3$ matrix-valued field $M(x)$,
chosen to maximize 
\begin{equation}
       R_M = \sum_{x,\m} \mbox{Tr}[M^T(x) U_{A\m}(x) 
         M(x+\hat{\m})] 
\end{equation}
($U_A$ is the link in the adjoint representation)
with the constraint that $M(x)$ is orthogonal ``on average.''
\begin{equation}
     {1\over \V} \sum_x M^T(x) M(x) = I
\end{equation}
This essentially introduces an adjustable weighting; $M(x)$ is allowed
to be small where the fit to a thin vortex is bad.

   $R_M$ is maximized (uniquely!) by finding the three lowest eigenvalues,
and corresponding eigenfunctions, of the lattice Laplacian eigenvalue
equation (no sum over $a=1,2,3$)
\begin{equation}
      \sum_y \jgD_{ij}(x,y) f^a_{j}(y) = \l_a f^a_i(x) 
\end{equation}
where
\begin{eqnarray}
\lefteqn{  \jgD_{ij}(x,y)  = }
\non \\
& &  - \sum_{\mu}\left( 
         [U_{A\m}(x)]_{ij}\delta_{y,x+\hat\mu}
       + [U_{A\m}(x)]_{ji}\delta_{y,x-\hat\mu}\
      -  2\delta_{xy}\delta^{ab}\right).
\end{eqnarray}
There are standard numerical algorithms which solve this equation,
and the matrix field formed by
\begin{equation}
       M_{ab}(x) = f_a^b(x)
\end{equation}
where $b=1,2,3$ denotes the three eigenvectors with lowest eigenvalues,
is the matrix field maximizing $R_M$.

  We then map
$M_{ab}(x)$ to the nearest SO(3) matrix
\begin{equation}
 [g_A(x)]_{ij} = \tilde{f}_i^j(x)
\end{equation}
which also satisfies a Laplacian equation
\begin{equation}
      \sum_y \jgD_{ij}(x,y) \tilde{f}^a_{j}(y) = \L_{ac}(x) \tilde{f}^c_i(x)
\end{equation}
This is equivalent to finding the SO(3)-valued matrix field closest to
$M_{ab}(x)$ which is also a local
maximum of $R_{dmc}$.  The mapping is accomplished in two steps: 

\begin{enumerate}
\item Polar decomposition
\begin{equation}
         M(x) = \pm \Omega(x) P(x)
\end{equation}
to find the SO(3) matrix $\Omega$ which is closest to $M$.
\item  Relaxation $\Omega \ra g_A$ to the nearest 
local maximum of 
\begin{equation}
       R_{dmc} = \sum_{x,\m} \mbox{Tr}[g_A^T(x) U_{A\m}(x) 
         g_A(x+\hat{\m})] 
\end{equation}
\end{enumerate}
Finally, transform $U_\m(x) \ra {}^gU_\m(x)$. This prescription for mapping
an arbitrary gauge field configuration to an (in principle unique) 
point on the gauge orbit is called {\bf direct Laplacian
center gauge}.  We apply center projection
\begin{equation}
       Z_\m(x) = \mbox{signTr}[{}^gU_\m(x)]
\end{equation}
as before, 
to locate the P-plaquettes forming thin vortices (known as ``P-vortices'')
on the projected lattice.

\subsection{Center Dominance}

  The most important property to check with the new procedure is
center dominance, since this is where problems were found numerically, in 
refs.\ \cite{BKP} and \cite{KT}, with the previous methods.

\begin{figure}[h!]
\centerline{\scalebox{1.0}{\includegraphics{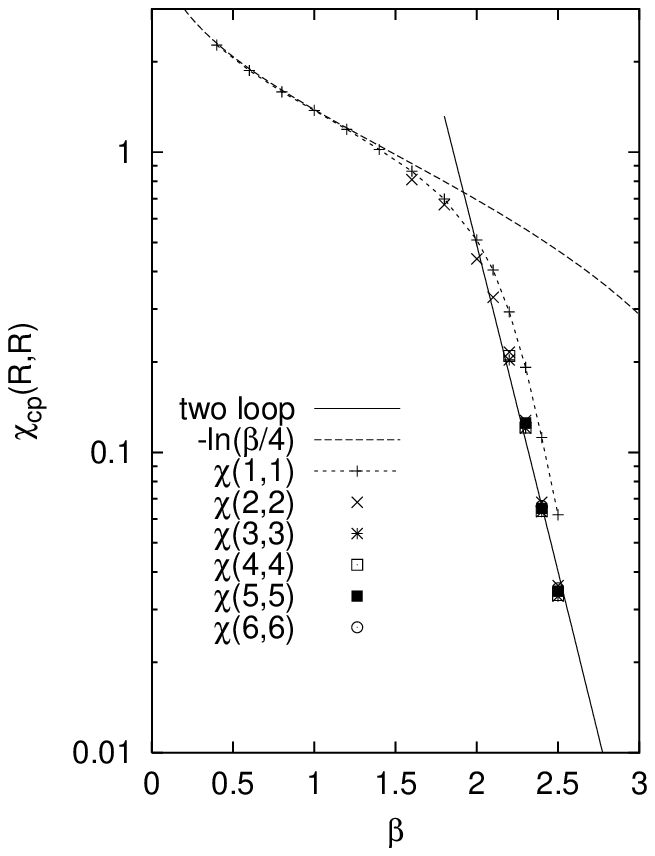}}}
\caption{%
Projected Creutz ratios of various sizes as a function of $\b$.}
\label{nchi}
\end{figure}

  The center vortex contribution to the asymptotic string tension
is extracted from
\begin{equation}
        W_{cp}(C) = \langle ZZ...Z \rangle
\end{equation}
in the center-projected configurations, obtained after direct
Laplacian gauge fixing.  Creutz ratios 
$\chi_{cp}[R,R]$ are plotted in Fig.\ \ref{nchi}.  The straight line is the
asymptotic freedom behavior, with 
$ {\sqrt{\s} / \Lambda} = 58$

\begin{figure}[h!]
\centerline{\scalebox{0.8}{\includegraphics{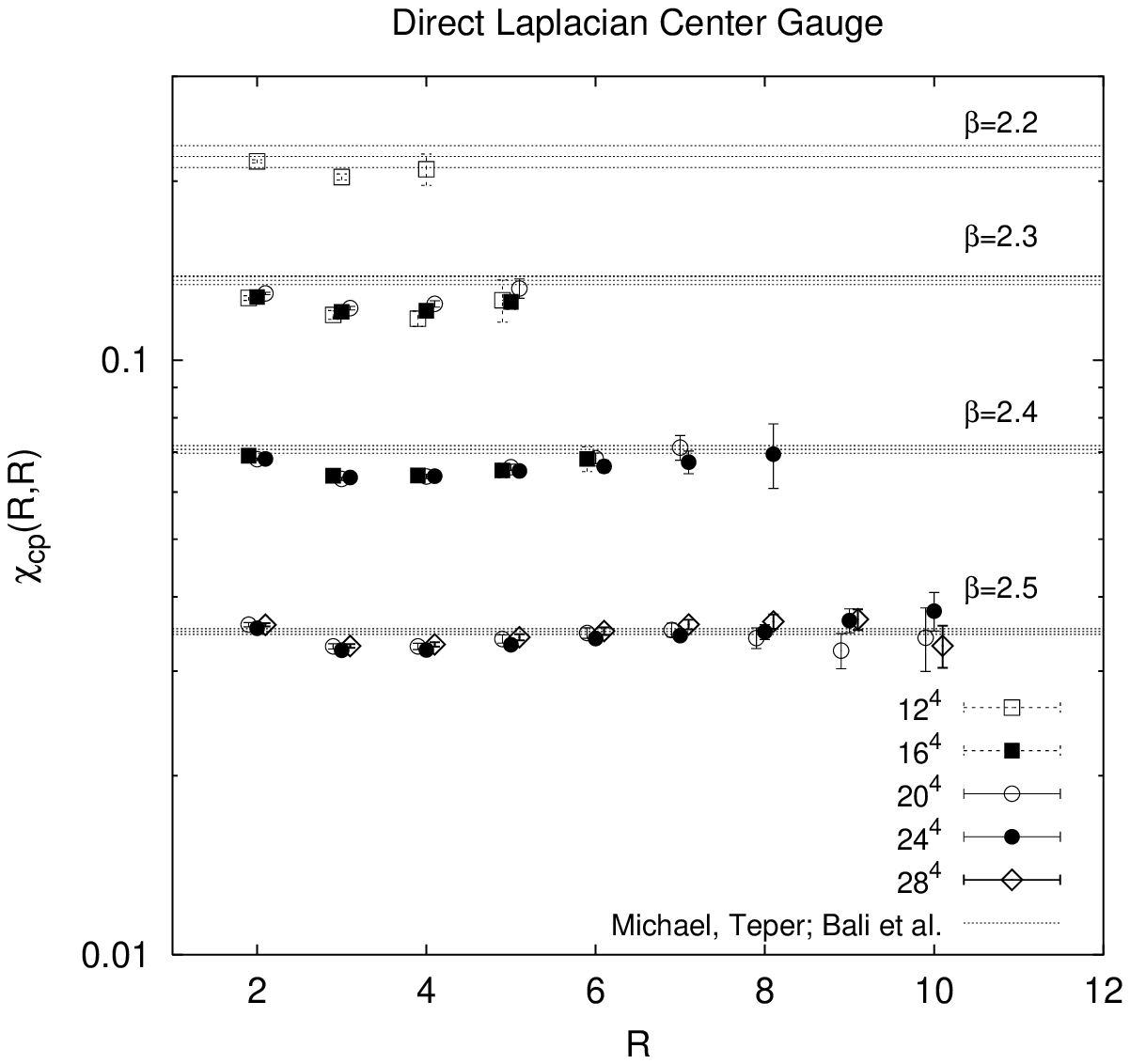}}}
\caption{%
Center projected Creutz ratios for various couplings
and lattice sizes. 
Straight lines show the values of
the asymptotic string tensions and corresponding error bars,
in the unprojected configurations, quoted in ref.\ 
\cite{Bali}.} 
\label{all_beta2}
\end{figure}

  Note that the Creutz ratios at each $\b$ are nearly the same
(with the exception of the $1\times 1$ loop), and this is known
as ``precocious linearity.''  This feature can also be clearly
seen in Fig.\ \ref{all_beta2}.  Here we also see
the property of center dominance: The
center-projected Creutz ratios $\chi_{cp}[R,R]$
at all $R \ge 2$ agree quite well with the full asymptotic string tension 
$\s$, extracted from unprojected Wilson loops, both at 
strong \emph{and} weak couplings

\begin{figure}[h!]
\centerline{\scalebox{0.8}{\includegraphics{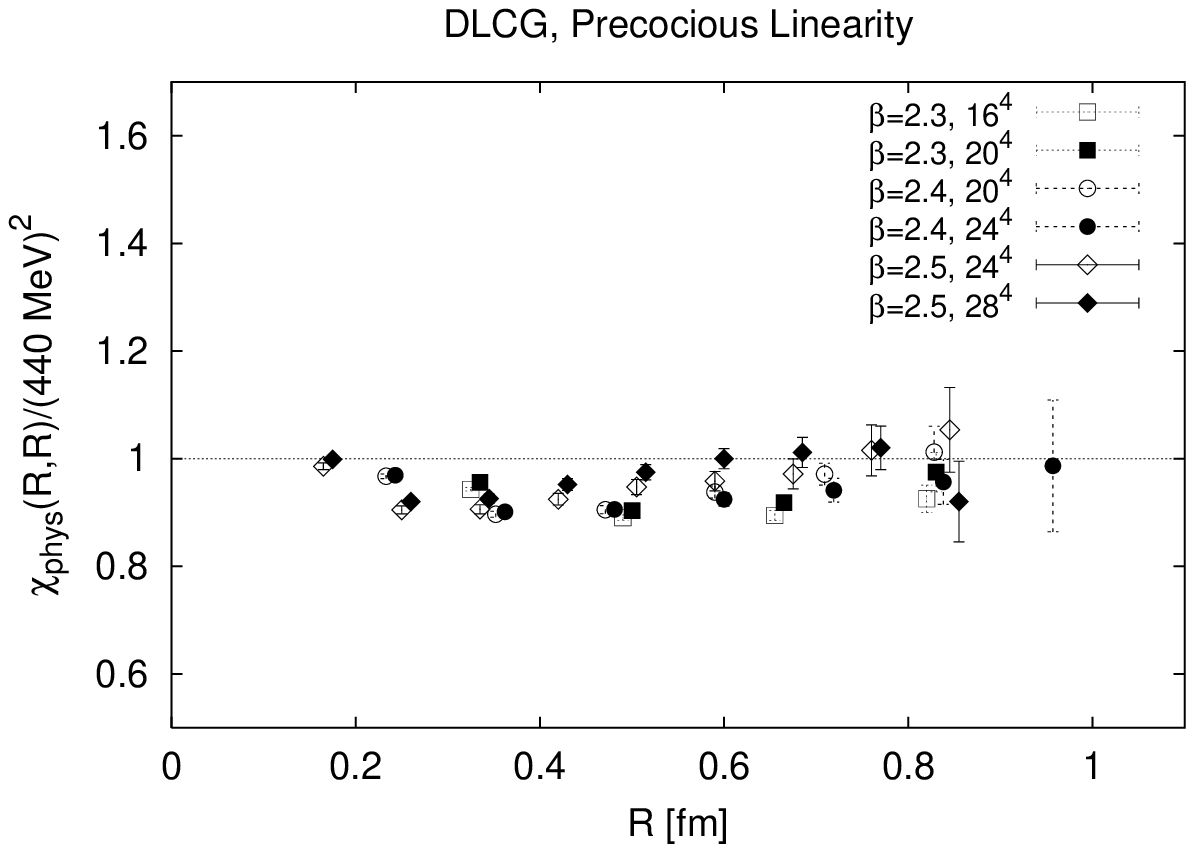}}}
\caption{%
Ratio of center-projected Creutz ratios $\chi[R,R]$
to the full asymptotic string tension in physical units vs.\ $R$
in physical units.} 
\label{chi_phys}
\end{figure}

  An alternative way of displaying our data is to
plot the ratio
\begin{equation}
   {\chi_{cp}[R,R] \over \s_{Lat}(\b)} \equiv 
         {\chi_{phys}[R,R]\over \s_{phys} }  
\end{equation}
as a function of $R$ in physical units,
where $\s_{Lat}$ is the asymptotic string tension in lattice units,
while the subscript ``phys'' refers to physical units.  In Fig.\
\ref{chi_phys} we
see that the ratios are not far from unity (center dominance) even
down to relatively small distance scales (precocious linearity).

\subsection{Other Tests}

   To check that P-vortices are really associated with center vortices
in the full configuration, and are not just artifacts of the projection,
it is necessary to verify their correlation with gauge invariant observables.
Such tests were invented in refs.\ \cite{Us,Jan98} and \cite{dFE}, and
it is important to repeat these with the new method.  Here I will only
touch on the results.
 
   We define the ``vortex-limited'' Wilson loops $W_n(C)$ 
to be loops evaluated on a 
subensemble of configurations,
selected such that precisely ~$n$~ P-vortices, in the 
corresponding center-projection, pierce the loop.  
Center projection is used only to select the data
set.  The loops themselves are evaluated using the full, 
{unprojected} link variables.  What we find numerically is that in the
limit of large loops,
\begin{equation}
       {W_n(C) \over W_0(C)} ~~ \longrightarrow ~~ (-1)^n
\label{WnW0}
\end{equation}
This is what one expects, if P-vortices locate thick center
vortices in the unprojected SU(2) lattice.

   It is also possible to measure the density of P-vortices as
a function of $\b$.  If $N_{vor}$ is the number of P-vortex plaquettes,
and $N_T$ is the total number of plaquettes on the lattice, 
then the density of P-vortices in lattice units is given by
\begin{eqnarray}
      p       & =& 
                  {N_{vor}\over N_T} = {N_{vor}a^2 \over N_T a^4} a^2
\non \\
              & =& 
               {\mbox{Total Vortex Area} \over 6
                  \times \mbox{Total Volume} } a^2
               =  {1\over 6} \rho a^2
\end{eqnarray}
where $\rho$ is the center vortex density (area per unit volume) 
in physical units. 
According to asymptotic freedom, if the P-vortices are locating
physical objects, we should have
\begin{equation}
     p = {\rho \over 6\Lambda^2} \left({6\pi^2 \over 11} \b \right)^{102/121}
            \exp\left[-{6\pi^2 \over 11}\b \right]
\label{jg-af}
\end{equation}
But also $p$ is related to the center-projected one-plaquette loop
\begin{equation}
     W_{cp}(1,1) =  (1-p)\times (+1) + p \times (-1) = 1-2p
\end{equation}
which allows us to calculate $p$ from center-projected plaquette values.
It is found that the the data for $p$ is quite consistent with the
asymptotic freedom result \rf{jg-af}.

   Finally, as suggested by de Forcrand and D'Elia \cite{dFE}, one
can remove center vortices from the
gauge-fixed configuration
by simply multiplying by the projected configuration
\begin{equation}
       {}^gU_\m(x) \ra {}^gU'_\m(x) = Z_\m(x)  {}^gU_\m(x)
\end{equation}
What we find, using the direct Laplacian gauge procedure, is that
confinement disappears when vortices are removed in this way.

\section{Conclusions}

   The conclusion we draw from all this is that center vortices in
the Yang-Mills vacuum are correctly located by the new procedure, and 
these vortices are responsible for the confining force.    
Direct Laplacian center gauge combined with center projection is found
to be an
improvement over the original approach to locating center vortices in
lattice configurations, and presumably works well even in the
continuum limit.  It should be a useful tool for the further study of
the center vortex ``spaghetti vacuum;'' a picture of the QCD vacuum
very similar to that proposed by Holger Nielsen and Poul 
Olesen back in 1978.

\section*{ Acknowledgements}

  Much of this work, and work that preceded it, was carried out
in Copenhagen at the Niels Bohr Institute, whose hospitality
I am happy to acknowledge.
This research is supported in part by
the U.S. Department of 
Energy under Grant No.\ DE-FG03-92ER40711.

\backmatter

\thispagestyle{empty}
\parindent=0pt
\begin{flushleft}
\mbox{}
\vfill
\vrule height 1pt width \textwidth depth 0pt
{\parskip 6pt

{\sc Blejske Delavnice Iz Fizike, \ \ Letnik~2, \v{s}t. 2,} 
\ \ \ \ ISSN 1580--4992

{\sc Bled Workshops in Physics, \ \  Vol.~2, No.~2}

\bigskip

Zbornik delavnic `What comes beyond the Standard model', 2000 in 2001

Zvezek 1: Zbornik ob 60. rojstnem dnevu Holgerja Bech Nielsena

Proceedings to the workshops `What comes beyond the Standard model',
2000 and 2001

Volume 1: Festschrift dedicated to the 60th birthday of 
Holger Bech Nielsen

\bigskip

Uredili in oblikovali Norma Manko\v c Bor\v stnik, Colin D. Froggat in 
Dragan Lukman 

Publikacijo sofinancira Ministrstvo za \v solstvo, znanost in \v sport 

Tehni\v{c}ni urednik Vladimir Bensa

\bigskip

Zalo\v{z}ilo: DMFA -- zalo\v{z}ni\v{s}tvo, Jadranska 19,
1000 Ljubljana, Slovenija

Natisnila Tiskarna MIGRAF v nakladi 100 izvodov

\bigskip

Publikacija DMFA \v{s}tevilka 1479

\vrule height 1pt width \textwidth depth 0pt}
\end{flushleft}



\begin{thebibliography}{}


\bibitem{becherandjoos1982} P. Becher and H. Joos (1982), \textit{The Dirac-K\" ahler equation and
Fermions on the Lattice.}, \textit{ Z. Phys. C-Part. and Fields} 
\textbf{15}, 343-365.
%
\bibitem{borstnikandmankoc1999} Anamarija Bor\v stnik and Norma Susana Manko\v c Bor\v
stnik (1999). \textit{Are Spins and Charges Unified? How Can One
Otherwise Understand Connection Between Handedness (Spin) and
Weak Charge?  Proceedings to the International Workshop on 
''What Comes Beyond the Standard Model, Bled,
Slovenia, 29 June-9 July 1998} 
Ed. by N. Manko\v c Bor\v stnik,
H. B. Nielsen, C. Froggatt, DMFA Zalo\v zni\v stvo 1999, p. 52-57,
hep-ph/9905357, and the paper in preparation. 

%
\bibitem{georgi1982} Howard Georgi (1982). \textit{ Lie Algebra in Particle
Physics},  The 
Benjamin/Cummings Publishing Company, Inc. Advanced Book
Program. 
%
\bibitem{ikemori1987} Hitochi Ikemori (1987). \textit{Superfield
Formulation of Relativistic Superparticle }, \textit{ Phys.Lett.}
\textbf{ 199}, 239-242.

%
\bibitem{kahler1962} Erich K\" ahler (1962), \textit{Der innere Differentialkalk\" ul}, 
\textit{Rend. Mat. Ser.V}, \textbf{21}, 452-523. 
%
\bibitem{mankoc1992} Norma Susana Manko\v c Bor\v stnik (1992).
\textit{Spin Connection as a Superpartner of a Vielbein},
\textit{ Phys. Lett.} \textbf{ B 292} 25-29. {From a World-sheet
Supersymmetry to the Dirac Equation } \textit{ Nuovo Cimento}
\textbf{ A  105}, 1461-1471.
%
\bibitem{mankoc1993} Norma Susana Manko\v c Bor\v stnik (1993).
\textit{Spinor and Vector Representations in Four Dimensional
Grassmann Space} \textit{ J. Math. Phys.} \textbf{ 34} 3731-3745. 
%
\bibitem{mankoc1994} Norma Susana Manko\v c Bor\v stnik (1994).
\textit{Spinors, Vectors and Scalars in Grassmann Space and
Canonical Quantization for Fermions and Bosons}, 
\textit{ Int. Jour. Mod. Phys.} \textbf{ A  9} 1731-1745;
\textit{ Unification of Spins and Charges in Grassmann Space}
\textit{ hep-th/9408002}; \textit{Qantum Mechanics in Grassmann
Space, Supersymmetry and Gravity} \textit{hep-th/9406083}.
%
\bibitem{mankoc1995} Norma Susana Manko\v c Bor\v stnik (1995).
\textit{Poincar\'e Algebra in Ordinary and Grassmann Space and
Supersymmetry } \textit{ J.
Math. Phys.} \textbf{ 36}, 1593-1601; \textit{Unification of
Spins and Charges in Grassmann Space} \textit{ Mod. Phys. Lett.}
\textbf{A 10}, 587-595; \textit{hep-th/9512050} 
%
\bibitem{mankoc1999} Norma Susana Manko\v c Bor\v stnik (1999,2001).
\textit{Unification of Spins and Charges in Grassmann Space},  
\textit{hep-ph/9905357}, 
\textit{In. J. of Theor. Phys.} \textbf{40} 315-337 
\textit{ Proceedings to the International Workshop ''What Comes
Beyond the Standard Model'', Bled,
Slovenia, 29 June-9 July 1998}, Ed. by N. Manko\v c Bor\v stnik,
H. B. Nielsen, C. Froggatt, DMFA Zalo\v zni\v stvo 1999, p. 20-29.
%
\bibitem{mankoc1997}
Norma Susana Manko\v c Bor\v stnik and Svjetlana Fajfer (1997),
\textit{Spins and Charges, the Algebra and Subalgebras of the
Group SO(1,14)}, \textit{ Nuovo 
Cimento} \textbf{B  112 }, 1637-1665; \textit{hep-th/9506175}.
%
\bibitem{mankocandnielsen1999} Norma Susana Manko\v c Bor\v stnik and Holger Bech
Nielsen (1999). \textit{Dirac-K\' ahler Approach Conencted to
Quantum Mechanics in Grassmann Space}, to appear in \textit{Phys.
Rev. D15}; \textit{hep-th/9911032},
\textit{ Proceedings to the International Workshop ''What Comes
Beyond the Standard Model'', Bled,
Slovenia, 29 June - 9 July 1998}, Ed. by N. Manko\v c Bor\v stnik,
H. B. Nielsen, C. Froggatt, DMFA Zalo\v zni\v stvo 1999, p. 68-73;
\textit{hep-ph/9905357;  hep-th/9909169}.
%
\bibitem{nielsenandninomija1981} Holger Bech Nielsen and M. Ninomija (1981).
\textit{ A No-go Theorem for Regularizing Chiral Fermions}
 \textit{ Phys. Lett.} \textbf{
B 105}, 219-223; \textit{ Nucl. Phys.  } \textbf{ B 185},
\textit{Absence of Neutrinos on a Lattice}, 20-40.
%
\bibitem{wessandbagger1983} Julius Wess and Jonathan Bagger (1983). \textit{
Supersymmetry and Supergravity}, Princeton Series in Physics,  
Princeton University Press, Princeton, New Jersey.   
%
\bibitem{mankocandnielsen2002} Norma Susana Manko\v c Bor\v stnik and Holger Bech
Nielsen (2002). \textit{Coupling constant unification in spin-charge unifying model
agreeing with proton decay measurement}, in preparation.
%
\bibitem{borstnikandmankocandnielsen2002} Anamarija Bor\v stnik and 
Norma Susana Manko\v c Bor\v stnik and Holger Bech
Nielsen (2002). \textit{Where do families of left handed weak charge doublets and right 
handed weak charge singlets come from?}, in preparation.
%
\end{thebibliography}

\begin{thebibliography}{99}
\bibitem{KN1}Z.Koba and H.B.Nielsen, Nucl.Phys.B10,633(1969)
\bibitem{KN2}Z.Koba and H.B.Nielsen, Nucl.Phys.B12,517(1969)
\bibitem{KN3}Z.Koba and H.B.Nielsen, Z.Physik 229,243(1969)
\bibitem{HBN}Y.Nambu, Proceedings of the International Conference on 
Symmetries and  Quark Models,edited by R.Chand.Gordan and Breach  
(1970) p.269;\\ H.B.Nielsen, in High Energy Physics, Proceedings 
of the 15th International Conference on High Energy Physics,\\ 
Kiev 1970, edited by V.Shelest(Naukova,Dunika,Kiev,\\USSR, 
1972);\\ L.Susskind, Phys.Rev.Letts.23, 545 (1969) 
\bibitem{PF}P.H.Frampton in "Dual Resonance Models",(Frontiers in Physics, no.5),
W.A.Benjamin, Inc (1974).
\bibitem{NO} H.B.Nielsen and P.Olesen, Nucl.Phys.B61,45(1973)
\bibitem{AFKP} M.Axenides, E.G.Floratos, S.Komineas and L.Perivolaro\\poulos,
"Metastable Ringlike Semitopological Solitons", Phys.Rev.Letts 86, 
4459 (2001)
\bibitem{lp92}T.D. Lee and Y. Pang, Phys. Rept. {\bf 221} (1992) 251
\bibitem{c85}S. Coleman, Nucl. Phys. {\bf B262} (1985) 263
\bibitem{bs00}
R.~Battye and P.~Sutcliffe, 
Nucl.\ Phys.\  {\bf B590}, 329 (2000) [hep-th/0003252]. 
\bibitem{lpwww}S. Komineas and L. Perivolaropoulos,
http://leandros.chem.demokritos.gr/qballs,
http://leandros.chem.demokritos.gr/qballs/3Dplots.html 
\bibitem{Witten:1985eb}
E.~Witten, 
Nucl.\ Phys.\  {\bf B249}, 557 (1985). 
\bibitem{hht88}
D.~Haws, M.~Hindmarsh and N.~Turok, 
Phys.\ Lett.\  {\bf B209}, 255 (1988). 
\bibitem{d64} G.H. Derrick, J. Math Phys. {\bf 5}, 1252 (1964)
\bibitem{k97}
A. Kusenko, Phys. Lett. {\bf B404} (1997) 285, hep-th/9704073 
\bibitem{akpf00}
M.~Axenides, S.~Komineas, L.~Perivolaropoulos and E.~Floratos, 
Phys.\ Rev.\  {\bf D61}, 085006 (2000) [hep-ph/9910388]. 
\bibitem{Faddeev:1997zj}
L.~Faddeev and A.~J.~Niemi, 
Nature {\bf 387}, 58 (1997) [hep-th/9610193]. 
\bibitem{Perivolaropoulos:2000gn}
L.~Perivolaropoulos, ``Superconducting semilocal stringy (Hopf) 
textures,'' in {\it ESF Network Workshop Les Houches, France, 
1999} hep-ph/9903539 \\ 
 L.~Perivolaropoulos and 
T.~N.~Tomaras, 
Phys.\ Rev.\  {\bf D62}, 025012 (2000) [hep-ph/9911227]. 
\bibitem{kk99}S.~Kasuya and M.~Kawasaki,
``Q-ball formation through Affleck-Dine 
mechanism,''hep-ph/9909509. 
\end{thebibliography}

\begin{thebibliography}{99}
\bibitem{abj} 
S. Adler, Phys. Rev. {\bf 177}, 2426 (1969); 
J. S. Bell and R. Jackiw, Nuovo Cimento, {\bf 60A}, 47 (1969); 
R. A. Bertlmann, 'Anomalies in quantum field theory', 
Oxford, UK: Clarendon (1996) and references therein. 
\bibitem{ms}D. Melikhov and B. Stech, hep-ph/0108165. 
\bibitem{dz} A. D. Dolgov and V. I. Zakharov, Nucl. Phys. {\bf B27}, 525 (1971); 
V. I. Zakharov, Phys. Rev. {\bf D42}, 1208 (1990). 
\bibitem{m} D. Melikhov, Phys. Lett. {\bf B380}, 363 (1996).  
\bibitem{mr} I. V. Musatov and A.V. Radyushkin, Phys. Rev. {\bf D 56}, 2713 (1997).
\bibitem{bb} P. Ball and V. M. Braun, Phys. Rev. {\bf D 58}, 094016 (1999).  
\bibitem{bmns}M. Beyer, D. Melikhov, N. Nikitin, and B. Stech, 
Phys. Rev. {\bf D64}, 094006 (2001).
\end{thebibliography}

\begin{thebibliography}{99}
\bibitem{maltoni} M. Maltoni, V.A. Novikov, L.B. Okun, A.N. Rozanov, M.I. Vysotsky: Extra quark-lepton generations and precision measurements,
Phys. Lett. \textbf{B476}, (2000), 107--115.
\newline
M. Maltoni: Ph.D. thesis, hep-ph/0002143, Ch. 5, (2000).
\bibitem{okun} V.A. Novikov, L.B. Okun, A.N. Rozanov, M.I. Vysotsky: Extra generations and discrepancies of electroweak precision data,
hep-ph/0111028.
\bibitem{gunion} J.F. Gunion, D. McKay, H. Pois: A Minimal Four-Family Supergravity Model, Phys. Rev. \textbf{D53}, (1996), 1616--1647.
\bibitem{pirogov} H. B. Nielsen, A. V. Novikov, V. A. Novikov, M. I. Vysotsky: Higgs potential bounds on extra quark-lepton generations, Phys. Lett. 
\textbf{B374}, (1996), 127--130. 
\newline
Yu. F. Pirogov, O.V. Zenin: Two-loop renormalization group restrictions on the standard model and the fourth chiral family,
Eur. Phys. J. \textbf{C10}, (1999), 629--638.
\bibitem{L3} L3 Coll: Search for Charginos with a Small Mass Difference to the Lightest Supersymmetric Particle at $\sqrt{s} = 189 GeV$,
Phys. Lett. \textbf{B482}, (2000), 31--42.
\bibitem{lepton} OPAL Coll., G. Abbieudi {\it et al}: Search for Unstable Heavy and Excited Leptons at LEP2, Eur. Phys. J. \textbf{C14}, (2000),
73--84.
\bibitem{hou} W. S. Hou, R. Stuart: Higgs-boson production from decays of the fourth-generation $B$ quark, Phys. Rev. \textbf{D43}, (1991),
3669--3682.
\bibitem{quark} CDF Coll., T. Affolder {\it et al}: Search for a Fourth-Generation Quark More Massive than the $Z^0$ Boson in $p\overline{p}$ Collisions
at $\sqrt{s} = 1.8 TeV$, Phys. Rev. Lett. \textbf{84}, (2000), 835--840.
\bibitem{susy} S. P. Martin: A Supersymmetry Primer, hep-ph/9709356.
\bibitem{string} C.-H. Chen, M. Drees, J. F. Gunion: A Non-Standard String/SUSY Scenario and its Phenomenological Implications, Phys. Rev. \textbf{D55},
(1997), 330--347.
\bibitem{long} DELPHI Coll., P. Abreu {\it et al}: A Search for Heavy Stable and Long-Lived Squarks and Sleptons in $e^+e^-$ Collisions at Energies from
130 to 183 GeV, Phys. Lett. \textbf{B444}, (1998), 491--502.
\bibitem{pdg} D. E. Groom {\it et al}: Particle Data Group Coll., Eur. Phys. J. \textbf{C15}, (2000), 1.
\bibitem{boer} W. de Boer, G. Burkart, R. Ehret, W. Oberschulte-Beckmann, V. Bednyakov, S. G. Kovalenko: Combined Fit of Low Energy Constraints
to Minimal Supersymmetry and Discovery Potential at LEPII, hep-ph/9507291.
\bibitem{ross} D. M. Ghilencea, G. G. Ross: Precision prediction of gauge couplings and the profile of a string theory, Nucl. Phys. 
\textbf{B606}, (2001), 101.
\bibitem{polonsky} P. Langacker, N. Polonsky: Uncertainties in Coupling Constant Unification, Phys. Rev. \textbf{D47}, (1993), 4028 - 4045.
\bibitem{invisible} ALEPH, DELPHI, L3 and OPAL Coll: Searches for Invisible Higgs bosons: Preliminary combined results using LEP data collected
at energies up to $209 GeV$, hep-ex/0107032.
\end{thebibliography}

\begin{thebibliography} {99}
\bibitem{gov}  Fairlie D.B., Govaerts  J. and  Morozov A., Universal Field Equations
with Covariant Solutions, {\it Nuclear Physics B 373} (1992) 214-232.
\bibitem{baker} Baker L.M. and  Fairlie D.B., Companion Equations for Branes, (1999) , {\it Journal of Mathematical Physics} {\bf 41} (2000) 4284-4292.\\
 Baker L.M. and  Fairlie D.B., Hamilton-Jacobi equations and Brane associated Lagrangians, {\it Nuclear Physics} {\bf B596} (2001) 348-364.
\bibitem{fai} Fairlie D.B The Multi-field Complex Bateman  Equation,\hfill\break {\bf hep-th 0106003} to appear in {\it Letters in Math. Phys} (2001).

\bibitem{chaundy} Chaundy T. , {\it The Differential Calculus} Oxford University Press (1935) p. 328.
\bibitem{leznov}Fairlie  D.B. and Leznov  A.N.,
 The Complex Bateman Equation, {\it Letters in Math. Physics} {\bf 49}(1999) 213-216.\\
 Fairlie D.B. and  Leznov A.N.,  The Complex Bateman equation in a space of arbitrary dimensions, {\it Journal of Mathematical Physics} {\bf 42} (2001) 453-462.
\bibitem{mulvey} Fairlie D.B. and  Mulvey J.A., Integrable Generalisations of the
\hfill\break 2-dimensional Born Infeld Equation,  {\it J. Phys} {\bf A27} (1994) 1317-1324.


\bibitem{arik} Arik M., Neyzi F., Nutku Y., Olver P.J. and Verosky J., Multi-Hamiltonian structure of the Born Infeld equations, {\it J.Math. Phys} {\bf 30} (1989) 1338-1344. 
\bibitem{intrev}Fairlie D.B., Integrable Systems in Higher Dimensions 
{\it Quantum Field Theory, Integrable Models and Beyond} 
Editors. T. Inami and R. Sasaki {\it Progress of Theoretical Physics Supplement}
 {\bf 118} (1995) 309-327.
\bibitem{dbf} Fairlie D.B.,Dirac-Born-Infeld Equations, {\it Phys .Lett.} {\bf B456}, (1999) 141-146.

\bibitem{leznov2}Leznov A.N., {\it Journal of Nonlinear Mathematical Physics}  {\bf 8:(1)} (2001)1-7;also 
'Integrable Hierarchies and Modern Physical Theories NATO Science Series ed.
Henrik Aratun and A.Sorin,` Mathematics, Physics and Chemistry-{\bf V0l. 18} (2001)
\end{thebibliography}

\begin{thebibliography}{99}

\bibitem{Schafer}
An introduction to nonassociative
algebras, Richard D. Schafer.
New York, Academic Press, 1966.

\bibitem{J}
P. Jordan, Z. Phys. {\bf 80,} 285 (1933).

\bibitem{JVW}
P. Jordan, J. Von Neumann and E.P. Wigner, Ann. Math. {\bf 35,} 29
(1934).

\bibitem{A}
A.A. Albert, Ann. Math. {\bf 35,} 65 (1934).

\bibitem{GunaydinGursey}
M.~Gunaydin and F.~Gursey,
J.\ Math.\ Phys.\  {\bf 14}, 1651 (1973).

\bibitem{Costa}
Nonassociative algebra and its applications
 : the fourth international conference /
edited by R. Costa ... [et al.].
New York : Marcel Dekker, c2000.

\bibitem{Lohmus}Nonassociative algebras in physics /
J. L$\tilde{o}$hmus, E. Paal, and L. Sorgsepp,
Palm Harbor, FL : Hadronic Press, 1994.

\bibitem{Grossman}
B.~Grossman, T.~W.~Kephart and J.~D.~Stasheff,
Commun.\ Math.\ Phys.\  {\bf 96}, 431 (1984)
[Erratum-ibid.\  {\bf 100}, 311 (1984)].

\bibitem{Fubini:1985jm}
S.~Fubini and H.~Nicolai,
Phys.\ Lett.\ B {\bf 155}, 369 (1985).

\bibitem{Duff:1990wu}
M.~J.~Duff and J.~X.~Lu,
Phys.\ Rev.\ Lett.\  {\bf 66}, 1402 (1991).

\bibitem{Harvey:1991eg}
J.~A.~Harvey and A.~Strominger,
Phys.\ Rev.\ Lett.\  {\bf 66}, 549 (1991).

\bibitem{Bruck}
A survey of binary systems,
R. H. Bruck,  in
Ergebnisse der Mathematik und ihrer Grenzgebiete Berlin, Heft 20,
Springer, 1958.

\bibitem{DK}
"Latin squares and their applications," J. D\'{e}nes and A. D. Keedwell.
London : English Universities Press, 1974.

\bibitem{LM}
"Discrete mathematics using Latin squares,"Charles F. Laywine and Gary
L. Mullen, New York : Wiley, 1998.

\bibitem{Heydayat}
Orthogonal arrays : theory and
applications,
A.S. Heydayat, N.J.A. Sloane, J. Stufken.
Springer series in statistics, Springer, New York, 1999.

\bibitem{Pickert}
Projektive Ebenen, Pickert, G.
Die Grundlagen der mathematischen
Wissenschaften in Einzeldarstellungen Bd. 80,
Berlin, Springer, 1955.

\bibitem{Bar-Natan}
Nonassociative Tangles, by D. Bar-Natan in
Geometric topology : 1993 Georgia International
Topology Conference, University of Georgia,
Athens, Georgia / William H. Kazez, editor.
Providence, R.I. : American Mathematical Society :
International Press, 1997


\end{thebibliography}

\begin{thebibliography}{99}
\bibitem{goldstein} H.\ Goldstein, {\em Classical Mechanics} (Addison-Wesley,
   1950)
\bibitem{ll} L.D.\ Landau and E.M.\ Lifshitz, {\em Fluid Mechanics} (Pergamon
   Press, 1959)
\bibitem{feynman} R.P.\ Feynman, in: {\em Progr.\ in Low Temp.\ Phys.,} ed.\
              C.J.\ Gorter (North Holland; Amsterdam, 1955), Vol.I, ch.2
\bibitem{dirac} P.A.M.\ Dirac, Proc.\ Roy.\ Soc.\ A33 (1931) 60 \\
            G.\ 't Hooft, Nucl.\ Phys.\ B79 (1974) 276 \\
            A.M.\ Polyakov, JETP Lett.\ 20 (1974) 194
\bibitem{abrikosov} F.\ London, in {\em Superfluids} (J.\ Wiley \& Sons,
                N.Y.\ 1950) 152 \\
                A.A.\ Abrikosov, Sov.\ Phys.\ JETP 5 (1957), 1174 \\
                H.B.\ Nielsen and P.\ Olesen, Nucl.\ Phys.\ B61 (1973) 45
\end{thebibliography}

\begin{thebibliography}{99}

\bibitem{PRY1} J.L. Petersen, J. Rasmussen and M. Yu, 
 \NP{B 457} (1995) 309
\bibitem{PRY2}J.L. Petersen, J. Rasmussen and M. Yu, \NP{B 457} (1995) 343
\bibitem{PRY3}J.L. Petersen, J. Rasmussen and M. Yu, in 
 {\em Proceedings of the workshop Gauge Theories, Applied Supersymmetry and
 Quantum Gravity, Leuven July 1995, Leuven Notes in Mathematical and 
 Theoretical Physics, Vol. 6}, Eds. B. de Wit {\em et al} (Leuven 1996)
 hep-th/9510059
\bibitem{PRY4}J.L. Petersen, J. Rasmussen and M. Yu, Nucl. Phys. 
{\bf B} (Proc. Suppl.)
 {\bf 49} (1996) 27
\bibitem{PRY5} J.L. Petersen, J. Rasmussen and M. Yu,
\NP{B 481} (1996) 577
\bibitem{PRY6}J.L. Petersen, J. Rasmussen and M. Yu, in 
{\em Proceedings of Inauguration
 Conference of the Asia Pacific Center for Theoretical Physics (APCTP),
 Seoul, Korea, 1996}
\bibitem{PRY7} J.L. Petersen, J. Rasmussen and M. Yu, \NP{B 502} (1997) 649
\bibitem{JR} J. Rasmussen, hep-th/9610167, Ph.D. thesis 
  (The Niels Bohr Institute) 
\bibitem{KK} V.G. Kac and D.A. Kazhdan, Adv. Math. {\bf 34} (1979) 97
\bibitem{MFF} F.G. Malikov, B.L. Feigin and D.B. Fuks,
Funkt. Anal. Prilozhen {\bf 20} (1986) 25
\bibitem{BF} D. Bernard and G. Felder, \CMP{127} (1990) 145
\bibitem{ATY} H. Awata, A. Tsuchiya and Y. Yamada, \NP{B 365} (1991) 680;
 H. Awata, Prog. Theor. Phys. Suppl. {\bf 110} (1992) 303
\bibitem{FGPP} P. Furlan, A.Ch. Ganchev, R. Paunov and V.B. Petkova,
 \PL{B 267} (1991) 63;\\
 P. Furlan, A.Ch. Ganchev, R. Paunov and V.B. Petkova,
 \NP{B 394} (1993) 665;\\
 A.Ch. Ganchev and V.B. Petkova, \PL{B 293} (1992) 56
\bibitem{Dot} Vl.S. Dotsenko, \NP{B 338} (1990) 747;\\
 Vl.S. Dotsenko, \NP{B 358} (1991) 547
\bibitem{JLP-A} K. Amoto, J. Math. Soc. Japan {\bf 39} (1987) 191
\bibitem{SV} V.V. Schechtman and A.N. Varchenko, \IM{106} (1991) 139
\bibitem{FIM} B. Feigin and F. Malikov, Adv. Sov. Math. {\bf 17} (1993) 15,
hep-th/9306137;\\
K. Iohara and F. Malikov, Mod. Phys. Lett. {\bf A 8} (1993) 3613
\bibitem{FV} G. Felder and A. Varchenko, hep-th/9502165, preprint;\\
G. Felder and C. Wieczerkowski, hep-th/9411004, preprint
\bibitem{FF} B.L. Feigin and E.V. Frenkel, Usp. Mat. Nauk. {\bf 43} (1988) 227 
(in Russian), Russ. Math. Surv. {\bf 43} (1989) 221;\\
 B.L. Feigin and E.V. Frenkel, \CMP{128} (1990) 161;\\
 B.L. Feigin and E.V. Frenkel,  Lett. Math. Phys. {\bf 19} (1990) 307;\\
 B.L. Feigin and E.V. Frenkel, in {\em Physics and Mathematics of Strings},
 Eds. L. Brink {\em et al.} (World Scientific, 1990); \\
 E. Frenkel, {\em Free field realizations in representation theory and 
 conformal field theory}, preprint hep-th/9408109 
\bibitem{An} O. Andreev, Int. J. Mod. Phys. {\bf A 10} (1995) 3221;\\
 O. Andreev, \PL{B 363} (1995) 166
\bibitem{AY} H. Awata and Y. Yamada, \MPL{A7} (1992) 1185
\bibitem{FM} B. Feigin and F. Malikov, Lett. Math. Phys {\bf 31} (1994) 315
\bibitem{FM95} B. Feigin and F. Malikov, preprint q-alg/9511011
\bibitem{Wak} M. Wakimoto, \CMP{104} (1986) 60
\bibitem{BO} M. Bershadsky and H. Ooguri, \CMP{126} (1989) 49
\bibitem{KZ} V. Knizhnik and A. Zamolodchikov, \NP{B 247} (1984) 83
\bibitem{CF} P. Christe and R. Flume, \NP{B 282} (1987) 466
\bibitem{FZ} V.A. Fateev and A.B. Zamolodchikov, \SJNP{43} (1986) 657
\bibitem{DF} Vl.S. Dotsenko and V.A. Fateev, \NP{B 240}{[}FS12{]} (1984) 312;\\
 Vl.S. Dotsenko and V.A. Fateev, \NP{B 251}{[}FS13{]} (1985) 691
\bibitem{MR} A.C. McBride and G.F. Roach (eds.) {\em Fractional Calculus}
(Pitman Advanced Publishing Program) (Boston 1985);\\
S.G. Samko, A.A. Kilbas and O.L. Marichec, {\em Fractional Integrals and
Derivatives,} Gordon and Breach, Science Publishers (1993)
\bibitem{GMMOS} A. Morozov, JETP Lett. {\bf 49} (1989) 345;\\
 A. Gerasimov, A. Marshakov, A. Morozov, M. Olshanetskii, and S. Shatashvili, 
 Int. J. Mod. Phys. {\bf A 5} (1990) 2495
\bibitem{BMP} P. Bouwknegt, J. McCarthy and K. Pilch, \PL{B 234} (1990) 297;\\
 P. Bouwknegt, J. McCarthy and K. Pilch, \CMP{131} (1990) 125;\\
 P. Bouwknegt, J. McCarthy and K. Pilch, Prog. Theor. Phys. Suppl. {\bf 102}
 (1990) 67;\\
  P. Bouwknegt, J. McCarthy and K. Pilch in {\em Strings and symmetries 1991,}
 eds. N. Berkovits {\em et al.}, (World Scientific, Singapore, 1992)
\bibitem{KOS} M. Kuwahara and H. Suzuki, \PL{B 235} (1990) 52;\\
 M. Kuwahara, N. Ohta and H. Suzuki, \PL{B 235} (1990) 57;\\
 M. Kuwahara, N. Ohta and H. Suzuki, \NP{B 340} (1990) 448;\\
 N. Ohta and H. Suzuki, \NP{B 332} (1990) 146
\bibitem{Ito0}K. Ito, \PL{B 252} (1990) 69
\bibitem{Ito} 
 K. Ito and Y. Kazama, Mod. Phys. Lett. {\bf A 5} (1990) 215;\\
 K. Ito and S. Komata, Mod. Phys. Lett. {\bf A 6} (1991) 581
\bibitem{Ku} G. Kuroki, \CMP{142} (1991) 511
\bibitem{Tay} W. Taylor IV, LBL-34507, hep-th/9310040, Ph.D. thesis
\bibitem{deBF} J. de Boer and L. Feh\'er, Mod. Phys. Lett. {\bf A 11} (1996) 
 1999;\\
 J. de Boer and L. Feh\'er, {\em Wakimoto realizations of current algebras: an 
 explicit construction},  
 LBNL-39562, UCB-PTH-96/49, BONN-TH-96/16, hep-th/9611083, preprint
\bibitem{KW} V.G. Kac and M. Wakimoto, Proc. Natl. Acad. Sci. 
 USA {\bf 85} (1988) 4956;\\
 V.G. Kac, and D.A. Kazhdan, Adv. Ser. Math. Phys., Vol. 7 (World Scientific, 
 1989), p. 138
\end{thebibliography}

\begin{thebibliography}{99}
\bibitem{2a}
H.B.Nielsen, "Dual Strings. Fundamental of Quark Models", in:       \\
{\it Proceedings of the XYII Scottish University Summer School in Physics},
St.Andrews, 1976, p.528.
\bibitem{2c}
H.B.Nielsen, D.L.Bennett, N.Brene, {\it Recent Developments in Quantum
Field Theory}, Amsterdam, 1985, p.263;\\
D.L.Bennett, H.B.Nielsen, I.Picek, Phys.Lett. {\bf B208}, 275 (1988).
\bibitem{2f}
H.B.Nielsen, Acta Physica Polonica, {\bf B20}, 427 (1989).
\bibitem{2g}
H.B.Nielsen, N.Brene, Phys.Lett. {\bf B233}, 399 (1989);
Phys. Mag. {\bf 12}, 157 (1989); Nucl.Phys. {\bf B359}, 406 (1991).
\bibitem{2k}
C.D.Froggatt, H.B.Nielsen, {\it Origin of Symmetries}, Singapore:
World Scientific, 1991.
\bibitem{10na}
L.Nottalle, Int.J.Mod.Phys. {\bf A7}, 4899 (1992);\\
L.Nottale, {\it Fractal Space-Time and Microphysics},
Singapore: World Scientific, 1993; \\
L.Nottalle and J.Schneider, J.Math.Phys. {\bf 25}, 1296 (1984).
\bibitem{32a}
H.P.Nilles, Phys.Reports {\bf 110}, 1 (1984).
\bibitem{33a}
P.Langacker, N.Polonsky, Phys.Rev. {\bf D47}, 4028 (1993); ibid, {\bf D49},
1454 (1994); ibid, {\bf D52}, 3081 (1995).
\bibitem{17p}
D.L.Bennett and H.B.Nielsen, Int.J.Mod.Phys. {\bf A9}, 5155 (1994);
ibid, {\bf A14}, 3313 (1999).
\bibitem{18p}
L.V.Laperashvili, Phys. of Atom.Nucl. {\bf 57}, 471 (1994);
ibid, {\bf 59}, 162 (1996).
\bibitem{19p}
L.V.Laperashvili and H.B.Nielsen, Mod.Phys.Lett. {\bf A12}, 73 (1997).
\bibitem{20p}
L.V.Laperashvili and H.B.Nielsen, "Multiple Point Principle and Phase
Transition in Gauge Theories", in: {\it Proceedings of the International
Workshop on "What Comes Beyond the Standard Model"}, Bled, Slovenia,
29 June - 9 July 1998; Ljubljana 1999, p.15.
\bibitem{21p}
L.V.Laperashvili and H.B.Nielsen, Int.J.Mod.Phys. {\bf A16}, 2365 (2001).
\bibitem{22p}
L.V.Laperashvili, H.B.Nielsen and D.A.Ryzhikh,
Int.J.Mod.Phys. {\bf A16}, 3989 (2001);\\
L.V.Laperashvili, H.B.Nielsen and D.A.Ryzhikh, Yad.Fiz. {\bf 65} (2002).
\bibitem{35}
C.D.Froggatt, G.Lowe, H.B.Nielsen, Phys.Lett. {\bf B311}, 163 (1993);
Nucl.Phys. {\bf B414}, 579 (1994); ibid {\bf B420}, 3 (1994);\\
C.D.Froggatt, H.B.Nielsen, D.J.Smith, Phys.Lett. {\bf B358}, 150 (1996);\\
C.D.Froggatt, M.Gibson, H.B.Nielsen, D.J.Smith, Int.J.Mod.Phys. {\bf A13},
5037 (1998);\\
H.B.Nielsen, C.D.Froggatt, "Masses and Mixing Angles and Going beyond
the Standard Model", in {\it{Proceedings of the International
Workshop on "What Comes Beyond the Standard Model?"}}, Bled, Slovenia, 29 June -
9 July 1998: Ljubljana 1999, p.29.
\bibitem{36}
C.D.Froggatt, L.V.Laperashvili, H.B.Nielsen, "SUSY or NOT SUSY",
{\it "SUSY98"}, Oxford, 10-17 July 1998; hepnts1.rl.ac.uk/susy98/.
\bibitem{37}
H.B.Nielsen, Y.Takanishi, Nucl.Phys. {\bf B588}, 281 (2000);
ibid, {\bf B604}, 405 (2001); Phys.Lett. {\bf B507}, 241 (2001);
hep-ph/0011168, hep-ph/0101181, hep-ph/0101307.
\bibitem{38}
L.V.Laperashvili, "Anti--Grand Unification and Phase Transitions at the
Planck Scale in Gauge Theories", in: "Frontiers of
Fundamental Physics", Proceedings of Forth International Symposium, Hyderabad,
India, 11-13 December, 2000. Ed. B.G.Sidharth.
\bibitem{1s}
K.Wilson, Phys.Rev. {\bf D10}, 2445 (1974).
\bibitem{10s}
J.Jersak, T.Neuhaus and P.M.Zerwas, Phys.Lett. {\bf B133}, 103 (1983);
Nucl.Phys. {\bf B251}, 299 (1985);   \\
H.G.Everetz, T.Jersak, T.Neuhaus, P.M.Zerwas, Nucl.Phys. {\bf B251},
279 (1985).
\bibitem{13p}
G.Parisi, R.Petronzio, F.Rapuano, Phys.Lett. {\bf B128}, 418 (1983);\\
E.Marinari, M.Guagnelli, M.P.Lombardo, G.Parisi, G.Salina, Proceedings Lattice '91', Tsukuba 1991, p.278-280;
Nucl.Phys.Proc.Suppl. {\bf B26}, 278 (1992).
\bibitem{9s}
G.Bhanot, Nucl.Phys. {\bf B205}, 168 (1982);
Phys.Rev. {\bf D24}, 461 (1981);
Nucl.Phys. {\bf B378}, 633 (1992).
\bibitem{14p}
M.L\"uscher, K.Symanzik, P.Weisz, Nucl.Phys. {\bf B173}, 365 (1980);\\
C.Surlykke, {\it On Monopole Suppression in Lattice QED}, preprint NBI (1994).
\bibitem{13s}
T.Suzuki, Progr.Theor.Phys. {\bf 80}, 929 (1988);      \\
S.Maedan, T.Suzuki, Progr.Theor.Phys. {\bf 81}, 229 (1989).
\bibitem{20s}
S.Coleman and E.Weinberg, Phys.Rev. {\bf D7}, 1888 (1973);\\
S.Coleman, in: {\it Laws of Hadronic Matter}, edited by
A.Zichichi, Academic Press, New York, 1975.
\bibitem{21s}
M.Sher, Phys.Rept. {\bf 179}, 274 (1989).
\bibitem{24p}
G. 't Hooft, Nucl.Phys. {\bf B190}, 455 (1981).
\bibitem{1a}
M.B.Voloshin and K.A.Ter-Martirosyan, {\it Theory of Gauge Interactions
of Elementary Particles,} Energoatomizdat, 1984.
\bibitem{40a}
P.A.Kovalenko, L.V.Laperashvili, Phys. of Atom.Nucl. {\bf 62}, 1729 (1999).
\bibitem{40}
S.Rajpoot, Nucl.Phys.Proc.Suppl. {\bf 51A}, 50 (1996).
\bibitem{41}
D.Zwanziger, Phys.Rev. {\bf D3}, 343 (1971);\\
R.A.Brandt, F.Neri, D.Zwanziger, Phys.Rev.{\bf D19}, 1153 (1979).
\bibitem{42}
F.V.Gubarev, M.I.Polikarpov, V.I.Zakharov,
Phys.Lett.{\bf B438}, 147 (1998).
\bibitem{43}
L.V.Laperashvili, H.B.Nielsen, Mod.Phys.Lett. {\bf A14}, 2797 (1999).

\end{thebibliography}

\begin{thebibliography}{99}

\bibitem{hoppe}J. Hoppe, Ph. D. thesis, MIT (1982); J. Hoppe, Int. J. Mod.
Phys. {\bf A4}(1989)5235.

\bibitem{floratos} E. G. Floratos, J. Iliopoulos, and G. Tiktopoulos,
Phys. Lett. {\bf B217}(1989)285.

\bibitem{fairlie} 
D. Fairlie, P. Fletcher, and G. Zachos, Phys. Lett. {\bf B218}(1989)203;
D. Fairlie and G. Zachos, Phys. Lett. {\bf B224}(1989)101;
D. Fairlie, P. Fletcher, and G. Zachos, J. Mat. Phys. {\bf 31}(1990)1088.

\bibitem{bars}I. Bars, Phys. Lett. {\bf B245}(1990)35.

\bibitem{krauth}W. Krauth and M. Staudacher, Phys. Lett. {\bf B435}(1998)350.
\bibitem{hotta}J. Hotta, J. Nishimura and A. Tsuchiya, Nucl. Phys. {\bf B545}(1999)543
\bibitem{austing}P. Austing and J. F. Wheater, hep-th/0101071.

\bibitem{us}K. N. Anagnostopoulos, J. Nishimura, and P. Olesen,
hep-th/0012061.
\bibitem{alvarez}O. Alvarez, Phys. Rev. {\bf D24}(1981)440.
\bibitem{po} P. Olesen, Phys. Lett. {\bf B160}(1985)144.
\end{thebibliography}

\begin{thebibliography}{99}

\bibitem{holger}  Ziro Koba, Holger Bech Nielsen,
 MANIFESTLY CROSSING INVARIANT PARAMETRIZATION OF N MESON AMPLITUDE.
 Nucl.Phys.B12:517-536,1969

\bibitem{aage}A. Kraemmer, H. Nielsen, and L. Susskind, Nucl. Phys. B28,
34 (1971);


\bibitem{karliner} M. Karliner, I. Klebanov and L Susskind, Size And
Shape Of Strings
 Int.J.Mod.Phys. A3 (1988) 1981

\bibitem{tooft}  G. 't Hooft  A Planar Diagram Theory For Strong
Interactions, Nucl.Phys. B72 (1974) 461

\bibitem{olesen}Holger Bech Nielsen, P. Olesen,
 VORTEX LINE MODELS FOR DUAL STRINGS,
 Nucl.Phys.B61:45-61,1973

\bibitem{confinejoe}Joseph Polchinski, Matthew J. Strassler,
 The String Dual of a Confining Four-Dimensional Gauge Theory,
  hep-th/0003136

\bibitem{hardjoe} Joseph Polchinski , Matthew J. Strassler
 HARD SCATTERING AND GAUGE / STRING DUALITY.
 hep-th/0109174

\bibitem{thorn}  R.R. Metsaev, C.B. Thorn, A.A. Tseytlin,
Light-cone Superstring in AdS Space-time,
hep-th/0009171

\end{thebibliography}

\begin{thebibliography}{99}
\bibitem{vys1}
I. Shapiro, Phys. Rev. Lett. {\bf 13}, 789 (1964); see also S. Weinberg,
{\it Gravitation and Cosmology}, Wiley, 1972.
\bibitem{vys2}
L.B. Okun, Mod. Phys. Lett. {\bf A15}, 1941 (2000).
\bibitem{vys3}
L.D. Landau and E.M. Lifshits, {\it Teoriya Polya},
Moscow, Nauka, 1967 (in Russian); {\it Classical Field Theory},
 Pergamon Press 1961 (in English).

\bibitem{MTW} C.W. Misner, K.S. Thorne, J.A. Wheeler,
{\it Gravitation},  W. H. Freeman \& Co., 1971.

\bibitem{lightm}
A.P. Lightman, W.H. Press, R.H. Price,  S.A. Teukolsky,
{\it Problem Book in Relativity and Gravitation},    Princeton U., 1975.


\bibitem{vys4}
L.D. Landau  and E.M. Lifshits, {\it Hydrodynamics} (in
Russian), Moscow, Nauka, 1986.


\bibitem{lanliffm} L.D. Landau  and E.M. Lifshits,
 {\it  Fluid Mechanics}, Oxford, Pergamon Press 1982.


\bibitem{vys6}
R. Kubo, {\it Statistical Mechanics}, North-Holland 1965.
\bibitem{tang}
F.R. Tangherlini, Schwarzschild field in n-dimensions and the
dimensionality of space problem, Nuovo Cimento, {\bf 27}, 636 (1963).
\bibitem{MP}
R.C. Myers and M.J. Perry, ``Black holes in higher dimensional
spaces'', Ann. Phys. {\bf 172}, 304 (1986);


\end{thebibliography}

\begin{thebibliography}{xx}
%
\bibitem{CV} H. B. Nielsen and P. Olesen, Nucl. Phys. B160 (1979) 380
\bibitem{tHooft} G. 't Hooft, Nucl. Phys. B138 (1978) 1;
G. Mack, in {\sl Recent Developments in Gauge Theories}, edited by
G. 't Hooft et al. (Plenum, New York, 1980).
\bibitem{Us}  L.~Del Debbio, M.~Faber, J.~Greensite, and 
  {\v S}.~Olejn\'{\i}k, Phys. Rev. D55 (1997) 2298, hep-lat/9610005.
\bibitem{Jan98} L.~Del Debbio, M.~Faber, J.~Giedt, J.~Greensite, and 
  {\v S}.~Olejn\'{\i}k, Phys. Rev. D58 (1998) 094501, hep-lat/9801027.
\bibitem{ER} M. Engelhardt and H. Reinhardt, Nucl. Phys. B567 (2000) 249,
hep-th/9907139.
\bibitem{BKP} V. Bornyakov, D. Komarov, and M. Polikarpov, Phys. Lett.
B497 (2001) 151, hep-lat/0009035.
\bibitem{Remarks} M.~Faber, J.~Greensite, and 
  {\v S}.~Olejn\'{\i}k, hep-lat/0103030.
\bibitem{Alex} C. Alexandrou, M. D'Elia, and Ph. de Forcrand, Nucl. Phys. 
Proc. Suppl. 83 (2000) 437, hep-lat/9907028.
\bibitem{Pepe} Ph. de Forcrand and M. Pepe, Nucl. Phys. B598 (2001) 557,
hep-lat/0008016.
\bibitem{Vink} J. Vink and U. Weise, Phys. Lett. B289 (1992) 122, 
hep-lat/9206006.
\bibitem{KT} T. Kovacs and E. Tomboulis, Phys. Lett. B463 (1999) 104.
\bibitem{Bali} C. Michael and M. Teper, Phys. Lett. B199 (1987) 95; \\
G. Bali, C. Schlichter, and K. Schilling, Phys. Rev. D51 (1995) 5165,
hep-lat/9409005.
\bibitem{dFE} Ph. de Forcrand and M. D'Elia, Phys. Rev. Lett. 82 (1999)
4582, hep-lat/9901020.
\end{thebibliography}
\end{document}